\documentclass[%
 aip,
 jmp,%
 amsmath,amssymb,
 reprint,%
 twocolumn,%
 author-year,%
]{revtex4-1}

\usepackage{graphicx}
\usepackage{dcolumn}
\usepackage{bm}
\usepackage{threeparttable}
\usepackage{booktabs}

\usepackage{graphicx,times}
\usepackage{subfigure}
\usepackage{amsmath}
\usepackage{mathrsfs}
\usepackage{cases}
\usepackage{bm}
\usepackage{longtable}
\usepackage{hyperref}
\usepackage{epstopdf}
\usepackage{amsmath,bm}
\usepackage{amssymb}
\usepackage{natbib}
\usepackage{morefloats}
\usepackage{multirow}
\usepackage{array}
\usepackage{verbatim}
\usepackage[stable]{footmisc}
\usepackage{stackrel}
\usepackage{graphicx}
\usepackage{dcolumn}
\usepackage{bm}

\def\aap{Astronomy \& Astrophysics}
\def\aapr{Astronomy and Astrophysics Reviews}
\def\apj{The Astrophysical Journal}
\def\apjl{The Astrophysical Journal Letters}
\def\apjs{The Astrophysical Journal Supplement Series}
\def\apss{Astrophysics and Space Science}
\def\araa{Annual Review of Astronomy \& Astrophysics}
\def\grl{Geophysics Research Letters}
\def\jgr{Journal of Geophysical Research}
\def\mnras{Monthly Notices of the Royal Astronomical Society}
\def\nat{Nature}
\def\physrep{Physics Reports}
\def\prd{Physical Review D}
\def\pre{Physical Review E}
\def\prl{Physical Review Letters}
\def\qjras{Quarterly Journal of the Royal Astronomical Society}
\def\solphys{Solar Physics}

\def\ssr{Space Science Reviews}

\newcommand{\ba}{{\bf a}}
\newcommand{\bE}{{\bf E}}

\newcommand{\bX}{{\bf X}}

\newcommand{\br}{{\bf r}}

\newcommand{\bx}{{\bf x}}

\newcommand{\bv}{{\bf v}}

\newcommand{\bB}{{\bf B}}

\newcommand{\bJ}{{\bf J}}

\newcommand{\bS}{{\bf S}}
\newcommand{\bR}{{\bf R}}

\newcommand{\grad}{{\mbox{\boldmath $\nabla$}}}

\newcommand{\brho}{{\mbox{\boldmath $\rho$}}}

\newcommand{\bxi}{{\mbox{\boldmath $\xi$}}}
\newcommand{\bzed}{{\bf 0}}

\newcommand{\bdot}{{\mbox{\boldmath $\cdot$}}}
\newcommand{\btimes}{{\mbox{\boldmath $\times$}}}
\newcommand{\bdots}{{\mbox{\boldmath $:$}}}

\newcommand{\bSigma}{{\mbox{\boldmath $\Sigma$}}}

\newcommand{\boeps}{{\mbox{\boldmath $\varepsilon$}}}
\newcommand{\btimescom}{{\stackbin{\rm \btimes}{,}}}

\newcommand{\wt}{\widetilde}

\begin{document}


\title[Turbulent Reconnection]{3D Turbulent Reconnection: Theory, Tests \& Astrophysical Implications}


\author{Alex Lazarian}
\email{lazarian@astro.wisc.edu}
\affiliation{Department of Astronomy, University of Wisconsin, 475 North Charter Street, Madison, WI 53706, USA}

\author{Gregory L. Eyink}
\affiliation{Department of Applied Mathematics \& Statistics, The Johns Hopkins University, Baltimore, MD 21218, USA}
\affiliation{Department of Physics \& Astronomy, Johns Hopkins University, 3400 N. Charles Street, Baltimore, MD 21218, USA}

\author{Amir Jafari}
\affiliation{Department of Physics \& Astronomy, Johns Hopkins University, 3400 N. Charles Street, Baltimore, MD 21218, USA}

\author{Grzegorz Kowal}
\affiliation{Escola de Artes, Ci\^encias e Humanidades, Universidade de S\~ao Paulo, Av. Arlindo B\'ettio, 1000 - Vila Guaraciaba, S\~ao Paulo - SP, CEP: 03828-000, Brazil}

\author{Hui Li}
\affiliation{Los Alamos National Laboratory, Los Alamos, NM 87545, USA}

\author{Siyao Xu}
\affiliation{Department of Astronomy, University of Wisconsin, 475 North Charter Street, Madison, WI 53706, USA}

\author{Ethan T. Vishniac}
\affiliation{Department of Physics \& Astronomy, Johns Hopkins University, 3400 N. Charles Street, Baltimore, MD 21218, USA}

\date{\today}

\begin{abstract}
 Magnetic reconnection, topological change in magnetic fields, is a fundamental process in magnetized plasmas. 
It is associated with energy release in regions of magnetic field annihilation,
but this is only one facet of this process. Astrophysical fluid flows normally have very large Reynolds
numbers and are expected to be turbulent, in  agreement with observations. In strong turbulence
magnetic field lines constantly reconnect everywhere and on all scales, thus making magnetic reconnection an
intrinsic part of the turbulent cascade. We note in particular that this is inconsistent with the usual practice of regarding magnetic field lines as persistent dynamical elements. A number of theoretical, numerical, and observational studies starting with 
the Lazarian \& Vishniac 1999 paper proposed that 3D turbulence
makes magnetic reconnection fast and that magnetic reconnection and turbulence are intrinsically connected.
In particular, we discuss the dramatic violation of the textbook concept of magnetic flux-freezing in the presence of
turbulence. We demonstrate that in the presence of turbulence the plasma
effects are subdominant to turbulence as far as the magnetic reconnection is concerned. The latter fact justifies
an MHD-like treatment of magnetic reconnection on all scales much larger than the relevant plasma scales. We
discuss numerical and observational evidence supporting the turbulent reconnection model. In particular, we demonstrate
that the tearing reconnection is suppressed in 3D and, unlike the 2D settings, 3D reconnection induces turbulence
that makes magnetic reconnection independent of resistivity. We show that turbulent reconnection dramatically affects
key astrophysical processes, e.g. star formation, turbulent dynamo, acceleration of
cosmic rays. We provide criticism of the 
concept of ``reconnection-mediated turbulence'' and
explain why turbulent reconnection is very different from enhanced 
turbulent resistivity and hyper-resistivity,  and why the latter have fatal conceptual flaws. 
\end{abstract}

\pacs{Valid PACS appear here}
\maketitle

\begin{quotation}

\end{quotation}

\section{Problem of Magnetic Reconnection in Astrophysics}

Magnetic fields are ubiquitous in astrophysical systems and
critically affect the dynamics and properties of
magnetized plasmas over an extended range of scales. These scales are typically much larger than any relevant plasma scales, e.g., the ion inertial length. The magnetic field 
of the Earth's magnetosphere, the case-study of many {\it in situ} measurements, is an exception
in the astrophysical context in terms of the involved scales. In what follows, we will discuss magnetic fields
at much larger scales.

One textbook concept related to magnetic fields, widely used in astrophysical studies, is that magnetic fields
in highly conducting plasmas are  nearly perfectly frozen into the fluid
and retain their topology for all time \citep{Alfven:1942, Parker:1979}. This concept of
flux-freezing is at the heart of many theories, e.g., the theory of 
star formation in magnetized interstellar medium.

At the same time, there is ample evidence that magnetic fields in astrophysical systems
do change their topology, solar flares being a classical example of such a process
\cite[see e.g.][]{Parker:1970, Lovelace:1976, PriestForbes:2002}.  The studies on the process enabling such a
topology change, i.e., magnetic reconnection, were historically motivated by observations of the solar corona
\citep{Innes_etal:1997, YokoyamaShibata:1995, Masuda_etal:1994}, which also created a number of
misconceptions. First of all, magnetic reconnection is frequently associated only with the
processes of conversion of magnetic energy into heat. In addition, this
understanding influenced attempts to find very peculiar conditions for flux conservation
violation. For instance, in a fundamental study by \cite{PriestForbes:2002}, a number of examples of magnetic
configurations that produce fast reconnection  are discussed. The attempts to accelerate the reconnection rate for 
special plasma conditions, e.g., in plasmas with very small collision
rates \cite[see e.g.][]{Shay_etal:1998, Drake:2001, Drake_etal:2006, Daughton_etal:2006,
UzdenskyKulsrud:2006,  Bhattacharjee_etal:2003,
ZweibelYamada:2009, Yamada_etal:2010} follow the same logic of seeking special circumstances where the magnetic flux-freezing is violated.

 We feel that the above treatment of magnetic reconnection is not generic or widely applicable. 
 Solar flares \citep{Sturrock:1966} are just one vivid example of
reconnection activity. In reality,
magnetic reconnection is a ubiquitous process taking place in various
astrophysical environments, both collisionaless and collisional \cite[see][]{ShibataMagara:2011}.  For instance, magnetic reconnection is a part of large-scale dynamo acting in stellar interiors
\citep{Parker:1993, Ossendrijver:2003}.
Magnetic reconnection is also required to make the theory of 
magnetohydrodynamic (MHD) turbulence \citep{GoldreichSridhar:1995} self consistent \citep{JafariVishniac:2019}, etc.  

The problem of magnetic reconnection is not limited to explaining its typically fast rates. In fact, it is often overlooked that the
observations of solar activity indicate that the reconnection rates can significantly vary. 
This presents serious problems for theories that are based on relating the rate of
magnetic reconnection to the peculiar properties of plasmas. These properties do not
change, for instance, as the flux gets accumulated prior to a solar flare and gets annihilated
during the flare. This may suggest that magnetic reconnection should have a sort of trigger.

It is also worth noting that for decades a paradoxical situation existed with the treatment of astrophysical magnetic reconnection. 
On the one hand, the core of the reconnection community claimed numerous restrictions either on magnetic field 
configurations or the properties of reconnecting plasmas. On the other hand, the practitioners of numerical astrophysics
used their MHD codes to simulate various astrophysical settings without worrying about the aforementioned restrictions. At the same time, the question of the actual effect of magnetic reconnection on the validity of numerical studies was somehow avoided. 

The absence of an effective communication between the communities is surprising.  For instance, for many years it was considered
that in order for reconnection to be fast, it must take place in collisionless environments. 
However, if collisionless
reconnection were the only way to make reconnection rapid, then numerical MHD
simulations of many astrophysical processes including those of the interstellar
medium (ISM), which is collisional, are in error.  Fortunately for numerical practitioners,
the observations of
collisional solar photosphere indicate that the reconnection is fast in these
environments as well (see \S \ref{sec:observations1}). 

In this review, we provide evidence that, in fact, turbulence determines the rate of
reconnection in realistic 3D astrophysical systems.  The concept of 3D turbulent reconnection was introduced by
\citet[][henceforth LV99]{LV99}. This model was followed by several subsequent theoretical studies, for instance, \citet[][henceforth ELV11]{Eyink_etal:2011}, \cite{Eyink2015}, and \cite{Jafari_etal:2018}. The LV99 theory has been supported by numerical simulations in non-relativistic 
\cite[see][]{Kowal_etal:2009, Kowal_etal:2012a, Eyink_etal:2013, Beresnyak:2017, Oishi_etal:2015, Kowal_etal:2017}, 
as well as relativistic settings 
\cite[see][]{Takamoto2016, Takamoto:2018}. 
In addition, different pieces of observational evidence,
discussed in the review, support the predictions of the 3D turbulent reconnection theory of LV99.

To understand the essence of LV99 model, one can recall that the problem of magnetic reconnection in astrophysical systems can be viewed as the problem related to the scale disparity. Reconnection occurs on very large scales, while the dissipation processes take place at the smallest plasma scales, which are set by, e.g., resistivity. LV99 theory shows how  turbulence, as a multi-scale phenomenon, can indeed resolve the problem of scale disparity as well as other problems that plague the traditional reconnection models. 

While the idea of turbulent reconnection is nearly coeval with the ideas of 2D collisionless Hall-MHD reconnection, the latter has been  substituted more recently by the ideas of tearing/plasmoid reconnection (Shibata \& Tanuma 2001). In fact, 2D numerical work has already shown that magnetic reconnection can be accelerated significantly by the tearing instability \citep{Loureiro_etal:2007, Lapenta:2008, Daughton_etal:2009a,
Daughton_etal:2009b, Bhattacharjee_etal:2009, Cassak_etal:2009}.

In a sense, this model has some similarities with the turbulent reconnection. Indeed, it departs from the Sweet-Parker Y-point reconnection but allows instabilities to evolve within the current sheet. We will discuss how in realistic 3D configurations this type of instability transfers to turbulent reconnection.

One should keep in mind that we deal with 3D turbulent reconnection on large scales at which an 
MHD-like approximation is applicable. In particular, we consider scales much larger than the ion Larmor radius or skin depth, where an ideal Ohm's law is generally agreed to be valid. 
At the same time, usual tests of reconnection are based on the {\em in situ} measurements of magnetospheric reconnection. This type of
 reconnection occurs on scales comparable to the ion inertial length and therefore is atypical for the large-scale reconnection that occurs in most astrophysical systems. For the reconnection explored at larger scales, e.g., in the solar wind
\citep{gosling2012magnetic, Lalescu_etal:2015},
the measured properties of reconnection are consistent with the general expectations based on the LV99 theory (see \S \ref{sec:observations2}). 

Occasionally, the process of magnetic reconnection is understood in a very narrow sense, e.g. magnetic reconnection necessarily in low-$\beta$ plasmas with the magnetic energy much larger than the thermal energy. Here we consider turbulent reconnection as a generic process, which is applicable to plasmas of arbitrary $\beta$ and for arbitrary arrangements of reconnecting magnetic fluxes.   
 
Our review is devoted to the astrophysical reconnection, although the magnetic reconnection is a fundamental physical process and therefore our conclusions can be applied also to the laboratory reconnection. We discuss the theory and observations of astrophysical turbulence in \S II. The LV99 model of 3D turbulent
reconnection and its extensions are described in \S III and \S IV. There we explain how turbulent reconnection and flux-freezing violation in turbulent fluids are interrelated. The numerical tests of turbulent reconnection and the testing of flux freezing violation in turbulent fluids  are provided in \S V. Both testing with MHD codes and Hall-MHD codes are discussed. The initial tests of selected ideas of turbulent reconnection  with 3D kinetic simulations are presented in \S VI. The observational evidence supporting turbulent reconneciton is briefly discussed in \S VII. The magnetic reconnection in high-Prandtl number medium, reconnection in relativistic fluids, and reconnection in the presence of whistler turbulence  are discussed in \S VIII. Astrophysical implications of reconnection for turbulent dynamo, star formation, as well as for particle acceleration and gamma ray bursts are covered in \S IX. In \S X, we discuss the differences between our model of turbulent reconnection and other ideas on the effects of turbulence on reconneciton and resistivity. We explain the reasons for our disagreement with (i) the currently-popular  ``reconnection-mediated turbulence'' idea  and (ii) the concepts of turbulent resistivity and turbulent ambipolar diffusion. There we also outline the problems of the mean field approach to studying reconnection, compare turbulent and tearing reconnection, and explain why reconnection rates are not expected to have a ``universal'' value of $0.1$ Alfven velocity. We present our summary in \S XI.

\section{Turbulence as a natural state of astrophysical fluids and its MHD Description}
\label{sec:turbulence}

\subsection{Observational Evidence}
\label{sec:turbulence1}

Observations of the diffuse warm ISM reveal a Kolmogorov spectrum of electron
density fluctuations \cite[see][]{Armstrong_etal:1995, ChepurnovLazarian:2010}. Similar
 spectral slopes of supersonic velocity fluctuations are measured using Doppler shifted spectral lines
\cite[see][for a review]{Lazarian:2009}, as well as through the {\it in situ} measurements of the solar wind fluctuations \citep{Leamon_etal:1998}.  The evidence of turbulence being ubiquitous comes from 
non-thermal broadening of spectral lines as well as measures obtained by other
techniques \cite[see][]{Burkhart_etal:2010}. This fact is not
 surprising as magnetized astrophysical plasmas generally have very large
Reynolds numbers. Indeed, the length scales involved  are large, while the 
motions of charged particles in the direction perpendicular to magnetic fields
are of the order of the Larmor radius.  Plasma flows at these high Reynolds numbers are prey
to numerous linear and finite-amplitude instabilities. This induces  turbulent
motions readily.

The precise origin of the plasma turbulence differs from instance to instance. 
It is sometimes driven by an external energy source, such as
supernova explosions in the ISM \citep{NormanFerrara:1996, Ferriere:2001}, merger events and
active galactic nuclei outflows in the intercluster medium (ICM)
\citep{Subramanian_etal:2006, EnsslinVogt:2006, Chandran:2005}, and baroclinic
forcing behind shock waves in interstellar clouds.  In other cases, 
turbulence is spontaneous, with available energy released by a rich array of
instabilities, such as magneto-rotational instability (MRI) in accretion disks
\citep{BalbusHawley:1998, JafariVishniac2018disks}, kink instability of twisted flux tubes in the solar
corona \citep{GalsgaardNordlund:1997, GerrardHood:2003}, etc.  In all these 
mentioned cases, turbulence is not driven by reconnection.  Nevertheless, we would like to
stress that the driving of turbulence through the energy release in
the reconnection zone is important for the transfer from laminar to turbulent reconnection.  All in all, whatever its origin is, the signatures of
plasma turbulence are seen throughout astrophysical media.

\begin{figure}[!t]
\centering
\includegraphics[width=0.48\textwidth]{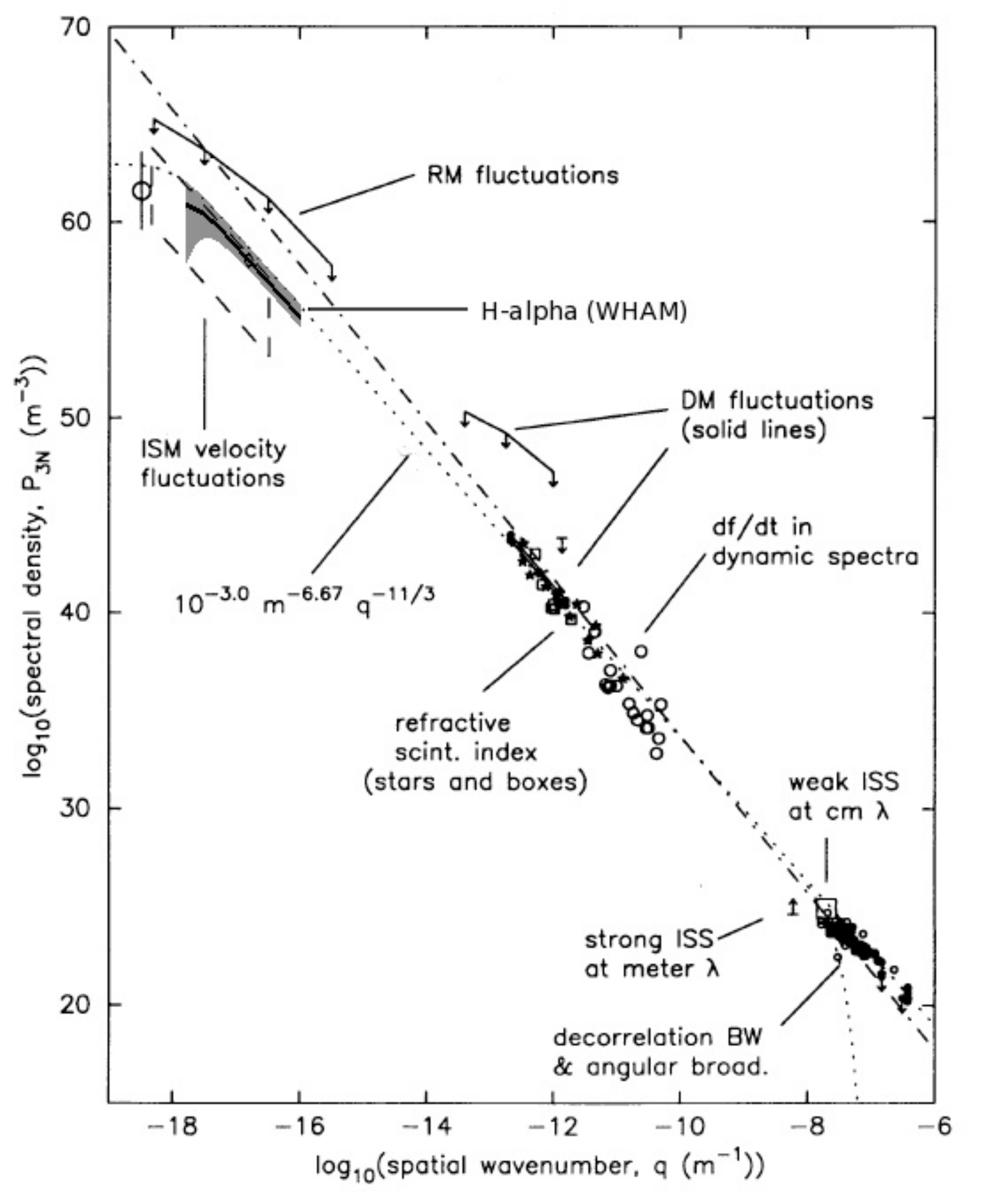}
\caption{Big power law in the sky from \cite{Armstrong_etal:1995} extended to scale of parsecs using WHAM data. From \cite{ChepurnovLazarian:2010}, \copyright~AAS. Reproduced with permission.
\label{fig1}}
\end{figure}

Indeed, observations show that turbulence is ubiquitous in all astrophysical
plasmas.  The spectrum of electron density fluctuations in Milky Way is
presented in Figure~\ref{fig1}, as one dramatic piece of evidence for cosmic turbulence. 
Similar spectra are reported in \cite{Leamon_etal:1998} and \cite{Bale_etal:2005} for solar wind, \cite{Padoan_etal:2006} for molecular clouds, and \cite{VogtEnsslin:2005} for the ICM.
New techniques for studying turbulence, e.g. the Velocity Channel Analysis (VCA) and Velocity Coordinate Spectrum (VCS) techniques \citep{LazarianPogosyan:2000, LazarianPogosyan:2004, LazarianPogosyan:2006} have provided important insight into the velocity spectra of turbulence in molecular clouds \cite[see][]{Padoan_etal:2006,Padoan_etal:2010}, Galactic and extragalactic atomic hydrogen \citep[][, see also the review by \cite{Lazarian:2009}, where a compilation of velocity and density spectra obtained with contemporary HI and CO data is presented]{StanimirovicLazarian:2001, Chepurnov_etal:2010, Chepurnov_etal:2015}.

\subsection{Basic Phenomenology of MHD turbulence}
\label{sec:turbulence2}

It is generally accepted that an MHD-like description of turbulence is applicable to the motions of magnetized plasmas at sufficiently large scales \cite[e.g. see discussion in][]{Schekochihin_etal:2009}.  Standard MHD 
or a multi-species variant is valid at length-scales much larger than the collisional mean-free-path of the plasma. However, even in a nearly collisionless but magnetized plasma, ``kinetic MHD'' or ``kinetic reduced-MHD'' equations are satisfied at scales much larger than the ion gyroradius 
\citep{kulsrud1983mhd,kunz2015inertial}. Since collisions in those plasmas are too infrequent 
to bring the particle distributions to Maxwellian, the pressure tensor 
is anisotropic and given not by a thermodynamic equation of state but instead by 1D kinetic
equations along particle guiding line-centers. This subtlety is irrelevant for the incompressible shear-Alfv\'{e}n wave modes, which nearly uncouple from the kinetic/compressible modes 
and satisfy the well-known ``reduced MHD'' (RMHD) equations \footnote{
 For relativistic turbulence
 \cite[see][]{Takamoto2017}
 the transfer of energy to fast modes is more important.}. Shear-Alfv\'{e}n waves are 
often energetically dominant, as in the solar wind where they contain about 90\% of the 
total energy of plasma and fields \cite[see][]{Dobr80}. For this reason, we shall generally assume that MHD 
is an accurate model at sufficiently large scales or at least a good working approximation 
within which to explore the effects of turbulence on large-scale reconnection. 

We shall invoke below in our discussion of strong MHD turbulence 
the \citet[][henceforth GS95]{GoldreichSridhar:1995} theory of Alfv\'enic turbulence.
We believe that the empirical evidence largely supports GS95 over its competitors
and gives a good description of MHD turbulence (apart from the phenomenon of small-scale intermittency). 
We wish to avoid here any controversy regarding the ``correct'' phenomenology, however. 
It may be shown following LV99, Appendix D, that the  predictions on turbulent reconnection 
are qualitatively independent of the particular phenomenology adopted and, in particular, do not depend 
sensitively upon the predicted spectral slopes. Therefore, our use of GS95 theory may be regarded 
by its persistent skeptics as a mere convenience, since it is the simplest model that incorporates the 
small-scale anisotropy due to a local magnetic field.

 \subsubsection{Derivation of the MHD turbulence relations}
 
 MHD turbulence was traditionally viewed as the process dominated by the Alfv\'en wave interactions. Below we follow this path to show how to obtain GS95 relations and go beyond them
 \cite[see also][]{Cho_etal:2003}.

 For Alfv\'enic perturbations the relative perturbations of velocities and magnetic fields are related in the
following way:
\begin{equation}
\frac{\delta B_l}{B}=\frac{\delta B_l}{B_L}\frac{B_L}{B}=\frac{u_l}{u_L} M_A=\frac{u_l}{V_A},
\label{same}
\end{equation}
where $B_l$ is the perturbation of the magnetic field $B$ at the scale $l$, $B_L$ is the perturbation of the magnetic field at the injection scale $L$, while $u_l$ is the velocity fluctuation at the scale $l$ in the turbulent flow with energy injected with velocity $u_L$.

To understand Alfv\'en wave damping it is advantageous to present MHD turbulence as a superposition of colliding Alfv\'en wave packets. 
Consider such packets with parallel scales $l_{\|}$ and perpendicular scales $l_{\bot}$. The system of reference with respect to the local 
magnetic field is what is relevant when dealing with MHD turbulence. In the rest of the discussion we operate with $l_\|$ and $l_\bot$ 
that are defined in the frame of wavepackets moving along the local magnetic field. 
The wave packet collision induces a change
\begin{equation}
 \Delta E \sim (du^2_l/dt) \Delta t,
 \label{init}
 \end{equation}
 where the first term is the change of the energy of a packet as it interacts with 
 the oppositely moving wave packet. The time of the interaction should be identified with the time of the passage of
 the given wave packet through the oppositely moving packet of the size $l_{\|}$.  Thus the interaction time is given by 
 $\Delta t \sim l_{\|}/V_A$. 
 
 The characteristic rate of cascade is due to the change of the structure of the oppositely moving wave packet. The 
 latter has the rate $u_l/l_{\bot}$. As a result Eq. (\ref{init}) gives
 \begin{equation}
  \Delta E 
 \sim {\bf u}_l \cdot \dot{\bf u}_l\Delta t
 \sim  (u_l^3/l_{\perp}) (l_{\|}/V_A),
 \label{change}
\end{equation}
The fractional energy change per collision is $\Delta E/E$, which measures the strength of the nonlinear interaction:
\begin{equation}
  f \equiv \frac{\Delta E}{u^2_l}
                           \sim \frac{ u_l l_{\|} }{ V_A l_{\perp} }.
                         \label{fraction}
\end{equation}
In Eq. (\ref{fraction}) $f$ measures the ratio of the rate of shearing of the wave packet
$u_l/l_{\bot}$ to the rate of the propagation of the wave packet $V_A/l_{\|}$.
Two cases can be identified. For $f\ll 1$ the shearing rate is much smaller than the propagation rate and  
the cascade process is a random walk. This means that
\begin{equation}
\aleph=f^{-2},
\label{aleph}
\end{equation}
which is the number of steps required for O(1) cascade of energy and therefore the cascade time is 
\begin{equation}
t_{cas}\sim \aleph \Delta t .
\label{tcas}
\end{equation}
For $\aleph\gg1$, the turbulence corresponds to the {\it weak MHD turbulence}. Naturally, $\aleph$ cannot be smaller than unity. 
The limiting case of $\aleph\approx 1$ corresponds to {\it strong MHD turbulence}. 

Consider first the case of weak MHD turbulence, which requires $V_A\ll u_L$ and very 
high-frequency oscillations.  This cascade thus proceeds by resonant 
3-wave Alfv\'enic interactions that satisfy $\omega_1 +  \omega_2  = \omega_3$ 
in addition to $ {\bf k}_1 + {\bf k}_2  = {\bf k}_3,$ 
where ${\bf k}$ and $\omega$ denote wavevector and angular frequency, respectively. 
Because of the dispersion relation $\omega = s V_Ak_\|$ with $s=\pm 1$ for shear-Alfv\'en 
waves travelling parallel and anti-parallel to magnetic fields and because only   
counter-propagating waves interact, one finds that $k_{2,\|}=0$ and $k_{1,\|}=k_{3,\|}.$
Thus resonantly interacting wavepackets have unchanging $k_{\|}$ but $k_\perp$ increases by 
distortions through collisions.  The decrease of $l_{\bot}$ while $l_{\|}$ does not change 
signifies the increase of the energy change per collision. This eventually makes $\aleph$ of the order of unity
and the cascade transitions to strong MHD turbulence. In this case one gets
\begin{equation}
u_l l_{\bot}^{-1}\approx V_A l_{\|}^{-1},
\label{crit1}
\end{equation}
which signifies the cascade time being equal to the wave period $\sim \Delta t$. It was hypothesized 
by \cite{GoldreichSridhar:1995} that the further decrease of $l_{\bot}$ entails
the corresponding decrease of $l_{\|}$ to keep Eq. (\ref{crit1}) satisfied, which is the 
{\it critical balance condition}.  The ability to change $l_{\|}$ by non-resonant strong cascade means 
that the frequencies of interacting waves increase. However, this increase  has a natural limit because 
a too high frequency would restore the resonance condition $\omega_1 +  \omega_2  = \omega_3$ 
and this would shut off the parallel cascade. Thus, the strong turbulent cascade has a tendency
to operate at marginal resonance, where the critical balance relation (\ref{crit1}) is satisfied. 

The cascade of turbulent energy satisfies the relation 
\citep{Bat53}:
\begin{equation}
\epsilon\approx u_l^2/t_{cas}=const,
\label{cascading}
\end{equation}
which for the hydrodynamic cascade provides
\begin{equation}
\epsilon_{hydro}\approx u_l^3/l\approx u_L^3/L=const,
\end{equation}
where the relation $t_{cas}\approx l/u_l$ is used.

For the weak cascade $\aleph \gg 1$ 
\begin{equation}
\epsilon_w \approx \frac{ u_l^4} {V_A^2 \Delta t (l_{\bot}/l_{\|})^2} \approx \frac{u_L^4 \ell_\|}{V_A \ell_\perp^2}
\approx \frac{u_L^4}{V_A L}
\label{eps_weak}
\end{equation}
where Eqs. (\ref{cascading}) and (\ref{tcas}) are used (see LV99). The isotropic turbulence 
injection at scale $\ell_\|\simeq \ell_\perp\simeq L$ results in the third relation in Eq. (\ref{eps_weak}). 
Taking into account that $l_{\|}$ is constant for weak turbulence, it is easy to see that 
Eq. (\ref{eps_weak}) provides
\begin{equation}
u_l\sim u_L (l_{\bot}/L)^{1/2},
\label{u_weak}
\end{equation}
which is different from the hydrodynamic $\sim l^{1/3}$ scaling.\footnote{Using the relation $k E(k) \sim u_k^2$ it is easy to show that the spectrum of weak turbulence is $E_{k, weak}\sim k_{\bot}^{-2}$ 
\cite[LV99, ][]{Galtier_etal:2000}.} 

For {\it sub-Alfv\'enic turbulence} injected isotropically at length scale $L$ with $u_L<V_A$ 
the transition to the strong regime, i.e. $\aleph\approx 1,$ happens at the scale (LV99)
\begin{equation}
l_{trans}\sim L(u_L/V_A)^2\equiv L M_A^2.
\label{trans}
\end{equation} 
Therefore, weak turbulence has a limited,  i.e. $[L, L M_A^2]$, inertial range and transits at $l_{trans}$ into strong turbulence. The velocity at  the transition follows
from $\aleph\approx 1$ condition given by Eqs. (\ref{aleph}) and (\ref{fraction}): 
\begin{equation}
u_{trans}\approx V_A \frac{l_{trans}}{L}\approx V_A M_A^2.
\label{vtrans}
\end{equation}
The scaling for the strong turbulence range at $l<l_{trans}$ can be easily obtained. Indeed, the turbulence 
cascades over one wave period, which according to Eq. (\ref{crit}) is equal to $l_{\bot}/u_l$.  
Substituting the latter in Eq. (\ref{cascading}) one gets
 \begin{equation}
 \epsilon_s\approx \frac{u_{trans}^3}{l_{trans}}\approx \frac{u_l^3}{l_\bot}=const. 
 \label{alt_sub2}
 \end{equation}
The latter is analogous to the hydrodynamic Kolmogorov cascade in
the direction perpendicular to the local direction of the magnetic field. This strong MHD turbulence cascade starts at $l_{trans}$ and
it has the injection velocity given by Eq. (\ref{vtrans}). This provides the another way to get the scaling 
relations for velocity in strong Alfv\'enic turbulence (LV99)
\begin{equation}
u_{l}\approx V_A \left(\frac{l_{\bot}}{L}\right)^{1/3} M_A^{4/3}.
\label{vll}
\end{equation}
which can be rewritten in terms of the injection velocity $u_L$ as 
\begin{equation}
 u_{l}\approx u_L \left(\frac{l_{\bot}}{L}\right)^{1/3} M_A^{1/3}.
\label{alternative}
\end{equation} 
Substituting this in Eq. (\ref{crit1}) we get the relation between the parallel and perpendicular
scales of the eddies (LV99):
\begin{equation}
l_{\|}\approx L \left(\frac{l_{\bot}}{L}\right)^{2/3} M_A^{-4/3}.
\label{Lambda1}
\end{equation}
The relations Eq. (\ref{Lambda1}) and (\ref{vll}) reduce to the GS95 scaling for trans-Alfv\'enic turbulence if 
$M_A\equiv 1$. 

In the opposite case of {\it super-Alfv\'enic turbulence}, i.e. for $u_L > V_A,$ the scales close
to the injection scale are essentially hydrodynamic
as the influence of magnetic forces is minimal. Thus the velocity is Kolmogorov
\begin{equation}
u_l=u_L (l/L)^{1/3}.
\label{u_hydro}
\end{equation}
The cascade changes its nature at the scale 
\begin{equation}
l_{A}=LM_A^{-3}, 
\end{equation}
for which the turbulent velocity is equal to the Alfv\'en velocity \citep{Lazarian2016}.  
The cascade rate for $l<l_A$ is:
\begin{equation}
\epsilon_{superA}\approx u_l^3/l\approx V_A^3/l_A=const. 
\label{super_alt1}
\end{equation}
This case likewise reduces to trans-Alfv\'enic turbulence, but 
with $l_A$ acting as the injection scale. At scales $\ell<l_A$ 
\begin{equation}
l_{\|}\approx L \left(\frac{l_{\bot}}{L}\right)^{2/3} M_A^{-1},
\label{Lambda}
\end{equation}
and
\begin{equation}
u_{l}\approx u_{L} \left(\frac{l_{\bot}}{L}\right)^{1/3}. 
\label{vl}
\end{equation}
The wave description of strong MHD turbulence is very productive in the case when the turbulence 
is imbalanced, i.e. the flow of wave-energy from one direction along field-lines significantly exceeds the flow 
from the other direction. An example of imbalanced MHD theory that agrees with numerical 
simulation is the theory by \citet{BeresnyakLazarian:2008} \cite[see also][]{BL10}. 

\subsubsection{Eddy description of MHD turbulence}

In our discussion above we revealed the analogy between the Alfv\'enic and Kolmogorov turbulence. This analogy can be extended in the case of strong Alfv\'enic turbulence. In fact, the GS95 relations can be naturally obtained if one assumes that such turbulence presents a collection of eddies that are mixing the conducting fluid perpendicular to the direction of the magnetic field present at the location of the eddies. This description can be justified in the presence of fast reconnection that we are describing in the review and this supports our claim that the MHD turbulence and turbulent magnetic reconnection are intrinsically connected. 

As it was described in LV99, the reconnection of magnetic fields associated with an eddy takes place within the eddy turnover time (see also \S \ref{sec:model}). In this situation, the random driving is preferentially exciting the motions that induce minimal bending of magnetic field lines, i.e. the hydrodynamic type motions of eddies that mix magnetized fluid perpendicular to the direction of magnetic field. This is the picture of MHD turbulence that follows from LV99 description. The key concept in the eddy description is that magnetic perturbations are measured in terms of the local direction of the magnetic field, not the mean field. The concept that is missing in GS95. and this causes a lot of confusion with different authors attempting to test the GS95 scale-dependent anisotropy in the frame of mean magnetic field. In fact, the latter anisotropy only exists in the local frame of the eddies, while the anisotropy in the reference frame of mean magnetic field is scale-independent and is determined by the anisotropy of the largest eddies \cite[see][]{Cho_etal:2003}.

Incidentally, within the description suggested in LV99, it is clear that the gradients of both velocities and magnetic field are expected to be perpendicular to the local magnetic field direction. Therefore by tracing the directions of the gradients one can trace the magnetic field direction in turbulent media. This idea is at the foundations of the new technique of magnetic field tracing that employs gradients of intensities within spectroscopic channel maps \citep{LazarianYuen:2018a}, gradients of synchrotron intensity \citep{Lazarian_etal:2018} and gradients of synchrotron polarization \citep{LazarianYuen:2018b}. The magnetic field tracing by the new technique provides results in good agreement with polarization data \cite[see][]{Hu_etal:2019}, which is an indirect way of confirming both the ubiquitous presence of the MHD turbulence in interstellar medium as well as a successful confirmation of the theoretical predictions.   

Adopting the picture of eddies perpendicular to the local magnetic field of the eddies, one can claim that in terms of perpendicular motions we expect to have the Komogorov picture, i.e. $u_\ell\sim l_{\bot}^{1/3}$. The critical balance relations in the picture of the eddies also follow naturally. Indeed, as the eddy provides mixing of magnetic field lines in the perpendicular direction over the time scale $l_{\bot}/u_{\ell}$ it sends an Alfv\'en wave with the same period. This is exactly 
\begin{equation}
l_{\|}^{-1}V_A\sim \ell_{\bot}^{-1} u_l,
\label{crit}
\end{equation}
where $l_{\|}$ and $l_{\bot}$
are eddy axes parallel and perpendicular to the {\it local} direction of
magnetic field, the magnetic field of the eddy. Note that this requires a more sophisticated ways to analyse numerical simulations. For instance, a way of determining of local magnetic field in numerical simulations using structure functions is discussed in \cite{ChoVishniac:2000}. In fact,  the notion of local magnetic field is the
essential part of the modern understanding of Alfv\'enic turbulence and it was
added to the MHD turbulence theory by the later studies \cite[LV99,][]{ChoVishniac:2000,
MaronGoldreich:2001}.  This notion is natural within the picture of LV99 turbulent reconnection. Indeed, if one considers the aforementioned eddy picture, the use of local magnetic field means 
that the eddies are only affected by the magnetic field around them and do not know anything about the mean magnetic field obtained by averaging over the entire turbulent volume. 

It is assumed in GS95 that the energy is injected at the scale $L$ with the injection velocity equal to the Alfv\'en velocity. 
If the injection velocity $u_L$ is smaller than $V_A,$ then the eddies at small scale get more elongated.
For $M_A<1$ the eddy description is valid only for the strong turbulence, i.e. starting with the scales less than $l_{trans}=LM_A^2$. Taking $l_{trans}$ as the injection scale of the turbulence and the velocity at this scale $V_A M_A^2$ (see Eq. (\ref{vtrans})) as the injection velocity one can repeat the arguments about the correspondence of the time scales of parallel and perpendicular motions to obtain the relations given by Eq.(\ref{vll}) and (\ref{Lambda1}).

\subsubsection{Compressible MHD turbulence}

The compressible MHD turbulence can be approximated by the superposition of 3 cascades of basic MHD modes, namely, Alfv\'enic, slow and fast. The original idea of the decomposition of low amplitude MHD perturbations into the modes can be traced to \citep{Dobr80}, while the practical realization for the actual transAlfvenic turbulence was first performed in \citep{ChoLazarian:2002}.  We, however, will concentrate on Alfv\'{e}nic part of the MHD turbulent cascade, as this component is most important for the theory of reconnection that we consider in the review. In non-relativistic MHD turbulence, the backreaction of slow and fast magnetosonic modes \citep{ChoLazarian:2002,
ChoLazarian:2003, KowalLazarian:2010} on Alfv\'enic cascade is insignificant
\citep{ChoLazarian:2002, GoldreichSridhar:1995, LithwickGoldreich:2001}.
Therefore in non-relativistic MHD turbulence the relations above obtained for Alfv\'enic turbulence are not much affected by the presence of compressible modes. 

In the rest of this section we discuss how the partial ionization and relativistic effects change the picture of Alfv\'enic turbulence.

\subsection{MHD turbulence in a partially ionized medium}
\label{sec:turbulence3}

In partially ionized astrophysical plasmas in e.g., the early universe, cold interstellar phases, protoplanetary disks, solar atmosphere, 
the MHD turbulence is subjected to the damping effects due to the presence of neutrals. 
Since we are interested in the turbulent reconnection associated with Alfv\'{e}nic turbulence,
here we focus on the damping of Alfv\'{e}nic turbulence. 

\begin{figure}[ht]
\centering
 \includegraphics[width=8cm]{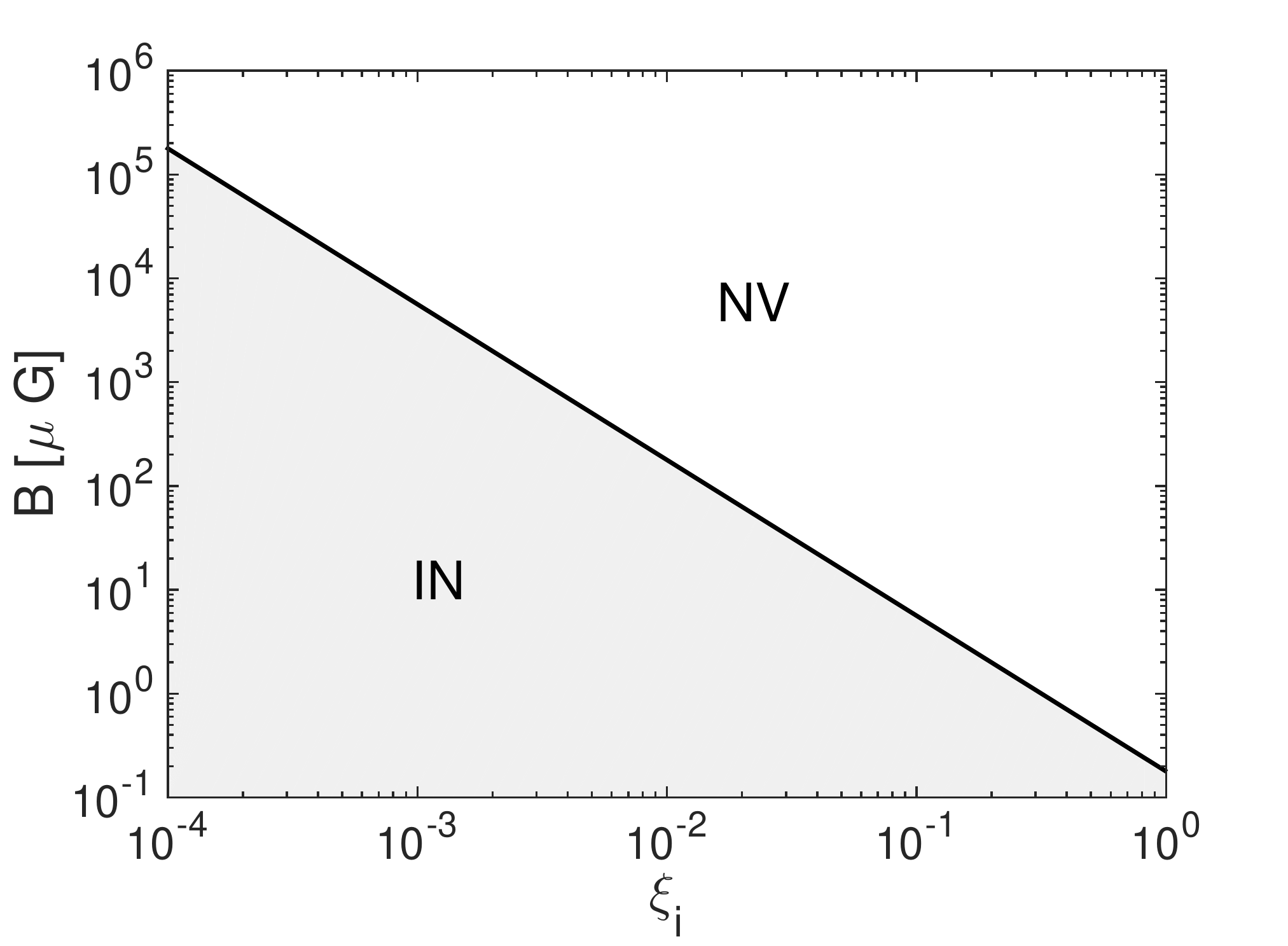}
\caption{The parameter space for determining the relative importance between IN and NV in the case of sub-Alfv\'{e}nic turbulence. From \citet{Xu2017}, \copyright~IOP Publishing \& Deutsche Physikalische Gesellschaft. Reproduced by permission of IOP Publishing. \href{https://creativecommons.org/licenses/by-nc-sa/3.0/}{CC BY-NC-SA}.}
\label{fig2}
\end{figure}

The damping effects in a partially ionized medium mainly arise from the collisional friction between ions and neutrals, i.e., 
ion-neutral collisional damping (IN) and the viscosity in neutral fluid, i.e., neutral viscous damping (NV)
\citep{LithwickGoldreich:2001, Lazarian_etal:2004}.
Their relative importance depends on the turbulence and plasma parameters
\citep{XLY14,Xuc16,Xu2017}.
In super-Alfv\'{e}nic turbulence with the turbulent energy larger than the magnetic energy, 
IN dominates over NV if the condition 
\begin{equation}
  \xi_n \nu_{ni}^2 \nu_n L u_L^{-3} < 0.5
\end{equation}
is satisfied, where $\xi_n$ is the neutral fraction, $\nu_{ni}$ is the neutral-ion collisional frequency, $\nu_n$ is the neutral viscosity, 
$L$ is the driving scale of turbulence, and $u_L$ is the turbulent velocity at $L$. 
In sub-Alfv\'{e}nic turbulence with the turbulent energy smaller than the magnetic energy, under the condition
\begin{equation}\label{eq: subcring}
  \xi_n \nu_{ni}^2 \nu_n L u_L^{-3} M_A^{-1} < 0.5, 
\end{equation}
IN is more important than NV. 
Here $M_A = u_L / V_A$ is the Alfv\'{e}n Mach number and $V_A$ is the Alfv\'{e}n speed. 
An illustration of the above condition in sub-Alfv\'{e}nic turbulence can be found in Fig. \ref{fig2}, where 
the typical parameters of the interstellar turbulence are used
\citep{Xu2017}.
\citet{Xuc16}
found that in the warm neutral medium of the ISM and the solar chromosphere, 
NV plays a significant role in damping MHD turbulence. 

When the damping effects are significant, MHD turbulence is suppressed. 
The damping scale where the MHD turbulence is dissipated can be determined by comparing the turbulent energy cascade rate 
with the damping rate. 
Different from the damping scale of linear MHD waves, 
the damping scale of MHD turbulence has parallel and perpendicular components, $k_{\text{dam},\|}$ and $k_{\text{dam},\perp}$, due to the anisotropy of MHD turbulence. 
In the frame of local magnetic field, i.e., the magnetic field averaged over the length scale of interest, 
magnetic fields are mixed by turbulent motions in the perpendicular direction, 
and the perturbed magnetic fields have wave-like motions in the parallel direction. 
For strong MHD turbulence, the lifetime of waves is equal to the eddy turnover time. 
As the turbulence anisotropy increases toward small length scales, 
when the damping scale of MHD turbulence is much smaller than $L$, 
it can be approximated by its perpendicular component in terms of magnitude.

Under the damping effects of both IN and NV, a general expression of the damping scale is  
\citep{Xu2017}
\begin{subequations}
\label{eq: dssup2dam}
\begin{align}
 & k_{\text{dam},\|}=\frac{-(\nu_n+\frac{V_{Ai}^2}{\nu_{in}})+\sqrt{(\nu_n+\frac{V_{Ai}^2}{\nu_{in}})^2+\frac{8V_A\nu_n l_A}{\xi_n}}}{2\nu_n l_A},  \label{eq: supkpar}\\
 & k_\text{dam}=k_{\text{dam},\|} \sqrt{1+l_A k_{\text{dam},\|}}
\end{align}
\end{subequations}
for super-Alfv\'{e}nic turbulence, and 
\begin{equation}
\begin{split}
\label{eq: subgsbc}
& k_{\text{dam},\|}=\frac{-(\nu_n+\frac{V_{Ai}^2}{\nu_{in}})+\sqrt{(\nu_n+\frac{V_{Ai}^2}{\nu_{in}})^2+\frac{8V_A\nu_n LM_A^{-4}}{\xi_n}}}{2\nu_n L M_A^{-4}}, \\
& k_\text{dam}=k_{\text{dam},\|} \sqrt{1+LM_A^{-4} k_{\text{dam},\|}}
\end{split}
\end{equation}
for sub-Alfv\'{e}nic turbulence, where $V_{Ai}$ is the Alfv\'{e}n speed of the ionized gas, $\nu_{in}$ is the ion-neutral collisional frequency, 
and $l_A = L M_A^{-3}$ in super-Alfv\'{e}nic turbulence.

When IN is the dominant damping effect, the damping is related to the coupling state between ions and neutrals. 
The most severe damping arises in the weakly coupled regime in a weakly ionized medium, 
where neutrals decouple from ions, but ions are still coupled with neutrals due to their 
frequent collisions with surrounding neutrals. 
In this situation, 
the (perpendicular) damping scale for super- and sub-Alfv\'{e}nic turbulence can be simplified as 
\citep{Xu2017}
\begin{equation} \label{eq: mtnisupds}
     k_{\text{dam,IN,sup},\perp}=\bigg(\frac{2\nu_{ni}}{\xi_n}\bigg)^{\frac{3}{2}}L^{\frac{1}{2}}u_L^{-\frac{3}{2}}
\end{equation}
and 
\begin{equation}\label{eq: mtnisubds}
   k_{\text{dam,IN,sub},\perp}=\bigg(\frac{2\nu_{ni}}{\xi_n}\bigg)^{\frac{3}{2}}L^{\frac{1}{2}}u_L^{-\frac{3}{2}}M_A^{-\frac{1}{2}}.
\end{equation}
Its relation to the neutral-ion decoupling scale is 
\begin{equation}\label{eq: damdec}
      k_\text{dec,ni} = \Big(\frac{2}{\xi_n}\Big)^{-\frac{3}{2}} k_\text{dam,IN} 
\end{equation}
for both super- and sub-Alfv\'{e}nic turbulence.  
The difference between the decoupling and damping scales has been confirmed by two-fluid MHD simulations 
\citep{Burk15}.
The above relation shows that the decoupling of neutrals from ions plays a key role in IN damping of MHD turbulence.

After neutrals decouple from ions, the hydrodynamic turbulence carried by neutrals is only subject to viscous damping, and its NV damping scale is 
smaller than the IN damping scale of MHD turbulence in ions. 
This differential damping of neutrals and ions can account for the observed linewidth difference of coexistent neutrals and ions in molecular clouds 
\citep{LH08,XLY14}.

When NV dominates over IN, the damping scale is approximately 
\citep{Xu2017}
\begin{equation}\label{eq: mtnvsupds}
   k_{\text{dam,NV,sup},\perp} = \Big(\frac{\xi_n}{2}\Big)^{-\frac{3}{4}} \nu_n^{-\frac{3}{4}} L^{-\frac{1}{4}}u_L^\frac{3}{4}
\end{equation}
for super-Alfv\'{e}nic turbulence, and 
\begin{equation}\label{eq: mtnvsubds}
k_{\text{dam,NV,sub},\perp} =\Big(\frac{\xi_n}{2}\Big)^{-\frac{3}{4}} \nu_{n}^{-\frac{3}{4}} L^{-\frac{1}{4}}u_L^\frac{3}{4} M_A^\frac{1}{4}
\end{equation}
for sub-Alfv\'{e}nic turbulence. 
MHD turbulence is damped at the NV damping scale. On the length scales smaller than the NV damping scale and larger than the 
IN damping scale, magnetic fluctuations persist, and the magnetic energy has a spectral form $k^{-1}$. 
This ``new regime of MHD turbulence'' in the sub-viscous range was first analytically predicted by 
\citet{Lazarian_etal:2004}
and further confirmed by MHD simulations 
\citep{CLV_newregime,CLV03}.
The $k^{-1}$ spectrum of magnetic energy in the sub-viscous range has also been observed in dynamo simulations in the 
nonlinear stage of dynamo 
\citep{Hau04,Sch04}.

 \subsection{Relativistic MHD turbulence}
 \label{sec:turbulence4}

Due to its numerical and theoretical simplicity, MHD turbulence in relativistic force-free regime has been studied first. Relativistic force-free formalism can be used for a system, such as the magnetosphere of a pulsar or a black hole, in which the magnetic energy density is much larger than that of matter.

Scaling relations for relativistic Alfvenic MHD turbulence were first derived by 
\citet{ThompsonBlaes:1998}. The predictions of these theory in terms of the spectral slope and anisotropy coincide with those in GS95 theory. These relations were first numerically tested by 
\citet{Cho:2005}
who performed a numerical simulation of a decaying relativistic force-free MHD turbulence with numerical resolution of $512^3$ and calculated energy spectrum and anisotropy of eddy structures.  After $t = 3$, the energy spectrum with $E\sim k^{5/3}$  was shown to decrease in amplitude without changing its slope. The anisotropy was also measured and was shown to be consistent with the GS95 expectations, i.e. $k_{\|}\sim k_{\bot}^{2/3}$.  Thus in terms of Alfvenic turbulence the correspondence between the non-relativistic theory and its relativistic counterpart is excellent. Interestingly enough, the relativistic simulations of imbalanced Alfvenic turbulence, i.e. the turbulence with the excess of the energy flux moving in one direction, performed in 
\citet{ChoLazarian:2014}
provided results in agreement with the theory of non-relativistic imbalanced turbulence in 
\citet{BeresnyakLazarian:2008}.

\begin{figure}
\centering
\includegraphics[width=0.48\textwidth]{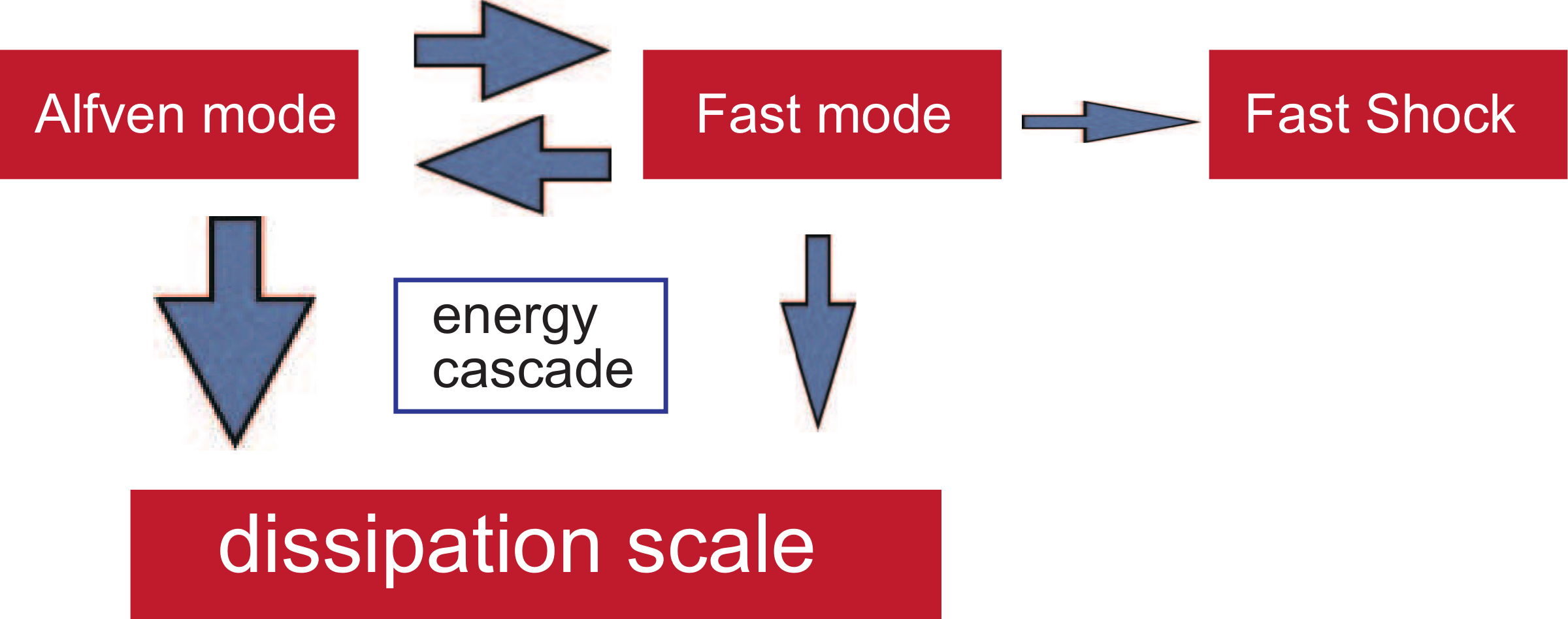}
\caption{A schematic picture of the energy transfer in high-$\sigma$ turbulence. From \cite{Takamoto2017}.}
\label{fig3}
\end{figure}

The situation is somewhat different in the case of relativistic compressible theory. The initial studies of relativistic MHD turbulence \cite[see][]{Zhang_etal:2009,Inoue_etal:2011,BeckwithStone:2011,2012ApJ...744...32Z,ZrakeMacFadyen:2013,GarrisonNguyen:2015} 
have not revealed much differences between the non-relativistic and relativistic turbulence. However, more recent studies in 
\citet{Takamoto2016,Takamoto2017}
provided the decomposition of the turbulent motions into Alfven, slow and fast modes. The comparison of the obtained results with the analogous decomposition into modes in non-relativistic case in 
\citet{ChoLazarian:2002,ChoLazarian:2003}
and 
\citet{KowalLazarian:2010}
revealed significant differences in terms of the coupling of the compressible and incompressible motions in the two cases. These differences are illustrated by Figure \ref{fig3} where the transfer of energy between Alfven and fast modes is shown in the relativistic case. Note, that the corresponding transfer of energy is suppressed for non-relativistic compressible turbulence. In terms of anisotropy, this energy transfer makes fast modes more similar to Alfven ones. We find that this effect of energy transfer is important for describing turbulent relativistic reconnection.

\subsection{Turbulence: Summary of basic facts}
\label{turb_summary}

For our discussion of turbulent reconnection it is important turbulence is ubiquitous in astrophysical environments and 
that Alfvenic modes usually represent the dominant component of MHD turbulent cascade.
These modes are present in collisional and collisionless environments. The turbulence in most astrophysical settings has the sources different from the magnetic reconnection. However, it is difficult to understand the properties of turbulent fluids if magnetic reconnection does not happen on the dynamical time of the fluid motions.

The partial ionization of plasmas changes the properties of MHD turbulence increasing the scale at which MHD turbulence is damped. The relativistic and non-relativistic MHD turbulence have significant similarities, although they show a difference in the coupling of fast and Alfven modes. More detailed description of the theory and implications of MHD turbulence can be found in a book by \cite{BeresnyakLazarian:2019}.

\section{Analytical model of turbulent reconnection}
\label{sec:model}

\subsection{Derivation of LV99 reconnection rate}

The 3D model of turbulent reconnection proposed in LV99 provides a natural generalization of the classical Sweet-Parker model \citep{Parker:1957, Sweet:1958}. Thus we remind the reader first the basic properties of this model. 

In the Sweet-Parker model, \textbf{two regions} with uniform {\it laminar} magnetic fields are separated by thin current sheet. The magnetic fields are frozen in through the two regions and the violation of the flux freezing takes place over a thin slot $\delta$, the latter being determined by Ohmic diffusion. Thus the speed of
reconnection is determined roughly by the resistivity divided by the sheet thickness,
i.e.
\begin{equation}
V_{rec1}\approx \eta/\Delta.
\label{eq.1}
\end{equation}
Obviously, the smaller $\Delta$ the larger the reconnection speed. Small $\Delta$, however, prevents plasmas from leaving the reconnection region. 

The Sweet-Parker model deals with the {\it steady state reconnection} and therefore the plasma in the diffusion region must be
ejected from the edge of the current sheet at the Alfv\'{e}n speed, $V_A$. As a result,
the reconnection speed is limited by the conservation of mass condition
\begin{equation}
V_{rec2}\approx V_A \Delta/L_x,
\label{eq.2}
\end{equation}
where $L_x$ is the lateral extend of the current sheet. 

The conditions imposed by Eq. (\ref{eq.1}) and (\ref{eq.2}) are acting in the opposing directions as far as the thickness of of the reconnection region $\Delta$ is concerned. As a result, the Sweet-Parker reconnection cannot be fast as it faces a contradictory requirements. Therefore the compromise speed of reconnection is the Sweet-Parker reconnection
speed that is reduced from the Alfv\'{e}n speed by the square root of the Lundquist
number, $S\equiv L_xV_A/\eta$, i.e.
\begin{equation}
V_{rec, SP}=V_A S^{-1/2}.
\label{SP}
\end{equation}
The Lundquist number in astrophysical conditions is enormous, it can be $10^{16}$ or larger. In this situation, the Sweet-Parker reconnection rate is absolutely negligible. 

Due to the disparity between $L_x$ and $\Delta$ the Sweet-Parker model cannot explain most of the observed astrophysical reconnection events, e.g. Solar flares. A way to deal with this problem was suggested by \cite{Petschek:1964}, who speculated that in special conditions 
the reconnection configurations may have magnetic field lines converging to the reconnection zone at a sharp angle making $L_x$ and $\Delta$ comparable. The corresponding configuration of magnetic field is known as X-point reconnection as opposed to the Sweet-Parker extended current sheet Y-type reconnection. The research exploring X-point reconnection was the dominant trend in the attempts of obtaining fast reconnection for decades. However, the major problem of such configurations is to preserve the X-point configuration over astrophysically large scales. 

\begin{figure}
\centering
\includegraphics[width=0.48\textwidth]{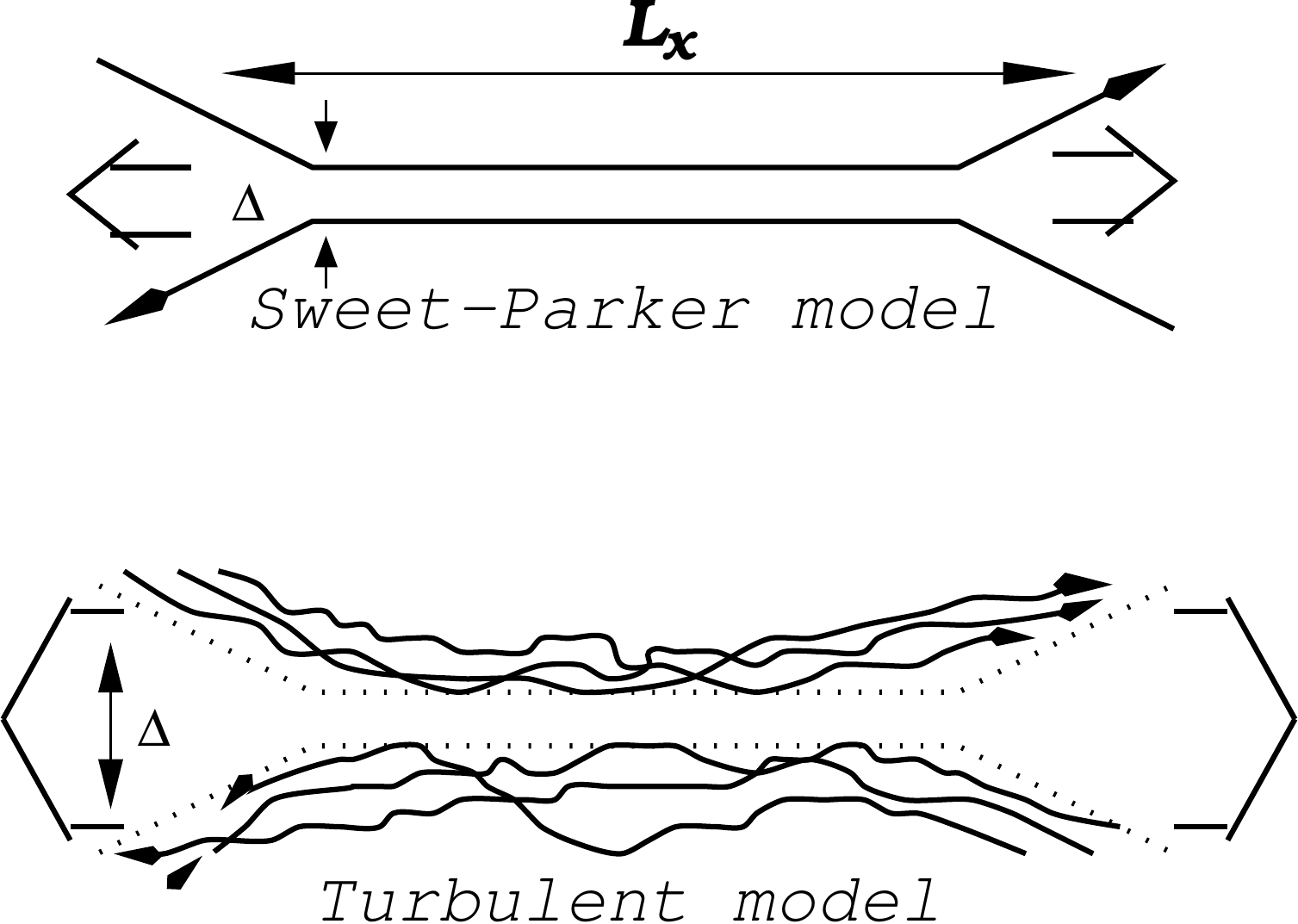}
\caption{{\it Upper plot}: Sweet-Parker model of reconnection.  The thickness of the outflow $\Delta$ is limited by Ohmic diffusivity that takes place on microscopic scales.  The disparity between this scale and the astrophysically large scale $L_x \gg \Delta$ makes the reconnection slow. 
{\it Lower plot}: LV99 model of turbulent reconnection model. The width of the
outflow $\Delta$ is determined by macroscopic field line wandering and it can be comparable with the scale $L_x$ for trans-Alfvenic turbulence.
\cite{LazarianVishniac:2000}.}
\label{fig:recon}
\end{figure}

The direction taken by LV99 was to modify the Y-type reconnection in order to make $\Delta$ macroscopic and even astrophysically large. Naturally, this is impossible with plasma effects. However, LV99 proposed that turbulence can induce magnetic field wandering that can do the job.
The corresponding model of magnetic reconnection is illustrated by
Figure~\ref{fig:recon}.

Similar to the Sweet-Parker model, the turbulent model developed in LV99 deals with a generic
configuration, that arises naturally as magnetic flux tubes crossing each other make
their way one through another.  The outflow thickness $\Delta$ is determined by the large-scale magnetic field wandering. This wandering depends on the level of turbulence. As a result, the rate of reconnection in LV99 model is determined by the level of MHD turbulence. The LV99 reconnection rates is fast, i.e. its rate do not depend on the fluid/plasma resistivity. However, the rates of turbulent reconnection can change significantly depending on the level of turbulence. This is an important prediction of LV99 theory which make it different from the theories that, for instance, appeal to plasma effects for describing fast reconnection. 

To quantify the reconnection rate in LV99 model one should use the scaling
relations for Alfv\'{e}nic turbulence from \S~\ref{sec:turbulence}. In fact, it is reconnection at low Alfven Mach numbers that corresponds to the classical representation of magnetic reconnection flux tubes. This, in fact required LV99 to generalize the GS95 MHD theory for arbitrary Alven Mach numbers. 

As the first step consider the problem of magnetic field wandering, as this magnetic wandering determines the thickness of outflow $\Delta$. We may note that magnetic wandering determined in LV99 is not only important for magnetic reconnection. In fact, it is a fundamental property of turbulent magnetic field and it is important for different problems, e.g. problems of heat transfer (Narayan \& Medvedev 2001, Lazarian 2006) and cosmic ray propatation (Yan \& Lazarian 2008, Lazarian \& Yan 2014). 

Using the MHD turbulence relations one can describe the diffusion of magnetic field lines. Indeed, a bundle of field lines confined within a region of width $y$ at some particular point
spreads out perpendicular to the mean magnetic field direction as magnetic field lines are traced  in either direction.  The rate of field
line spread is given by (see more in \S \ref{Rich_disper})
\begin{equation}
\frac{d\langle y^2\rangle}{ds} \sim \frac{\langle y^2\rangle}{\lambda_\parallel},
\end{equation}
where $\lambda_{\|}^{-1}\approx \ell_{\|}^{-1}$, $\ell_{\|}$ is the parallel
scale. The corresponding transversal scale, $\ell_{\perp}$, is $\sim \langle
y^2\rangle^{1/2}$, and $x$ is the distance measured along the 
magnetic field direction.  As a result, using equation (\ref{Lambda}) one gets
\begin{equation}
\frac{d\langle y^2\rangle}{ds} \sim L_i\left(\frac{\langle y^2\rangle}{L_i^2}\right)^{2/3}
\left(\frac{u_L}{V_A}\right)^{4/3}
\label{eq:diffuse}
\end{equation}
where the substitution $\langle y^2\rangle ^{1/2}$ for $\ell_{\perp}$ is used.  This
expression for the magnetic field dispersion is applicable only when $y$ is small
enough for us to use the strong turbulence scaling relations, i.e.
when $\langle y^2\rangle < L_i^2(u_L/V_A)^4$. 

For scales less that $L_i$ it follows from Eq.(\ref{eq:diffuse}) that the mean squared magnetic field wandering is equal to:
\begin{equation}
    \langle y^2\rangle\sim \frac{s^3}{L_i} \left(\frac{u_L}{V_A}\right)^4, s<L_i
    \label{eq:superdiff}
\end{equation}
which provides a superdiffusive dependence of 
$\langle y^2\rangle\sim s^3$, which was one of the major findings in LV99 as far as the properties of magnetic field in turbulence are concerned. Due to the above scaling the probability of wandering magnetic field line to return back to the reconnection layer is very small.

The dependence of $\langle y^2\rangle\sim M_A^4$ is significant as it shows that if the turbulent kinetic energy is a small portion of magnetic energy of the fluid, the magnetic field lines are nearly straight. This provides a smooth transition to the slow reconnection in the absence of turbulence and, as we discuss later, to the flux freezing being a fair approximation to the low $M_A$ fluids.

 For describing magnetic field wandering over scales larger than the turbulence injection scale, i.e. $s\gg L_i$ magnetic field wandering obeys the usual random walk scaling. The corresponding random walk step is $s/L_i$ with the mean squared displacement per step equal to
$L_i^2(u_L/V_A)^4$.  As a result,
\begin{equation}
\langle y^2\rangle^{1/2}\approx (L_i s)^{1/2} (u_L/V_A)^2~~~x>L_i,
\label{eq:diffuse3}
\end{equation}
which provides the other limit for the magnetic field wandering.
While the diffusive regime of magnetic field lines given by Eq. (\ref{eq:diffuse3}) was well known in cosmic ray literature \cite[see][]{Schlickeiser2002}, the superdiffusive one was introduced in LV99\footnote{This resulted in several important developements not only for reconnection but for the understanding of cosmic ray propagation and acceleration \cite[see][]{LazarianYan:2014}.}.

Using Eqs. (\ref{eq:superdiff}) and (\ref{eq:diffuse3}) one can describe the thickness of the outflow for the reconnection of the anti-parallel magnetic fields in turbulent media, i.e. in the setting generalizing Sweet-Parker reconnection without guide field for turbulent media. Measuring $s$ along the horizontal $x$-axis along the magnetic field reversal and combining Eqs. (\ref{eq:superdiff}) and (\ref{eq:diffuse3}) it is possible to derived the
thickness of the outflow $\Delta$ in the aforementioned
two regimes. Then, using the mass conservation Eq. (\ref{eq2}) it is straightforward to obtain
\begin{equation}
V_{rec} \approx V_A\min\left[\left(\frac{L_x}{L_i}\right)^{1/2},
\left(\frac{L_i}{L_x}\right)^{1/2}\right]
M_A^2,
\label{eq:lim2a}
\end{equation}
where $V_A M_A^2$ is proportional to the turbulent eddy speed. As we discussed earlier, the obtained  reconnection
rate vary depending on Alfven Mach number $M_A$ and for $M_A\sim 1$ can represent a large fraction of the Alfv\'{e}n speed in the case when $L_i$ and $L_x$
are not too comparable.

Similar to the Sweet-Parker picture, the contribution of the guide field does not change the dynamics of magnetic reconnection. Figure \ref{fig:recon} presents a $XY$ plane of cross section of the reconnection region perpendicular to the guide field $B_z$ shared by two fluxes. The guide field should be ejected together with plasma from the reconnection region. Indeed, both random velocity and magnetic field can be decomposed into components in the direction of the guide field and perpendicular to it. The field wondering in the direction perpendicular to the guide field is affected by the velocity perturbations in the same direction. The guide field does not oppose to the corresponding bending of magnetic field. The outflow velocity is, however, is determined only by the Alfven velocity corresponding to the perpendicular component of magnetic field. In fact, for the sake of clarity, it could be right to change $M_A$ for $M_{A,XY}$, where the latter denotes the Alfven Mach number corresponding to the Alfven velocity in the plane perpendicular to the guide field. This notation of the Alfven Mach number is not common and we will not use it therefore. 

It was communicated to us by Andrey Beresnyak that the independence of the reconnection rate on the guide field can be also justified from the point of Reduced MHD (RMHD) approach (see Beresnyak 2012). Formally, the RMHD is applicable for describing incompressible fluids when the perturbations are much smaller than $V_A$. However, in practical cases RMHD is well applicable for many collisionless systems, e.g. Solar wind (see Schekochihin et al. 2009), and its practical validity is good even for transAlfvenic turbulence (Beresnyak \& Lazarian 2009). Within the RMHD the dynamics of turbulent motions is not changed if the ratio $B_0/l_{\|}$ stays the same. If no constraines are imposed on the parallel scale of perturbations $l_{\|}$ they increases with the increase of the guide field $B_0$ as was demonstrated in numerical simulations in Beresnyak (2017). The dynamics in terms of perpendicular perturbations $l_{\bot}$ and field wondering that these perturbations entailed increasing the reconnection outflow stayed the same. 

For distances along magnetic field less than the injection scale $L_i$ using the definition of cascading rate for strong MHD turbulence given by Eq. (\ref{alt_sub2}), we can rewrite the expression for magnetic field wandering (see Eq. (\ref{eq:superdiff}))
as 
\begin{equation}
    \langle y^2 \rangle \sim \epsilon_s\left(\frac{s}{V_A}\right)^3
    \label{wantering_strong}
\end{equation}
from which it is clear that the increase of $V_A$ can be naturally compensated by the increase of the parallel scale along the magnetic field. 

For the scales larger than the injection scale, the Eq. (\ref{eq:diffuse3}) can be rewriten as 
\begin{equation}
    \langle y^2\rangle \sim \frac{L_i^2 s}{V_A^3}\epsilon_w,
    \label{wandering_weak}
\end{equation}
where the expression for the cascading of the weak MHD cascade given by Eq.(\ref{eps_weak}) is used. It is also obvious that the simultaneous increase of the parallel scale of both injection and turbulent motions and the Alfven velocity does not change the magnetic field wandering. 

For the formal applicability of RMHD one should assume that the guide field $B_z$ is much stronger than the reconnecing magnetic field $B_x$. The relevant setting corresponds to reconnection of two flux tubes intersecting at a small angle $\alpha$. The corresponding setting was discussed in LV99 for the reconnection induced by individual eddies within MHD turbulence and it was shown there that the reconnection happens within one eddy turnover time. Here we consider a more generic situation of reconnecting flux tubes with an arbitrary turbulence driving. 

In the above setting, $B_x\approx B_z\sin \theta/2$ and the perturbations at the injection scale $L_i$ result in the perturbations in the $XY$ plane with the scale $L_{i,XY}\sim L_i \theta/2$. Similarly the path $x$ measured along the magnetic flux tubes projects into the path $x_{XY}\approx x\theta/2$ in the $XY$ plane. If one considers the dynamics of magnetic field lines in the $XY$-plane the speed of propagation of Alfvenic perturbations along $x$-axis is $V_A\theta/2$ and therefore in this approximation it is obvious that the growth of the outflow region $\Delta$ as seen in $XY$-plane does not depend on the angle $\theta$ between reconnecting fluxes, i.e. magnetic reconnection proceeds independently of the value of guide field, but depends only on the $B_x$ field and the properties of turbulence in $XY$-plane.

In numerical experiments described in \S \ref{sec:testing} the properties of driving as well as value of the reconnecing component of magnetic field were fixed in the $XY$-plane. Therefore the magnetic field wandering for such settings can be described by Eq. (\ref{wantering_strong}) and (\ref{wandering_weak}) with $V_A$ arising only from $B_x$ and $L_i$ being the injection scale of turbulence in the $XY$-plane. Correspondingly, Eq. (\ref{eq:lim2a}) should use $M_A\approx \delta B_x/B_x$, where $\delta B_x$ is a turbulent perturbation of $B_x$ at the injection scale $L_i$.

The LV99 theory of turbulent reconnection has provided a number of predictions very different from the contemporary theories of fast magnetic reconnection. First of all, the turbulent reconnection was identified as a generic process that takes place everywhere in the magnetized conducting fluid. A violation of flux freezing implicitly follows from that. Second, LV99 does not prescribe a given the reconnection rate, but provides an expression that shows that the speed of reconnection vary depending on the level of turbulence. This can explain the variability of the energy release in reconnection events observed in e.g. Solar flares. Indeed, many of the observation processes require that the rate of reconnection changes. In LV99 theory this is controlled by the level of turbulence that can vary. For instance, the low rate of turbulent reconnection can explain how magnetic flux can be accumulated prior to a Solar flare. Third, the independence on the plasma parameters makes the turbulent reconnection a generic reconnection process in the astrophysical settings. 

While the earlier theories of turbulent reconnection were focused X-point reconnection, LV99 showed that Y-type reconnection can be fast. The tearing reconnection that was developed later (see \S \ref{sec:alternatives1}) presented another version of the fast Y-type reconnection. Similar to LV99 it departed from the regular structures prescribed by Sweet-Parker and Petscheck reconnection models. 

Being Y-type fast reconnection the LV99 necessarily involved a significant volume of magnetized fluid in the reconnection process. This volume reconnection changes the nature of the energetic particle acceleration as it is discussed in \S \ref{sec:implications3}

\subsection{Flux freezing violation: Richardson dispersion }
\label{ssec: rdffvio}

\subsubsection{Alfv\'en flux freezing}

Since the seminal work of \cite{Alfven:1942}, the ``flux freezing principle'' has become a powerful tool to estimate solutions of MHD equations 
in a diverse range of problems \citep{Parker:1979,Kulsrud:2005}. Besides its utility in plasma physics, in astrophysics as well it has found many applications, for example, in explaining the low angular momentum of stars and the spiral structure of magnetic field lines in the solar wind. Despite its success in addressing diverse problems, however, there have been also fatal failures \cite[see][]{Burlaga_etal:1982, KO2012, Richardsonetal:2013, Eyink2015}. For instance, assuming flux freezing, the magnetic topology in a high-conductivity magnetized fluid should not change rapidly, whereas observations of solar flares and Coronal Mass Ejections (CMEs) indicate otherwise. Also, in star formation, the magnetic pressure of in-falling magnetized material should be strong enough to prevent gravitational collapse altogether, if magnetic flux-freezing held with good accuracy. Finally, if magnetic field were frozen into the fluid in small-scale astrophysical dynamos, the tangled structure of the generated magnetic field would quench its further growth.

In order to account for these difficulties with magnetic flux-freezing, one might find it tempting to appeal to additional effects beside collisional resistivity. 
For instance, one might consider additional terms, such as Hall effect and electron pressure anisotropy, in the generalized Ohm's law (electron momentum equation) 
\citep{burch2016electron, cassak2016inside}. This does not answer the question, unfortunately, how magnetic reconnection events can be fast in fluids with high rates of collisions. Indeed, both observations and theory imply fast magnetic reconnection in the solar photosphere, and other environments such as collisional parts of accretion disks, which cannot be explained by appealing to collisionless mechanisms \citep{chae2010new, singh2011chromospheric}. The problem persists even for weakly collisional systems where an ideal MHD-type description should be valid at length-scales much larger than the ion gyroradius, above which flux-freezing is often assumed to hold. However, observations indicate that magnetic structures with length scales much larger than ion gyroradius  reconnect very rapidly, e.g. at the heliospheric current sheet in the solar wind \citep{gosling2012magnetic}. It is unclear how collisionless mechanisms at scales below the ion cyclotron radius can lead to fast magnetic reconnection on much larger scales. This is the essence of the ``scale problem'' in magnetic reconnection \citep{ji2019major}. 

While the necessity of fast magnetic reconnection was widely accepted, most of the studies assumed that the fast reconnection happens for very special configurations of magnetic fields. This way the community could reconcile at least some of the observational evidence of fast reconnection and the concept of magnetic flux freezing. LV99, as discussed previously, identified magnetic reconnection as an intrinsic property of MHD turbulent cascade implying the violation of the flux freezing principle \cite[see][for an explicit statement about the flux freezing in turbulent fluids]{VishniacLazarian:1999}. 

Below we provide a description of the Richardson dispersion, the phenomenon that manifests a gross violation of the flux freezing in magnetized turbulent fluids. This numerically confirmed phenomenon (see \S \ref{ssec: doffvio}) provides another way of re-deriving LV99 rate of turbulent reconnection. A more rigorous approach to the flux freezing violation is discussed in \S \ref{sec:level1}.

\subsubsection{Time dependent Richardson dispersion}

Another way to understand the violation of flux freezing in turbulent fluids is related to the concept of Richardson dispersion. This process initially introduced in hydrodynamic turbulence carries over to MHD turbulence. 
\citet{richardson1926atmospheric}
empirically analyzed the dispersion of the volcanic ashes during the eruption and formulated the law that can be derived using hydrodynamic turbulence theory. Note, however, that  the corresponding theory was formulated much later by 
\citet{Kol41}. 
Using this theory one can predict that the separation between two particles obeys the equation $d/dt[l(t)]\sim v(l)\sim \alpha l^{1/3}$, where $\alpha$ is proportional to a cubical root of the energy cascading rate. The solution of this equation $l^2 \sim t^3$   describes the Richardson 
dispersion, i.e. the mean square separation between particles increasing in proportion to the cube of time. The accelerated growth of the separation is due to the fact that the larger the separation, the larger the eddies that induce the separation. In Kolmogorov turbulence the larger eddies have the larger velocity dispersion.
An interesting feature of this solution above is that the provides the type of fast separation even if the initial separation of particles is zero. Formally this means the violation of Laplacian determinism. Mathematically the above paradox is resolved by accounting to the fact that turbulent field is not differentiable\footnote{The Kolmogorov velocity field is Holder continuous, i.e. $|v(r_1) - v(r_2)| \lesssim C|r_1 - r_2|^{1/3.}$} and therefore it is not unexpected that the initial value problem does not have a unique solution. In any realistic physical settings the turbulence is damped at a non-zero scale $l_{min}$, which makes the paradox less vivid. At the scales less than the $l_{min}$ the separation of points is not stochastic and it follows the Lyapunov growth
\cite[see][]{Lazarian2006}. 
However, at the scales larger $l_{min}$ but smaller than the injection scale of turbulence $L_{max}$, i.e. over the inertial range of turbulence, the explosive growth of separation is still present. 

\begin{figure}
\centering
\includegraphics[width=0.48\textwidth]{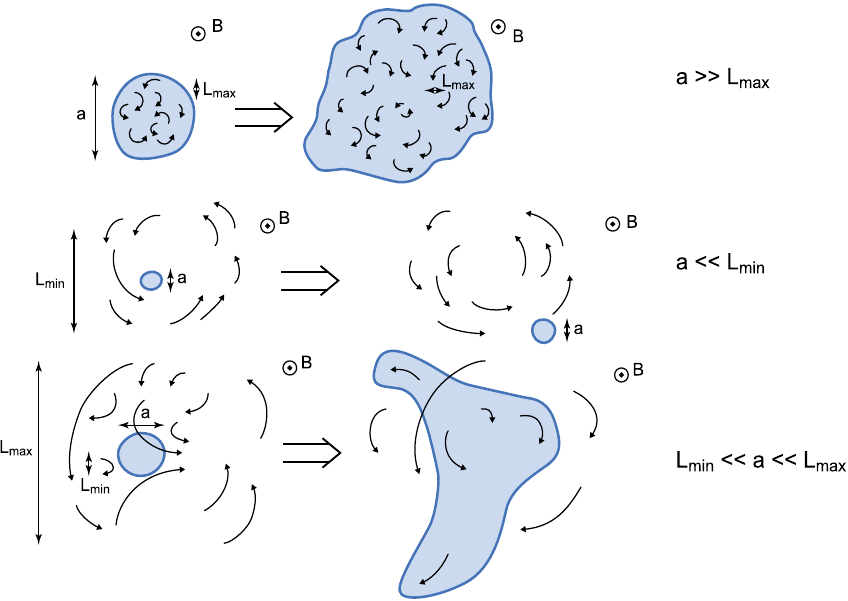}
\caption{Illustration of  diffusion for different regimes. Magnetic field ${\bf B}$ is perpendicular to the eddies. Upper plot corresponds to the diffusion on scales larger than the maximal size of turbulent eddies; Middle plot corresponds to the diffusion at scales less than the damping scale of turbulence; Lower plot corresponds to a Richardson dispersion case, i.e. when the dispersion is measured for fluid elements with separations within the inertial range of turbulent motions. From \cite{Lazarian:2014}.}
\label{fig4}
\end{figure}

In terms of motions perpendicular to the local direction of magnetic field, the scaling of velocities in GS95 picture corresponds to the Kolmogorov one. Therefore we do expect that the plasma particles to separate in accordance to the Richardson law over the inertial range. The situation is illustrated by Figure~\ref{fig4}.  The magnetic field is shown to be perpendicular to the plane of the picture. A cloud of test particles undergoes an ordinary diffusion for $a\gg L_{max}$, undergoing Lyapunov separation of field lines for $a<l_{min}$ and exhibiting Richardson dispersion over the inertial range $[L_{max}, l_{min}]$. Obviously, the Richardson-type dynamics is not possible without magnetic field lines disentangling themselves via fast reconnection. This corresponds to the notion in LV99 that turbulent reconnection is required for turbulent magnetized fluid to preserve fluid-type behavior.
We note that if reconnection were slow the intersecting magnetic field flux tubes would form unresolved knots. This inevitably would arrest the fluid motions, making the fluid with dynamically important magnetic field Jello or felt-like substance. The spectrum of magnetic fields would be very modified in this case with a significant increase of energy at small scales. Only acoustic type waves can propagate in this substance and definitely no present-day numerical simulations of MHD type can describe it. This picture of slow reconnection in turbulent media contradicts both to numerical simulations and observations of interstellar turbulence (see \S \ref{sec:turbulence1} and \S \ref{sec:turbulence2}). The turbulent reconnection, on the contrary, predicts the free evolution of eddies that produce Kolmogorov-type cascade for eddies perpendicular to local direction of magnetic field as discussed in \S \ref{sec:turbulence2}. 

The Richardson dispersion is a numerically proven phenomenon (see \S \ref{ssec: doffvio}) and it was used in ELV11 to re-derive the LV99 rates of reconnection. 

Consider first the Sweet-Parker reconnection. The Ohmic diffusion induces the
mean-square displacement across the reconnection layer  in a time $t$ given by
\begin{equation}
\langle y^2(t)\rangle \sim \lambda t.
\label{diff-dist}
\end{equation}
where $\lambda=c^2/4\pi\sigma$ is the magnetic diffusivity. As in our previous treatment the magnetized plasmas is
ejected out of the sides of the reconnection layer of length $L_x$ at a
velocity of order $V_A$.  Thus, the time that the lines can spend in the
resistive layer is the Alfv\'{e}n crossing time $t_A=L_x/V_A$. Therefore, the reconnected field lines
 are separated by a distance
\begin{equation}
\Delta = \sqrt{\langle y^2(t_A)\rangle} \sim \sqrt{\lambda t_A} = L_x/\sqrt{S},
\label{Delta}
\end{equation}
where $S$ is Lundquist number.  Combining Eqs. (\ref{eq.2}) and (\ref{Delta})
one the Sweet-Parker result, $v_{rec}=V_A/\sqrt{S}$.

The difference with the turbulent case is that instead of Ohmic diffusion 
the Richardson dispersion dominates (ELV11).  In this case the mean
squared separation of particles follows the Richardson law, i.e. $\langle |x_1(t)-x_2(t)|^2 \rangle\approx
\epsilon t^3$, where $t$ is time, $\epsilon$ is the energy cascading rate.
For subAlfv\'{e}nic turbulence $\epsilon\approx u_L^4/(V_A L_i)$ (see LV99) and
therefore one can write
\begin{equation}
\Delta\approx \sqrt{\epsilon t_A^3}\approx L(L/L_i)^{1/2}M_A^2
\label{D2}
\end{equation}
where it is assumed that $L<L_i$.  Combining Eqs. (\ref{eq.2}) and (\ref{D2})
it is easy to obtain
\begin{equation}
v_{rec, LV99}\approx V_A (L/L_i)^{1/2}M_A^2.
\label{LV99}
\end{equation}
in the limit of $L<L_i$.  Similar considerations allowed ELV11 to re-derive the LV99
expression for $L>L_i$, which differs from Eq.~(\ref{LV99}) by the change of the
power $1/2$ to $-1/2$.

\subsubsection{Field line wandering}
\label{Rich_disper}

Magnetic fields produce a new effect compared to the Richardson dispersion in hydrodynamics. In magnetized fluids one can trace magnetic line separation. This effect has been discussed for decades in the cosmic ray literature as the cause of the perpendicular diffusion of cosmic rays in the galactic magnetic field 
\citep{Jokipii1973}. It worth mentioning that the correct quantitative description of the effect obtained only in LV99. Earlier papers assumed that the diffusive separation of magnetic field lines, which is the effect taking place only if magnetic field lines are separated by more than the turbulence injection scale $L$. It was shown in LV99 that at scales less than $L$ the magnetic field lines separate superdiffusively. This changes a lot in terms of both the propagation and the acceleration of cosmic rays \cite[see][]{LazarianYan:2014}.

\begin{figure}
\centering
\includegraphics[width=0.48\textwidth]{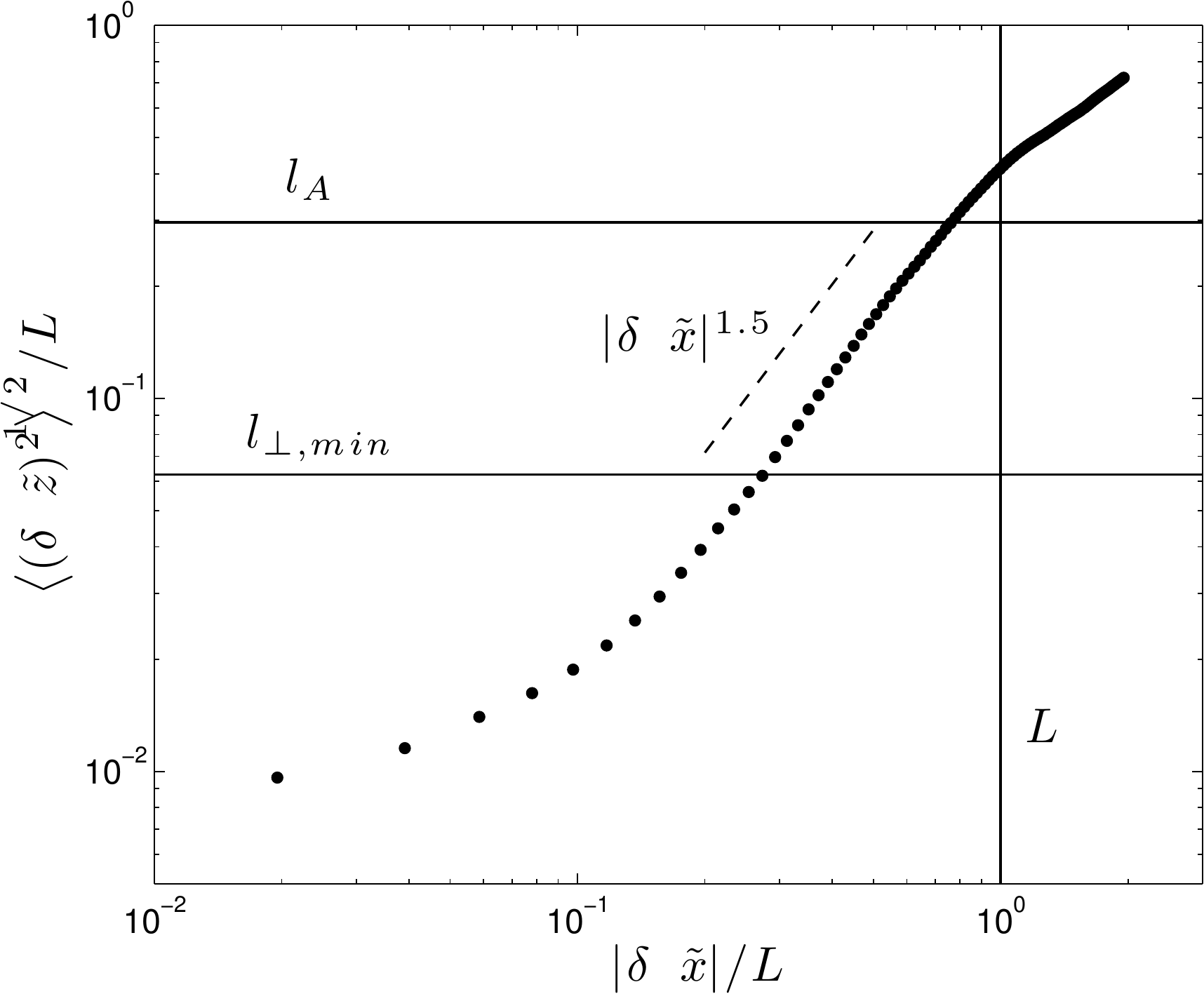}
\caption{Tracing magnetic fields with charged particles.  The mean squared separation of particle trajectories $\langle (\delta z)^2\rangle^{1/2}$ as the particles move along along the magnetic field is measured in the $x$-direction.  
From \citet{XY13}, \copyright~AAS. Reproduced with permission.}
\label{fig5}
\end{figure}

Dealing with magnetic reconnection in \S \ref{sec:model} we have discussed the separation of magnetic field lines. 
Figure \ref{fig5} illustrates the spread of magnetic field lines in the perpendicular direction in the situation when magnetic field lines are traced by particles moving along them. This figure confirms the prediction for the magnetic field line separation in Eq. (\ref{eq:superdiff}), i.e.   that the squared separation between the magnetic field lines increases in proportion the cube of the distance measured along the magnetic field lines (see \ref{fig5}). 

\begin{figure}
\centering
\includegraphics[width=0.48\textwidth]{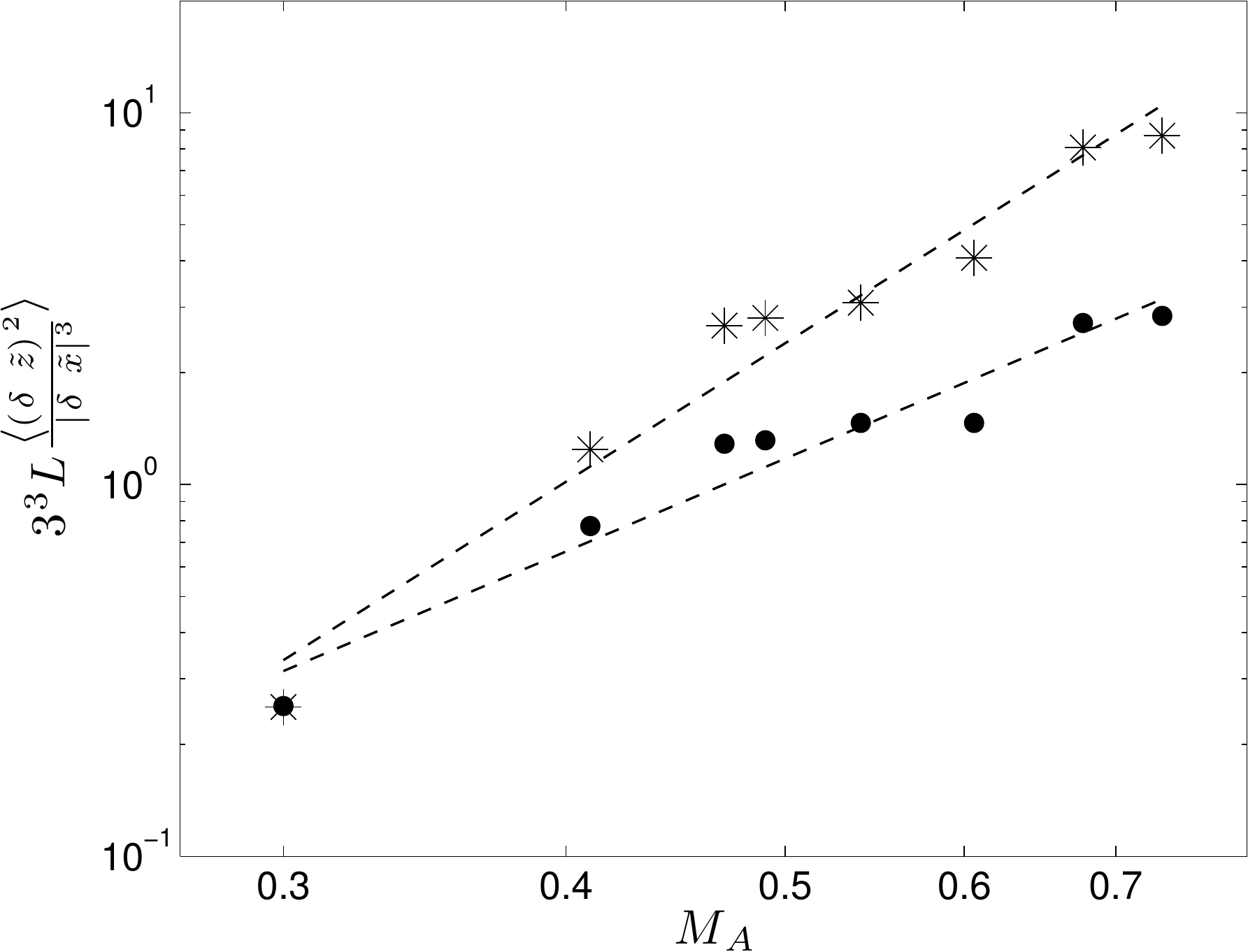}
\caption{The dependence of the magnetic field line separation 
on the Alfven Mach number $M_A$ as measured by particles streaming freely along the magnetic field lines. The fit for the asterisks corresponds to the $\sim M_A^{-3.8}$ dependence, very close to the predicted one.
From \citet{XY13}, \copyright~AAS. Reproduced with permission.}
\label{figmaxl13}
\end{figure}

The other important prediction in Eq. (\ref{eq:superdiff}) is the change of the square of the separation between the magnetic field lines increases with the forth power of Alfven Mach number, i.e. $\sim M_A^4$. The successful testing of this prediction is demonstrated in Fig. \ref{figmaxl13} that is taken from \citet{XY13}. The asterisks correspond to the calculations obtained using the parallel particle velocity. This way of calculating the separation provides a better description of magnetic field wandering. The fitted dependence is $M_A^{-3.8}$, which is very different from the $M_A^2$ dependence traditionally accepted in the cosmic ray literature. 

We note that the magnetic field field tracing with particles is not the only way to numerically explore the magnetic field separation. In fact, the analytical predictions of magnetic field line superdiffusion were also explored numerically by direct tracing of magnetic field lines first with low resolution in \cite{Lazarian_etal:2004} and with higher resolution in \citet{B13b}.

Taken together, these numerical testing provides a solid confirmation of the predictions given by Eq. (\ref{eq:superdiff}). It is significant that the turbulent magnetic field wandering induced by strong MHD turbulence provides the separation of magnetic field lines $\delta_{lines}^2 \sim s^3$, where the distance $s$ along the instantaneous realization of a magnetic field line.  Thus for sufficiently large $s$  all parts of the volume of magnetized fluid get connected. In other words, the entire volume becomes accessible to particles moving along magnetic field lines. This is important for a wide variety of astrophysical processes, e.g. the cosmic ray propagation and acceleration \citep{YanLazarian:2008, LazarianYan:2014}, heat propagation \citep{NarayanMedvedev:2001, Lazarian2006}.

 We would like to stress again that magnetic wandering cannot be understood using the notion of magnetic field lines preserving their identify in turbulent flows.  Magnetic field lines wandering in the turbulent magnetized flow constantly reconnect inducing the exchange of plasmas and magnetic field. In view of this, it is interesting to recall the considerations on the non-trivial nature of astrophysical magnetic field lines that can be found in a prophetic book by 
 \citet{Parker:1979}. 
 There it was mentioned that a magnetic field line may be traced through a significant part of the galaxy. We add that this is only possible through constant reforming of the magnetic field lines via reconnection.

We note that there is a formal analogy between the Richardson diffusion that predicts that the mean squared separation between particles $\delta_{particles}^2$ growth with time as $t^3$ and the mean squared separation between the magnetic field lines $\delta_{lines}^2$ that grows with the distance $s$ as $s^3$. Therefore, rather than use a new term to describe the latter dependence theoretically predicted in LV99, it was decided in ELV11 to term this new type of dispersion, as {\it the Richardson diffusion in space}, which keeping the term  ``Richardson dispersion'' for the original time-dependent process empirically discovered in \cite{richardson1926atmospheric}.

\subsection{Applicability of LV99 approach}
\label{sec:model3}

The derivation of LV99 reconnection rate assumes that all magnetic energy is converted to kinetic energy of
the outflow.  This may not be true when $ M_A$ is large. In this case, the
energy dissipation in the reconnection layer becomes non-negligible and the decrease of the outflow velocity is expected (see ELV11).  

The corresponding effect of turbulent dissipation can be estimated from steady-state energy
balance in the reconnection layer:
\begin{equation}
\frac{1}{2}v_{out}^3 \Delta = \frac{1}{2}V_A^2 v_{ren} L_x - \varepsilon L_x \Delta,  \label{Ebal}
\end{equation}
where kinetic energy of the outflow is balanced against magnetic
energy transported into the layer minus the energy dissipated by turbulence.
The turbulent dissipation is given by $\varepsilon =
u_L^4/V_A L_i$ for sub-Alfv\'enic turbulence \cite{Kraichnan:1965}. Dividing
(\ref{Ebal}) by $\Delta=L_x v_{rec}/v_{out}$, one obtains
\begin{equation}
     v_{out}^3 = V_A^2 v_{out} - 2 \frac{u_L^4}{V_A} \frac{L_x}{L_i}.
\end{equation}
 The corresponding solutions are easy to get
by introducing the ratios $f=v_{out}/V_A$ and $r= 2 {\cal M}_A^4 (L_x/L_i)$ which
are the measures of the outflow speed and the energy
dissipated by turbulence in units of the available magnetic energy. This gives
\begin{equation}
    r = f - f^3.
\label{cubic}
\end{equation}
When $r=0$, the only solution of (\ref{cubic}) with $f>0$ is $f=1,$ which is the 
the LV99 estimate $v_{out}=V_A$ for ${\cal M}_A\ll 1$.  For  larger
values of $r,$ $f\simeq 1-(r/2)$, in agreement with the formula $f=(1-r)^{1/2}$
that follows from Eq.~(65) in ELV11. This implies a decrease in $v_{out}$ as
compared to $V_A$.  Note that formula (\ref{cubic}) can be used when  $r$ is not too large. The
largest value of $r$ for which a positive, real $f$ exists is $r_{max} =
2/\sqrt{27}\approx 0.385$. For $r_{max}$ one gets $f_{min} =
1/\sqrt{3}\approx 0.577$. This implies that the LV99 approach is limited to sufficiently small
$ M_A$.  For $L_x \simeq L_i$, one may consider values of
$ M_A$ up to $0.662$.  Therefore, the original LV99 approach is accurate for $M_A\lesssim 1$. 

While the accuracy of LV99 expressions can be decreases due to the additional dissipation of the turbulent energy, the predicted phenomenon of fast reconnection arising from the turbulence persists for any $M_A$. In fact, one can see that the effect of the reduced outflow velocity 
{\it increases} the reconnection rate.  The reason is that field-lines now spend
a time $L_x/v_{out}$ exiting from the reconnection layer, greater than assumed
in LV99 by a factor of $1/f$.  This implies a thicker reconnection layer
$\Delta$ due to the longer time-interval of Richardson dispersion in the layer,
greater than the LV99 estimate by a factor of $(1/f)^{3/2}$.  The net reconnection speed
$v_{rec}=v_{out}\Delta/L_x$ is thus larger by a factor of $(1/f)^{1/2}$.  The
increased width $\Delta$ more than offsets the reduced outflow velocity
$v_{out}$.  At even larger $M_A$ magnetic
fields at large scale are weak that they are easily bent and twisted by the fluid turbulent motions. However, at the scales smaller than $l_A$ we expect the original LV99 treatment to be accurate again.

\subsection{LV99 model: summary of facts}
\label{LV99_summary}

The model of turbulent reconnection suggested in LV99 departed from the earlier attempts to explain astrophysical magnetic reconnection as something that required special conditions. On the contrary, it suggested that reconnection takes place everywhere in realistically turbulent fluids. Moreover, the transition to turbulent reconnection is inevitable as the 3D outflow from the reconnection region gets turbulent for sufficiently large Lundquist numbers. 

The model identified turbulent magnetic field wandering as the process that opens up the reconnection region outflow making reconnection fast, i.e. independent of fluid resistivity or microscopic plasma effects. The derived dependences for the field wandering were successfully tested with the numerical simulations. 

The model provided analytical expressions relating the reconnection rate with the level of turbulence. It also provided the relations between the thickness of the reconnection layer and the level of turbulence. This explains that rate of magnetic reconnection may vary significantly and also provides ways for quantiatative numerical and observational testing of the model.

\section{\label{sec:level1} Violation of Flux-Freezing, Renormalization and the Scale Problem}

In \S \ref{ssec: rdffvio} we discussed the issue of flux freezing on the basis of the simplified treatment of Richardson dispersion, treating the latter more like an empirical fact and using the corresponding scaling relations to re-derive the LV99 reconnection rates. However, the phenomenon of Richardson dispersion presents a much more fundamental property of magnetic fields in turbulent fluids. Below we discuss the violation of flux freezing on the basis of a more rigorous theoretical treatment. 

\subsection{Spontaneous stochasticity}
\label{sec:violation1}

 The conventional flux-freezing principle has not only problems with explaining observations. It also encounters serious theoretical difficulties. This can be exemplified by naively extending the concept to turbulent MHD flows with both small resistivity and small viscosity. To understand the subtlety of 
such singular limits, consider the problem of incompressible fluid turbulence. 
As is well-known, the kinetic energy dissipation rate, in a fluid with viscosity $\nu$ and velocity field $\bf u$, does not vanish in the limit of vanishing viscosity 
(e.g., see \cite{sreenivasan1984scaling, sreenivasan1998update}). 
On the contrary, $\lim_{\nu\rightarrow 0}\nu |\nabla {\bf u}|^2 >0$ which implies a ``dissipative anomaly''. This in turn requires that velocity gradients must blow up, with $\nabla{\bf u}\rightarrow \infty$ in the limit $\nu\rightarrow 0$. In fact, as Onsager pointed out in a prescient 
work (see \S\ref{renormalizing} below), the velocity field in a turbulent fluid must develop H{\"older}-type singularities in the limit as viscosity tends to zero 
\citep{onsager1949statistical, eyink2006onsager, eyink2018review}. 
These singularities render all hydrodynamic equations containing velocity gradients naively undefined. Extending Onsager's analysis to MHD plasmas, turbulent solutions are found to suffer several dissipative anomalies which require diverging gradients of all MHD fields, e.g., velocity field, magnetic field, density etc. (see e.g., 
\cite{mininni2009finite, caflisch1997remarks, eyink2018cascades}. 
For example, magnetic-field gradients blow up in the limit as magnetic diffusivity $\eta$ tends to zero, with $\nabla{\bf B}\rightarrow \infty$ as $\eta\rightarrow 0$. Because velocity fields in particular become non-smooth and singular 
as $\nu\to 0$, the magnetic field lines cannot freeze into the fluid in a standard deterministic sense. 
Instead a non-deterministic frozen-in property known as ``stochastic flux freezing'' (\S\ref{flux-freezing}) has been established 
by \cite{eyink2009stochastic, eyink2011stochastic}, 
which is intrinsically random in nature due to the roughness of the advecting velocity field 
\citep{eyink2010fluctuation} 
and is associated with the {\it spontaneous stochasticity} of Lagrangian particle trajectories 
\citep{bernard1998slow, Eyink_etal:2011, Eyink_etal:2013}. 

The latter concept is crucial and must be briefly explained. In essence, ``spontaneous stochasticity'' means that stochastic 
Lagrangian particle trajectories remain non-unique and indeterministic (random) for turbulent flows even in the limit of vanishing viscosity and molecular diffusivity 
\citep{bernard1998slow}. 
This phenomenon is directly related to Richardson super-ballistic behavior 
\citep{richardson1926atmospheric} (see \S \ref{ssec: rdffvio}). 
In a turbulent flow, therefore, infinitely-many magnetic field lines are advected to each point in space even in the ideal limit and must be averaged to obtain the resultant magnetic field \citep{eyink2011stochastic, Eyink_etal:2013}. 
Due to Richardson dispersion, in fact, even a tiny breakdown of magnetic flux freezing at small scales will be explosively amplified into larger scales in the turbulent inertial range 
\cite[][see also \S\ref{slippage} below]{Eyink2015}. The rates of such rapid amplification are non-vanishing in the ideal limit \citep{eyink2011stochastic, Eyink_etal:2011}.  
This breakdown of conventional magnetic flux freezing has already been verified in numerical simulations of resistive MHD turbulence at high conductivity 
\cite[e.g.][]{Eyink_etal:2013}.

To appreciate the importance of (asymptotic) singularities in MHD, we may note that without differentiability of fields even fundamental concepts 
such as the notion of field line need to be reconsidered. Taking magnetic field $\bB(\bx)$ as an example, its {\it field line}  through point 
$\bx$ is conventionally defined as the integral curve $\bxi(s;\bx)$ whose tangent vector at any point is parallel to the vector field at that point, 
so that, parameterized with arclength $s,$  
\begin{equation} d\bxi(s)/ds = \bB(\bxi(s))/|\bB(\bxi(s)|, \ \bxi(0)=\bx. \label{xi-eq} \end{equation} 
If the magnetic field satisfies the following inequality for some constant $C>0$;
$$ | \bB(\bx) - \bB(\bx') | \leq C| \bx - \bx|^h $$
with $h=1$ (Lipschitiz condition), then there exists a unique field-line $\bxi(s;\bx)$ passing through point $\bx$; however, if this condition
holds merely with $0<h<1$ (H\"older continuity), then there are generally infinitely many such integral curves (see \cite{JafariVishniac:2019}; \cite{dynamicsJV2019}; \cite{SecondJVV2019}). If the Lipschitz condition is not satisfied and hence the uniqueness theorem cannot be applied for a given magnetic field, what would a magnetic field line mean after all? 
When resistivity is tiny but non-zero, then $\bB$ is differentiable and indeed equation (\ref{xi-eq})  has unique solutions. However, 
these solutions exhibit such extreme sensitivity to initial conditions that the distance $|\bxi(s;\bx')-\bxi(s;\bx)| $ between field lines   
through neighboring points becomes {\it independent of}  $\bx'-\bx$ at distances $s\gg |\bx'-\bx|$ along the lines 
(LV99).\footnote{We provide more detail on this effect in \S \ref{ssec: rdffvio}.} 
Such explosive separation of magnetic field lines is not observed if the field lines exhibit standard ``deterministic chaos'', where memory 
of the initial separation is preserved.  Exactly analogous difficulties plague the concepts of ``Lagrangian 
fluid particle'' or ``fluid trajectory'', which are ill-conditioned and asymptotically non-unique if the fluid velocity 
$\bv(\bx,t)$ becomes spatially non-Lipschitz as $\nu\to 0.$  

The singularities associated with dissipative anomalies lead to subtle effects whose consistent description 
requires great care, otherwise contradictory conclusions are obtained. Similar difficulties were encountered already many 
decades ago in quantum field theory and condensed matter physics, which could be alleviated by mathematical 
methods collectively known as regularization and renormalization. As we review in \S\ref{renormalizing} below, such methods 
can be applied also to plasma turbulence by exploiting a spatial coarse-graining operation, which removes 
divergences and leads to meaningful equations. Physically speaking, ``coarse-graining'' simply means 
averaging quantities in a volume in space. Indeed what is measured in the laboratory as magnetic or velocity fields 
at a point in space and time is actually always an average over a small volume around that point during a time interval. 
Paradoxes are avoided by developing a theory based upon such quantities that can be measured in principle 
rather than idealized pointwise fields that are strictly unobservable. Such renormalization introduces an 
``observation scale'' or ``resolution scale'' which is entirely arbitrary. However, obviously, no objective physical fact 
can depend upon this arbitrary scale.  In the language of quantum field theory and statistical physics, this principle is called ``renormalization-group invariance'' \citep{gross1976applications}. 

\subsection{ Renormalization of MHD Equations and Singularities}\label{renormalizing}

We illustrate Onsager's renormalization-type arguments in one of the simplest models with 
realistic astrophysical applications,  the standard MHD  
equations for a compressible, magnetized, single-component fluid with mass density $\rho,$
velocity $\bv,$ magnetic field $\bB,$ and internal energy density $u$:  
$$ \partial_t\rho+\grad\bdot(\rho\bv)=0,  $$
\begin{eqnarray*}
&& \partial_t(\rho\bv)+\grad\bdot\bigg[\rho\bv\bv-\frac{1}{4\pi}\bB\bB+\left(p+\frac{1}{8\pi}B^2\right){\bf I}   \cr 
&& \hspace{150pt} -2\mu\bS-\zeta \Theta{\bf I}\bigg]=\bzed, 
\end{eqnarray*}
$$ \partial_t\bB =\grad\btimes(\bv\btimes\bB-c\bJ/\sigma), \quad \grad\bdot\bB=0,  $$
\begin{eqnarray*} 
&& \partial_t(\frac{1}{2}\rho v^2+\frac{1}{8\pi}B^2+u)\cr
&& \hspace{30pt}+\grad\bdot\bigg[(u+p+\frac{1}{2}\rho v^2)\bv+\frac{1}{4\pi}\bB\btimes(\bv\btimes\bB)  \cr
&& \hspace{60pt}-\kappa \grad T -2\mu \bS\bdot\bv-\zeta \Theta\bv - \frac{c}{4\pi\sigma}\bB\btimes\bJ \bigg]=0. 
\end{eqnarray*} 
Here $\bf I$ is the identity tensor and $p=p(\rho,u)$ is the thermodynamic equation of state for the pressure and 
\begin{eqnarray*}
&& S_{ij}=\frac{1}{2}\left(\frac{\partial v_i}{\partial x_j}+\frac{\partial v_j}{\partial x_i}-\frac{2}{3}(\grad\bdot\bv)\delta_{ij}\right), 
\quad \Theta=\grad\bdot\bv, \cr
&& \hspace{60pt} \quad \bJ=c(\grad\btimes\bB)/4\pi,  
\end{eqnarray*} 
with shear viscosity $\mu,$ bulk viscosity $\zeta,$ thermal conductivity $\kappa,$ and electrical conductivity $\sigma$ (or magnetic diffusivity $\eta=c^2/4\pi\sigma$). See 
\cite{landau2013electrodynamics}, 
Chapter VIII, \S 65-66.

The local balance of mechanical energy for the above model is given by the equation: 
\begin{eqnarray*} 
&& \partial_t(\frac{1}{2}\rho v^2+\frac{1}{8\pi}B^2)+\grad\bdot\bigg[(p+\frac{1}{2}\rho v^2)\bv+\frac{1}{4\pi}\bB\btimes(\bv\btimes\bB) \cr
&& \hspace{120pt}  -2\mu \bS\bdot\bv-\zeta \Theta\bv- \frac{c}{4\pi\sigma}\bB\btimes\bJ \bigg]\cr
&&\qquad\qquad =p(\grad\bdot \bv)-2\mu|\bS|^2-\zeta \Theta^2 - \eta (\grad\btimes\bB)^2/4\pi. 
\end{eqnarray*}
Just as for the case of incompressible, neutral fluids studied by Onsager, turbulent solutions 
are characterized by {\it dissipative anomalies} \cite[see, e.g.][]{mininni2009finite} in which 
$$ 2\mu|\bS|^2,\ \zeta \Theta^2,\ \eta (\grad\btimes\bB)^2
\nrightarrow 0, \quad {\rm as} \ \mu,\zeta,\kappa,\eta\to 0.
$$
Similar anomalous dissipation must then appear also in the 
balances of internal energy and entropy. Such anomalies obviously require diverging space-gradients 
$|\grad\bv|,\, |\grad \bB|\to\infty$
for $\mu,\zeta,\kappa,\eta\to 0$, or a ``violet catastrophe'' in the words of 
\cite{onsager1945distribution}.  

To obtain a well-defined dynamical description in this singular ideal limit, one must regularize these UV divergences. 
For example, one may employ a ``coarse-graining'' or ``block-spin'' regularization, defined as 
$$ \overline{\bB}_\ell(\bx,t) = \int d^3r\, G_\ell(\br)\, \bB(\bx+\br,t) $$ 
and likewise for $\bv,$ $\rho,$ $u,$ etc. with $G_\ell(\br)=\ell^{-3} G(\br/\ell)$ for some positive, smooth,
rapidly decaying kernel $G(\brho)$ normalized as $\int d^3\rho\, G(\brho)=1.$ Coarse-grained gradients 
such as $\grad\overline{\bB}_\ell(\bx,t)$ necessarily remain finite as $\mu,\zeta,\kappa,\eta\to 0,$ their 
UV divergences removed. It should be emphasized again that such a coarse-graining is purely passive and corresponds 
to observing the flow ``with spectacles off'' so that eddies of size $<\ell$ cannot be resolved.
The basic principle of renormalization-group invariance is that this regularization scale $\ell$ 
is completely arbitrary and no objective physics can depend upon it. Because the coarse-graining 
operation commutes with space-time derivatives, the effective equations of motion for eddies of 
size $>\ell$ are exactly the same as the original MHD equations with just an additional overbar
$\overline{()}_\ell$. 

However, even when there are dissipative anomalies in the microscopic or ``bare'' balance equations
for conserved quantities, the dissipative transport terms 
such as $\grad\btimes\overline{(\eta\grad\btimes\bB)}_\ell$ in the coarse-grained MHD equations  
can be shown to vanish for $\mu,\zeta,\kappa,\eta\ll 1$ at fixed $\ell$ 
or for $\ell\gg 1 $ at fixed $\mu,\zeta,\kappa,\eta$. These dissipative terms 
are thus {\it irrelevant}  in the renormalization-group sense. It follows, when $\mu,\zeta,\kappa,\eta\ll 1,$
that there exists an ``inertial-range'' of length-scales $\ell$  
at which the ideal MHD equations hold in the ``coarse-grained sense'', i.e. that: 
\begin{equation}  \partial_t\overline{\rho}_\ell+\grad\bdot \overline{(\rho\bv)}_\ell=0,  \label{eq1} \end{equation} 
\begin{equation} \partial_t \overline{(\rho\bv)}_\ell 
+\grad\bdot\overline{\left[\rho\bv\bv-\frac{1}{4\pi}\bB\bB+\left(p+\frac{1}{8\pi}B^2\right){\bf I}\right]}_\ell=\bzed,  
 \label{eq2} \end{equation} 
\begin{equation}  \partial_t\overline{\bB}_\ell  =\grad\btimes \overline{(\bv\btimes\bB)}_\ell, \quad \grad\bdot\overline{\bB}_\ell=0 
 \label{eq3} \end{equation} 
\begin{eqnarray} 
&& \partial_t \overline{\left(\frac{1}{2}\rho v^2+\frac{1}{8\pi}B^2+u\right)}_\ell \cr 
&& +\grad\bdot\overline{\left[(u+p+\frac{1}{2}\rho v^2)\bv+\frac{1}{4\pi}\bB\btimes(\bv\btimes\bB)\,\right]}_\ell =0. 
 \label{eq4} \end{eqnarray} 
Taking the limit $\mu,\zeta,\kappa,\eta\to 0$ these coarse-grained equations hold for all $\ell>0,$ 
which is equivalent to the standard mathematical notion of a ``weak''/``distributional'' solution, 
or what Onsager called a ``more general description'' than standard differential equations.  

The validity of the ideal MHD equations in this generalized sense in not sufficient, however, to guarantee 
that naive balances of mechanical energy, internal energy and entropy will hold. To show this clearly, 
the above ideal equation can be cosmetically rewritten by introducing the {\it density-weighted Favre-average} 
$$    \wt{f}_\ell = \overline{(\rho\, f)}_\ell/\overline{\rho}_\ell $$
and {\it pth-order cumulants} $\overline{\tau}_\ell(f_1,f_2,...,f_p)$ and $\wt{\tau}_\ell(f_1,f_2,...,f_p)$ for the two 
types of coarse-graining averages. With these definitions, the effective equations of motion 
for scales $>\ell$ become: 
\begin{equation} \partial_t\overline{\rho}_\ell+\grad\bdot (\overline{\rho}_\ell\wt{\bv}_\ell)=0, 
 \label{eq5}\end{equation} 
\begin{equation} \partial_t(\overline{\rho}_\ell\wt{\bv}_\ell)+\grad\bdot\left[\overline{\rho}_\ell\wt{\bv}_\ell\wt{\bv}_\ell+\overline{p}_\ell{\bf I} 
+{\bf T}_\ell\right]=\frac{1}{c}\overline{\bJ}_\ell\btimes\overline{\bB}_\ell, 
 \label{eq6}\end{equation} 
\begin{equation} \grad\bdot\overline{\bB}_\ell=0, \qquad \partial_t\overline{\bB}_\ell =\grad\btimes(\overline{\bv}_\ell\btimes\overline{\bB}_\ell+
\boeps_\ell),   \label{eq7}\end{equation} 
and 
\begin{eqnarray} 
&& \partial_t\left(\frac{1}{2}\overline{\rho}_\ell |\wt{\bv}_\ell|^2+\frac{1}{8\pi}|\overline{\bB}_\ell|^2+\overline{u}_\ell+ \delta \overline{u}_\ell\right) + \cr 
&&\grad\bdot\bigg[\left(\frac{1}{2}\overline{\rho}_\ell |\wt{\bv}_\ell|^2+\overline{u}_\ell+\delta\overline{u}_\ell+\overline{p}_\ell\right)\wt{\bv}_\ell
+\frac{1}{4\pi}\overline{\bB}_\ell\btimes(\overline{\bv}_\ell\btimes\overline{\bB}_\ell) \cr
&& \hspace{75pt} 
+{\bf T}_\ell\bdot \wt{\bv}_\ell -\frac{c}{4\pi}\boeps_\ell\btimes\overline{\bB}_\ell +{\bf J}_\ell^h\bigg]=0.
 \label{eq8} \end{eqnarray} 
Here we have introduced the {\it turbulent stress tensor} 
$$ {\bf T}_\ell=\overline{\rho}_\ell \wt{\tau}_\ell(\bv,\bv)-\frac{1}{4\pi}\overline{\tau}_\ell(\bB,\bB) 
+\frac{1}{8\pi}\overline{\tau}_\ell(B_i,B_i){\bf I}, $$
{\it turbulent electromotive force (emf)}
\begin{equation}\label{emf}
\boeps_\ell=\overline{\tau}_\ell(\bv\btimescom\bB):=\overline{(\bv\btimes\bB)}_\ell-\overline{\bv}_\ell\btimes\overline{\bB}_\ell, \end{equation}
{\it turbulent internal energy}  
$$\,\,\delta \overline{u}_\ell = \frac{1}{2}\overline{\rho}_\ell \wt{\tau}_\ell(v_i,v_i)+\frac{1}{8\pi}\overline{\tau}_\ell(B_i,B_i) $$
and {\it turbulent enthalpy transport}
\begin{eqnarray*}
 \bJ_\ell^h &=&   \overline{\tau}_\ell (h,\bv)-\overline{h}_\ell\overline{\tau}_\ell(\rho,\bv)/\overline{\rho}+\frac{1}{2}\bar{\rho}_\ell\tilde{\tau}_\ell(v_i,v_i,\bv) \cr
 && \hspace{-10pt} + \frac{1}{4\pi}\left(\overline{\tau}_\ell(\bB,\bB)-\overline{\tau}_\ell(B_i,B_i){\bf I}\right)\bdot\overline{\tau}_\ell(\rho,\bv)/\overline{\rho} \cr
 && \hspace{-10pt} + \frac{1}{4\pi}\left(\overline{\tau}_\ell(\bv,\bB)-\overline{\tau}_\ell(v_i,B_i){\bf I}\right)\bdot\overline{\bB}_\ell
 +\frac{1}{4\pi}\overline{\tau}_\ell(\bB\btimescom (\bv\btimescom\bB))  \cr 
\end{eqnarray*} 
with $h=u+p$ the enthalpy per volume. 

Although equations (\ref{eq5})-(\ref{eq8}) are exactly equivalent to the coarse-grained ideal MHD equations (\ref{eq1})-(\ref{eq4}), they appear ``non-ideal''. Integrating 
out the small-scale eddies has led to ``renormalized'' equations 
with new terms ${\bf T}_\ell,$ $\boeps_\ell,$ $\delta \overline{u}_\ell,$ $\bJ_\ell^h$ that would not 
appear in naive ideal MHD equations for the resolved fields $\overline{\rho}_\ell,$ $\wt{\bv}_\ell,$ 
$\overline{\bB}_\ell,$  $\overline{u}_\ell$ at length-scales $>\ell.$ 
The precise condition for validity 
of the ideal equations (\ref{eq1})-(\ref{eq4}) at length-scale $\ell$ is that the standard dissipative 
transport terms are negligible compared with these new terms. For example, the 
ideal magnetic induction equation (\ref{eq7}) is valid precisely when 
\begin{equation}  |\boeps_\ell | \gg  |\overline{(\eta\grad\btimes\bB)}_\ell |   \label{inert-cond} \end{equation} 
One of Onsager's key insights was that contributions from coarse-graining such as $\boeps_\ell$ 
can be expressed entirely in terms of {\it space-increments} \cite[see][]{eyink2006breakdown, eyink2018cascades}, such as 
$$ \delta\bB(\br;\bx,t):= \bB(\bx+\br, t)-\bB(\bx, t). $$

It follows that 
\begin{equation}
 |\boeps_\ell | \sim  \delta v(\ell) \delta B(\ell), \quad  |\overline{(\eta\grad\btimes\bB)}_\ell | \sim \eta \delta B(\ell)/\ell. 
 \label{eps-dB} 
\end{equation} 
In that case, the condition (\ref{inert-cond}) becomes
$$ \ell \delta v(\ell)/\eta \gg  1, $$ 
which precisely defines the inertial-range of scales $\ell$ for which the ideal induction equation is valid. This condition can be restated as $\ell\gg \ell_\eta$ where $\ell_\eta$ is the resistive dissipation 
length specified by $\ell_\eta \delta v(\ell_\eta)\simeq \eta.$  Similar 
conditions guarantee validity of all of the other ideal equations as well
\cite[see][]{eyink2018cascades}. 

From the coarse-grained ideal MHD equations rewritten in this manner, one obtains a 
balance equation for {\it resolved mechanical-energy}: 
\begin{eqnarray} 
&& \partial_t\left(\frac{1}{2}\overline{\rho}_\ell |\wt{\bv}_\ell|^2+\frac{1}{8\pi}|\overline{\bB}_\ell|^2\right)  \cr 
&& \hspace{50pt} +\grad\bdot\bigg[\overline{p}_\ell\,\overline{\bv}_\ell+\frac{1}{2}\overline{\rho}_\ell|\wt{\bv}_\ell|^2\wt{\bv}_\ell
+\frac{1}{4\pi}\overline{\bB}_\ell\btimes(\overline{\bv}_\ell\btimes\overline{\bB}_\ell) \cr 
&& \hspace{150pt} +{\bf T}_\ell\bdot\wt{\bv}_\ell -\frac{c}{4\pi}\boeps_\ell\btimes\overline{\bB}_\ell\bigg] \cr 
&& \hspace{100pt} = \overline{p}_\ell\grad\bdot\overline{\bv}_\ell - Q_\ell^{flux}. 
 \label{eq9}\end{eqnarray} 
where the renormalization has introduced new apparently non-ideal terms. In particular, the 
right side of this balance contains the {\it turbulent energy flux}
$$ Q_\ell^{flux} = -{\bf T}_\ell\bdots\grad\wt{\bv}_\ell -\boeps_\ell\bdot\overline{\bJ}_\ell
-\overline{\tau}_\ell(\rho,\bv)\bdot \frac{1}{\overline{\rho}_\ell}\left(\frac{1}{c}\overline{\bJ}_\ell\btimes\overline{\bB}_\ell-\grad\overline{p}_\ell\right)$$
which represents nonlinear cascade of mechanical energy to unresolved scales $<\ell$. 
The energy lost from resolved mechanical modes reappears as total unresolved energy 
$\overline{u}_\ell^*=\overline{u}_\ell+\delta \overline{u}_\ell,$ where $\overline{u}_\ell$ is the thermal kinetic 
energy of microscopic particles and $\delta \overline{u}_\ell$ represents mechanical energy in 
unresolved turbulent eddies at scales $<\ell.$  The balance equation for this total unresolved 
energy or {\it intrinsic internal energy} can be obtained by combining Eq.(\ref{eq9}) 
with coarse-grained energy conservation to give 
$$ \partial_t\overline{u}_\ell^* + \grad\bdot (\overline{u}_\ell^*\wt{\bv} + {\bf J}_\ell^u)= -\overline{p}\grad\bdot\overline{\bv} + Q_\ell^{flux}$$
where ${\bf J}_\ell^u=\bJ_\ell^h+\overline{p}\,\overline{\tau}(\rho,\bv)/\overline{\rho}$ and where $Q_\ell^{flux}$ is the same 
term that appears in Eq.(\ref{eq9}) but with the opposite sign. 

If mechanical energy is dissipated in the limit $\mu,\zeta,\kappa,\eta\to 0,$ then such dissipation 
will be apparent even to a ``myopic'' observer who sees only eddies $>\ell$ and the dissipation rate
must become independent of $\ell$ for $\ell\to 0.$ From estimates of the form given by Eq.(\ref{eps-dB}) it follows 
that $Q_\ell^{flux}$ can be expressed entirely in terms of {\it space-increments} as discussed above, therefore  $Q_\ell^{flux}$ can be non-vanishing as $\ell\to 0$
only if the solution fields develop a critical degree of singularity. For example, suppose that for all 
space-time points $(\bx,t)$
\begin{equation}\label{Holder}
| \delta\bB(\br)|\leq  (const.) B_{rms} \;\Big({|{\bf{r}}|\over L}  \Big)^{h_B},
\end{equation}
and similarly for $\delta\bv(\br)$ with some exponent $h_v,$ etc. Since 
\begin{equation}
\boeps_\ell\bdot\overline{\bJ}=O\left(\frac{\delta v(\ell)(\delta B(\ell))^2}{\ell}\right) = 
O(\ell^{h_v+2h_B-1}), 
\end{equation}
this term, which represents cascade of magnetic energy, 
can remain non-vanishing as $\ell\to 0$ only if 
$$ h_v+2 h_B\leq 1. $$
Similarly, the first term in $Q_\ell^{flux}$ which represents cascade of kinetic energy can 
remaining non-vanishing only if $3h_v\leq 1$ and the third term, which generalizes 
to MHD the ``baropycnal work'' of 
\cite{aluie2013scale},  
requires at least one of the inequalities  
$h_\rho+h_v+h_B\leq 1$ or  $h_\rho+h_v+h_p\leq 1$ in order to be non-vanishing. 
These conditions taken altogether imply that $h_v<1$  as  $\mu,\zeta,\kappa,\eta\to 0$
and thus the velocity field {\it cannot} remain Lipschitz in the limit. 

It should be clear from the previous discussion that the balances of other quantities 
besides mechanical energy can suffer from dissipative anomalies. In particular, 
the induction equation (\ref{eq7}) contains the turbulent electromotive force $\boeps_\ell$
which not only drives cascade of magnetic energy but also vitiates freezing 
of the resolved magnetic field $\overline{\bB}_\ell$ to the resolved velocity $\overline{\bv}_\ell.$ 
If instead the magnetic field is referred to the Favre-averaged velocity $\wt{\bv}_\ell,$
then violation of flux-freezing is due to the combination $\boeps_\ell-(1/\overline{\rho}_\ell)\overline{\tau}_\ell(\rho,\bv)\btimes\overline{\bB}_\ell.$ 
Because generally $\boeps_\ell\bdot\overline{\bB}_\ell\neq 0,$ it is not possible to express the 
turbulent emf as $\boeps_\ell=\Delta \bv_\ell\btimes\overline{\bB}_\ell $ for any choice of vector field $\Delta \bv_\ell,$
which would allow one to regard lines of $\overline{\bB}_\ell$ as moving with a velocity $\overline{\bv}_\ell^*=\overline{\bv}_\ell
+\Delta \bv_\ell.$ Thus, {\it it is not true, as is often stated, that ``because the ideal Ohm's law holds 
in the turbulent inertial-range, therefore magnetic field-lines are frozen-in at those scales.''}  The error 
in this commonplace statement is that no precise operational meaning is ever given to the
claim of ``flux-freezing in the inertial-range of scales'', so that an experimentalist could 
in principle test it. As known from exalted areas of physics such as relativity and quantum mechanics,
the failure to formulate such claims in operationally meaningful terms can lead to paradoxes and 
inconsistencies. If one adopts the natural interpretation that the magnetic field $\overline{\bB}_\ell$ 
of the eddies $>\ell$ be frozen-in to the velocity fields $\overline{\bv}_\ell$ or $\wt{\bv}_\ell$ at those 
same scales, then the statements are certainly false.

We discuss this later in \S \ref{sec:alternatives2} in relation to the concept of ``reconnection-mediated turbulence'' which implies that magnetic reconnection takes place only at some small scales where tearing gets important. It is obvious from the discussion above that magnetic reconnection takes place over the entire inertial range rather than some chosen small scales where dissipative processes take place. The reconnection at all scales of the turbulent cascade is necessary to explain the Richardson dispersion (see \S \ref{ssec: rdffvio}) and the numerical simulations demonstrating the gross violation of the classical flux freezing (see \S \ref{ssec: doffvio}. In fact, this idea is implicitly at the very core of the LV99 theory. 

A clear-minded skeptic might argue that such violations of flux-freezing in the turbulent 
inertial-range are mere artefacts of the coarse-graining and not ``real''. In particular, coarse-graining 
could be applied even to a laminar ideal MHD flow and would lead to apparent violations of flux-freezing 
at length-scales $\ell\gg L_\nabla,$ the ``gradient-length'' over which the MHD solution fields sensibly vary. 
If such violations are real, then, invoking renormalization-group invariance, they must be observed
for {\it all} resolution scales. An experimentalist who measured the velocity and magnetic fields in a 
laminar ideal MHD flow would find that, as the length-scales resolved by measurements were refined more, 
standard flux-freezing would hold better. However, in a turbulent MHD flow, deviations from 
flux-freezing predictions would persist as the experimentalist improved measurement resolution
down to smaller length-scales $\ell$ within the inertial-range, even though ideal MHD holds there in 
the ``coarse-grained sense''. Eventually, the experimentalist would resolve down to the resistive 
scale $\ell_\sigma$ and would find the same violations of flux-freezing predictions, but now due 
to Ohmic electric fields. We note, that the assumption that the turbulence is not damped up to the Ohmic diffusion scale is not a necessary one. In \S \ref{ssVSR} we show that fast reconnection is possible in the situation when viscosity is larger than resistivity.

Considering the infinite-Reynolds limit  $\mu,\zeta,\kappa,\eta\to 0$, the 
ideal MHD equations would be valid at {\it all} scales $\ell$ and the violations of flux-freezing would be 
due entirely to the turbulent emf $\boeps_\ell.$  
\cite{eyink2006breakdown} have shown 
that magnetic flux conservation can be violated at an instant of time for an arbitrarily small length scale 
$\ell$ in the absence of any microscopic plasma non-ideality only if at least one of the following necessary 
conditions is satisfied: (i) non-rectifiability of advected loops; (ii) unbounded velocity or magnetic fields; or 
(iii) existence of singular current sheets and vortex sheets that intersect in sets of large enough dimension. 
Both conditions (i) and (iii) can be expected to hold in MHD turbulence. 

As we now review, this conclusion that flux-freezing can be violated in turbulent solutions 
of ideal MHD is corroborated by two entirely independent arguments. The first of these
appeals to the phenomenon of spontaneous stochasticity of Lagrangian fluid particle trajectories 
for non-Lipschitz velocity fields. The second argument invokes the spontaneous stochasticity 
of the magnetic field-lines themselves and the concept of ``slippage'' of field-lines relative to 
the plasma fluid.    

\subsection{Conventional and Stochastic Flux-Freezing}\label{flux-freezing}

The flux-freezing properties of smooth, laminar solutions of the ideal MHD equations are well-known. 
One can write the induction equation for magnetic field $\bf B$ as 
$$D_t\bB= (\bB\bdot\grad)\bv-\bB(\grad\bdot\bv)+\eta \triangle \bB$$ 
where $D_t\equiv \partial_t+{\bf u.\nabla}$. Assuming that the limiting solution 
for $\eta\to 0$ remains smooth, this equation may be combined with mass balance 
$D_t\rho+\rho(\grad\bdot\bv) =0$ to yield 
$$
D_t\Big({ {\bf B}\over \rho}\Big)=\Big( {{\bf B}\over \rho}\Big)\bdot\grad\bv, $$
which succinctly expresses flux-freezing. The latter equation is solved explicitly 
by the Lundquist formula,
$$ \bB(\bx,t) = \left. \frac{\bB_0(\ba)\bdot \grad_a\bX(\ba,t)}
{{\rm det}\,(\grad_a\bX(\ba,t))}\right|_{\bX(\ba,t)=\bx},$$
where $\bX(\ba,t)$ is the position at time $t$ of the Lagrangian fluid particle 
that was at point $\ba$ at time $t_0,$ so that 
$$ \frac{d}{dt}\bX(\ba,t) = \bv(\bX(\ba,t),t), \quad \bX(\ba,t_0)=\ba. $$
The mass-density ratio is represented by the determinant 
$\left.{\rm det}\,(\grad_a\bX(\ba,t))\right|_{\bX(\ba,t)=\bx} = \rho(\ba,t_0)/\rho(\bx,t).$ 

It is less widely known that analogous {\it stochastic flux-freezing} properties hold 
even if magnetic diffusivity is non-vanishing, or $\eta>0.$ In that case, 
one must consider stochastic Lagrangian trajectories that satisfy 
$$ d\tilde{\bX}(\ba,t) = \bv(\tilde{\bX}(\ba,t),t) dt + \sqrt{2\eta} \, d\tilde{{\bf W}}(t)$$
where $\tilde{{\bf W}}(t)$ is a 3D Brownian motion that represents the effects of magnetic diffusivity. 
The exact solution of the magnetic induction equation is then represented by the {\it stochastic 
Lundquist formula} 
\begin{equation}
 \bB(\bx,t) = \overline{\left. \frac{\bB_0(\ba)\bdot \grad_a\tilde{\bX}(\ba,t)}
{{\rm det}\,(\grad_a\tilde{\bX}(\ba,t))}\right|_{\tilde{\bX}(\ba,t)=\bx}}, 
\label{stochLund} \end{equation}
where the overbar $\overline{(\cdot)}$ represents an average over the Brownian motion $\tilde{{\bf W}}(t)$
or, equivalently, a path-integral over the stochastic Lagrangian trajectories $\tilde{\bX}(t)$ themselves
\cite[see][]{eyink2009stochastic, eyink2011stochastic}. 
If all solution fields remain smooth in the limit 
$\eta\to 0,$ then the ensemble of stochastic Lagrangian trajectories $\tilde{\bX}(\ba,t)$ collapses 
to the unique deterministic Lagrangian trajectory $\bX(\ba,t)$ and conventional flux-freezing is 
recovered in the limit.   

This is not the case for turbulent solutions of ideal MHD obtained in the joint limit $\mu,\zeta,\kappa,\eta\to 0.$
As we have observed in \S\ref{renormalizing}, the velocity field $\bv(\bx,t)$ cannot remain uniformly Lipschitz  
if there is anomalous energy dissipation in that limit. In that case, there is no longer a unique Lagrangian trajectory 
for each fluid element in the infinite-Reynolds limit. Therefore, one may wonder with such an infinite number of 
limiting particle trajectories, which trajectory will be selected in the joint limit $\mu,\zeta,\kappa,\eta\to 0$? 
The surprising answer is that Lagrangian trajectories, under such conditions, can remain random---the phenomenon 
of  {\it spontaneous stochasticity} 
\citep{bernard1998slow,chaves2003lagrangian}. 
This effect resembles 
``spontaneous symmetry-breaking'' in condensed matter physics and quantum field theory, where below a critical temperature 
(such as the Curie temperature for a magnet) the mean order parameter remains non-vanishing even as the external
symmetry-breaking field is taken to zero.   Spontaneous stochasticity differs greatly from the usual randomness in 
turbulence theory associated to an ensemble of velocity fields, because the Lagrangian fluid 
trajectories remain non-unique and stochastic for a {\it fixed} velocity field $\bv(\bx,t)$ and {\it fixed} particle position $\ba$ 
at the initial time $t_0.$ The Laplacian determinism expected for classical dynamics breaks down because of 
the explosive separation of particles undergoing turbulent Richardson dispersion.  This remarkable prediction 
is supported by direct empirical evidence for turbulent MHD flows 
\cite[for details see][]{eyink2011stochastic,Eyink_etal:2013} (see also \S \ref{ssec: doffvio}). 

In consequence of this surprising remnant or persistent randomness of Lagrangian trajectories, the 
stochastic flux-freezing relation given by Eq. (\ref{stochLund}) for non-ideal MHD solutions does {\it not} reduce 
to conventional flux-freezing in the limit $\mu,\zeta,\kappa,\eta\to 0.$ Instead, flux-freezing in such regimes 
remains valid in only a statistical sense associated to spontaneous stochasticity of the fluid trajectories 
\cite[][see also \cite{jafari2018introduction}]{eyink2011stochastic}. 
Magnetic flux conservation in MHD turbulence
at large kinetic and magnetic Reynolds numbers, thus holds neither 
in the conventional sense nor is entirely broken. In 
ELV11
this stochastic flux-freezing relation was employed, together with GS95 predictions
for turbulent Richardson dispersion, to rederive the predictions of LV99 for magnetic reconnection rates 
in the presence of MHD plasma turbulence.  The detailed predictions depend upon the phenomenological 
theory of MHD turbulence which is adopted, but the qualitative features are independent of such assumptions 
and, in particular, the prediction that reconnection rates are independent of the precise values of microscopic resistivity or, similarly, of any  effective ``anomalous resistivity'' arising from any other microscopic plasma processes.  

\subsection{ Field-Line Stochasticity and Slippage}\label{slippage}

The original arguments of LV99 rested instead upon the random wandering of magnetic field-lines $\bxi(s;\bx)$ 
with increasing arclength $s$ as they move away from their initial position $\bx.$ Anticipating the concept of 
spontaneous stochasticity, LV99 realized that field-lines would disperse explosively, so that separations 
between neighboring field-lines $\bxi(s;\bx),$ $\bxi(s;\bx')$ would become independent of the initial separation
$|\bx'-\bx|$ with increasing $s.$  This phenomenon allows field-lines from widely separated space regions to wander into the same thin current sheet with thickness governed by $\eta,$ even in the limit as $\eta,$ $\nu\to 0.$
In this way, LV99 argued that violations of flux-freezing by tiny amounts of resistivity can be enhanced to macroscopic scales. 
\cite{boozer2018fast} 
has emphasized that similar wandering due to deterministic chaos of field-lines 
can lead to reconnection rates nearly independent of resistivity, but ubiquitous plasma turbulence can 
produce strictly fast reconnection with no dependence on resistivity whatsoever.  

We may note, however, a formal inconsistency in the argument of LV99 in the limit of vanishing resistivity. 
On the one hand, they used
GS95 phenomenology of MHD turbulence to estimate the mean-square dispersion of field-lines  
as proportional to $\varepsilon (s/v_A)^3,$ where $\varepsilon$ is the energy dissipation per mass
and $v_A$ is the Alfv\'en speed. On the other hand, GS95 theory predicts that the Ohmic electric field 
$\bR = (\eta/c)\grad\btimes\bB$ has a magnitude $R\sim (\eta/c) B_{rms}/(\eta^3/\varepsilon)^{1/4}
\propto \eta^{1/4} \to 0$ as $\eta\to 0.$ Thus, Ohm's law 
$ \bE + (\bv/c)\btimes\bB = \bR$
is predicted by GS95 phenomenology to become {\it pointwise ideal} in the limit $\eta,$ $\nu\to 0,$
with not only the width of current sheets but also the magnitude of the non-ideal Ohmic electric field
vanishing in that limit. How can magnetic reconnection persist in the zero-resistivity limit when non-ideal 
electric fields vanish uniformly? One obvious answer is that GS95 theory ignores small-scale intermittency 
which can produce current sheets with thickness $\propto \eta$  and thus Ohmic fields that persist 
as $\eta \to 0$. However, a deeper solution to this puzzle lies in the recognition that magnetic reconnection 
is possible even at points where non-ideal electric fields tend to zero!    

A complete analysis of this problem was presented by 
\cite{Eyink2015} 
who studied the 
rate at which magnetic-field lines ``slip'' through a non-ideal plasma, or, more accurately, 
the rate at which magnetic connections between plasma elements change due to non-ideality. 
As argued already by 
\cite{axford1984magnetic}, 
such change of magnetic connections is the basal 
effect that underlies all other reconnection phenomena (diffusion of plasma through separatrices, 
topology change, conversion of magnetic energy to kinetic energy, etc.) This effect was quantified 
in 
\cite{Eyink2015} 
by the {\it slip velocity} $\Delta {\bf{w}}_\perp(s;\bx)$, which is the relative velocity 
between the magnetic field line $\bxi(s;\bx)$ and the fluid velocity $\bv$ at a distance $s$ in arclength 
away from the plasma element $\bx$ which that particular field-line threads. A simple equation was 
obtained in 
\cite{Eyink2015} 
for the growth of slip-velocity $\Delta {\bf{w}}_\perp(s;\bx)$ in arclength $s$ 
along the field-line: 
\begin{eqnarray}\notag
{d\over ds}\Delta {\bf{w}}_\perp&-&\Big[ ( \grad_{\boldsymbol{\xi}}\hat{\bf{B}})^\top-(\hat{\bf{B}}\hat{\bf{B}})\bdot 
(\grad_{\boldsymbol{\xi}}\hat{\bf{B}})          \Big]\bdot \Delta{\bf{w}}_\perp\\\label{Eyink10}
&=&-{c(\grad\times{\bR})_\perp\over |{\bf{B}}|}.
\end{eqnarray}
where $\hat{\bB}(\bx,t)=\bB(\bx,t)/|\bB(\bx,t)|$ denotes the magnetic field direction and subscript ``$\perp$''
the vector component perpendicular to $\bB.$ Because this linear ODE can have solutions 
$\Delta {\bf{w}}_\perp(s;\bx)\neq {\bf 0}$ only if the righthand side is non-zero, the above expression indicates that 
flux-freezing holds if and only if $(\grad\btimes \bR)_\perp={\bf 0}$. In fact, the relation $\hat{\bf{B}}\times (\grad\btimes {\bf{R}})={\bf 0}$
has long been known as the general condition for flux freezing 
\citep{newcomb1958motion}. 
The quantity 
$\bSigma=-c(\grad\times{\bR})_\perp/|{\bf B}|$ represents the rate of development of slip-velocity per unit 
length of field-line and was dubbed in 
\cite{Eyink2015} 
the {\it slip-velocity source}. This expression has 
a topological interpretation as well, arising from the relation $\partial_t\hat{\bB}=(\partial_t\bB)_\perp/|\bB|$ \citep{JafariVishniac:2019, dynamicsJV2019}. 

The key point in the resolution of the ``vanishing Ohmic electric-field puzzle'' is that,  even if non-ideal electric fields $\bR\to {\bf 0}$, 
the corresponding slip-velocity source $\bSigma$ can remain non-vanishing or even diverge to infinity! 
For example, within GS95 phenomenology of MHD turbulence, $\Sigma\sim \eta/(\eta^3/\varepsilon)^{1/2}
\propto \eta^{-1/2}\to\infty$ as $\eta\to 0.$  Thus, the original argument of LV99 is self-consistent 
if GS95 phenomenology is employed throughout. More importantly, the above arguments show that 
the most intense current sheets due to small-scale intermittency are {\it not} required to produce flux-freezing 
breakdown and that field-lines even far from those strong sheets will slip through the background turbulence
at very sizable rates. The slip velocities will certainly be greatest at the most intense current sheets but, because 
of their rarity, more slippage may occur over time due to the weaker turbulence background. 
\cite{Eyink2015} 
argued that such turbulent line-slippage accounts for observed deviations from the Parker spiral 
model in the solar wind. This line-slippage is the same phenomenon which was termed ``reconnection diffusion'' introduced 
in \cite{Lazarian:2005} and numerically tested in \cite{Santos-Lima_etal:2010}. This process was invoked to explain the removal of magnetic flux in star-formation (see more in \S \ref{ssec: redsf}).

One may roughly translate the quantitative arguments of \cite{Eyink2015}, discussed above, into a more qualitative statement by saying that if the magnetic field is not perfectly frozen into the turbulent fluid, i.e., in the presence of stochastic flux freezing, the field should slip through the fluid. The differential equation governing this field-fluid slippage is given by Eq.(\ref{Eyink10}) with the source term $(\nabla\times{\bf B})_\perp/B$. Interestingly, it turns out that this term also acts as a source in the evolution equation of the unit vector $\hat {\bf B}$, which in turn implies that $(\nabla\times{\bf B})_\perp/B$ has a topological interpretation. In fact, it is simple calculus to see that the derivative of the unit vector $\hat{\bf{B}}={\bf{B}}/|{\bf{B}}|$ is given by 

  \begin{equation}\notag
  \partial_t \hat {\bf{B}}=\Big({\partial_t{\bf{B}}\over B} \Big)_\perp,
 \end{equation}
where $(.)_\perp$ denotes the perpendicular component with respect to ${\bf B}$. The evolution equation of $\hat{\bf B}$, therefore, is easily obtained using the induction equation:
  \begin{equation}\label{zap10}
  \partial_t \hat {\bf B}-{\nabla\times({\bf u}\times{\bf B})_\perp \over B}=-{(\nabla\times {\bf B})_\perp\over B}.
   \end{equation}

\cite{JafariVishniac:2019} used the relationship between the source term $(\nabla\times{\bf B})_\perp/B$ and the field's topology, implied by Eq.(\ref{zap10}) above, in order to define a measure for the spatial complexity, or self-entanglement, of the magnetic field $\bf B$ as

\begin{equation}\label{stochasticity-rate}
S (t) ={1\over 2} ( 1- \hat{\bf B}_l.\hat{\bf B}_L)_{rms},
\end{equation}
 where $\hat{\bf B}_l={\overline{\bf B}}_l/B_l$ and $\hat{\bf B}_L={\overline{\bf B}}_L/B_L$ with $\overline{\bf B}_l$ and $\overline{\bf B}_L$ as the coarse-grained, or renormalized, components at arbitrary scales $l$ and $L$. If we take $L$ of order of the system size and $l\ll L$ in the inertial range, $\overline{\bf B}_l({\bf x}, t)$ will be the average field of a fluid parcel of size $\sim l$ while $\overline{\bf B}_L({\bf x}, t)$ can be thought of as the average field of a fluid parcel of scale $L\gg l$. The former plays the role of a local field whereas the latter acts as a global field, therefore, $1- \hat{\bf B}_l.\hat{\bf B}_L$ is in fact a local measure of the deviation of $\overline{\bf B}_l$ from $\overline{\bf B}_L$. The RMS value of this quantity measures the global spatial complexity of the field (for a mathematical treatment of magnetic topology and complexity see \cite{dynamicsJV2019}). Magnetic reconnection, field-fluid slippage and magnetic topology change can be studied statistically using the function $S(t)$ and its time derivative $\partial_t S(t)$ \citep{JafariVishniac:2019, SecondJVV2019}. It also turns out that $S(t)$ is related to Richardson diffusion of magnetic field in turbulence; see \cite{Jafari_etal:2019}.

\subsection{\label{sec:level4}Irrelevance of Small Scale Physics in Large Scale Turbulent Reconnection}

Although the previous arguments were presented in the context of resistive MHD, they are not 
restricted to plasmas with high collision rates but apply with equal force to inertial-range turbulence in collisionless plasmas,
such as in the solar wind. The main difference is that in the {\it generalized Ohm's law} 
of a collisionless plasma, 
$$ \bE + \frac{1}{c}\bv\btimes\bB = \bR, $$
the non-ideal electric field $\bR$ contains many additional terms from the Hall effect, electron pressure anisotropy, 
electron inertia, etc. While these microscopic non-idealities play a crucial role in reconnection
of magnetic reversals near electron- and ion-scales, they are totally negligible and irrelevant to reconnection
of structures at scales $\ell\gg \rho_i,$ the ion gyroradius. This has been demonstrated in detail, 
starting with either the generalized Ohm's law 
\citep{Eyink2015} 
or with the full Vlasov-Maxwell-Landau kinetic equations 
\citep{eyink2018cascades}
In both cases, it was shown by a Favre coarse-graining analysis that 
the generalized Ohm's law at scales $>\ell$ takes the form 
$$ \wt{\bE}_\ell + \frac{1}{c}\wt{(\bv\btimes\bB)}_\ell = \wt{\bR}_\ell, $$
and that all of the coarse-grained non-ideal terms on the righthand side are damped out by powers 
$(\delta_i/\ell)$ and $(\delta_i/\ell)^2$ for $\ell\gg \delta_i,$ the ion skin depth. The non-ideal electric 
field $\wt{\bR}_\ell$ is thus irrelevant in the renormalization-group sense and an 
{\it ideal Ohm's law} holds in the coarse-grained sense 
$$ \wt{\bE}_\ell + \frac{1}{c}\wt{(\bv\btimes\bB)}_\ell = {\bf 0} $$
at inertial-range length-scales $\ell\gg \rho_i.$ 

Exactly as in our discussion of the MHD 
model, the validity of an ideal Ohm's law in this generalized sense does not imply that 
magnetic flux is frozen-in at inertial-range scales. Indeed, reconnection of magnetic structures 
many orders of magnitude larger than ion scales is observed in the turbulent solar wind
\citep{gosling2012magnetic,Eyink2015}
and such reconnection will be observed, obviously, 
whether ion-scale eddies are resolved or not! The principle of renormalization-group invariance
implies once again that the rate of reconnection of such large-scale structures must be 
independent of the resolution scale $\ell.$ Within the coarse-grained description, the 
ideal Ohm's law at scales $\ell\gg \rho_i$ can be rewritten as 
$$ \overline{\bE}_\ell + \frac{1}{c}\wt{\bv}_\ell\btimes\overline{\bB}_\ell = -\boeps^v_\ell-\boeps^n_\ell $$
where two apparently ``non-ideal'' terms now appear, one due to velocity fluctuations
$$ \boeps^v_\ell = \frac{1}{c}\wt{\tau}_\ell(\bv\btimescom\bB):
=\frac{1}{c}[\wt{(\bv\btimes\bB)}_\ell - \wt{\bv}_\ell\btimes\wt{\bB}_\ell] $$
and the other to density fluctuations 
$$ \boeps^n_\ell = \frac{1}{\overline{n}_\ell}[\overline{\tau}_\ell(n,\bE)+(\wt{\bv}_\ell /c) \btimes \overline{\tau}_\ell(n,\bB)]. $$ 
Here $n$ and $\bv$ are the number density and bulk velocity of the ions. Of course, these 
two terms are actually effects of the ideal Ohm's law and together they account for all of the 
reconnection and breakdown of flux-freezing at length-scales $\ell\gg \rho_i.$ Simple analytical estimates 
show indeed with $\boeps_\ell:= \boeps^v_\ell+\boeps^n_\ell$ that  $|\boeps_\ell|\gg |\wt{\bR}_\ell|$
and also $|(\grad\btimes \boeps_\ell)_\perp|\gg |(\grad\btimes\wt{\bR}_\ell)_\perp|$ so that 
reconnection and line-slippage at scales $\ell\gg \rho_i$ is due to ideal turbulence effects 
and not to microscopic plasma non-ideality. 

Just to be completely clear, this conclusion does {\it not } mean that large-scale turbulent reconnection
in a weakly collisional plasma such as the solar wind will be described by ideal MHD equations. 
Although an ideal Ohm's law holds,  there are not enough collisions to make the ion distribution function 
close to a local Maxwellian and to make the ion pressure tensor an isotropic function of local density 
and temperature. Thus, to provide a detailed quantitative description of large-scale reconnection
in the solar wind, one must resort either to the full Vlasov-Maxwell kinetic equations or to gyrofluid 
models that take advantage of the small ion gyroradius to develop closures for the ion pressure tensor.   
Ideal MHD should, however, be a suitable model to investigate some of the basic effects of turbulence
at inertial-range scales. Since the energetically dominant wave component of solar wind turbulence consists 
of incompressible shear-Alfv\'en modes, quantitative measures of field-line stochasticity and  turbulent Richardson dispersion should carry over with little change to this collisionless plasma environment. 

\subsection{Insight into physics of turbulent cascade}
\label{insight}

The explanation of the violation of textbook concept of flux freezing provided above is related to the properties of turbulence first discovered by Onsager (1949). In his prophetic paper he wrote that ideal equations indeed hold in a turbulent inertial range but ``the ordinary 
formulation of the laws of motion in terms of differential equations becomes inadequate and must be replaced by a more general description''. In other words, the {\it ideal equations} (Euler, MHD, etc.) holds in the inertial range in a {\it generalized 
sense}, but it does not hold in the standard, naive sense of partial differential equations. In particular, 
it does not hold in a sense that implies conclusions such as conservation of total kinetic energy for 
Euler or magnetic flux-freezing for ideal MHD.

This profound insight by Lars Onsager is not properly appreciated by the community. It is frequently assumed that the modes at scale $>L$
should satisfy the standard ideal equations, in the usual sense of partial differential equations, for about one eddy turnover time at that scale, $t_L~L/u_L$, until small scales $l_{\nu}$ are created where viscosity/resistivity (or microscpic plasma 
effects in a collisionless plasma) become non-negligible. This is not right way of thinking. In a turbulent flow all scales in the range $L>l>l_{\nu}$ are are already present at the initial instant. Low-pass filtering the fields to observe the scales $>L$ does not mean that that scales $<L$ have been 
physically removed. Turbulent motions of the cascade are physically present and cannot be disregarded. For instance, in the solar wind at every instant of time there is a broad-band spectrum with excitation at all scales. 

If one wants to know what is happening at a given scale $l>l_{\nu}$ low-pass filtering the fields at this scale is a purely passive operation of observing only the modes 
at length scales $>l$ and ignoring those at scales $<l$. But the modes at scales less than $l$ are still present and cannot be ignored. Therefore the dynamics at the sclae $l$ is not directly affected by the viscosity/resistivity/etc. in the equations for scales larger than $l$, but those equations are not the standard ideal equations. In the presentation above dealing with magnetic reconnection we wrote out these “ideal equations” 
in physical space as partial differential equations, but had to add  the “subgrid terms” that 
arise from the modes at scales less than $l$.  Those “subgrid terms” are exactly what one 
gets by coarse-graining the ideal PDE’s, i.e. one may say that the ideal equations hold 
“in the coarse-grained sense”.  This is equivalent (as pointed out by Lars Onsager) to saying that Fourier 
coefficients of the fields satisfy ordinary differential equations that follow from the ideal PDE’s.
However, none of these equivalent notions of “generalized solution”  have the usual implications 
of the ideal PDE's, such as conservation of total kinetic energy or conservation of magnetic-flux. 

The necessity of using {\it a more general description} is obvious from a trivial hydrodynamic example. If the large scales $>L$ in hydrodynamic turbulence 
{\it obeyed Euler equations} 
in the usual, naive sense, then the kinetic energy at scales $>L$ would be conserved in time. 
But it isn’t true. This is the easiest way to see that “obeying Euler equations” does not mean 
in this context the usual, naive thing. It is exactly this experimental observation that large eddies 
decay away, at a rate independent of viscosity, that motivated Onsager’s mathematical analysis.

In other words, turbulent cascade connects the scales and it is because of the motions associated with this cascade it is not possible to consider motions at large scales as ideal. The real world is decribed at appropriate macroscopic scales by non-ideal fluid or kinetic PDE’s with viscosity, 
resistivity,  collisions, etc. If you resolve all motions of a turbulent flow down to those smallest scales of validity, 
you will see solutions of the non-ideal PDE’s. All dissipation and reconnection arises from the nonideal terns 
within this “fine-grained” description. However, for many problems, including the problem of astrophysical magnetic reconnection, it is natural to consider only large scales $l\gg l_{\nu}$. At these scales the direct effect of non-ideal, e.g. resistivity, Hall, etc. terms are negligible. However, ``subgrid terms'' should be accounted for. With these terms the equations are valid in the {\it generalized sense} and we explicitly accounted for these terms in our quantiatative discussion above.

\subsection{Rigorous description of flux freezing violation: summary}
\label{flux_summ}

Turbulent reconnection predicts reconnection taking place at all scales in the turbulent media and this entails flux freezing violation. This section provides a rigorous description of the magnetic field behavior in turbulent fluids and demonstrates clearly that for considering the reconnection at scale $l$ one should account for the turbulent motions at the scale less than $l$. This vividly demonstrates and quantifies  the violation of the conventional flux freezing. In approach proves that the non-ideal plasma effects can be safely disregarded if magnetic reconnection is studied at scales much larger than the corresponding plasma scales.

This section provides strong support for conclusion in LV99, resolves a number of problems related to the earlier treatment of magnetic field in turbulent fluids and establishes solid mathematical foundations of the turbulent reconnection theory. It also clarifies the dynamics of magnetic field in turbulent media and suggests practical ways for evaluating the violation of the magnetic flux freezing.

\section{Testing of turbulent reconnection and flux freezing violation}
\label{sec:testing}

\subsection{Testing LV99 predictions with MHD codes}
\label{sec:testing1}

Numerical testings of reconnection in the presence of turbulence has a long
history.  Pioneering two-dimensional studies in a periodic box were done by
\cite{MatthaeusLamkin:1985, MatthaeusLamkin:1986}.  Although the turbulence was
only injected initially and not driven, it persevered whole simulation time.
The results indicated strong effect of turbulence presence on the reconnection
process.  For instance, the rates of production of reconnected magnetic islands
were insensitive to resistivity.  Unfortunately, the numerical setup precluded
calculation of the long-term reconnection speed and no studies were done on how
the properties of turbulence itself affect reconnection.  More recently,
\cite{Watson_etal:2007} studied the effects of small-scale turbulence on
two-dimensional reconnection, however, no significant effects of turbulence on
reconnection were observed.  Some possible explanations for such effects was
provided by the authors.  Of the most important is that the perturbations were
injected in the flow-driven merging of the super-Alfv\'enic regime, resulting in
a quick ejection of the fluctuations from the sheet and two-dimensional
configuration limiting the degree of freedom of magnetic field lines.

Dynamics of magnetic fields is three-dimensional by nature.  Any two-dimensional
configuration will significantly restrict the magnetic field freedom, making
impossible, for example, the conventional magnetic field line wandering.  Moreover,
two-dimensional and three-dimensional magnetized turbulence are very
different in nature (see the discussion in Eyink et al. 2011). Nevertheless, due to the computational
challenges and in the absence of any quantitative model to be tested, simulations
aimed at studying reconnection have been performed in two-dimensions for a long time.
For instance, the two-dimensional study in
\cite{MatthaeusLamkin:1986} stresses the importance of turbulence for modifying
the character of magnetic reconnection and specifies heating and transport as
the effect of particular significance, as well as formation of Petschek-type
X-point in two-dimensional turbulence.  \cite{KimDiamond:2001} showed that
the transport of magnetic flux is not enhanced to the reconnection zone.  At the
same time, field wandering described in LV99 is the essential feature of three-dimensional
reconnection. The dynamics of field lines is very different  in two dimensions (see Eyink et al. 2011). Therefore 2D processes cannot provide us with the guide to what is going on in the real astrophysical situations. 

\cite{Loureiro_etal:2009} studied two-dimensional turbulent reconnection
performing high-resolution simulations in a periodic box for Lundquist numbers
up to $5 \times 10^4$ concluding the existence of a critical threshold of the
injection rate above which the reconnection rate is enhanced.
\cite{Kulpa-Dybel_etal:2010} performed numerical simulations of the LV99
model in two dimensions showing that the relations
between the reconnection speed and turbulence properties are different than
those predicted by the LV99 model, and the reconnection is not fast in the
presence of turbulence, since it still depends on the resistivity. In any case, it is obvious that the difference in physics of magnetic turbulence in 2D and 3D makes two dimentional studies a subject of pure academic interest. The actual testing of turbulent reconnection requires one to use the numerical set up of the realistic 3D set up.

\begin{figure*}
 \centering
 \includegraphics[width=0.48\textwidth]{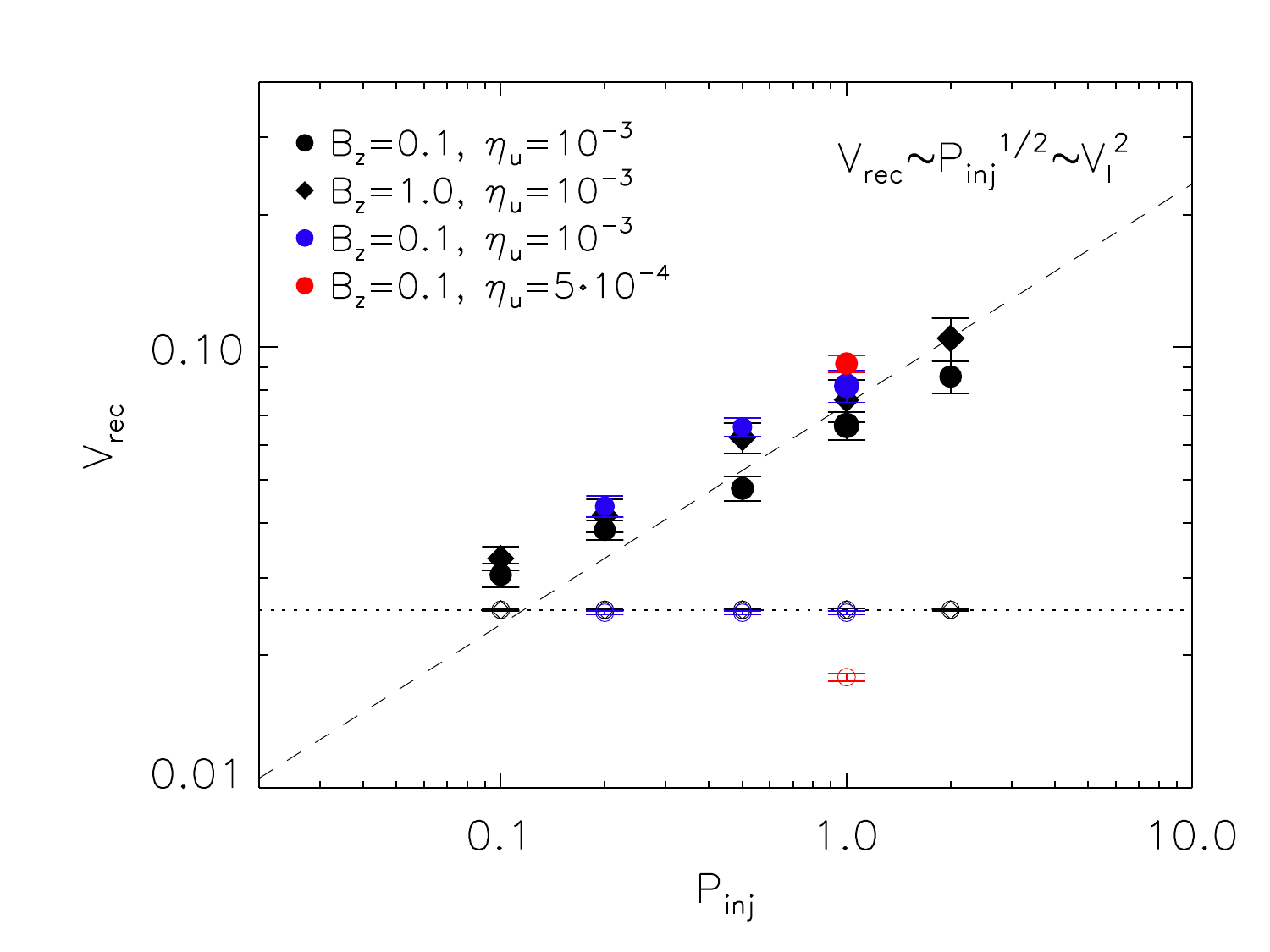}
 \includegraphics[width=0.48\textwidth]{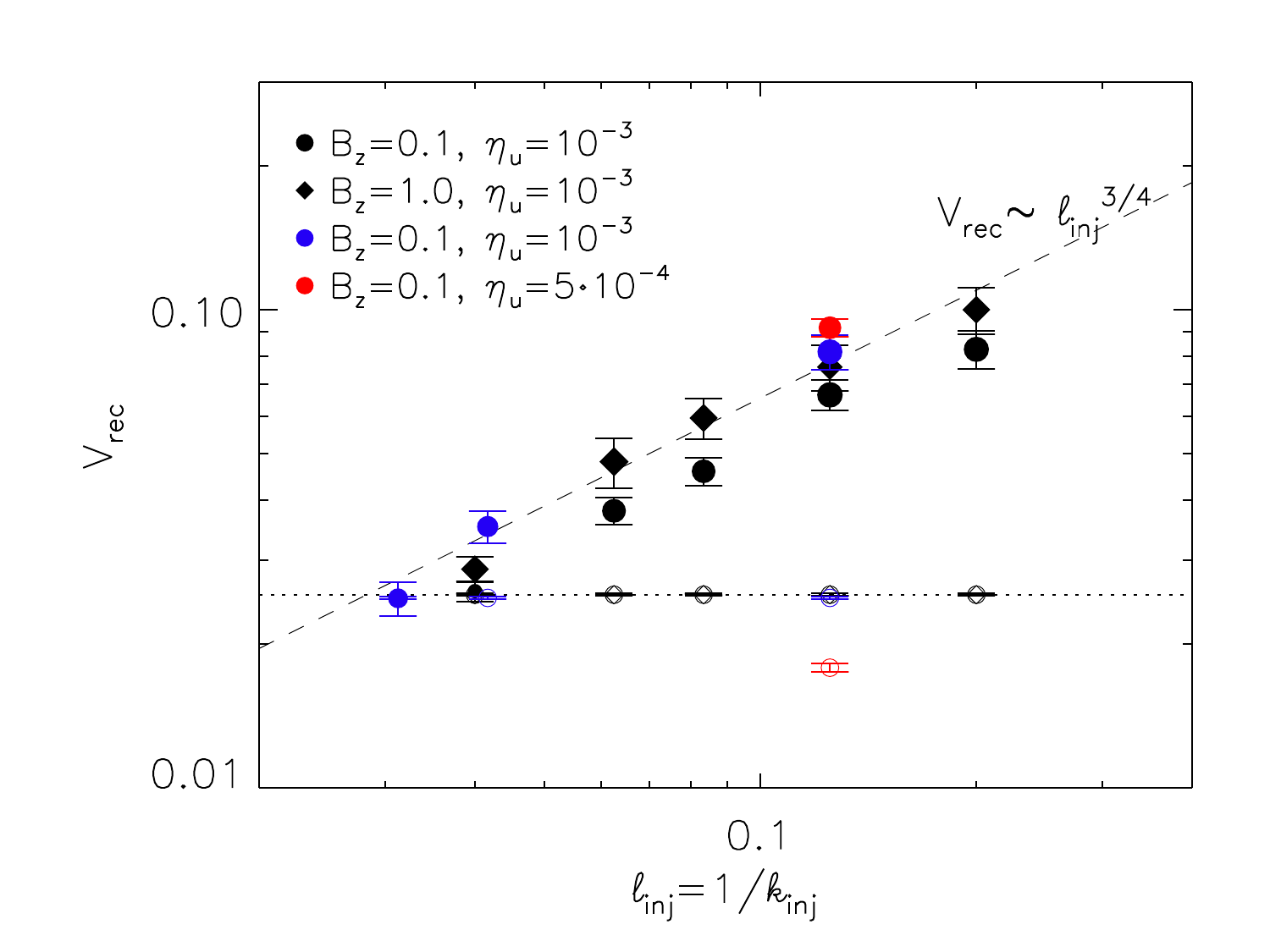}
 \caption{Dependence of the reconnection speed $V_{rec}$ on the injection
power $P_{inj}$ (left) and the injection scale $k_{inj}$ (right).  Different
colors correspond to different turbulence driving methods (black - Fourier driving in
velocity, blue and red - real space driving in velocity and magnetic field,
respectively).  The dotted line corresponds to the Sweet-Parker reconnection
rate for models with $\eta_{u}=10^{-3}$.  A unique red symbol shows the
reconnection rates for model with driving in velocity performed with higher
resolution ($512 \times 1024 \times 512$) and resistivity coefficient reduced to
$\eta_{u}=5\cdot10^{-4}$. Reprinted from \cite{Kowal_etal:2012a} under
\href{https://creativecommons.org/licenses/by/3.0/}{CC BY} license.}
 \label{fig:lv99_testing}
\end{figure*}

At the time of publication of the theoretical LV99 model of
turbulent reconnection, it was not possible to verify its predictions
immediately using numerical simulations. The model is three-dimensional
intrinsically, requiring significant computational resources and requires good
resolution to resolve all scales from large scales where turbulence is injected to
dissipation scales at which magnetic resistivity becomes important.  Even though the
relations between the reconnection rate and turbulence properties could be
tested using relatively low resolutions, any attempt to test the most
important prediction of the rate being independent on the magnetic dissipation
would be exposed to immediate critics.  Knowing that numerical simulations would
not reach astrophysical Lundquist numbers for decades, it was challenging to
undertake such a task.

\begin{figure}
 \centering
 \includegraphics[width=0.48\textwidth]{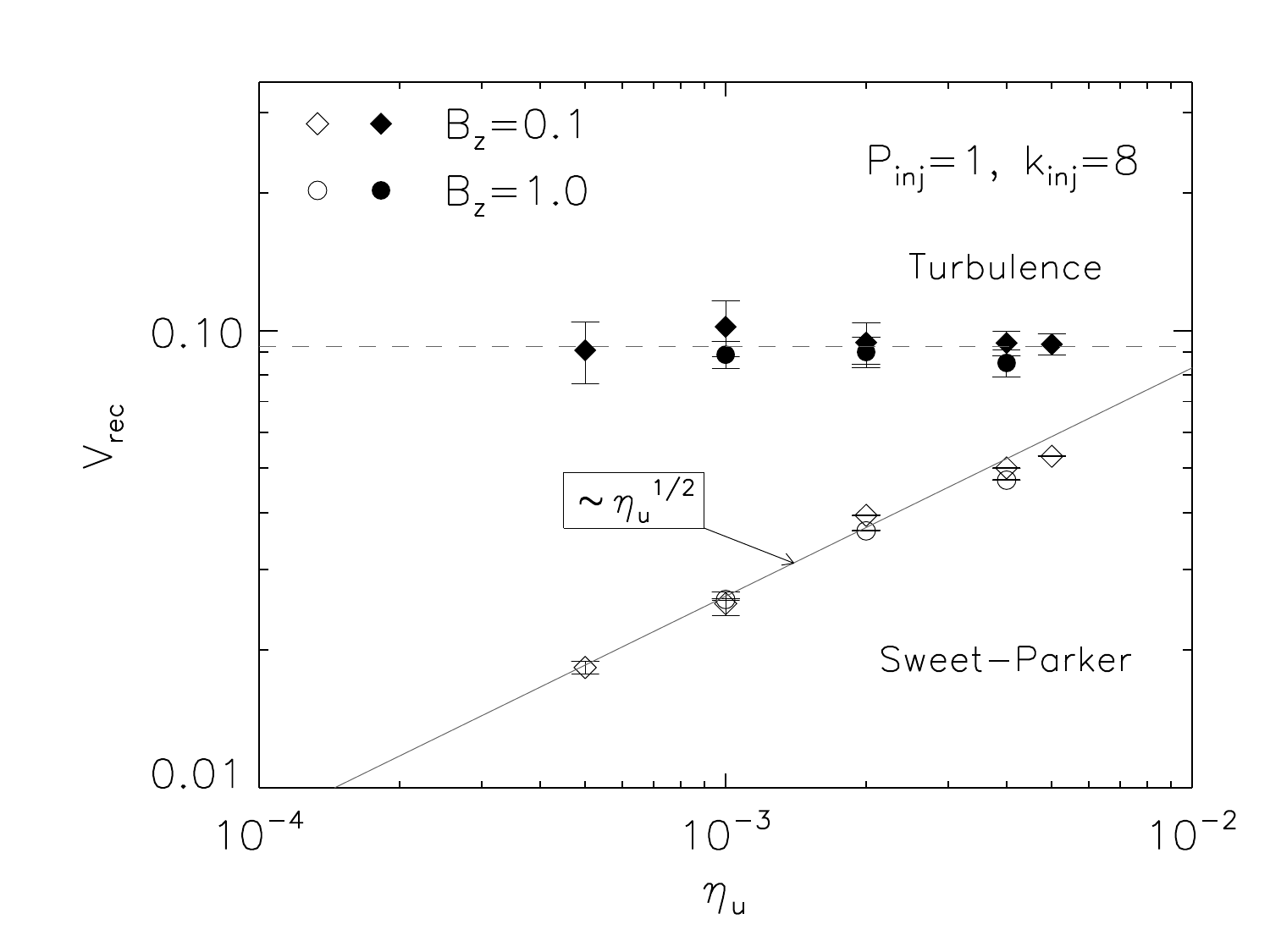}
 \caption{Dependence of the reconnection speed $V_{rec}$ on the resistivity
$\eta$ for models with weak (diamonds) and strong (circles) guide field.  From
\cite{Kowal_etal:2009}, \copyright~AAS. Reproduced with permission.}
 \label{fig:lv99_testing:eta}
\end{figure}

The first extensive testing of LV99
by means of numerical simulations was performed
in \cite{Kowal_etal:2009}, a decade after the publication of LV99.  The
tests were done using MHD compressible code in three-dimensional domain with
open boundary conditions in the subAlfv\'enic turbulence regime. Note, that for simplicity the periodicity was applied
along the guide field direction, i.e. $Z$-direction, but this does not affect our results, as there is no outflow/inflow along $Z$-direction. 

Additional tests in terms of different turbulence driving and exploring the parameter space have been performed in
\cite{Kowal_etal:2012a}. These pioneering provided new numerical approaches to exploring 3D magnetic reconnection. In particular, the ways of measuring 3D reconnection rate were introduced there.

The authors performed numerical models changing three
basic parameters, the power $P_{inj}$ and scale $k_{inj}$ of driving, and the
resistivity coefficient $\eta$. Figure~\ref{fig:lv99_testing} shows the measured
dependencies of the reconnection speed $V_{rec}$ on the turbulence parameters
$P_{inj}$ and $k_{inj}$.  Even though relatively low resolution was used, the
authors confirmed theoretical LV99 model predictions of reconnection rate
dependence on the turbulence parameters.  Most importantly, however, the
numerical studies of \cite{Kowal_etal:2009} have proven that the reconnection is
indeed fast in the presence of turbulence, since it does not depend on
the resistivity $\eta$, as shown in Figure~\ref{fig:lv99_testing:eta}. Moreover,
the same dependence was observed for models with weak and strong guide fields
$B_z$.  In \cite{Kowal_etal:2009}, the authors also tested the role of the
anomalous resistivity, and found that there is no essential dependence of the
reconnection rate on anomalous effects, as well.  Apart from the confirmation of these
important relations, \cite{Kowal_etal:2012a} reported a weak dependence of the
reconnection rate on viscosity, $V_{rec} \sim \nu^{-1/4}$, later studied in
more details in \cite{Jafari_etal:2018} in the context of large magnetic Prandtl
numbers.

As expected in the LV99 model, the current sheet is broad with individual
currents distributed widely within a three dimensional volume and the turbulence
within the reconnection region is similar to the turbulence within a
statistically homogeneous volume.  The structure of the reconnection region was
analyzed in more details by \cite{Vishniac_etal:2012} based on the numerical
work by \cite{Kowal_etal:2009}.  The results supported LV99 assumptions about
reconnection region being broad, the magnetic shear is more or less coincident
with the outflow zone, and the turbulence within it is broadly similar to
turbulence in a homogeneous system.

\subsection{Demonstration of flux freezing violation}
\label{ssec: doffvio}

As we discussed earlier the LV99 idea of turbulent reconnection is closely related to the gross violation of flux freezing in turbulent environments. As we discussed in \S \ref{ssec: rdffvio} Richardson dispersion is a vivid manifestation of the flux freezing violation in turbulent fluids. Richardson dispersion in space was derived in LV99 and it was used for predicting the rate of turbulent reconnection. 
Figure \ref{fig5} confirms the corresponding LV99 predictions. Indeed, it is clearly seen that over the inertial range the separation between following magnetic field lines is growing as $s^{3/2}$, where $s$ is the distance calculated along the magnetic field line. 

\begin{figure}
\centering
\includegraphics[width=0.48\textwidth]{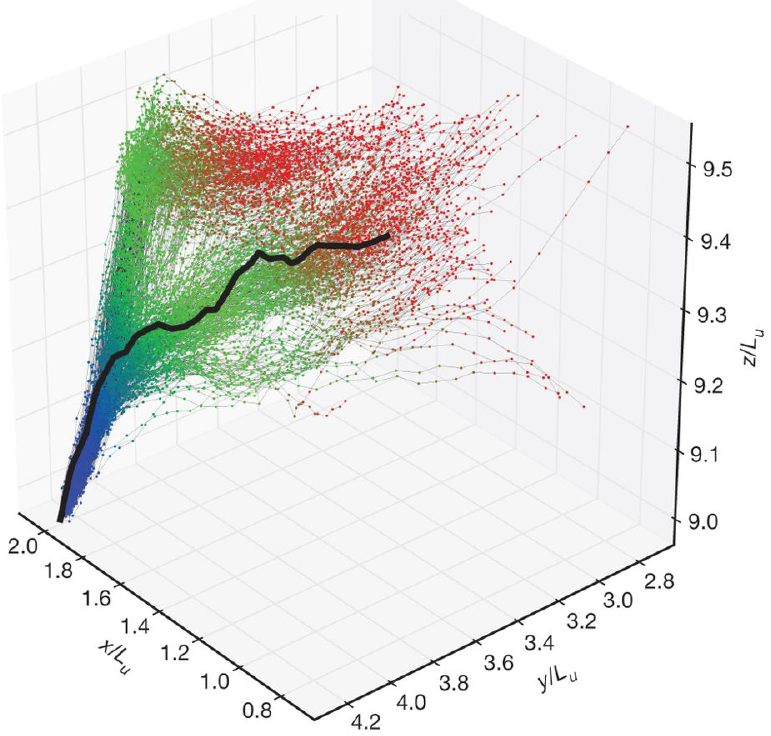}
\caption{Stochastic trajectories that arrive at a fixed point in the archived MHD flow, color-coded red, green, and blue from earlier to later times. From \cite{Eyink_etal:2013}.}
\label{fig11}
\end{figure}

A direct testing of the temporal Richardson dispersion of magnetic field-lines was performed recently in 
\citet{Eyink_etal:2013}.
For this experiment, stochastic fluid trajectories were tracked backward in time from a fixed point in order to determine which field lines at earlier times would arrive to that point and be resistively ``glued together''. For the flux-freezing situation the one to one mapping is expected for different times, but the Richardson dispersion prescribes quite a different behavior. To test what is right, many time frames of an MHD simulation were stored so that equations for the trajectories could be integrated backward. The results of this study are presented in Figure~\ref{fig11}. It shows the trajectories of the arriving magnetic field- lines, which are clearly widely dispersed backward in time, more resembling a spreading plume of smoke than a single ``frozen-in'' line. The quantitative results of the time-dependent separation correspond well to the expectations for the temporal Richardson dispersion. This numerical experiment also demonstrates that magnetic reconnection happens at every scale of the cascade as we discussed in \S \ref{insight}.

\subsection{\label{sec:level2b}Self-driven turbulent reconnection}

Although studies on reconnection with imposed driven turbulence have been done
for a few decades already, the question of self-driven reconnection was
approached by aims of numerical simulations within last several years only.  The
question was first raised in LV99 and further discussed
in \cite{LazarianVishniac:2009}.  It was realized, that due to the existence of
magnetohydrodynamic instabilities, turbulence could be generated directly by the
inhomogeneous current sheet, leaving the imposed external turbulence obsolete.
In the self-driven reconnection those inhomogeneouties of current sheet should
grow, resulting in variations of the current sheet thickness and the size and
strength of the outflow from the local reconnection regions, which interacting
develop turbulence.  While the reconnection feeds these interactions, so and the
turbulence should growth.  The developed turbulence, in return, influences the
reconnection process, resulting in its enhanced rate.

\begin{figure}[t]
\centering
\includegraphics[width=0.48\textwidth]{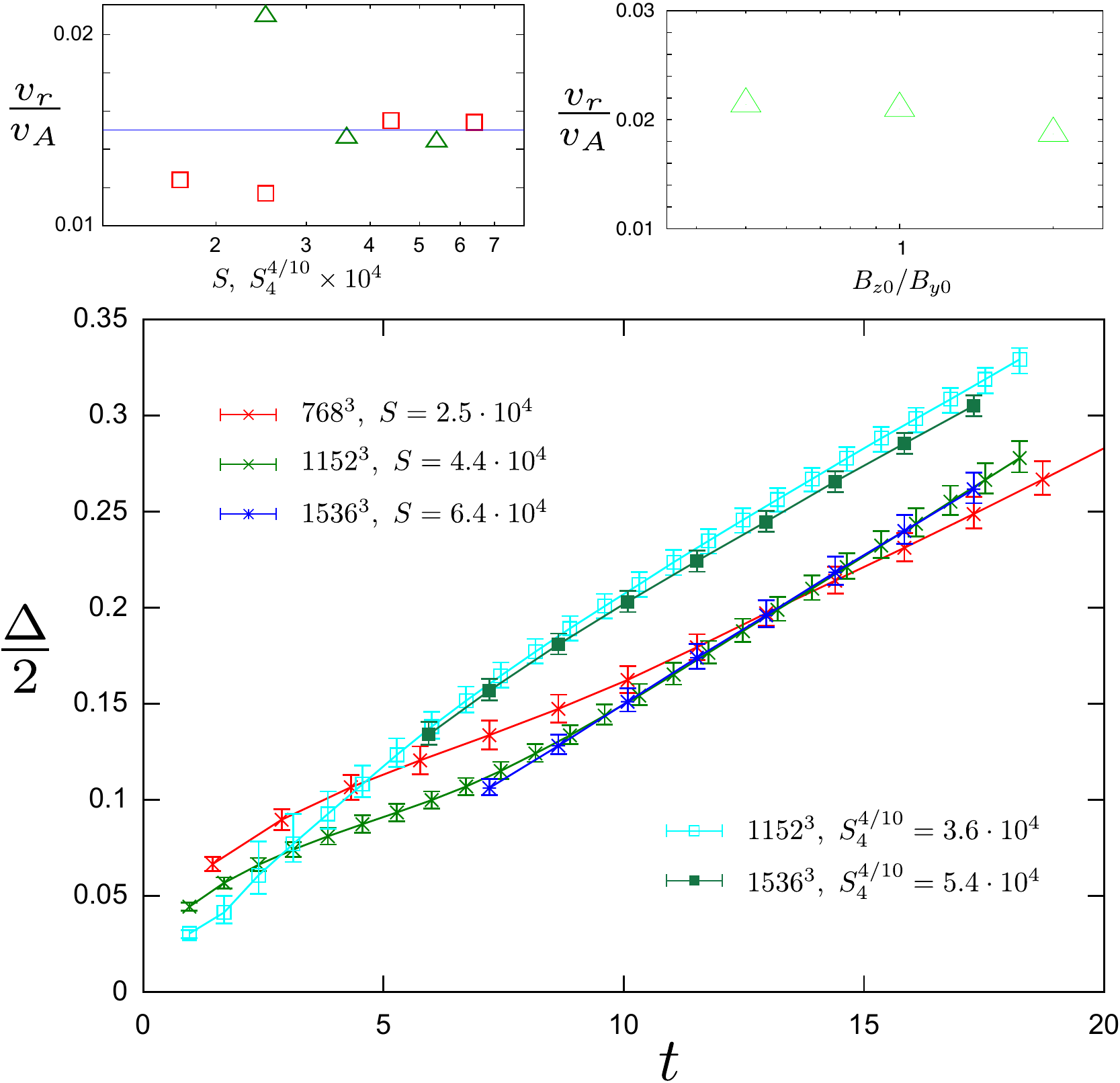}
\caption{The evolution of the current layer width
$\Delta$ (bottom) and the inferred reconnection speed as a function $S$ (top
left) and the ratio of $B_{z0}/B_{y0}$ (top right). From \cite{Beresnyak:2017},
\copyright~AAS. Reproduced with permission.}
\label{fig:beresnyak}
\end{figure}

The difficulty of studying self-driven reconnection comes from several factors.
First of all, in order for small amplitude fluctuations to not simply propagate
out of the box or dissipate, one has to use very small Ohmic resistivites $\eta
\ll 10^{-3}$, corresponding to large Lundquist numbers, $S \gg 10^{3}$.
However, the thickness of current density scales as $\sim \eta^{1/2}$, e.g. in
the Sweet-Parker model, indicating necessity of using very large resolutions in
numerical models. For instance, for Lundquist number $S = V_A L / \eta \sim
10^6$, assuming $V_A \sim 1$ and $L \sim 1$, the current sheet thickness would
be around around $10^{-3} L$, where $L$ is the longitudinal size if the current
sheet. Resolving such a small thickness numerically requires at least
resolutions of $10^4$ cells per unit length $L$. Moreover, the goal is to reach
a steady-state of self-driven reconnection, which cannot be achieved in a
periodic box, and one has to use numerical domains with open boundary
conditions, which proper implementation is far from trivial.  The steady-state
also cannot be achieved using very short runs, therefore, one has to simulate a
model for at least a few Alfv\'en times.  Finally, one would like to study
realistic three-dimensional configurations, which for the estimated resolutions
would require enormous computational resources.  There are only a few most
powerful supercomputers which could deal with such a huge task nowadays.

The numerical simulations with turbulence generated through  magnetic
reconnection were presented in preprint by Beresnyak (2013) with the final extended version of the paper published in \citet{Beresnyak:2017}.  
The studies were done in a
three-dimensional periodic box using spectral incompressible MHD simulations up
to resolution $1536^3$ with the highest Lunquist number of $S = 6.4\times10^4$. The incompressible simulations used correspond the limiting case of high beta plasma, i.e. plasma with sound velocity $\gg$ the Alfven one. While in 2D the location of the current sheet was remained in the same position in 3D the location of the original current sheet position was largely forgotten as turbulence was developing.  

The study reported magnetic reconnection that is very different from the magnetic reconnection studied earlier in 2D settings. First of all, the reconnection was observed to proceed at a steady rate, which is in contrast to the to 2D simulations \cite[see][]{Loureiro2012}, exhibiting significant time dependence. The flux ropes were observed in 3D, but, unlike magnetic islands in 2D, the flux ropes were turbulent inside. The number of these structures did not depend on the Lunquist number, contrary to the tearing reconnection described e.g. in \cite{Uzdensky_etal:2010}.

Clear turbulent structures
the kinetic energy of the order of $10^{-6}$ of the magnetic one within several
Alfv\'en times \cite[see Fig.~2 in][]{Beresnyak:2017}.  The reported
reconnection rate was measured to be $\sim 1.5\%$ of the Alfv\'en speed (see Fig.~\ref{fig:beresnyak}).  The reconnection rate was derived as $V_{rec}
= d\Delta/dt$, where $\Delta$ is the measured of the turbulent layer thickness and this result was justified using LV99 theory.\footnote{A somewhat different derivation of the growth of $\Delta$ is obtained in Lazarian et al. (2015).}. Analysing his data Beresnyak (2017) concluded that the expression for the LV99 reconnection rate $V_{rec}=C_{LV} \delta v_L^2/V_{A,x}$ should have the constant $C_{LV}$ very close to unity, with $C_{LV}=0.94$ being claimed.

\cite{Beresnyak:2017} also studied the properties of
generated turbulence, reporting turbulence consistent with the Goldreich-Sridhar
\citep{GoldreichSridhar:1995}, i.e. kinetic energy power slope $\sim -1.5 \div
-1.7$, and scale-dependent anisotropy $\lambda_\parallel/\lambda_\perp \sim
\lambda_\perp^{-1/3}$, although, the inertial range of compatible anisotropy was
relatively short.

Later works include numerical simulations of stochastic reconnection done by
\cite{Oishi_etal:2015} and \cite{HuangBhattacharjee:2016}, both using
compressible MHD approximation, in slightly different reconnection
configurations, however.
\cite{Oishi_etal:2015} used a numerical setup starting from the Sweet--Parker
setup with periodic box along the current sheet.  This setup is somewhat
unfortunate, since the Sweet--Parker configuration requires open boundary
conditions to allow inflow and outflow of the magnetic flux in order to maintain
steady-state.  Due to periodic boundary conditions, the growth of the
fluctuations at large scales may be attributed to the presence of the magnetic
field setup which is not initially in the equilibrium.  Nevertheless, they
performed several long time simulations with Lundquist numbers up to $S \sim 3.2
\times 10^6$ with the conclusion that for sufficiently large Lundquist numbers
$S \gtrsim 10^5$, the thickness of current sheet $\delta_\mathrm{SP}$ becomes
weakly dependent on $S$ (see their Figure~1).  However, it should be noted that
simulations with $S > 10^5$ ($\eta < 10^{-5}$) where done with the maximum
effective resolution $1024^3$, much lower than the resolutions needed to resolve
the thickness of current sheet estimated above, therefore, the conclusion might
be attributed to the numerical effects.  They also estimated the reconnection
rate using the decay rate of the magnetic energy and concluding that 3D MHD
stochastic reconnection can be fast without recourse to kinetic effects or
external turbulence. They did not, however, provide any studied of the
properties of reconnection-generated turbulence.

\begin{figure*}[t]
\centering
\includegraphics[width=0.96\textwidth]{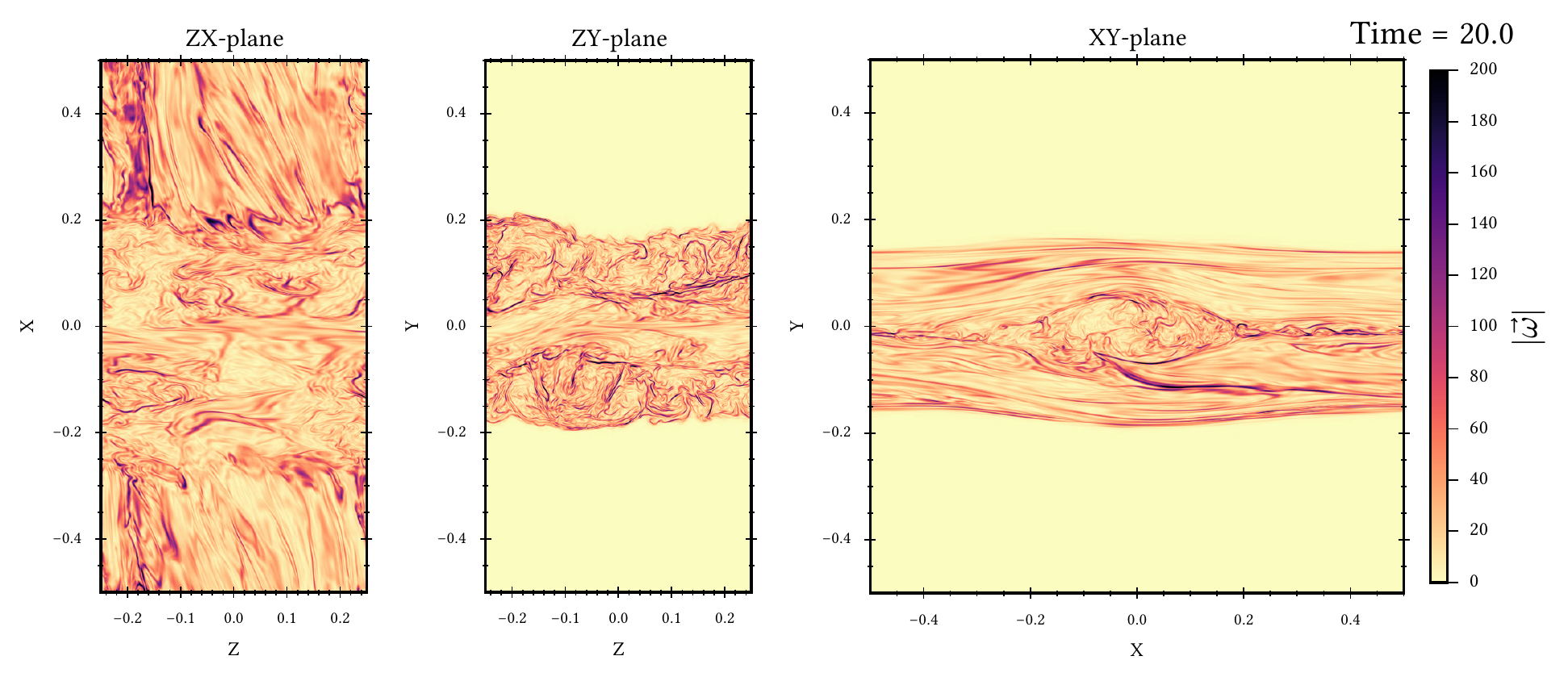}
\caption{Slices along three main planes, $ZX$ (left), $ZY$ (center), and
$XY$ (right) of the amplitude of vorticity $\omega = \nabla \times \vec{v}$
at time $t = 20.0V_A$ for a model with sound speed $a = 1.0$ and grid
size $h = 1/1024$. The turbulent region develops near the initial current
sheet and expands in the $Y$ direction. From \cite{Kowal_etal:2017},
\copyright~AAS. Reproduced with permission.
\label{fig:kowal:image}}
\end{figure*}

\cite{HuangBhattacharjee:2016} analyzed the statistics of velocity fluctuations
generated by reconnection also in the presence of the initial velocity noise,
and used, as a starting point, a global Sweet--Parker configuration, but with
non-periodic boundaries along the current sheet and a guide field comparable in
strength to the reconnecting component, which additionally prevented the
interactions of the waves.  Weak velocity perturbations were injected into such
a domain in order to understand their effect on the reconnection rate.  The
reported turbulent power spectra of kinetic energy with slopes $-2.5 \div -2.3$
and observed nearly scale independent anisotropy of velocity fluctuations, both
in contrast with the predictions of the Goldreich-Sridhar theory.  As pointed
out later by \cite{Kowal_etal:2017}, the power spectra and structure functions
calculated in \cite{HuangBhattacharjee:2016} from two-dimensional $XZ$-planes
and later averaged cause strong bias, resulting in steeper slopes and much
reduced anisotropy, when compared to the fully three-dimensional spectral and
structure function analysis.

While the study in \cite{HuangBhattacharjee:2016} support the LV99 expectation to the transfer of reconnection to the turbulent regime, some statements of these study contradict to the findings in all other numerical studies by performed by different groups around the world that also studied 3D reconnection. In particular, the statement of the similarity of 2D and 3D magnetic reconnection  in \cite{HuangBhattacharjee:2016} is at odds with the findings by other authors who report that in 3D the reconnection is evolving differently from that in 2D \cite[see][]{Oishi_etal:2015, Wang_etal:2015, Striani_etal:2016, Beresnyak:2017, Kowal_etal:2017, Takamoto:2018, WangYokoyama:2019, Kowal_etal:2019}. This statement in  
\cite{HuangBhattacharjee:2016} as well as their finding that the spectral slope and anisotropy of the obtained turbulence are different from expectations of MHD turbulence theory (see \S \ref{sec:turbulence2}) contradict to the LV99 expectations. Below we provide the results of numerical testing of these claims in \cite{HuangBhattacharjee:2016}.

\begin{figure}
\centering
\includegraphics[width=0.48\textwidth]{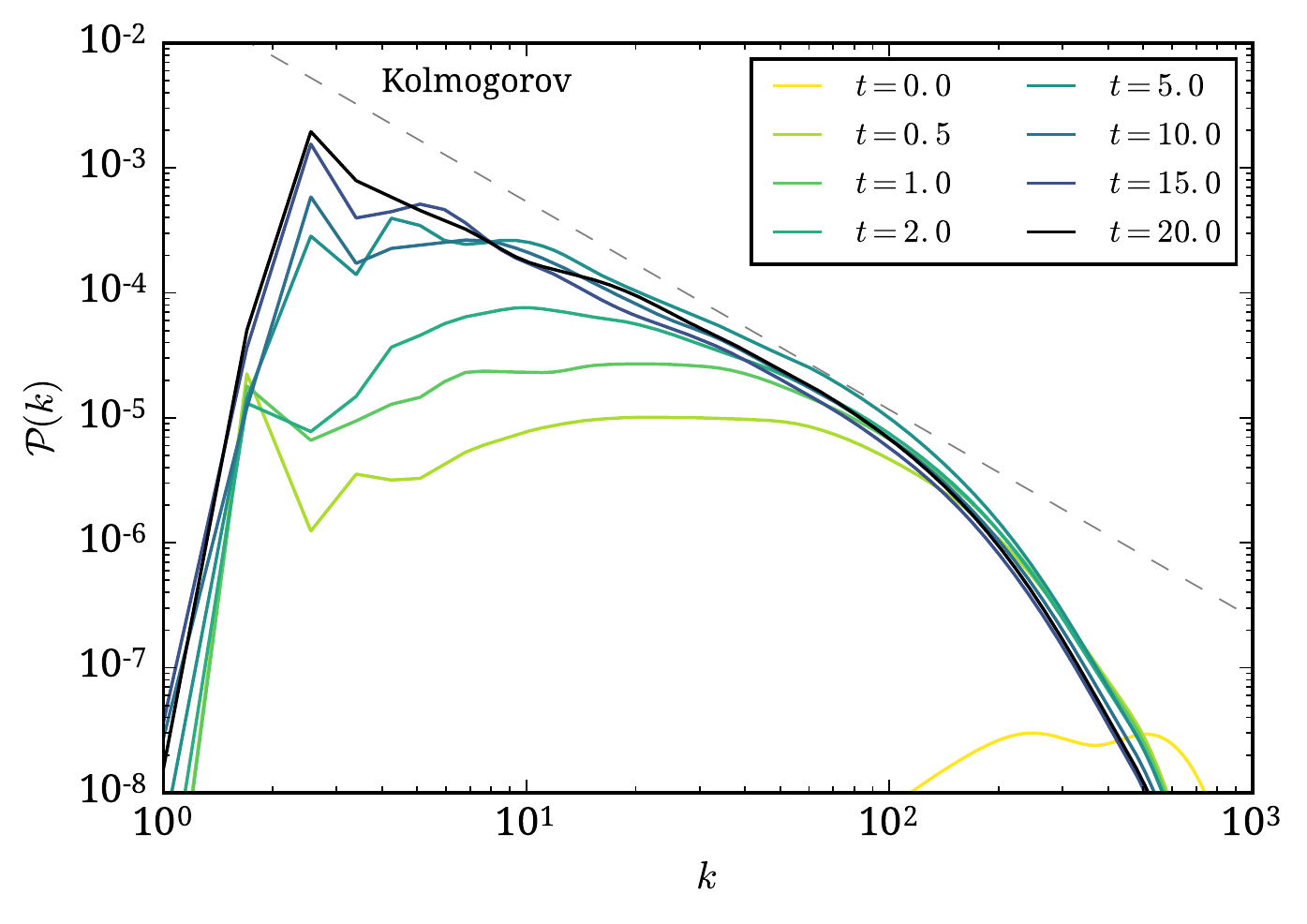}
\caption{Evolution of power spectra of velocity at different times. From
\cite{Kowal_etal:2017}, \copyright~AAS. Reproduced with permission.
\label{fig:kowal:spectra}}
\end{figure}

Below we discuss the most recent studies of the reconnection-driven turbulence in
\cite{Kowal_etal:2017}.  As the initial setup they used a uniform box with a
single discontinuity of the X-component of magnetic field placed in the middle
of the box along the XZ plane.  Although they used periodicity along the X and Z
directions, the box was open in the vertical direction, perpendicular to the
current sheet, and significantly elongated in Y, with physical dimensions of
$1.0 \times 4.0 \times 1.0$ in order to prevent any influence of the open
boundary conditions.  They let the turbulence grow from the initial noise of
amplitude $0.01$ applied to each velocity component for long runs up to $t = 20
t_A$.  Figure~\ref{fig:kowal:image} shows three main cuts across the domain of
the vorticity amplitude at the final time of one of the investigated models.

\begin{figure}
\centering
\includegraphics[width=0.48\textwidth]{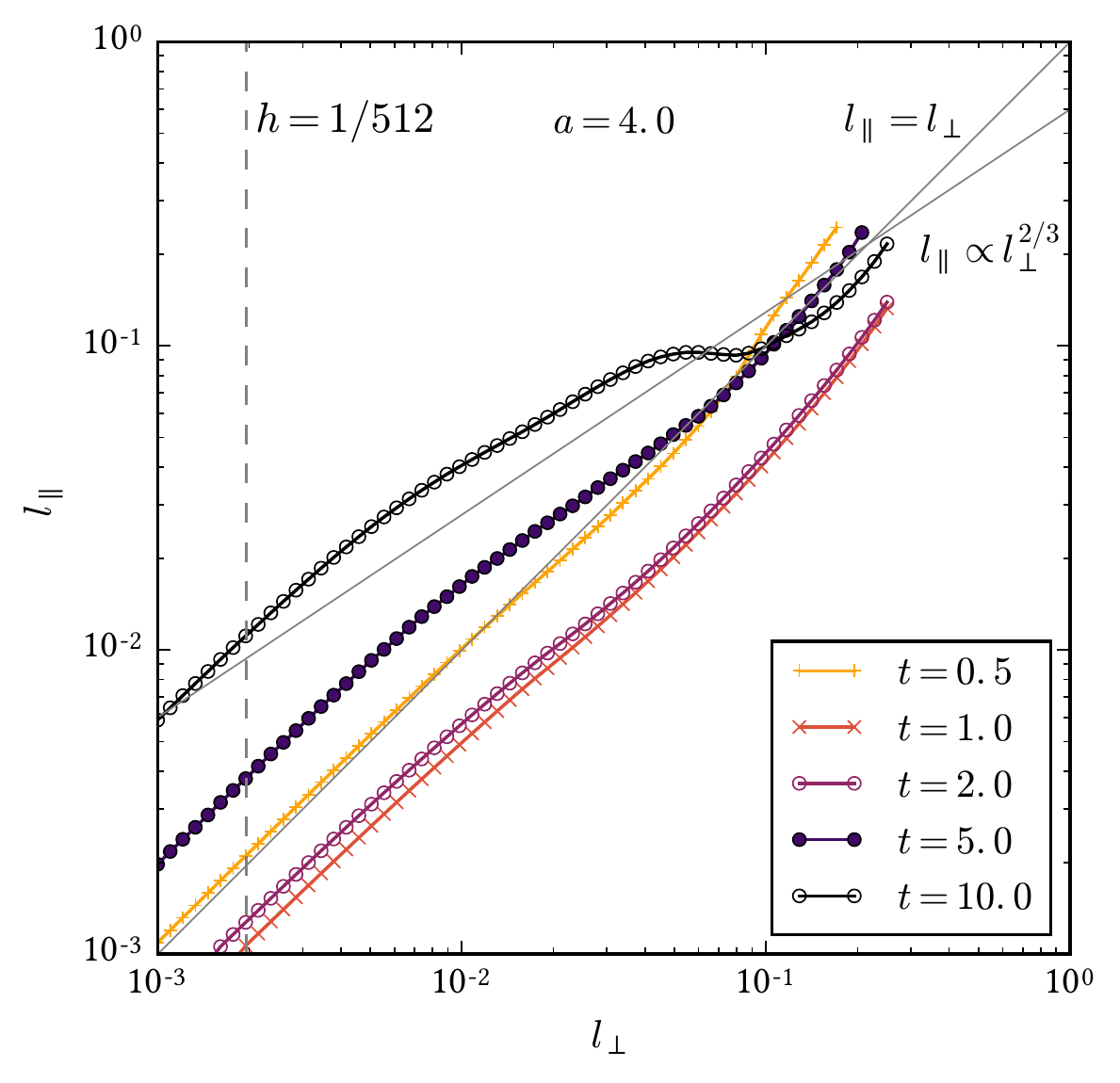}
\caption{Anisotropy scalings for total velocity for models with sound speed $a =
4$ obtained at different moments. For comparison, we show isotropic and strong
turbulence scalings (denoted by gray thin lines). From \cite{Kowal_etal:2017},
\copyright~AAS. Reproduced with permission.
\label{fig:kowal:anisotropy}}
\end{figure}

The choice of different vertical boundaries had some important
consequences. Periodicity in all three directions with two separated current
sheets imposed, as used in \cite{Beresnyak:2017} for instance, allows for the
possibility of horizontal large scale interactions through the deformations of
both current sheets.  These interactions, especially at later times, can
generate large scale motions in the perpendicular direction to the current sheet
providing additional to the reconnection energy input
\cite[see, e.g.][for a similar setup with many current sheets,
in which the current sheet deformations are well manifested]{Kowal_etal:2011}. 
In this case, such large-scale
interactions are not allowed, therefore the only energy input comes from the
small-scale reconnection events.  They, contrary to other studies, investigated
the dependence of the properties of developed turbulence on the $\beta$-plasma
parameter.  They reported Kolmogorov-like power spectra of generated velocity
fluctuations (see Fig.~\ref{fig:kowal:spectra}), which also presented
scale-dependent Goldreich--Sridhar anisotropy scaling at later times in large
$\beta$ models (shown in Fig.~\ref{fig:kowal:anisotropy}).  They claimed that
the turbulent statistics are similar to strong MHD turbulence, but are
significantly affected by the dynamics of reconnection induced flows, more
clearly manifested in the low-$\beta$ regime, where supersonic reconnection
outflows were present.  In high-$\beta$ regime the Goldreich--Sridhar anisotropy
scaling manifested at much earlier times, due to the fluctuations generated by
reconnection outflows propagating and interacting faster.  They also estimated
the reconnection rate to be twice as high as those estimated in
\cite{Beresnyak:2017}, which they justified by more diffusive code.

\begin{figure*}
 \centering
 \includegraphics[width=0.48\textwidth]{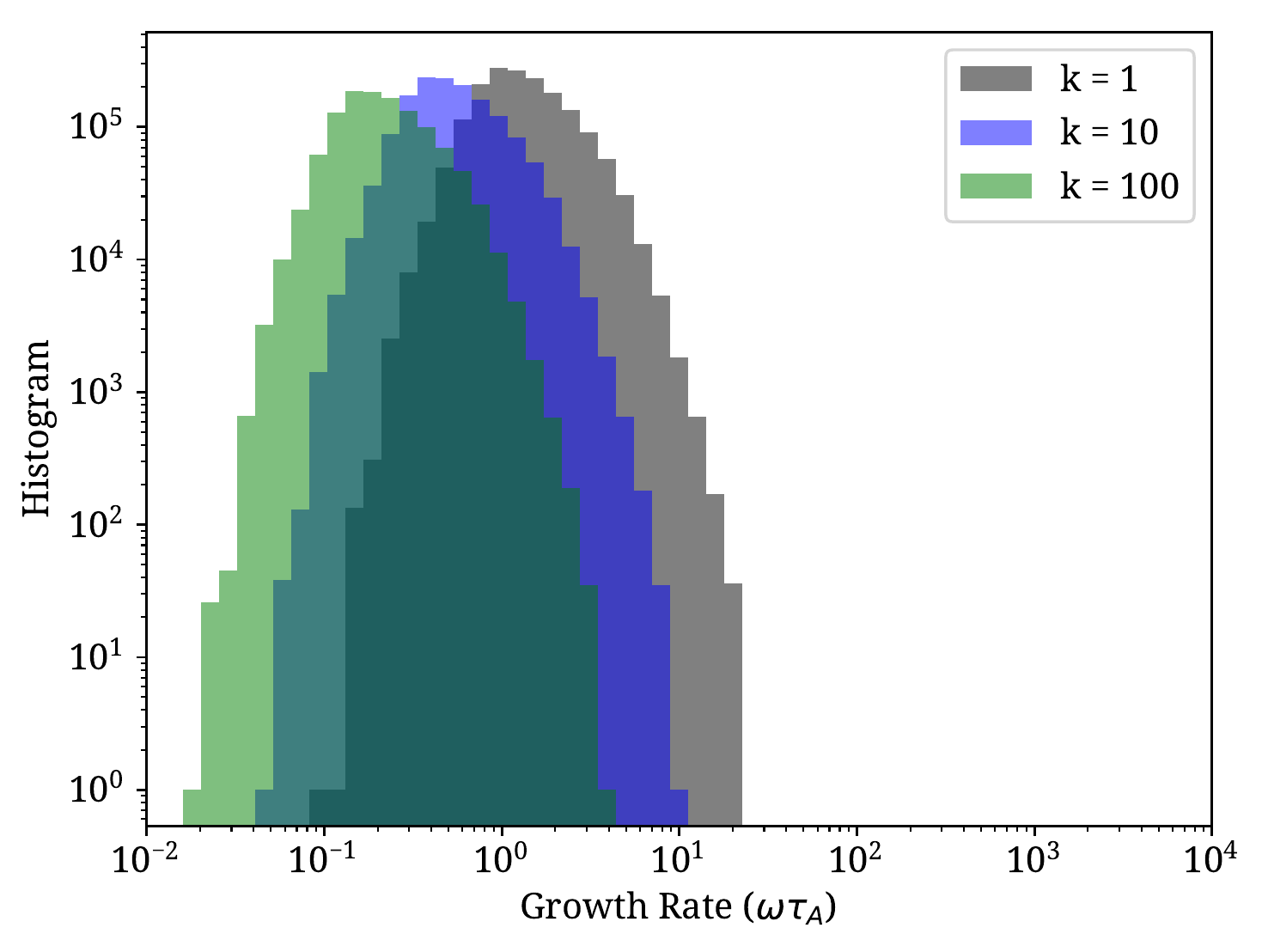}
 \includegraphics[width=0.48\textwidth]{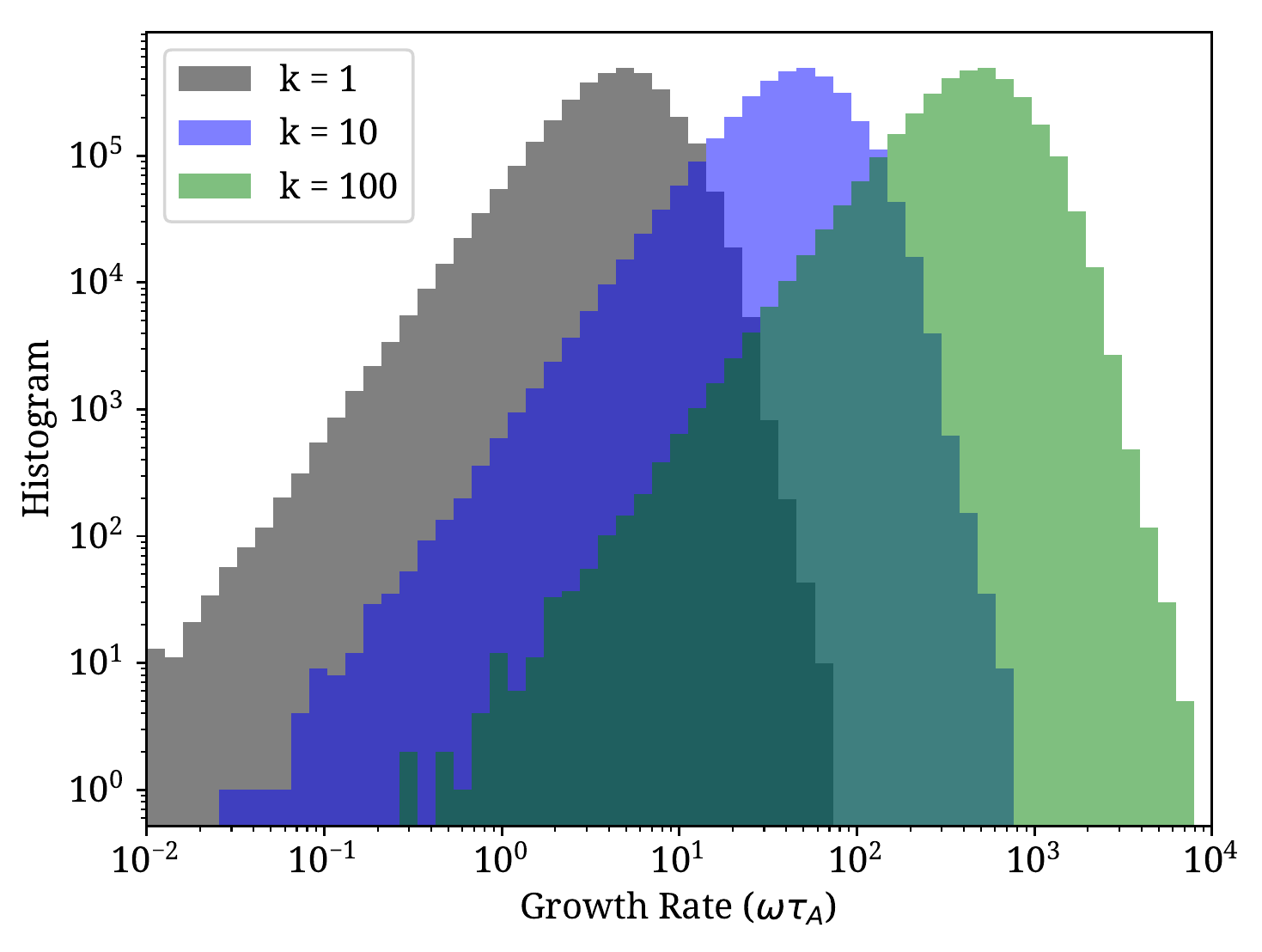}
 \caption{Distribution of the growth rate of the tearing mode (left) and
Kelvin-Helmholtz instability (right) for reconnection-driven turbulence
simulations by \cite{Kowal_etal:2019}. Distributions with different colors
correspond to three different perturbation wave numbers assumed (see legend).
From \cite{Kowal_etal:2019}.}
\label{Fig:Growth}
\end{figure*}

\cite{Kowal_etal:2019} studied what exactly drives turbulence in their
stochastic reconnection models.  They considered two instabilities, tearing mode
and Kelvin-Helmholtz instability.  For each analyzed model they determined the
magnetic (in the case of tearing mode) and velocity (in the case of
Kelvin-Helmholtz instability) shear location and performed detailed analysis of
the field profiles taking into account local geometry, i.e. the local coordinate
system oriented with respect to the shear plane.  Using such analysis, they
could determine a number of quantities important for the development of
instabilities.  For instance, in order to study the tearing mode, they
determined the local thickness of the current sheet, the length of the local
sheet, the strength of the transverse and guide components of magnetic field.
For Kelvin-Helmholtz instability, they determined the shear amplitudes and the
strength of the local magnetic field component which stabilized the instability,
or sonic and Alfv\'{e}nic Mach numbers.  These parameters allowed also to
estimate the growth rates for both instabilities.  

\begin{figure}
 \centering
 \includegraphics[width=0.48\textwidth]{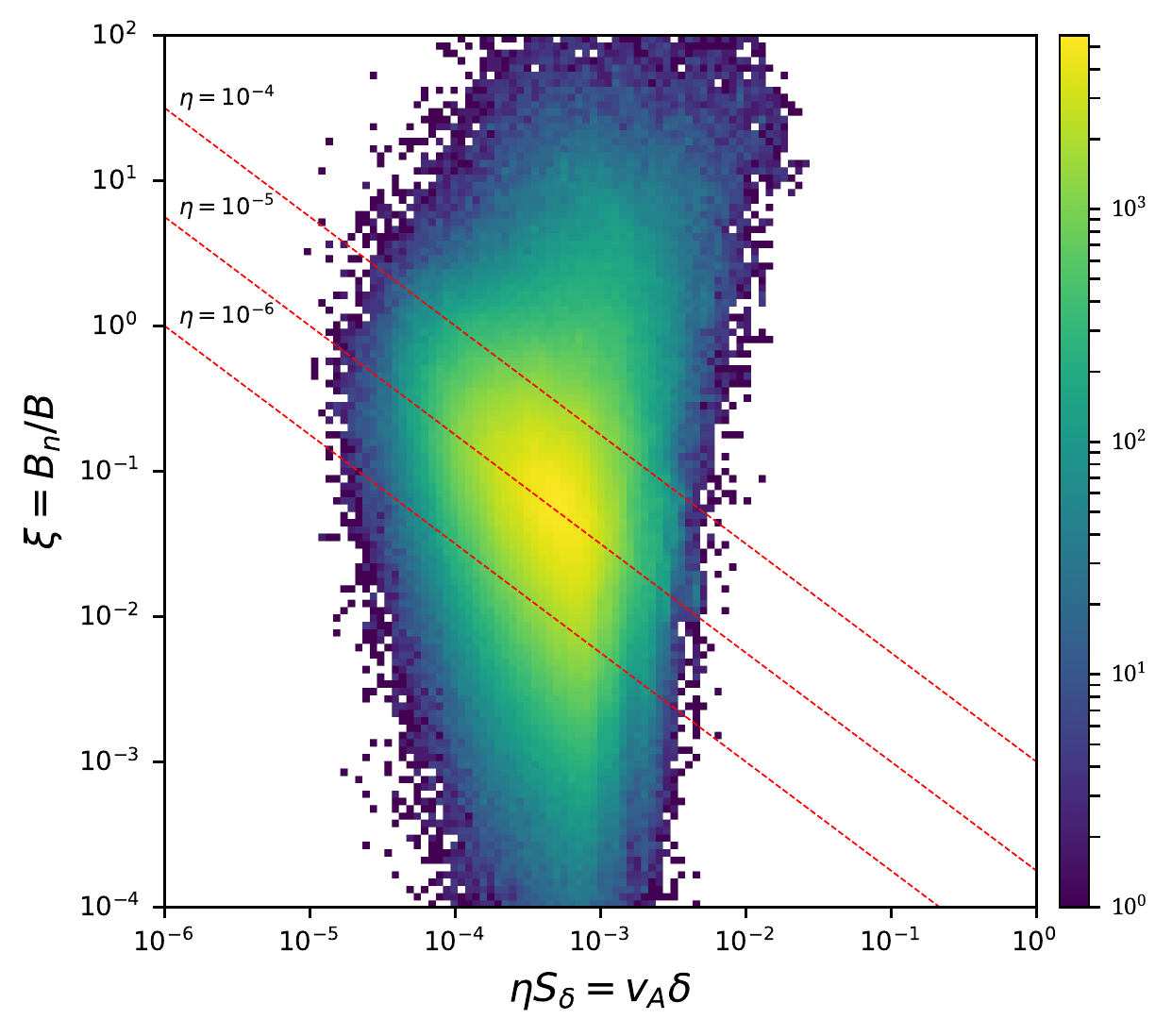}
 \caption{Distribution of $\xi = B_n / B$ as a function of $v_A \delta$. Since
the specific Lundquist number is $S_\delta = \frac{v_A \delta}{\eta}$, by
assuming explicit resistivity $\eta$ we can find the stability region limit.
Red lines correspond to three values of $\eta = 10^{-4}$, $10^{-5}$, $10^{-6}$.
The stable region lays above the red lines. From \cite{Kowal_etal:2019}.}
\label{Fig:Xi}
\end{figure}

The results indicated
tearing mode growth is significantly suppressed at all studied wave numbers. See
for example Figure~\ref{Fig:Growth}, where statistical distribution of the
growth rates for tearing mode  (left panel) and Kelvin-Helmholtz instability
(right panel) for three different perturbation wave numbers are shown.  Strong
suppression of tearing mode can be explained by the fact, that its growth rate
actually decreases with the wave number $k$, while for Kelvin-Helmholtz
instability it increases.  It could be also attributed to the presence of the
transverse component of the magnetic field, easily generated by turbulence.
Figure~\ref{Fig:Xi} shows distribution of the transverse component relative
strength $\xi = B_n / B$ vs $v_A \delta$.  It is well seen, significant number
of cells within current sheets are stable due to strong transverse component.

\begin{figure*}
 \centering
 \includegraphics[width=0.48\textwidth]{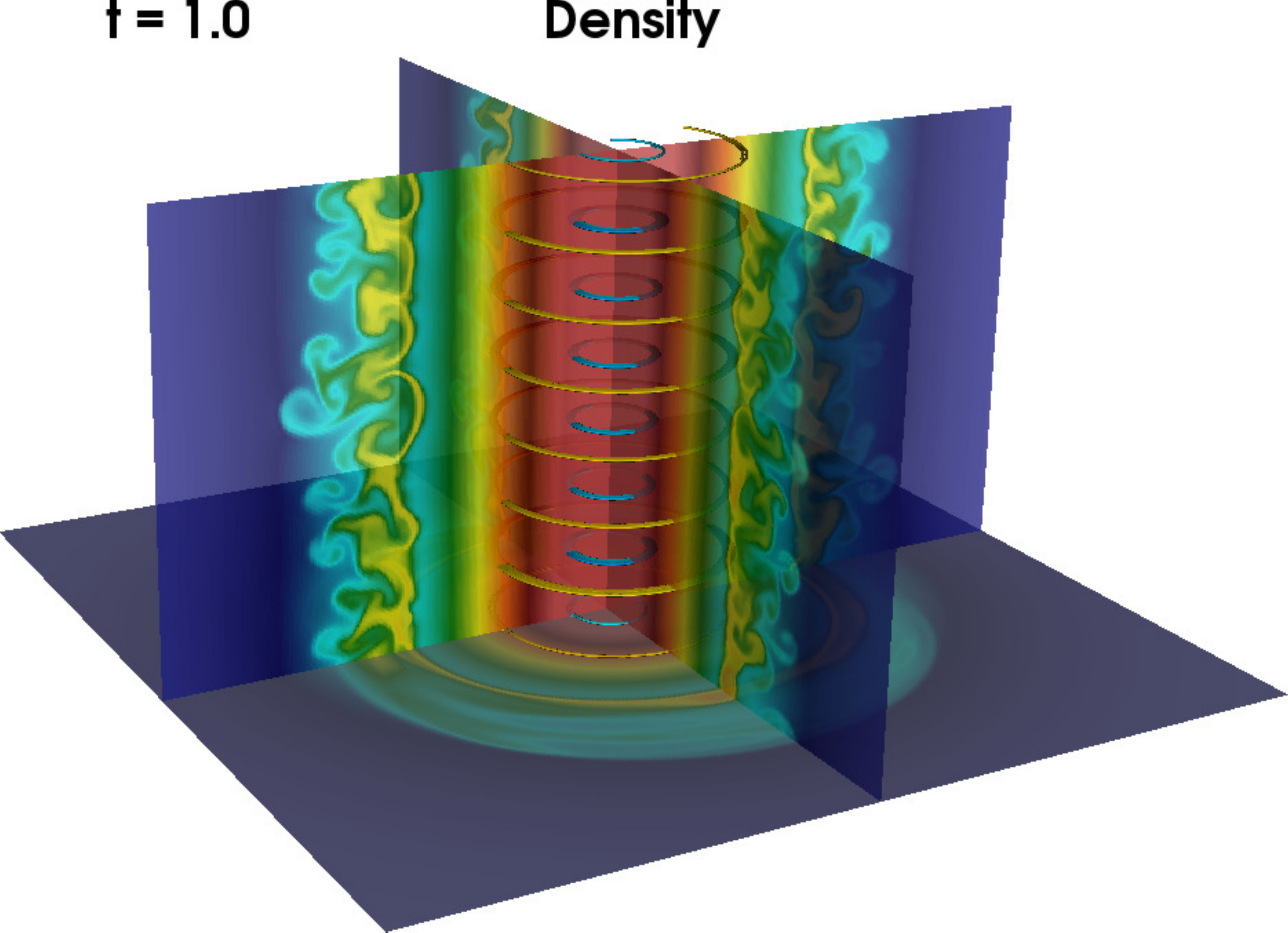}
 \includegraphics[width=0.48\textwidth]{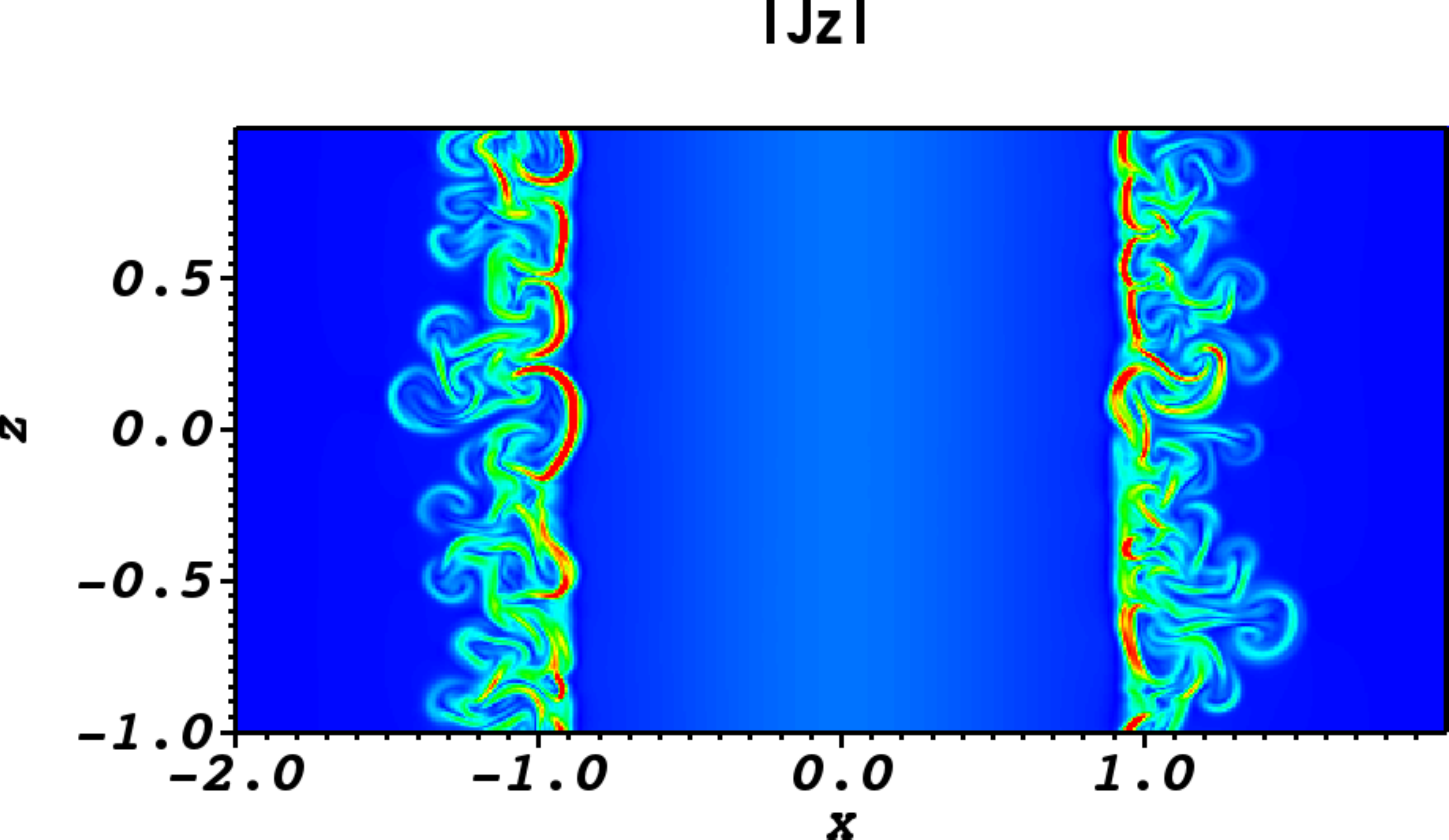}
 \caption{{\it Left}: Snapshot for cylindrical plasma colum with $S = 2.4 \times 10^4$ at
$t = 1.0 t_A$ showing the three-slice (pseudocolor rendering) of the density.
{\it Right}: 2D slice on the $YZ$  plane (at $x = 0)$ of
the current density. Reprinted from \cite{Striani_etal:2016} with permission of Oxford University Press.}
\label{Fig:Striani}
\end{figure*}

\begin{figure*}
 \centering
 \includegraphics[width=0.48\textwidth]{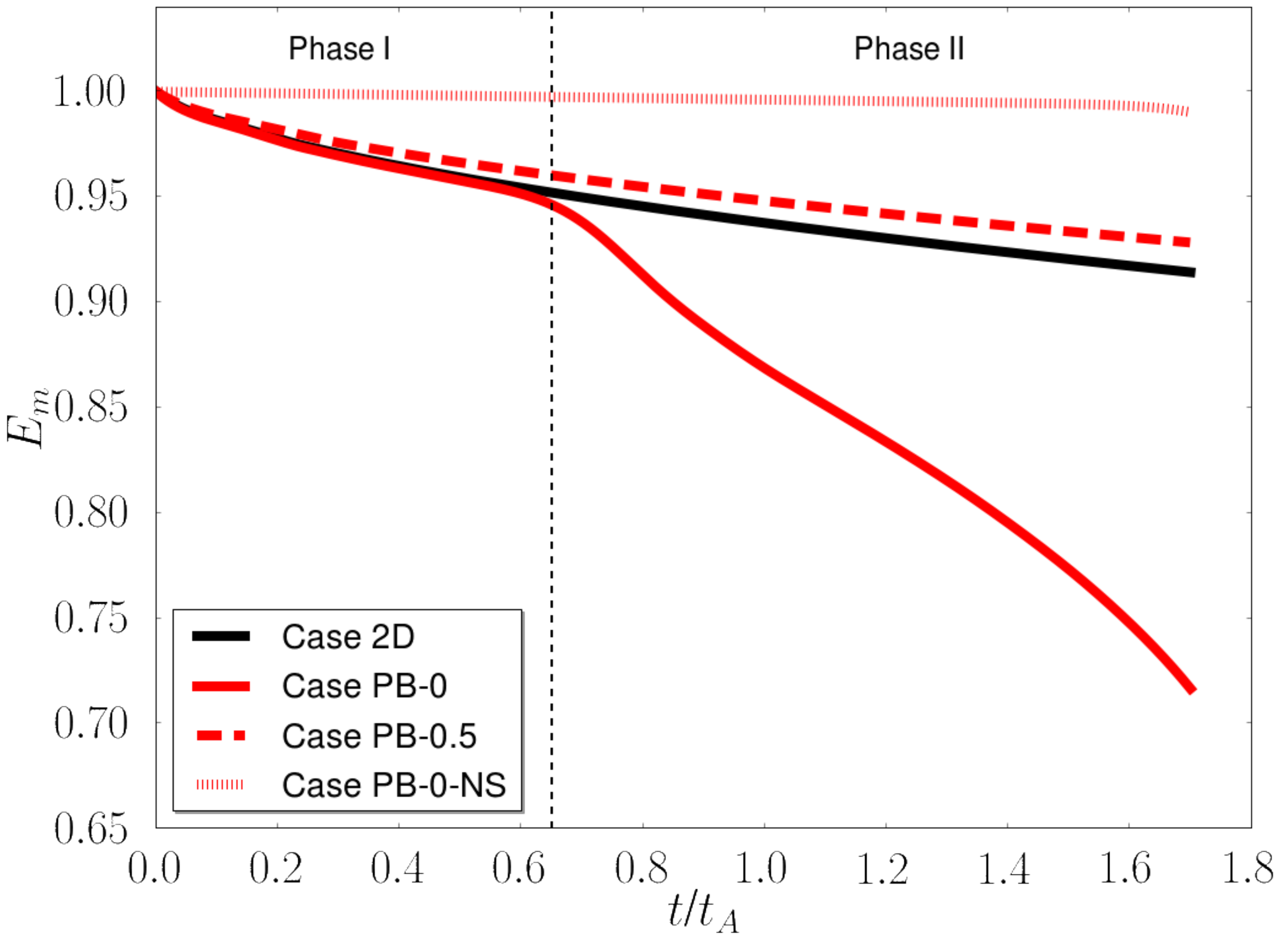}
 \includegraphics[width=0.48\textwidth]{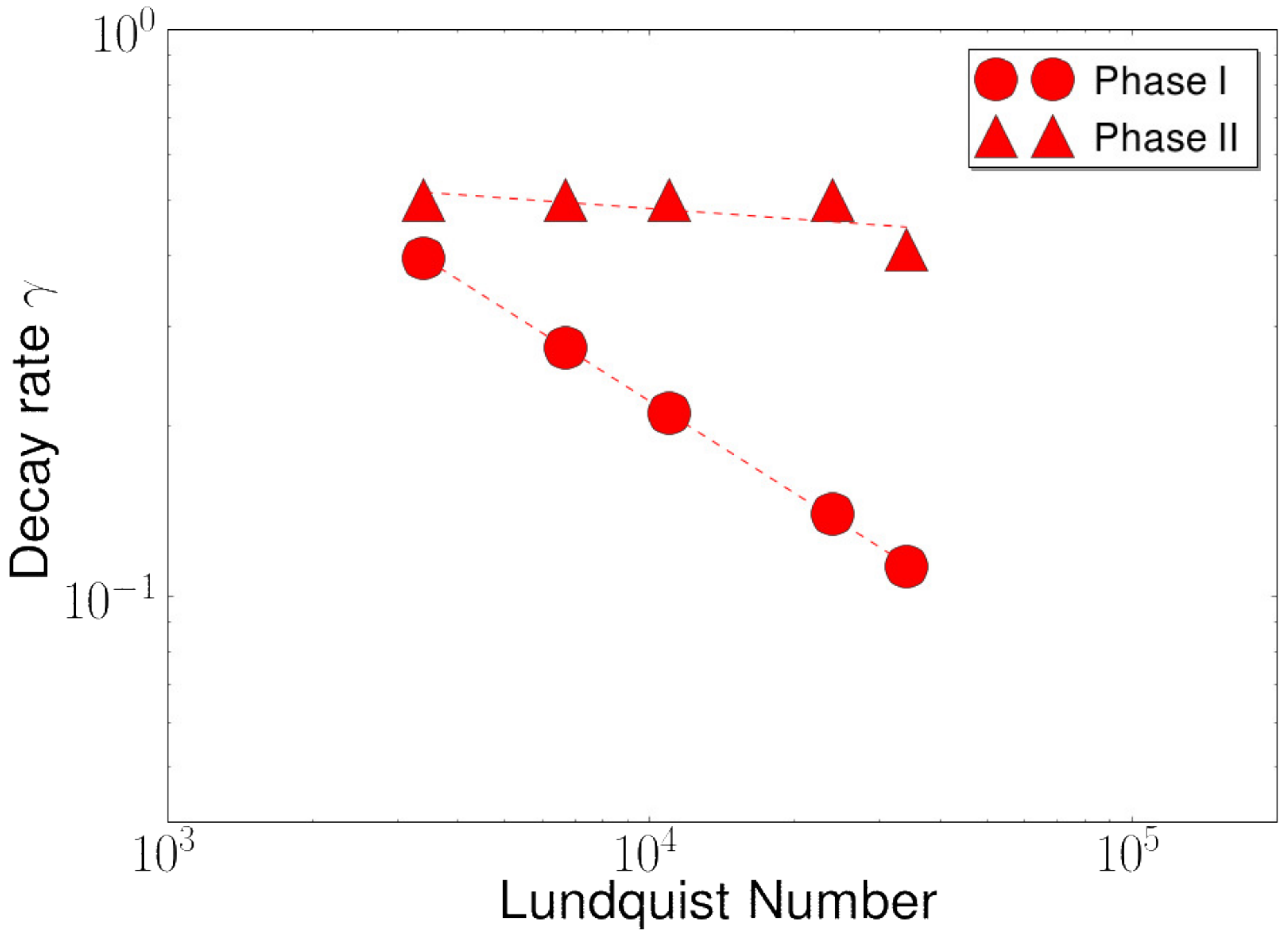}
 \caption{{\it Left}: Temporal evolution of the volume averaged magnetic energy normalized to the initial magnetic energy for case PB-0 (red solid line), case PB-0.5 (dashed line) and case PB-0-NS (dotted line), along with case two-dimensional (black).  For all cases the Lundquist number is $S = 2.4 \times 10^4$.  The black vertical line separates phase I and phase II for case PB-2D.  {\it Right}: Decay rate $\gamma$ of the magnetic energy as a function of $S$  for case PB-0. For each $S$ value, $\gamma$ is computed in {phase I} (circles) and {phase II} (triangles).  Note that the decay rate follows the Sweet-Parker scaling during phase I, and is nearly independent on $S$ in phase II. Reprinted from \cite{Striani_etal:2016} with permission of Oxford University Press.}
\label{fig:striani:vrec}
\end{figure*}

It is evident that the study in Kowal et al. (2019) support the expectation of the turbulent reconnection theory and disagrees with some of the conclusions in \cite{HuangBhattacharjee:2016}. In particular, the spectrum and anisotropy that are reported are consistent with the expectations of the MHD turbulence theory (see \S \ref{sec:turbulence2}). In addition, Kowal et al (2019) testifies that, in agreement with other earlier studies, the reconnection in 3D is different from that in 2D. In particular, the tearing modes that dominate the reconnection dynamics in 2D are replaced by perturbation induced by Kelvin-Helmholtz instability. In other words, the steady state self-driven reconnection is not driven by tearing/plasmoids. The issue why the results in \cite{HuangBhattacharjee:2016} are different requires further studies. One possibility is that the development of a true turbulent state requires more time that the current sheet was evolving in the simulations in \cite{HuangBhattacharjee:2016}. 

Studies of 3D self-driven turbulent magnetic reconnection with different initial conditions is important. In this respect, it is worth noting other studies of turbulence generating by stochastic
reconnection in different than the Harris current sheet configurations. For
example, turbulence developed through reconnection in a three-dimensional
cylindrical plasma column was investigated in \cite{Striani_etal:2016}. An
example visualization is shown in Figure~\ref{Fig:Striani}. The authors report a
fragmentation of the initial current sheet in small filaments, which leads to
enhanced reconnection rate, independent of the Lundquist number already at $S
\sim 10^{3}$. Moreover, they observe at the first phase, before the initial
perturbation develop turbulence, the classical Sweet-Parker dependence of the
reconnection rate on the resistivity. Once the turbulence grows, the
reconnection rate increases quickly and becomes independent of the Lundquist
number (see Fig.~\ref{fig:striani:vrec}). Apparently this study also shows that 2D and 3D self-driven reconnection are different.

\subsection{Testing turbulent reconnection with Hall-MHD code}
\label{Hall}

\begin{figure}
 \centering
 \includegraphics[width=0.48\textwidth]{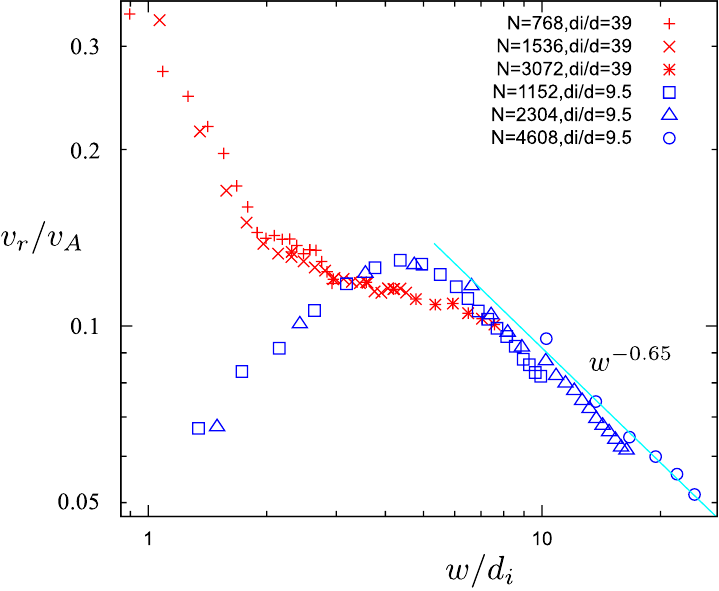}
 \caption{Normalized reconnection rate as a function of the reconnection region width. For relatively small width the reconneciton rate is $\sim 0.1 V_A$, but it decreases as the width increases. The value of $0.015 V_A$ reported in the earlier MHD simulations by \cite{Beresnyak:2017} is expected for $170 d_i$. Reprinted from \cite{Beresnyak:2018} under \href{https://creativecommons.org/licenses/by/3.0/}{CC BY} license.}
\label{fig:hall_recon}
\end{figure}

One may ask a question whether the magnetic reconnection that we described above for the MHD case both in the externally driven turbulence and in the case of self-driven turbulence is going to be different from the reconnection in actual plasmas. The LV99 theory gives a negative answer to this question and states that the microphysical plasma affects are not important at the scales much larger than the ion gyroradius or/and the ion skin depth $d_i=c/\omega_{pi}$, where $\omega_{pi}$ is the ion plasma frequency. This conclusion corresponds to both the common knowledge that the MHD turbulence properties that should not depend on the physics at the dissipation scale and to the accepted practise of representing magnetized flows on the scales much larger than $d_i$ using the MHD equations. However, the direct testing of the effects of plasma microphysics for magnetic reconnection on much larger scales is extremely challenging numerically. We address the approaches to this problem in the next section (\S \ref{sec:kinetic}). In this section we discuss simplified ways to perform this testing.

To test the LV99 prediction that plasma effects are not irrelevant for large scale turbulent reconnection  \cite{Kowal_etal:2009} performed the simulations of turbulent reconnection where the plasma effects were simulated through anomalous resistivity. No change of the reconnection was observed as the anomalous resistivity was varied. 

For the case of reconnection with self-generated turbulence \cite{Beresnyak:2018} performed Hall-MHD simulations with the resolution up to $2304\times 4608^2$. The simulations had different ratios of the ion-electron depth $d_i$ to the grid size $d$, namely, $d_i/d$ were chosen to be 9.5 and 39. The latter ratio approaches the ratio of the ion to electron skin depths, the former is used to obtain follow the scaling for larger current sheet widths. 

Figure \ref{fig:hall_recon} shows the dependence of the Hall-MHD reconnection rate $v_r$ as a function of the width $W$ of the reconnection layer. Similar to the earlier MHD reconnection simulations in \cite{Beresnyak:2017} the reconnection rate in the Hall-MHD simulations was measured as the increase of the reconnection layer thickness\footnote{A quantitative study of this measure of reconnection for the simulations with periodic boundary conditions and its relation to the LV99 theory is provided by \cite{Lazarian_etal:2015} in section 4(d) there.}
\begin{equation}
v_r=\frac{dw}{dt}.
\end{equation}
 It can be shown that the differences between the set-ups with different $d_i/d$ disappear when the width of the reconnection layer gets larger than $3.6 d_i$.

The current layer in the simulations was getting fully turbulent with magnetic fluctuations exhibiting whister turbulence (\S \ref{whistler}) below the $d_i$ scale. The original rate of reconnection in the set up was larger that in the case of MHD self-excited reconnection. However, this rate was decreasing with the increase of the width of the reconnection layer.  In other words, these findings support the notion that turbulent reconnection on the scales where one expects the MHD approximation to be valid \cite[see a discussion of this in][]{Eyink_etal:2011}, does not depend on the underlying microphysical plasma processes. This corresponds both to the physical intuition and to the accepted practice of using MHD for modeling the processes at scales much larger than the scales of the proton gyroradius. 

We note that while the main achievement of the study by \cite{Beresnyak:2018} was to demonstrate that the effects of Hall-MHD get negligible for turbulent reconnection at scales $\gg d_i$, the study is also of importance for the reconnection at sufficiently small scales. Such type of reconnection can take place e.g. in the Earth magnetosphere and the study demonstrates how the turbulent plasma-scale reconnection gradually transfers into the reconnection that is determined entirely by the MHD turbulence. We, however, expect that the reconnection will depend on the plasma $\beta$ and the corresponding studies would be useful.  

\subsection{MHD and Hall-MHD testing of turbulent reconnection: summary}

3D magnetic turbulent reconnection is notoriously difficult to study. The initial studies employed the externally driven turbulence and confirmed the analytical dependences for the reconnection rate predicted in LV99. Similarly, the violation of the conventional flux freezing predicted by the turbulent reconnection theory was also demonstrated in simulations with externally driven turbulence. 

How reconnection proceeds in the situation when no turbulence is initially present is the
 focus of the ongoing research on the spontaneous 3D reconnection. 
 This sort of reconnection is traditionally studied in 2D resulted and such studies in the MHD limit established the tearing/plasmoid reconnection model. However, the 3D simulations clearly show that for sufficiently large Lundquist numbers the reconnection layer in 3D gets turbulent and this makes the reconnection in 2D and 3D are very different as it is established in many independent studies. 3D studies are computationally costly, but, unlike 2D ones, they are astrophysically relevant. 
 
 Quantitative measurements of the properties of magnetic field in self-driven 3D reconnection are necessary. The first measurements of turbulence spectrum in the 3D reconnection layer, turbulence anisotropy, estimates of the growth rate of Kelvin-Helmholtz instability are very encouraging. These results support the predictions of turbulent reconnection theory. More studies of self-driven turbulent reconnection are necessary for exploring the phenomenon of flares induced by magnetic reconnection.

The work that shows that 3D self-driven reconnection naturally transfers to turbulent regime testifies that the turbulent reconnection is a generic type of astrophysical large-scale reconnection. At the same time, tearing and various plasma effects can be important only in special circumstances, e.g. in the transient regime before the system gets turbulent. For studies of 3D reconnection it is important to have high resolution. Otherwise, the numerical viscosity can suppress turbulence, completely distorting the real picture of astrophysical reconnection.

 A recent study of the self-driven turbulent reconnection with the Hall term confirms the prediction of the turbulent reconnection theory that small-scale plasma effects do not change the rate of large scale reconnection in turbulent fluids. This is a confirmation of a practically important conclusion in LV99 that the large scale processes in turbulent fluids can be simulated using MHD codes.

\section{Turbulent Reconnection with Kinetic Simulations}
\label{sec:kinetic}	

	
Kinetic simulations offer another perspective at understanding the 
role of turbulence on reconnection. 
As energy cascades to smaller scales,  the interactions between particles and
fields could become increasingly important, including many collisionless processes.
Fully kinetic simulations can track the motion of ions and electrons rigorously, 
this self-consistent ``closure'' is different from the resistivity and viscosity treatment 
often invoked in fluid studies. In fact, a direct comparison \citep[][]{Makwana2015}
between the compressible MHD and kinetic simulations of the turbulent cascade process has 
demonstrated the validity of using particle-in-cell (PIC) codes to study the nonlinear processes 
in the continuum limit at large scales. 

As described earlier in this review, 
it is instructive to quantify the field line diffusion behavior 
in kinetic simulations of reconnection. In addition to the often pre-existing turbulent
background and the collisionless tearing driving scales, 3D large-scale 
kinetic reconnection process also
contains several characteristic scales associated with ions and electrons that
naturally lead kinetic instabilities on small scales \cite[e.g.][]{BowersLi:2007, Daughton_etal:2011, Guo_etal:2015}, 
as well as hydrodynamic instabilities  at large scales \cite[see][]{Kowal_etal:2019}, can produce fluctuations on various scales, 
one might expect that the magnetic field diffusion at the microscopic plasma scales to be more complicated 
than its fluid counterpart. Below we describe some initial results based on 
one 3D PIC simulation of magnetic reconnection. It is very difficult to get fully turbulent fluid behavior for the outflow from the reconnection region. Therefore in what follows the study is focused on exploring the separation of the field lines, which is related to the Richardson dispersion  (see \S \ref{ssec: rdffvio}). 

\subsection{Numerical Method and Initial Set-up}
\label{subsec:num}
  
\begin{figure*}
\includegraphics[width=5in]{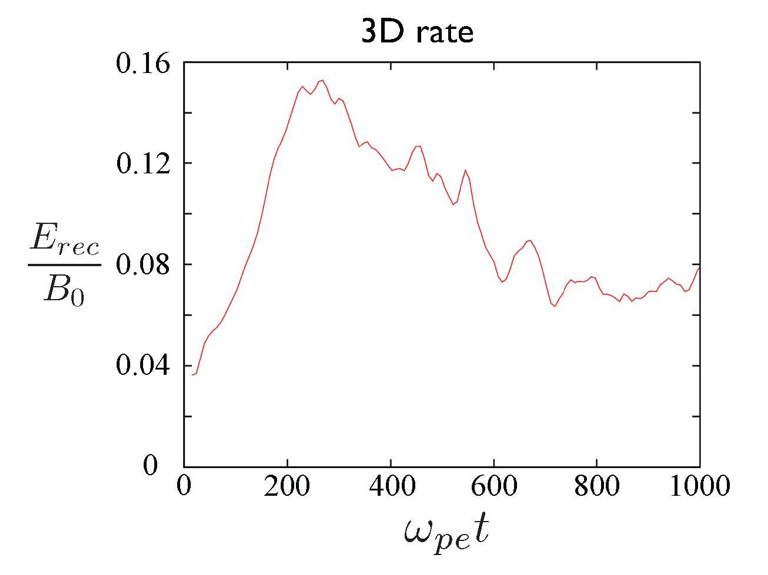}
\caption{The global reconnection rate derived from 3D reconnection. The initial turbulence
drastically shortens the early stage and causes the reconnection to start quickly.}
\label{F1}
\end{figure*}
	
\begin{figure*}
\includegraphics[width=6in]{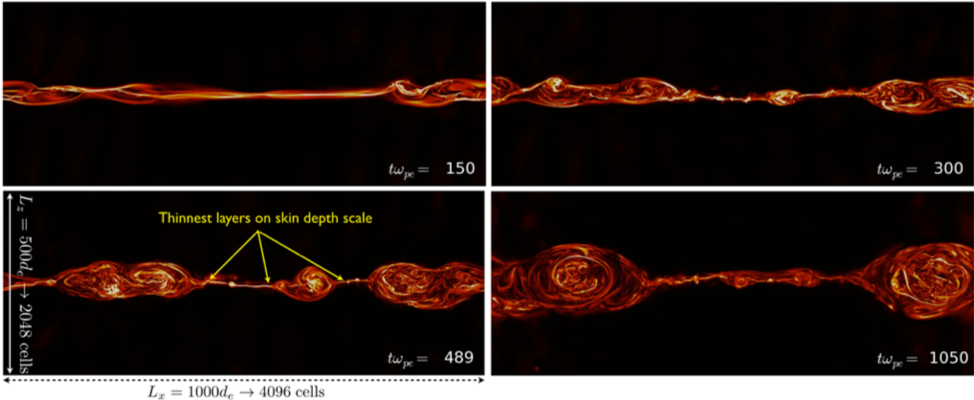}
\caption{Total current intensity $|J|$ evolution in the $\{x,z\}$ plane, 
showing the formation and merging of flux ropes in 3D and development of additional 
thin sheets. In addition to the initially injected turbulence, additional turbulence is produced
as the reconnection proceeds.}
\label{F2}
\end{figure*}

3D kinetic simulations were carried out using the VPIC code~\citep{2008PhPl...15e5703B}. 
VPIC is a first-principles electromagnetic charge-conserving code solving 
Maxwell's equations and the Vlasov equation in a fully relativistic manner. 
It has been carefully designed and optimized on recent peta-scale computers, 
enables multi-dimensional simulations with more than 1 trillion particles and 
5 billion numerical cells. Using the newly implemented Trinity machine at LANL, 
we have successfully run simulations that contains 5.2 trillion particles with 
about 17 billion cells. 

The simulation presented here \citep{guo2019} starts from a force-free current sheet with 
$\textbf{B}=B_0\mathrm{tanh}(z/\lambda)\hat{x} + B_0\mathrm{sech}(z/\lambda)\hat{y}$, 
where $B_0$ is the strength of the reconnecting magnetic field and 
$\lambda$ is the half-thickness of the current sheet. We assume a 
positron-electron pair plasma with $m_i/m_e=1$. Initially $\lambda= 12 d_e$ which
implies that the growth rate for collisionless tearing is quite small. Here, 
$d_e = d_i =c/\omega_{pe}=c/\sqrt{4\pi n_ie^2/m_e}$ is the electron/ion inertial length. 
The initial particle distributions are Maxwellian with uniform density $n_0$ 
and temperature $T_i=T_e=m_e c^2$. The magnetization parameter is 
$\sigma = (\Omega_{ce}/\omega_{pe})^2 = 100$. 
The domain size is $L_x\times L_z\times L_y=1000d_e\times 500d_e\times 500 d_e$. 
The grid sizes are $ N_x\times N_z\times N_y=4096\times2048\times 2048$. 
 The 150 particles per cell per species were used. For electric and magnetic fields, 
periodic boundaries along the $x$ and $y$ directions and 
perfectly conducting boundaries along the $z$-direction were employed. 
For particles, we employ periodic boundaries along the $x$ and 
$y$-direction and reflecting boundaries along the $z$-direction. 
The simulation lasts $\omega_{pe} t \approx 1050$, which is about the time
for an Alfv\'en wave to travel $L_x$. The timescale to traverse the current sheet
thickness, however,
$\omega_{pe} \tau = 2\lambda / c = 24$, is much shorter. 

Different from the earlier simulations, we inject an array of perturbations with 
different wavelengths with the total summed amplitude
$(dB/B_0)^2 \sim  0.1$ to initiate a background turbulence at the beginning of the simulation. 
(Injection of the initial perturbations follow the procedure described in
\cite{Makwana2015}.) These initially injected perturbations have wavelengths
longer than the initial thickness of the current sheet. 

\subsection{Analysis}
\label{subsec:ana}

Figure \ref{F1} depicts the overall global reconnection rate derived from this simulation,
and Figure \ref{F2} shows several snapshots of the total current density $|J|$ indicating the
evolution of many flux ropes (in 3D) and the development of subsequent thin current sheets. The reconnection rate is evaluated using the method described in 
\citep{Daughton_etal:2014},
where the mixing of electrons originating from separate sides of the initial current 
layer is used as a proxy to rapidly identify the magnetic topology and track the evolution of magnetic flux. 
It was found that, while the distribution and dynamics of reconnection X-points 
are strongly modified by the 
injected fluctuations and self-generated fluctuations due to secondary instabilities, 
the peak rate is on the order of 0.1 times of the Alfv\'en speed, 
consistent with previous 2D and 3D kinetic studies. The reconnection changes by about
a factor of 2 from its peak to its final quasi-steady rate. 

\begin{figure*}
\includegraphics[width=0.7\textwidth]{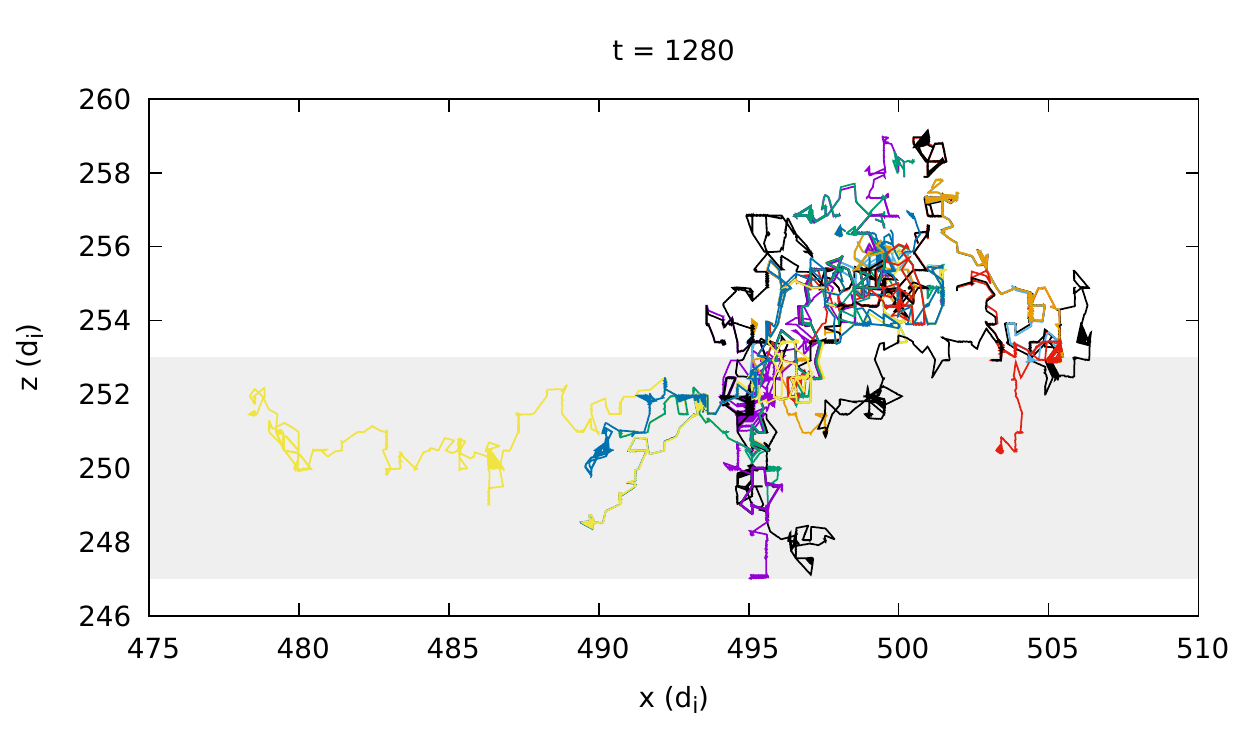}
\caption{Field line behavior traced in 3D but projected on the 
$\{x, z\}$ plane at $\omega_{pe} t = 136.3$, before the reconnection rate reaches its peak.
The initial locations are around $x=500$, $y=250$, $z=256~d_e$. Some magnetic field
lines have wandered through the current sheet, which is at $z = 250 \pm 3~d_e$ (the 
light shaded area).}
\label{F3}
\end{figure*}

To quantify the magnetic field diffusion during the reconnection,  
the following procedure was adopted: using an output at a particular time snapshot, 
the two field lines were traced by starting from a pair positions which are 
closely spaced (the typical initial separation is $s_0 = 0.01 d_i$). 
The separation between them $s (d_i)$ as
a function of field line path length $l (d_i)$ was calculated. Pairs at different
initial distances (above and below) from the initial current sheet layer center are chosen
randomly. 
Typically 1000 initial pairs to enhance the statistics were used. 
Figure \ref{F3} shows some sample magnetic field lines. There are 100 pairs
within $0.01 d_e$ of each other, whose starting 
locations are randomly chosen within a $2\times2\times2 d_e$ box that is $2 \lambda$ above
the middle plane of the initial current sheet at $x=500, y=250, z=256 d_e$. 
The time snapshot is at $t = 1280$ or 
$\omega_{pe} t = 126.3$, which is in the early stage of reconnection (before the peak,
cf. Figure \ref{F1}). These field lines are integrated for up to $500 d_e$ in total length
and we stop following them when they reach any simulation boundary. 
The initial turbulence is apparently causing a fraction of magnetic field lines to go cross
the current sheet. This could have interesting implications for initiating reconnection. 

\begin{figure*}
\begin{center}
\includegraphics[width=0.48\textwidth]{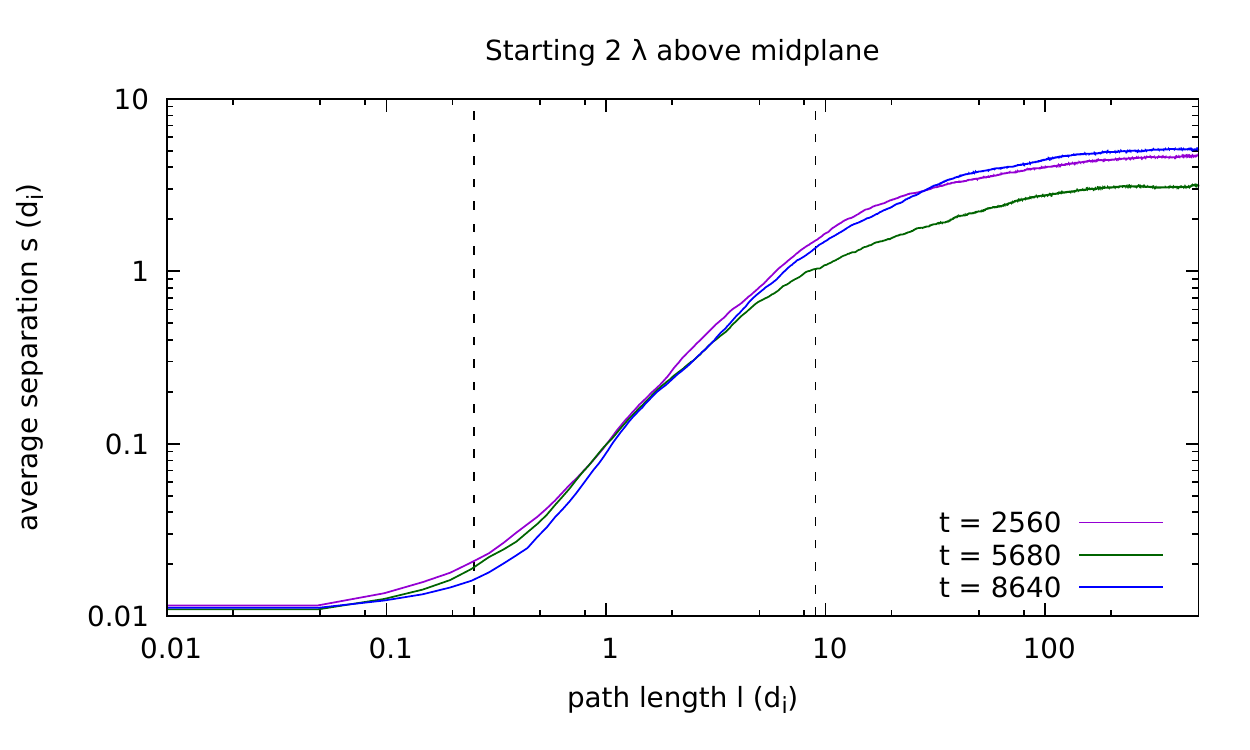}
\includegraphics[width=0.48\textwidth]{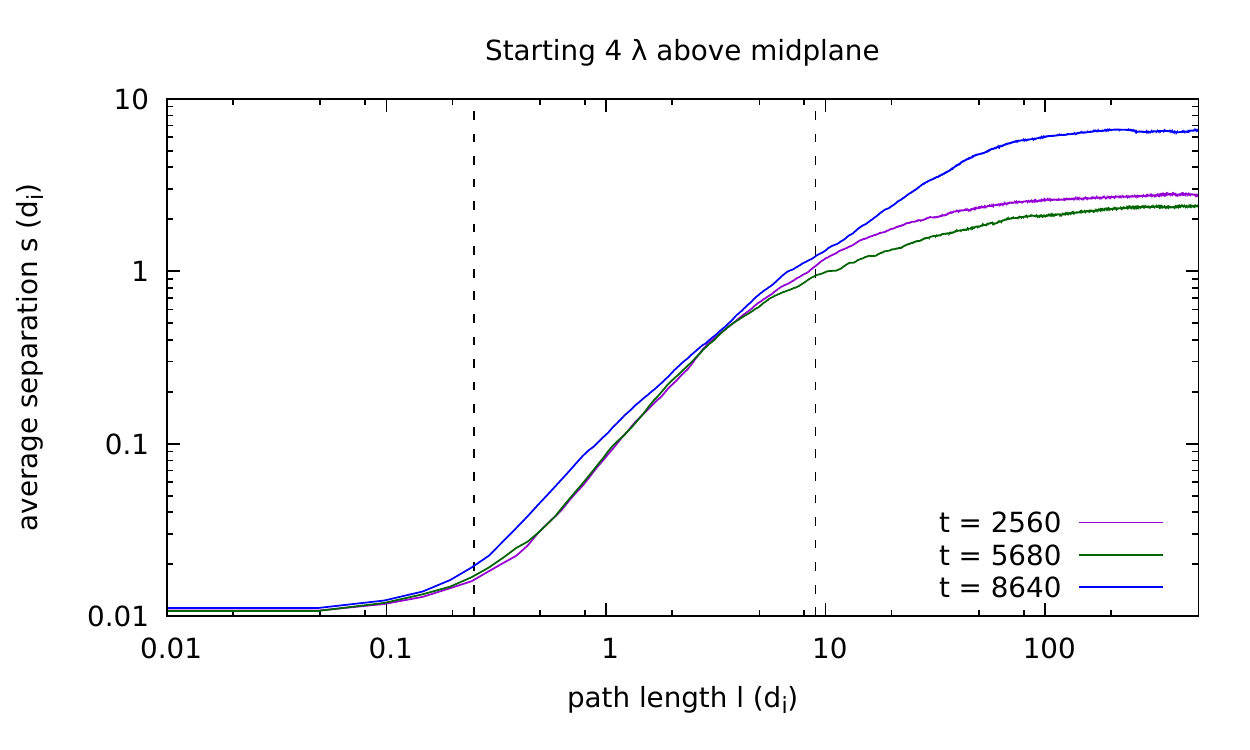}
\includegraphics[width=0.48\textwidth]{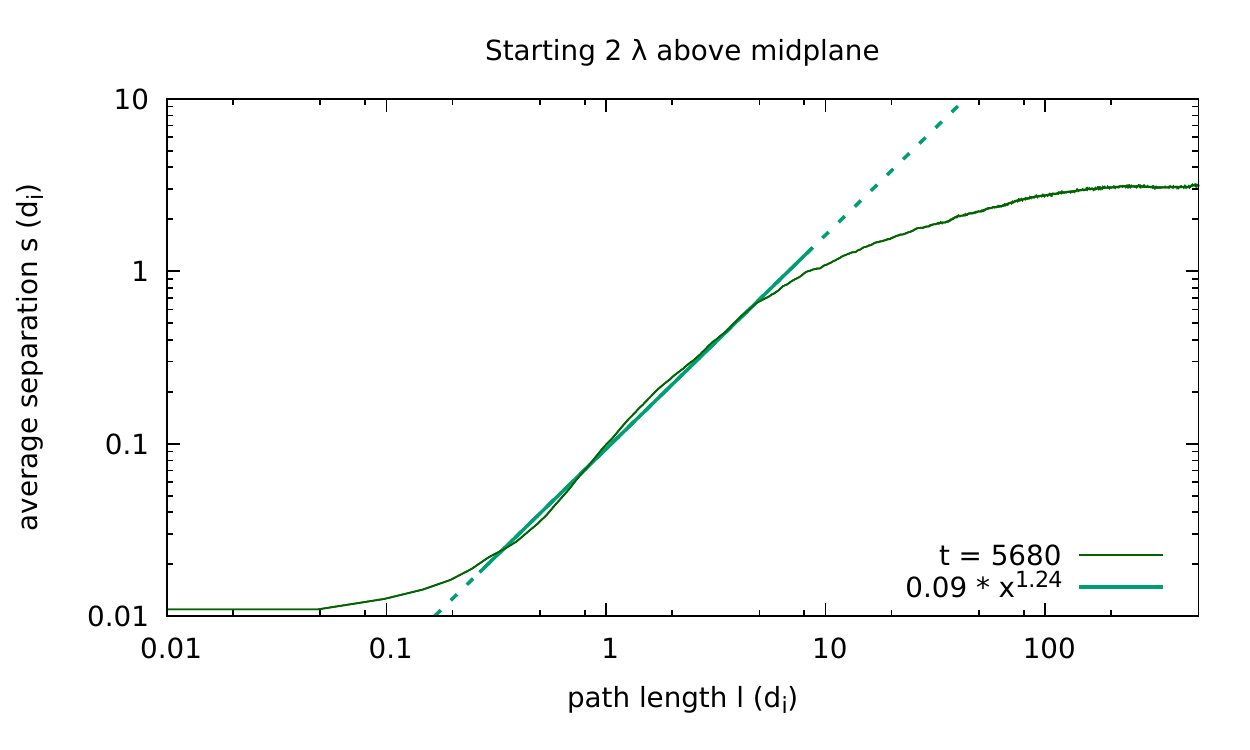}
\includegraphics[width=0.48\textwidth]{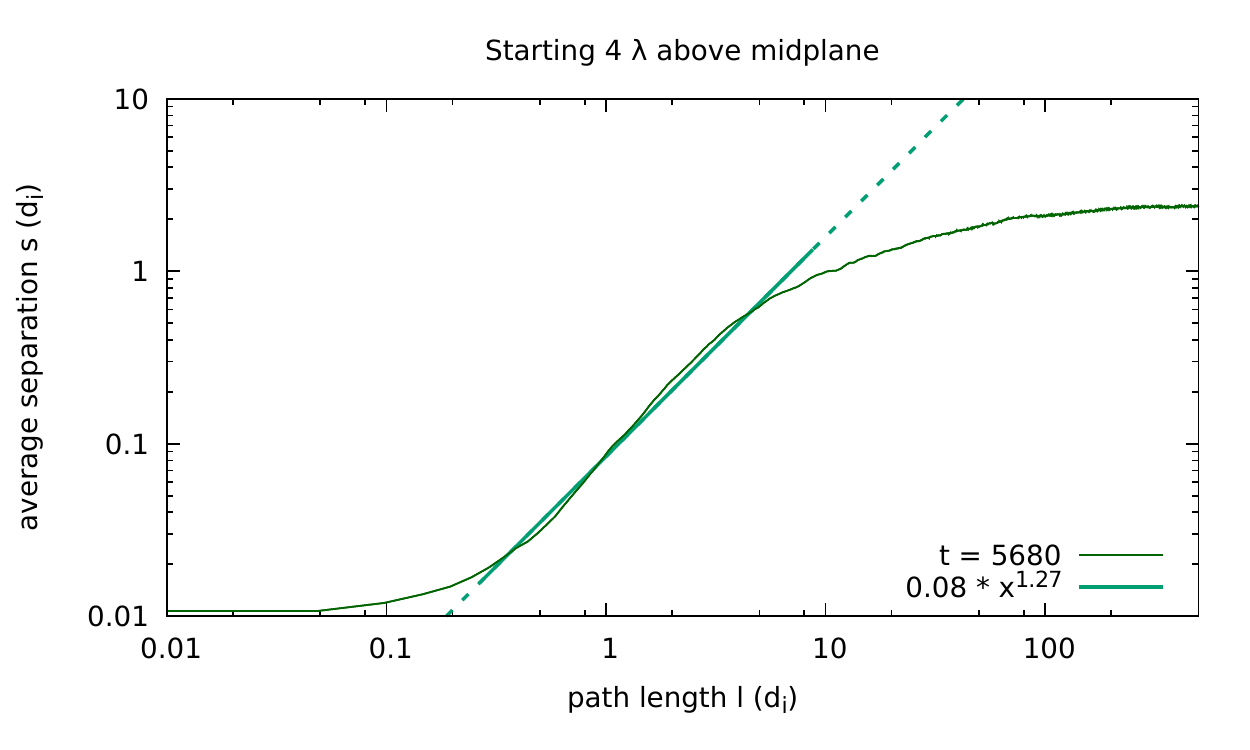}
\end{center}
\caption{Field line diffusion based on the 3D kinetic reconnection simulation. 
The top row shows the field line separation by starting from $2\lambda$ (left) 
and $4\lambda$ (right) above the 
initial current sheet central layer. The three curves in the plots are made using 
three time snapshots at $ t = 2560, 5680, 8640$, which correspond to 
$\omega_{pe} t = 252.6, 560.3,$ and $852.5$, respectively. They represent at-peak, 
post-peak and quasi-steady stages of reconnection (cf. Figure \ref{F1}), respectively.
The bottom row shows an approximate fit to the separation from two cases, 
depicting a power-law behavior over a part of the curves.}
\label{F4}
\end{figure*}

Figure \ref{F4} gives the details of magnetic field separation behavior. 
Overall, we find that magnetic field lines indeed separate from each other at a rate
faster than the regular diffusion process. The slopes from the power-law portion cluster
around $1.24 - 1.35$. In addition, we have chosen two initial heights 
($2 \lambda$ and $4 \lambda$) for tracing field lines and their behavior is quite similar. 
Furthermore, such analyses were done using three different time snapshots capturing the
at-peak and post-peak reconnection stages. 
Within the uncertainties of the fits, both the amplitudes and the exponents 
stay constant in time. 

The exponent ranging between $1.24 - 1.35$ suggests that magnetic field lines
exhibit super-diffusive behavior, consistent with the turbulent reconnection theory
as discussed earlier. The relatively narrow range of this exponent also suggests that
the initially injected turbulence might have strongly regulated the diffusion behavior,
though the turbulence produced by 3D kinetic reconnection could have impacted the
diffusive dynamics. More detailed analyses are needed to differentiate whether 
the injected and self-generated turbulence could lead to different field line diffusion behavior
and how they might interact with 
each other. 

\subsection{Kinetic simulations of turbulent reconnection: summary}
\label{subsec:sum}

By analyzing the magnetic field line behavior based on a large 3D kinetic simulation of
reconnection that is subject to an initial turbulence injection, it was found that: First, the overall
global reconnection rate reaches $0.1$ Aflv\'en speed, similar to previous kinetic simulations
without the initial turbulence injection. This is perhaps not surprising, 
given that the kinetic physics is probably still dominating the dynamics despite the relatively
large simulation box. Second, the presence of the initial turbulence, however, drastically
shortens the stage through which the peak reconnection rate is achieved. This suggests
the turbulence can accelerate the ``triggering'' process. 
Third, magnetic field lines exhibit super-diffusive 
behavior throughout the reconnection process, generally consistent with the 
super-diffusion theory (LV99, Eyink et al. 2011, Lazarian \& Yan 2014). as well as previous 3D MHD reconnection studies (see (see Kowal et al. 2017). 
More studies are needed to better understand the detailed dynamics as well as to 
delineate any possible different roles played by the initial background turbulence versus
the self-generated turbulence via reconnection.  

\section{Observations of Turbulent Reconnection}
\label{sec:observations}

Magnetic reconnection is notoriously difficult to study directly. In most cases one can see the consequences of the reconnection, e.g. heat release or energetic particles accelerated, rather than the actual picture of reconnection. For turbulent large scale reconnection the situation is really disadvantageous, as most of the in situ magnetic reconnection measurements have been done for the Earth magnetosphere where the thickness of the current sheets is comparable to the ion Larmor radius. In such settings one does not observe MHD-type turbulence and therefore magnetospheric measurements provide information about plasma reconnection dominating small scales rather than large scale reconnection relevant to most of astrophysical settings. Nevertheless, there have been both observations and measurements in the relevant parameter regime and those support the turbulent reconnection model.

\subsection{Solar Flares}
\label{sec:observations1}

Turbulent reconnection provides the most natural explanation for the variations of solar activity. Indeed, the reconnection speed that can be inferred from the observations of the magnetic activity in the solar atmosphere varies dramatically. This is difficult to account on the basis of any plasma reconnection process that prescribes a given reconnection rate. On the contrary, as we discussed earlier, the LV99 predictions suggest that the rate of magnetic reconnection vary with the level of turbulence and this level of turbulence can be affected by the reconnection itself. The latter provides a natural explanation of solar flares \cite[LV99,][]{LazarianVishniac:2009}. 

One can estimate the reconnection rates for solar reconnection assuming that turbulent velocity dispersion $U_{obs, turb}$  measured 
during the reconnection events is due to the strong MHD turbulence regime.  This is reasonable as we expect strong bending of magnetic field lines in the 
reconnection region. Therefore:
\begin{equation}
V_{rec}\approx U_{obs, turb} (L_{inj}/L_x)^{1/2},
\label{obs}
\end{equation}
Similarly, the thickness of the reconnection layer should be defined
as
\begin{equation}
\Delta\approx L_x (U_{obs, turb}/V_A) (L_{inj}/L_x)^{1/2}.
\label{delta_obs}
\end{equation}

For solar atmospheres the expected ratio of $\Delta$ to the proton gyroradius $\rho_i$ is $10^6$ or even larger. Therefore we expect to deal with the 
LV99 type of reconnection. 

The expressions given by Eqs.~(\ref{obs}) and (\ref{delta_obs}) should  be compared
with observations.  Using the corresponding data  in \cite{CiaravellaRaymond:2008} one can obtain the widths of the
reconnection regions  in the range from 0.08$L_x$ up to 0.16$L_x$
and the observed Doppler velocities in the units of $V_A$  of the
order of 0.1 ((see more in Lazarian et al. 2015). It is easy to see that these values agree well with
the predictions given by Eq.~(\ref{delta_obs}). In fact, the corresponding comparison was first done in \cite{CiaravellaRaymond:2008}
in order to observationally test the LV99 theory. However, the authors assumed that turbulence was isotropically driven, which 
was unrealistic for the solar flare case. While  \cite{CiaravellaRaymond:2008}
provide, within the observational uncertanities, a satisfactory quantitative agreement between their observations and the LV99 theory prediction, it is demonstrated in Lazarian et al. (2015) that this agreement can be further improved if the actual anisotropic driving is accounted for. 

A more recent study exploring the growth of the thickness of the reconnection layer of the microflaring loop is presented in \cite{ChittaLazarian:2019}. This study shows that the turbulent velocity measured using  Doppler-broadened lines and the width of the reconnection region are in a good agreement with the predictions of the LV99 theory.

A very distinct prediction in the LV99 theory is the spread of magnetic reconnection from an active reconnection region to the
adjacent reconnection regions. Indeed, the turbulence induced by the reconnection in one flaring region is expected to destabilize
the adjacent regions that are undergoing relatively slow reconnection and induce flares of reconnection in them. 
It is evident that one should not expect any similar effects in the models of plasma reconnection. 
 In \cite{Sych_etal:2009} and \cite{Sych_etal:2015} the authors observed the effect similar to one we just described. They explaining
quasi-periodic pulsations in observed flaring energy releases at an active
region above the sunspot and proposed that the wave packets arising from the
sunspots can trigger such pulsations via the LV99 mechanism. 

Plasmoids associated with tearing reconnection has also been observed \citep{ShibataTanuma:2001} and on the basis of our experience with self-driven
turbulent reconnection that we described in \S... we associate the plasmoid stage with the onset of turbulent reconnection.

\subsection{Reconnection in Solar Wind}
\label{sec:observations2}

\begin{figure*}
 \centering
 \includegraphics[width=4.5in]{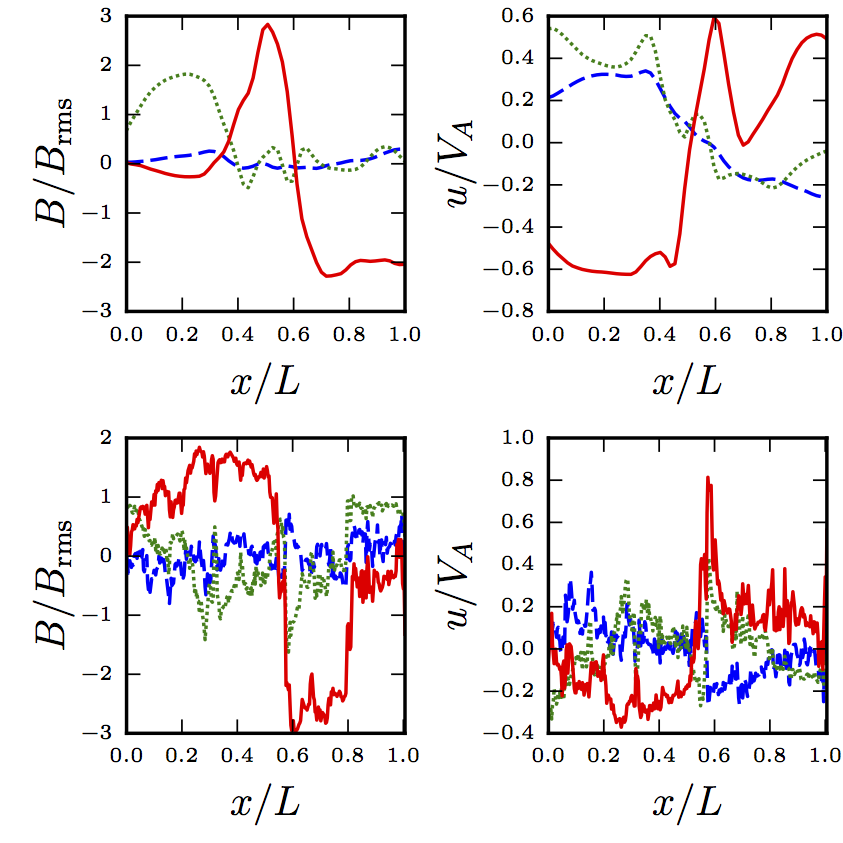}
 \caption{{\it(Upper panels)} Reconnection event in 3D MHD turbulence. \textit{(Lower panels)}. Reconnection event in High-Speed Solar wind.  The left panels show magnetic field components and the right panels velocity components. Both velocities and magnetic components are rotated into a local minimum-variance frame of the magnetic field. Red denotes the component of maximum variance. It is the reconnecting component.  The component of medium variance in green corresponds to guide-field direction. The minimum-variance direction depicted in blue is perpendicular to the reconnection layer.  From \cite{Lalescu_etal:2015}. }
\label{fig:lalescu1}
\end{figure*}

Reconnection throughout most of the heliosphere is expected to be similar to that in the Sun and therefore we do expect that the LV99 theory can be applicable to it.
However, for the measurement interpretation one should treat the small scale and large scale reconnection events separately. 
 For example, there are now extensive observations of strong narrow current sheets in the solar wind
\citep{gosling2012magnetic}. The most intense current sheets observed in the solar wind 
have widths of a few tens of the proton inertial length $d_i$ or
proton gyroradius $\rho_i$ (whichever is larger). The reconnection associated with these small-scale current sheets of
exibits exhausts with widths of a few hundreds of ion inertial lengths \citep{Gosling_etal:2007, GoslingSzabo:2008}.  Naturally, such
small-scale reconnection events  require collisionless physics for
their description \citep{Vasquez_etal:2007}. Within the LV99 picture these are small scale events representing the microphysics, but 
not much affecting the large scale reconnection. 

At the same time there are very large-scale reconnection events in the solar wind, often associated with interplanetary coronal mass
ejections and magnetic clouds or occasionally magnetic disconnection events at the heliospheric current sheet  \citep{Phan_etal:2009, gosling2012magnetic}.
These events have reconnection outflows with widths up to nearly $10^5$ of the ion $d_i$ and exhibit a prolonged, quasi-stationary regime
with reconnection lasting for several hours. Such large-scale reconnection is expected within the LV99 theory when very large flux-structures with
oppositely-directed components of magnetic field interact with each other in the
turbulent environment of the solar wind. The ``current sheet'' producing such large-scale reconnection in the LV99 theory contains itself many ion-scale,
intense current sheets embedded in a diffuse turbulent background of weaker current.

\begin{figure*}
 \centering
 \includegraphics[width=0.48\textwidth]{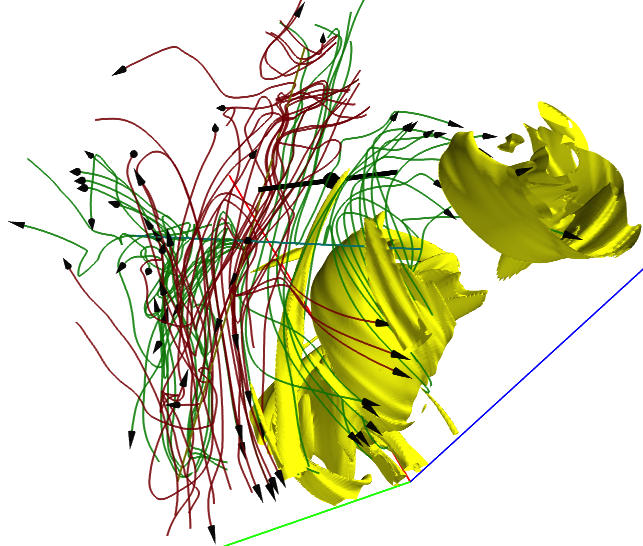}
 \includegraphics[width=0.48\textwidth]{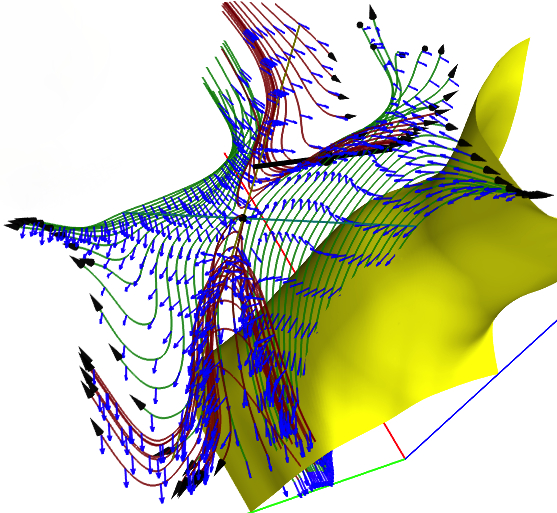}
 \caption{{\it Left}: Magnetic field iso-surface at half-maximum value 1.11 in given in yellow. B-lines sampled along the direction of minimal variance are given in green and in the directions of maximal variance are given in red.  {\it Right}: Same as figure on the left, but in terms of the averaged magnetic field.  The blue vectors on the field-lines are velocities measured in the local magnetic field frame. From \cite{Lalescu_etal:2015}. }
\label{fig:lalescu2}
\end{figure*}

The comparison between the reconnection events in High-Speed Solar Wind and those in MHD turbulence simulations was performed in \cite{Lalescu_etal:2015}. 
In the latter simulations
the magnetic field undergo Richardson diffusion as proven in \cite{Eyink_etal:2013} and therefore their turbulent reconnection corresponds to LV99 theory. Figure \ref{fig:lalescu1} shows that an inertial-range magnetic reconnection event in an MHD simulation is similar to those reconnection events observed in Solar Wind. The authors observed stochastic wandering of magnetic field lines in space, breakdown of conventional flux-freezing principle as a result Richardson dispersion of magnetic field lines, and also a broadened reconnection region containing several current sheets. The coarse-grain magnetic geometry in this study shown in Figure \ref{fig:lalescu2} appears as a large scale reconnection event in the solar wind, while the detailed picture of magnetic field lines exhibits a much more stochastic turbulent nature of the event.  \cite{Lalescu_etal:2015} concluded that the breakdown of conventional magnetic flux-freezing as an evidence of reconnection and also verified topology change in both fine-grained and coarse-grained magnetic fields. 

We believe that further studies of solar wind reconnection is a good way to test LV99 model as well as the related theories of flux freezing violation, slippage of magnetic field lines that we discussed in the review. In view of that we note, that LV99 better applicable to reconnection events more distant  from the Sun, because of the scales of the events increase. For
instance, reconnecting flux structures in the inner heliosheath could have sizes
up to $\sim$100 AU, much larger than the ion cyclotron radius $\sim10^3$ km  \cite[see][]{LazarianOpher:2009}.

\subsection{Heliospheric Current Sheet and Parker Spiral}
\label{sec:observations3}

The violation of flux freezing required by the model of turbulent reconnection can be also directly observable. For instance, \cite{Eyink2015} analyzed the data relevant to the region
associated with the broadened heliospheric current sheet (HCS), noticed its
turbulent nature and argued that LV99 magnetic
reconnection can explain the properties of the HCS. Naturally, more dedicated studies of the phenomenon are necessary.

The Parker's spiral model \citep{Parker:1958} of the interplanetary magnetic field remains as one of the applications of the magnetic flux-freezing. The deviations from this model have been observed, however. Using yearly averaged magnetic field strengths calculated from observations of Voyager 1 and 2, \cite{Burlaga_etal:2002} found a mean deviation of $-2\%$ between the observations and the Parker model. Yearly deviations were around $\pm 20\%$ with the maximum error (of about $-40\%$).  
In addition to magnetic field strengths studied by \cite{Burlaga_etal:2002}, an earlier work by \cite{Burlaga_etal:1982} considered also the magnetic field geometry which revealed notable deviations from the spiral model. This work found a different radial dependence for the radial and azimuthal magnetic field component from that predicted by the Parker model. These observations were based on the data collected by Voyager 1 and 2, while a comparison with earlier Pioneer 10 and 11 investigations during the less active solar activity period ($1972$ through $1976$) confirmed the spiral model modulo variability on timescales shorter than a solar rotation period. 

These early observations were recently confirmed by \cite{KO2012}. The extensive averaging used in the latter work eliminated the sizable fluctuations still observed in the daily averages of \cite {Burlaga_etal:1982}; see their Fig.1. \cite{KO2012} interpret their observations as due to a quasi-continuous magnetic reconnection, occurring both at the heliospheric current sheet and at local current sheets inside the IMF sectors. They present extensive evidence that most nulls of radial and azimuthal magnetic field components (i.e., potential reconnection sites) are not associated to the heliospheric current sheet. This interpretation is consistent with the results of \cite{Eyink2015} as magnetic slippage through the fluid as a result of pervasive turbulence in the near-ecliptic solar wind leads to a less tightly wound spiral and a stronger radial field strength than in the Parker model, as observed by \cite{KO2012}. This is intimately related to the reconnection diffusion proposed by  \cite{Lazarian:2005} and first studied numerically by  \cite{Santos-Lima_etal:2010}  (see more in \S \ref{ssec: redsf}).

In addition, \cite{Eyink2015} analyzed the data relevant to the region
associated with the broadened heliospheric current sheet (HCS), noticed its
turbulent nature and provided arguments on the applicability of LV99 magnetic
reconnection model to HCS. This seems to be a very promising direction of
research to study turbulent reconnection in action using in situ  spacecraft
measurements.

\subsection{Observational testing turbulent reconnection: summary}

Quantitative studies of magnetic reconnection from observations and in-situ spacecraft measurements are notoriously difficult. However, in the last 20 years a number of predictions of the turbulent reconnection theory has been successfully tested. For instance, the prediction of the relation between the level of turbulence and the width of the reconnection layer was confirmed with solar observations, the deviations of the Parker spiral and heliospheric current sheet were explained with the turbulent reconnection theory. Very importantly, the predicted effect of inducing magnetic reconnection by a Alfvenic perturbations coming from the neighoring reconnection event was observed.

The quantitative studies of the reconnection in Solar Wind showed that the reconnection events identified there have direct analogs in the reconnection events observed in numerical simuations of MHD turbulence. We believe that Solar Wind studies can provide the best testing ground for quantitative comparing the prediction of the theory of turbulent reconnection with the in situ measurements of reconnection. At the same time, the magnetospheric reconnection events cannot serve the same purpose as they happen in different regime that is dominated by plasma effects. The existing theory of turbulent reconnection cannot be directly applied for describing such reconnection events.

\section{Turbulent reconnection in special conditions}
\label{sec:special}

\subsection{Stochastic Reconnection For Large $Pr_m$}\label{ssVSR}

Partially ionized, turbulent regions of the Interstellar Medium (ISM) are often associated with a high magnetic Prandtl number $Pr_m$ \cite[see e.g.][]{Xu2017}. This regime of large $Pr_m$ is also prevalent in the laboratory fusion plasmas, different phases of star formation as well as in galaxies, protogalaxies and clusters which often are filled with hot and rarefied plasmas \citep{Jafari_etal:2018}. \cite{Lazarian_etal:2004} proposed a model for magnetic reconnection in partially ionized gases extending the arguments of stochastic reconnection (LV99) which was concerned with $Pr_m\simeq 1$. Another attempt, to study magnetic reconnection in large $Pr_m$ regime, was made by \cite{Lazarian_etal:2015} who considered the role of viscosity in the diffusion of field lines caused by neutrals in a partially ionized gas. To reach a better agreement with numerical simulations \citep{Kowal_etal:2009, Kowal_etal:2012a}, in a more recent work \cite{Jafari_etal:2018} have considered the effects of viscosity on fluid motions along magnetic field lines which is basically a generalization of stochastic reconnection (LV99) for viscous fluids.

Stochastic reconnection was proposed in LV99 for high $\beta$ plasmas with a magnetic Prandtl number of order unity, $Pr_m\simeq 1$. It predicts reconnection speeds of order the large scale turbulent eddy velocity. A modified version of stochastic model for $Pr_m>1$ in high $\beta$ plasmas, developed by \cite{Jafari_etal:2018}, showed that the width of the matter outflow layer and the ejection velocity are both unaffected by viscosity. Nevertheless, if $Pr_m>1$ viscosity may suppress small scale reversals near and below the viscous damping scale. In that case, there will be a threshold for the suppression of large scale reconnection by viscosity when the Prandtl numbers is larger than the root of the Reynolds number, that is $Pr_m>\sqrt{Re}$. For $Pr_m>1$ this leads to the spectral index $\sim-4/3$ for length scales between the viscous dissipation scale and eddies larger by roughly $Pr_m^{3/2}$. Here, we briefly review this regime of large magnetic Prandtl number studied in \cite{Jafari_etal:2018} \cite[see also][]{jafari2018introduction}.

The parallel wavelength at the viscous damping scale, $\lambda_{\parallel,d}=k_{\parallel, d}^{-1}$ can be obtained noting that $k_{\perp ,d}^{-1}\sim l_\perp(\nu/l_\perp u_T)^{3/4}$. The latter vertical damping scale comes from the Kolmogorov scaling for the viscous damping length scale, $l_d\sim \epsilon^{-1/4}\nu^{3/4}$ with $\epsilon\simeq u_T^3/l_\perp\simeq u_T^2V_A/l_\parallel$ where we have also used $l_\perp V_A=l_\parallel u_T$. If we define the injection velocity as $u_L=\sqrt{u_TV_A}$, then we can write $\epsilon\simeq u_T^2V_A/l_\parallel=u_L^4/(l_\parallel V_A)$. 

Analytical (see e.g., \cite{Lazarian_etal:2004}; \cite{Jafari_etal:2018}) and numerical (\cite{Cho_etal:2002}; \cite{Cho_etal:2003}; \cite{Schekochihin_etal:2005}) work indicate that a Prandtl number larger than unity will affect MHD turbulence in major ways. In MHD turbulence, there are two dissipation scales; the viscous damping scale $l_{d}$ and the magnetic diffusion scale, $l_{m}$. In fully ionized collisionless plasmas, both of these dissipation scales are very small. In partially ionized plasmas, they can be very different but always we have $l_{d}\gg l_{m}$ ([\cite{Jafari_etal:2018}]; [\cite{Lazarian_etal:2004}]). The kinetic energy is dissipated at viscous damping scale, $l_{d} \sim \nu^{3/4}\epsilon^{-1/4}$,
\begin{equation}
\label{89}
l_{d} \sim l _\perp\left ( {\nu\over l_\perp u_T}\right)^{3/4},
\end{equation}
where we have used the sub-Alfv\'enic energy transfer rate $\epsilon \sim u_L^4/(l_\parallel V_A)\simeq u_T^2V_A/l_\parallel$ ($u_L=\sqrt{u_TV_A}$) with critical balance condition (\cite{GoldreichSridhar:1995}; \cite{LV99}; \cite{Jafari_etal:2018}) $u_T l_\parallel\simeq l_\perp V_A$. However, at this scale magnetic energy does not dissipate as resistivity is smaller than viscosity by assumption. Instead, the magnetic damping scale is given by

\begin{equation}
\label{90}
l_{m} \sim l_\perp \left ( {\eta\over l_\perp u_T}\right)^{3/4}.
\end{equation}

The width of the  reconnection layers, created by the viscosity damped turbulence, can be determined using the eddy size at the viscous damping scale $l_{d}$. Consequently, the width of such a reconnection layer would be roughly $k_{\bot,d}^{-1}=l_{d}=\lambda_{\perp,d}$. The corresponding parallel length at the damping scale, $\lambda_{\parallel, \; d}$, can be obtained as

\begin{equation}\label{parlambda}
\lambda_{\parallel,d} \simeq  l_\parallel^{1/2}(u_TV_A)^{-1/4}\Big({V_A\over u_T}\Big)^{3/4}\nu^{1/2}.
\end{equation}

 Below $l_{d}$, the energy cascade rate $\tau_{cas}^{-1}$ is scale independent, and thus the constant magnetic energy transfer rate $b_l^2 \tau_{cas}^{-1}$ leads to $b_l\sim const.$;
\begin{equation}\label{power1}
E_b(k)\sim k^{-1}.
\end{equation}
The above expression is the magnetic power spectrum by the assumption that the curvature of field lines is almost constant $k_\parallel\sim const.$. This is consistent with a cascade driven by repeated shearing at the same large scale. Also, the component of the wave-vector parallel to the perturbed field does not change much and therefore any change in $k$ comes from the third direction \citep{Jafari_etal:2018}.
 
To find the kinetic spectrum, we may use the filling factor $\phi_l$ as the fraction of the volume that contains strong field perturbations with a scale $k^{-1}$ \citep{Lazarian_etal:2004}. Assuming a constant filling factor, $\phi_l\sim const.$, the kinetic spectrum is given by $E_v(k)\sim k^{-5}$. Otherwise, we find $E_v(k)\sim k^{-4}$ \citep{Jafari_etal:2018, jafari2018introduction}.

\begin{figure} 
\centering
\includegraphics[scale=.12]{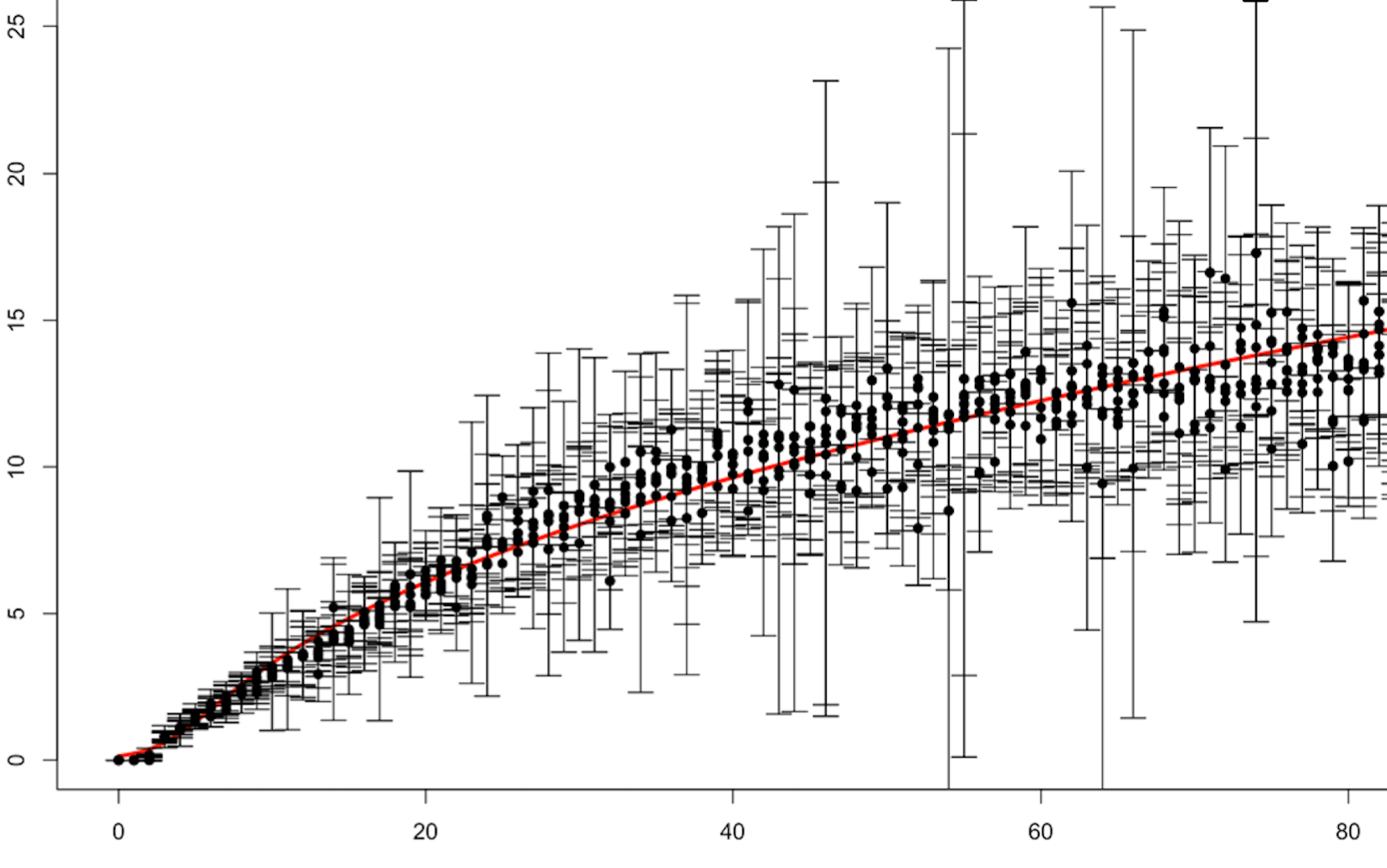}
\caption {Separation of magnetic field lines of \cite{Kowal_etal:2012a} compared with the analytical solution of eq.(\ref{general}) presented by the red line. The magnetic Prandtl number is of order $10$ \cite{Jafari_etal:2018}, \copyright~AAS. Reproduced with permission.}
\label{SeparationPlot}
\end{figure}

To find the reconnection speed in large magnetic Prandtl number regime, \cite{Jafari_etal:2018} used the following equation for the diffusion of field lines \cite[see also][]{jafari2018introduction}:

\begin{equation} \label{general}
{dy\over dx}\simeq {\eta/V_A\over y}+\frac{y}{ay^2+b y^{2/3}+c\sqrt{\nu}} .
\end{equation}
where $a=(l_\parallel/l_\perp^2) $, $b=l^{1/3}(V_A/u_T)^{2/3}$ and $c= l^{1/2}(u_TV_A)^{-1/4}(V_A/u_T)^{3/4}$ and finally $dx=V_A dt$. In order to compare to numerical data, we can use eq.(\ref{general}) with calibrated coefficients $a=a_1 l_\parallel/l_\perp^2$, $b=b_1 l_\parallel^{1/3}(V_A/u_T)^{2/3}$ and $c= c_1 l_\parallel^{1/2}(u_TV_A)^{-1/4}(V_A/u_T)^{3/4}$. The numerical factors $a_1 = 1.32$, $b_1 = 0.36$ and $c_1 = 0.41$ are constants introduced to match the solution to the available numerical data obtained by \cite{Kowal_etal:2012a} in which the injection power $P=u_T^2V_A/l_\parallel$ and the Alfv\'en speed $V_A$ are set to unity and the injection wavelength $k_{inj}=l_\parallel^{-1}=8$. For these simulations, numerically, we have $a=a_1/4$ and $b=b_1$ and $c=c_1$. Fig.(\ref{SeparationPlot}) shows a comparison of the solution of eq.(\ref{general}) with the numerical data obtained by \cite{Kowal_etal:2012a}. As mentioned before, we combine the solution $y=y(t)$  of eq.(\ref{general}) with the conservation of mass; $V_R=V_{eject} (\langle y^2 \rangle)^{1/2}/L_x$ where $V_{eject}$ is the ejection speed of plasma out of the reconnection zone. 

For $Pr_m<1$, one expects $V_{eject} \sim V_A$. However, for $Pr_m>1$, we may expect that the outflow is affected by viscosity if the viscous timescale is shorter than the outflow timescale, $\langle y^2 \rangle /\nu  < L_x/V_{eject}$. Using $\langle y^2\rangle/l_\perp^2 \sim L_x/l_\parallel $, $V_A l_\perp\simeq l_\parallel u_T$ and defining the Reynolds number as $Re=l_\perp u_T/\nu$ where $u_T$ is the velocity characteristic of largest strongly turbulent eddies, one finds
\begin{equation}\label{c_1}
Re< {V_A\over V_{eject}}.
\end{equation}

This condition can hardly be satisfied, hence viscosity is not expected to affect the outflow speed \citep{Jafari_etal:2018}. But it can still affect the reconnection rate by changing the outflow width. In fact, viscosity can affect reconnection process either by hindering the outflow from the reconnection zone or by narrowing the reconnection channel width \citep{jafari2018introduction}. If we assume that the perpendicular separation is completely inside the turbulence inertial cascade, then viscosity cannot not play any important role. That is if $y> l_{d}$ even for $x=L_x$, viscosity's effect will be negligible. In the dissipative range, the diffusion equation (\ref{general}) has the solution
\begin{equation}
\langle y^2 \rangle \simeq \Big({\eta\over V_A}  \Big)\lambda_{\parallel,d} \exp{\Big({L_x\over \lambda_{\parallel, d}} \Big)}.
\end{equation}
The condition $y \lesssim l_{d}$ then becomes 
\begin{equation}
 Re^{1/2} <\Big({ l_\parallel\over L_x} \Big) \Big[1+\ln ( Pr_m)\Big].
 \label{limit}
\end{equation}
where we have used $Re=l_\perp u_T/\nu$. Consequently, the outflow width, $y=\langle y^2\rangle^{1/2}$, is unaffected unless the magnetic Prandtl number is exponentially larger than the Reynolds number.

\begin{figure}
\begin{centering}
\includegraphics[width=0.48\textwidth]{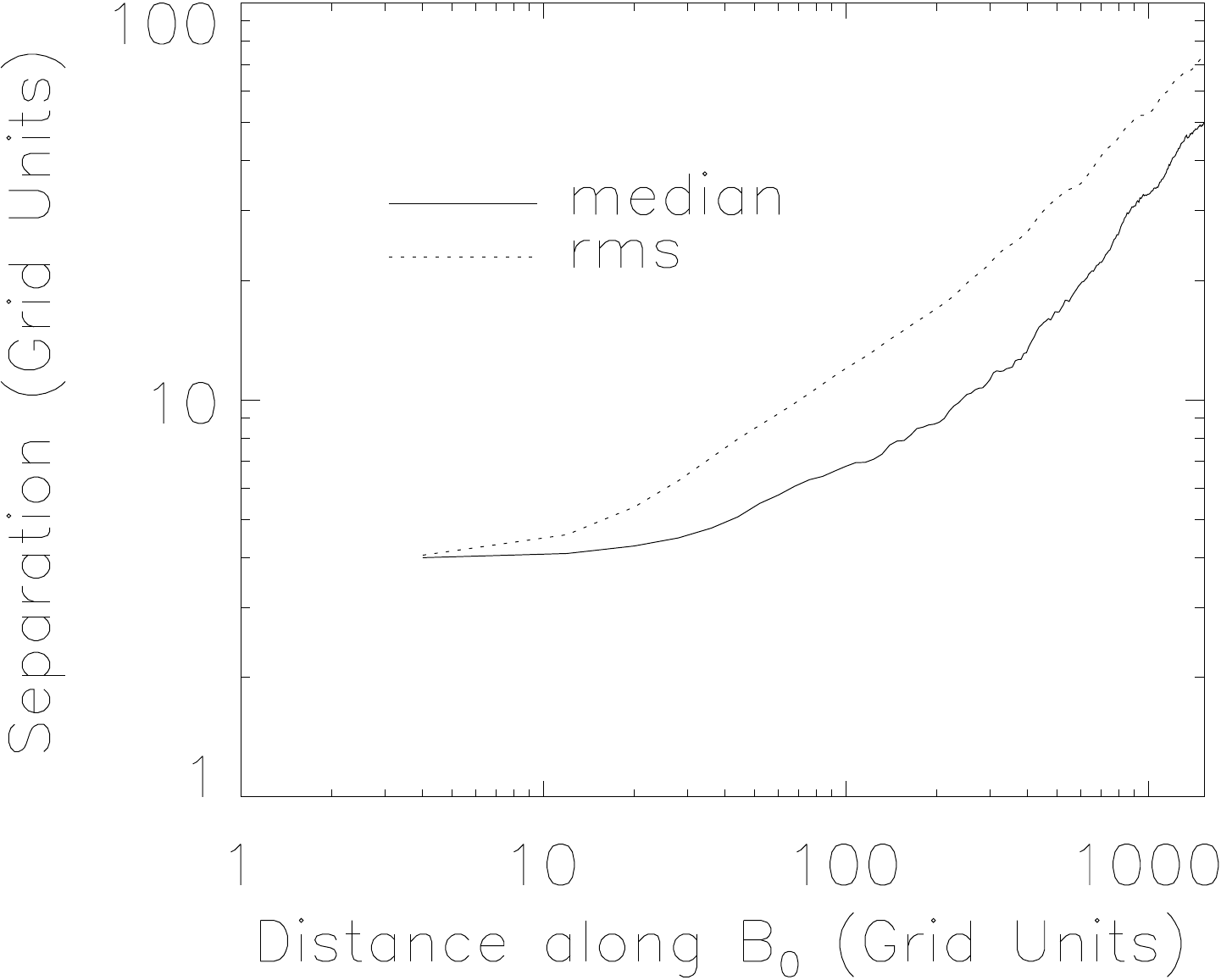}
\caption {The rms and median separation of magnetic field lines plotted versus the distance parallel to the mean field in the viscosity damped turbulence using a $384^3$ grid by \cite{Lazarian_etal:2004}. The median tracks the mean at a slightly smaller amplitude consistent with the exponential growth. This exponential growth in the dissipative regime can also be seen looking at the analytic result given by eq.(\ref{viscosity-damped}). From \cite{Lazarian_etal:2004}, \copyright~AAS. Reproduced with permission.}
\label{fig: lazrms}
\end{centering}
\end{figure}

For $Pr_m\gg 1$, one generally expects that viscosity will slow down the reconnection speed. 
Fig. \ref{SeparationPlot}
compares the general solution of eq.(\ref{general}) with the numerical simulations performed by \cite{Kowal_etal:2009} and  \cite{Kowal_etal:2012a}. Below the damping scale, eq.(\ref{general}) yields 
\begin{equation} \label{viscosity-damped}
y^2\sim {\eta\over V_A}\lambda_{\parallel,d}\Big[ \exp{(x/\lambda_{\parallel,d})}-1\Big],
\end{equation}
which, for $x\gg \lambda_{\parallel,d}$, becomes Lyapunov exponential growth whose exponent is inversely proportional to the square root of the viscosity. Fig.
\ref{fig: lazrms}
illustrates this exponential growth of field line separation in a set of numerical simulations preformed by \cite{Lazarian_etal:2004}.

Another way that viscosity may affect reconnection is through changing the local small scale reversals. It is a very stringent condition for viscosity to affect the large scale outflow. This may imply that viscosity will almost never be important, thus the question arises that if the small scale effect of viscosity can be significant.

Another effect of viscosity is associated with topologically distinct local reconnections at small scales. In stochastic reconnection, the global reconnection speed is the collective effect of numerous local reconnection events which occur simultaneously. This sets an upper limit on the global reconnection speed \cite{LV99}:
\begin{equation}
V_{rec}< N_{eddy} v_{rec,eddy},
\end{equation}
where $v_{rec,eddy}$ is the reconnection speed within an individual eddy and $N_{eddy}$ is the number of independent eddies along a field line. \cite{Jafari_etal:2018} obtained the following condition for viscosity to have an appreciable effect on local reversals:

 \begin{equation}
V_{rec} < v_{rec,eddy} \left({\eta\over\nu}\right)^{1/2} N_{eddy}\simeq u_T { Re^{1/4} Pr_m^{-1/2}\over 1+\ln(Pr_m)}{L_x\over l_\|}.
\label{plimit}
 \end{equation}

The implication is that for  $Pr_m=1$, the above condition is not an interesting limit. however, for $Re$ not very large and resistivity appreciably less than viscosity, it can be the controlling limit on reconnection (and not the outflow zone width).  In particular the dividing line between viscosity dominated reconnection and unimpeded reconnection is no longer overwhelmingly in favor of unimpeded reconnection. Therefore, the critical magnetic Prandtl number does not scale exponentially with the Reynolds number but close to its square root. The simple model outlined above, however, is based on the assumption of a large degree of intermittency below the Kolmogorov scale. This may not be very realistic. In addition, numerical simulations of the viscous regime have currently low resolutions and their results are affected by a damping scale very close to the eddy scale. The small Reynolds numbers accessible by numerical simulations at present may be the reason why they hardly show any effect associated with viscosity \citep{Jafari_etal:2018}.

\begin{figure*}
\includegraphics{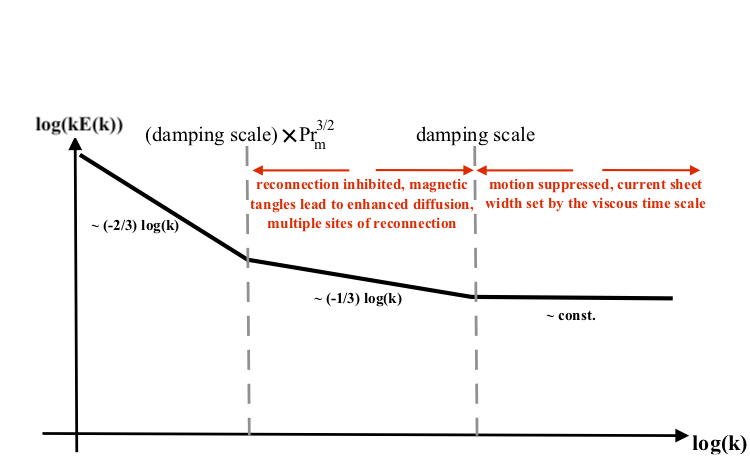}
\centering
\caption {\footnotesize {Magnetic fluctuation spectrum $E(k)k$ \cite{jafari2018introduction}: Below viscous damping scale, the hydrodynamic motions are suppressed and we have a flat spectrum for magnetic fluctuations. The width of the current sheet is controlled by viscosity in this regime. Above the viscous damping scale, and below the scale set by $(damping\;scale)\times Pr_m^{3/2}$, reconnection is inhibited while diffusion is enhanced by magnetic tangles. }}
\label{x}
\end{figure*} 

The magnetic structures are sheared below the viscous damping scale with a scale independent energy cascade rate and we find a power spectrum for the magnetic field as $E_b(k)\sim k^{-1}$: eq.(\ref{power1}). It follows that more power concentrates on the small scales than predicted by the Goldreich-Sridhar spectrum \citep{GoldreichSridhar:1995}. With a viscosity of order, or slightly larger than the resistivity, the  the outflow width remains independent of the small scale physics, which is similar to the LV99 model. Even with viscosity appreciably larger than resistivity, the width of the outflow region is unaffected unless the magnetic Prandtl number is exponentially larger than the Reynolds number. The viscosity is even less important when the current sheet instabilities dominate over external turbulence. For significantly large Prandtl numbers, i.e., of order $\sqrt{Re}$, the magnetic field perturbations cannot relax and a flat magnetic power spectrum extends below the viscous damping scale; see Fig.\ref{x}.

\subsection{Turbulent reconnection in relativistic fluid}
\label{sec:special2}

The relativistic turbulent simulations in 
\citet{Takamoto_etal:2015}
(henceforth TIL15) 
confirmed a general similarity between the relativistic reconnection and the non-relativistic one. 
They demonstrated that the turbulent reconnection speed can be as large as $0.3 c$,
which thus enables highly efficient 
conversion of magnetic energy into kinetic motions and particle acceleration. 
The numerical results in TIL15 can be understood on the basis of the expression LV99 expression for turbulent reconnection. This expression can be written as
\begin{equation}
V_{rec}\approx V_A\min\left[\left({L_x\over l}\right)^{1/2},
\left({l\over L_x}\right)^{1/2}\right] \left(\frac{u_L}{V_A}\right)^2,
\label{recon}
\end{equation}
where for sub-Alfvenic turbulent the energy cascading is given by expression (LV99):
\begin{equation}
\epsilon_{inj}\approx u_L^4/ lV_A.
\label{eps}
\end{equation}

However, for the relativistic case  
one should take into account both the density change in the relativistic plasma and the modification of turbulence properties in the relativistic regime. 
The former can be obtained from the energy flux conservation requirement.
With both analytical considerations and numerical simulations provided in TIL15, there is 
\begin{equation}
\frac{\rho_s}{\rho_i}\sim 1-\beta \left(\frac{u_L}{c_A}\right)^2,
\end{equation}
where $\beta$ was found in numerical simulations to be a function of plasma relativistic magnetization parameter $\sigma=B^2/4\pi\rho h c^2$, where $h=1+(4 (p/\rho c^2)$ is the specific enthalphy of the ideal relativistic gas, while  $\rho$ and $p$ are, respectively, density and pressure. 

The change of the outflow region $\Delta$ was shown to correspond to the original LV99 calcualtion 
but with injection $\epsilon_{inj}$ reduced compared to the value in non-relativistic case. 
This corresponds to the transfer of larger fraction of energy to the fast modes which are subdominant in inducing magnetic field wandering, as indicated by simulations in 
\citet{Takamoto2016,Takamoto2017}
With these modifications, the corresponding expression of the turbulent reconnection can be written as
\begin{eqnarray}
V_{rec, relativ.} & \approx & V_A \left(\frac{\rho_s}{\rho_i}\right) \left(\alpha \epsilon_{inj} l {V_A^3}\right)^{1/2} \nonumber \\
& \times & \min\left[\left({L_x\over l}\right)^{1/2},
\left({l\over L_x}\right)^{1/2}\right],
\label{relativ}
\end{eqnarray}
where $\alpha<1$ is the factor accounting for the decrease in the fraction of magnetic energy that induces magnetic field wandering. 

\begin{figure*}
\centering
\includegraphics[width=0.48\textwidth]{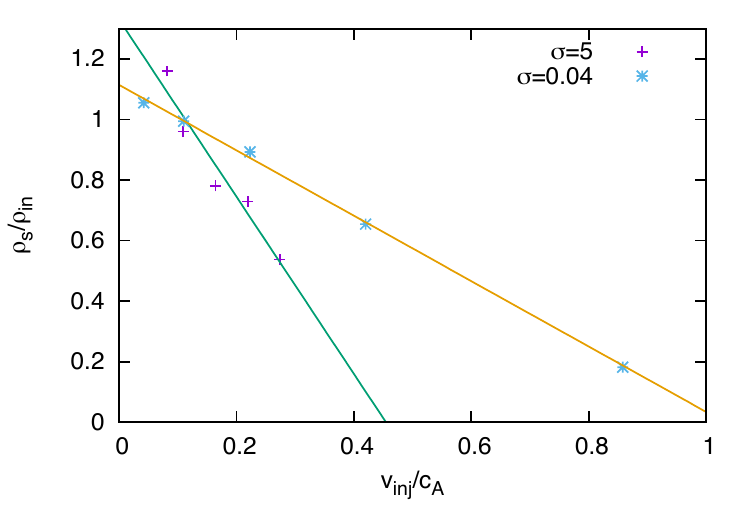}
\includegraphics[width=0.48\textwidth]{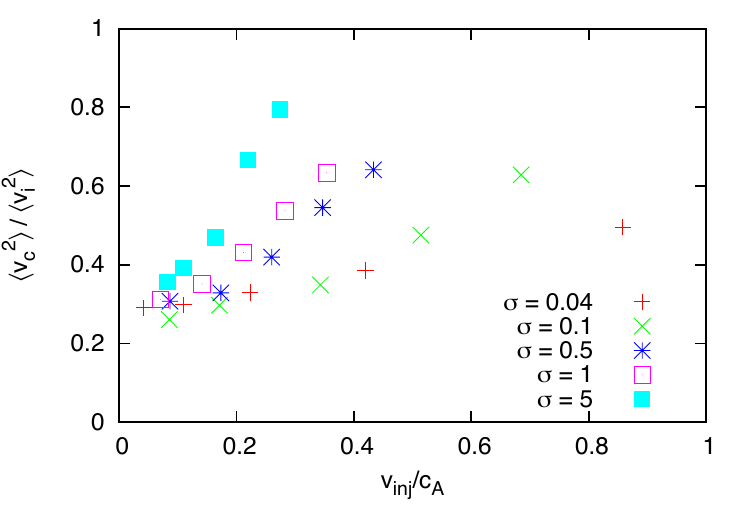}
\caption{{\it Left}: Variations of plasma density in relativistic reconnection. {\it Right}: Generation of compressible modes in relativistic reconnection. From \cite{Takamoto_etal:2015}, \copyright~AAS. Reproduced with permission.}
\label{fig25}
\end{figure*}

Figures \ref{fig25} illustrate the results of the numerical calculations in TIL15. These study shows that with the suggested modifications the LV99 theory reflects the major features of 
turbulent relativistic reconnection. 

It is evident from Eq. (\ref{relativ}) that the theory of relativistic turbulent reconnection does require further development in order to decrease the uncertainties related to magnetic turbulence in relativistic fluids. 
For the time being, it is important for our further discussion that qualitatively relativistic turbulent reconnection is similar to its non-relativistic counterpart. 
The existing numerical simulations in TIL15 provide us with guidance for studying the reconnection in GRBs. 
In particular, it is clear from the simulations that in high $\sigma$ flows, 
the turbulent reconnection speed only slowly changes with the injection velocity and does not depend on the guide magnetic field,
i.e., the common field component shared by the two reconnected fluxes.
The injection scale of the turbulence induced through the kink instability is likely to be comparable to the scale of the magnetic field flux tubes, 
i.e. $l\sim L_x$. 
In this situation, by extrapolating the results in TIL15, one can claim that the reconnection rate is larger than $0.1 c_A$, 
where the relativistic Alfv\'en speed $c_A$ is very close to the light speed.

\subsection{Whistler turbulence and reconnection}
\label{whistler}

In a turbulent environment, magnetic field wandering also takes place at scales below the proton Larmor radius. For instance, observations of the Earth's magnetosphere are frequently on scales less than the ion gyro-radius. At such scales electron MHD or the EMHD approximation can be relevant for describing turbulence and reconnection \cite[see][]{Mandt_etal:1994}.

The study of EMHD goes back more than 30 years \citep{Kingsep_etal:1987} and the turbulent spectrum  $E(k)\sim k^{-7/3}$ was explored more than 20 years ago \citep{Biskamp_etal:1996, Biskamp_etal:1999}. A quantitative description of anisotropy in EMHD turbulence anisotropy was studied by \cite{ChoLazarian:2004, ChoLazarian:2009}. 

We can derive the required scalings in analogy the approach that we discussed in \S \ref{sec:turbulence2} for the derivation of MHD turbulence scalings. The EMHD approximation applies below the ion inertial length $d_i = c/\omega_{pi}$, where $c$ is the speed of light and $\omega_{pi}$ is the ion plasma frequency. At these scales the ions constitute a smooth motionless background and electron flows carry all the current. (The ion inertial length is not the same as the ion Larmor radius, but differs by a factor of $\beta^{-1/2}$, where $\beta$ is the plasma $\beta$. Here we will neglect the difference.) As a result 
\begin{equation}
    {\bf v_e}=-\frac{{\bf J}}{ne}\sim \nabla \times {\bf B},
\end{equation}
where $n$ is the electron density and $e$ is the charge, respectively. In this limit the magnetic field is dragged by the motions of the electrons.

The magnetic energy cascades to smaller scales at a rate:
\begin{equation}
    {\cal E}=\frac{b_{l}^2}{t_{cascade}}=const,
    \label{whist}
\end{equation}
where $b_l$ is the magnetic perturbation at the scale $l$ and $v_l$ is the electron velocity. Finally $t_{cascade}\approx l/v_l$ is the cascading time at the scale $l$. Here the scale $l$ refers to the scale perpendicular to the large scale field. We note that Whistler waves have comparable amounts of energy in components parallel to and perpendicular to the mean field and the parallel current is comparable to the perpendicular current. The parallel wavevector component, $k_\|$, will in general be much smaller than the perpendicular component,$k_\perp$, but this does not effect estimates of the coherence time.  The energy cascade argument depends only on the perpendicular component. 
Substituting $t_{cascade}$ in Eq. (\ref{whist}) one gets $b_l\propto l^{2/3}$. 
The magnetic field wandering can be described as
\begin{equation}
\frac{d}{ds}y\sim \frac{b_l}{B_0},
\end{equation}
where $B_0$ is the mean magnetic field, $s$ is the distance alone the field lines and $y$ is the rms perpendicular distance between two field lines. Substituting the scaling of $b_l$ and using $y$ for $l$ one gets the perpendicular deviation of magnetic field lines
$y\sim s^3$. This is a larger exponent than classical Richardson divergence of $3/2$, which also applies to MHD turbulence.

Another way to understand this is to use the critical balance condition, which in this context means equating the cascade rate with the Whistler frequency
\begin{equation}
    \frac{1}{t_{cascade}}\sim \frac{d_i}{l}k_\| V_A.
\end{equation}
This implies that 
\begin{equation}
    k_\|\propto l^{-1/3},
\end{equation}
and that the divergence of field lines follows the rule that two field lines separated by an eddy will, on the average, increase their separation by the width of the eddy over a distance comparable to the length of the same eddy.

The speed of reconnection on a scale $l$ is 
\begin{equation}
    V_{rec}\sim v_l\sim d_i k_\| V_A\sim \frac{d_i}{l}\frac{b_l}{B}V_A,
\end{equation}
i.e. it actually increases on smaller scales. In terms of the strength of the energy cascade it is
\begin{equation}
    V_{rec}\sim \left(\frac{\cal E}{\rho}\frac{d_i^2}{l}\right)^{1/3}.
    \label{whistler_recon}
\end{equation}

Due to the limited range of whistler turbulence we do not expect this type of reconnection to play big role for large scale astrophysical reconnection. In fact, the existing Hall-MHD simulations of reconnection that show spectrum of whistler turbulence transfer for larger scale reconnection events to the turbulent reconnection in the MHD regime (see \S \ref{Hall}). However, we expect this type of reconnection to be present in the events that can be probed by in situ measurements in the Earth magnetosphere. Thus the numerical testing of Eq. (\ref{whistler_recon}) is of importance for understanding the magnetospheric physics. 

\subsection{Special regimes: summary}

The theory of turbulent reconnection was formulated for the non-relativistic reconnection in MHD regime. The numerical studies of the relativistic regime show that the reconnection in relativistic regime is rather similar to that in non-relativistic one. This arises from the fact that the properties of Alfvenic turbulence and therefore magnetic field wandering are similar in the two regimes. However, the effects of compressibility and the coupling of Alfven and fast modes are more important for the relativistic reconnection.

The reconnection at the high Prandtl number demonstrates that the turbulent reconnection is robust process and its rate is only marginally affected by the changes of the microscopic properties of the fluid. Indeed, the study shows that viscosity does not change the reconnection rate at large scales much larger than the viscous dissipation scale. We also note that the particular study is of practical interest, as high Prandtl number turbulence can represent some regimes of turbulence in partially ionized gas. 

Finally, the very idea of turbulent reconnection theory that magnetic field wandering regulates the outflow region and determines the reconnection rate carries over to scales less than the ion gyroradius. The corresponding whistler (or Kinetic Alfven wave) turbulence can also induce turbulent reconnection. Such reconnection may be important when reconnection events at the plasma scales are considered. Numerical testing of the predictions that we provide in this section would be of interest e.g. for magnetospheric reconnection research.

\section{Astrophysical implications of turbulent reconnections}
\label{sec:implications}

\subsection{Nonlinear turbulent dynamo and turbulent reconnection}
\label{sec:implications1}

Turbulent reconnection plays an important role for the theory of generating magnetic field by turbulent motions, i.e. the theory of turbulent dynamo. 
The turbulent dynamo has been recognized as the most promising mechanism to account for the cosmic magnetism 
\citep{Bran05}.
The dynamo process operating on the length scales smaller than the driving scale of turbulence, 
the so-called ``small-scale'' turbulent dynamo (SSTD), 
is especially important for the amplification and maintenance of the magnetic fields in the ISM and intracluster medium. 
Earlier studies on the SSTD have been restricted to the kinematic regime, 
where the magnetic field is dynamically insignificant and its effect on turbulent motions is negligible
\citep{Kaza68,KulA92}.
In the kinematic regime, the dynamo growth of magnetic fields can be analytically described by using the Kazantsev theory and considering 
the microscopic diffusion of magnetic fields, including the resistive diffusion and ambipolar diffusion
\citep{SchK02,KulA92,XL16,XuG19}.
However, the kinematic dynamo gives rise to the magnetic fields with the correlation scale comparable to the dissipation scale of 
magnetic fluctuations. 
This cannot explain the observed magnetic fields with the magnetic energy concentrated at a large correlation scale in diverse astrophysical media 
\citep{Leamon_etal:1998,VogtEnsslin:2005,BecM16}. 
The nonlinear dynamo, as confirmed by numerical simulations, accounts for not only the amplification of magnetic field strength but also the increase of 
its correlation length and thus is astrophysically important.

Analytical studies on the nonlinear dynamo is challenging because of the strong back-reaction of magnetic fields on turbulent motions. 
Numerical simulations reveal that the nonlinear dynamo has a universal linear-in-time growth of magnetic energy 
and the growth rate is only a small fraction of the turbulent energy transfer rate
\citep{Ryu08,CVB09, BJL09, Bere11}.
These numerical measurements can be used for testing the validity of analytical theories of nonlinear dynamo.

The efficiency of nonlinear dynamo is determined by the interaction between the dynamo stretching of magnetic fields by turbulent motions and 
the diffusion of magnetic fields. 
Earlier theoretical attempts to formulate the nonlinear dynamo focused on the microscopic diffusion of magnetic fields,
which is applicable to the kinematic regime. 
For instance, for the nonlinear dynamo in a conducting fluid, 
\citet{Sch02}  
considered the resistive diffusion as the dominant diffusion mechanism and concluded that the resulting magnetic energy spectrum peaks at the 
resistive scale. 
In the case of a partially ionized medium,
\citet{KulA92}
adopted the ambipolar diffusion as the dominant diffusion mechanism and found that the magnetic fields amplified by the nonlinear dynamo have 
the spectral peak at the ambipolar diffusion scale. 
More recently, 
\citet{Scho15}
introduced the ``effective magnetic diffusion'' 
\citep{Sub99}
to replace the microscopic diffusion used in the nonlinear dynamo. 
As a result, the magnetic energy spectrum peaks at the effective resistive scale. 
However, the theoretical predictions of these models on many characteristics including the dynamo efficiency, magnetic energy spectrum, and correlation length 
of magnetic fields are inconsistent with numerical findings. 

As shown in dynamo simulations, MHD turbulence is developed as a consequence of nonlinear dynamo with a Kolmogorov spectrum of magnetic fluctuations
\citep{Bran05,Bere11}. We discussed earlier that the violation of flux freezing predicted by the turbulent reconnection theory and supported by numerical simulations (see \S \ref{ssec: rdffvio}, \S \ref{sec:violation1}, \S \ref{ssec: doffvio}) in most cases exceeds orders of magnitude the magnetic diffusion that directly follows from non-ideal effects, e.g. resistivity and plasma effects. Therefore the process that was termed in Lazarian (2005) the reconnection diffusion (RD) (see \S \ref{ssec: redsf}) dominates over microscopic diffusion processes in MHD turbulence.  
Naturally, one should include the RD to properly model the nonlinear dynamo and this was done in 
\citet{XL16}. The latter paper
analytically derived the evolution of magnetic energy $\mathcal{E}$ for the nonlinear dynamo, 
\begin{equation}
   \mathcal{E} \sim \frac{3}{38} \epsilon t, 
\end{equation}
where $\epsilon$ is the scale-independent turbulent energy transfer rate. 
The linear dependence on time comes from the constant energy transfer rate of both hydrodynamic and MHD turbulence. 
The small fraction $3/38$ comes from the RD, 
which causes the inefficiency of nonlinear dynamo, and 
quantitatively agrees with numerical measurements
\citep{CVB09, Bere11}. Note that in the above equation, $3/2$ is used as the slope of the growing magnetic energy spectrum \citep{Kaza68}. 
If a much 
steeper slope $4$ is used \citep{eyink2010fluctuation,Krai67}, 
then an even smaller dynamo growth rate is expected.

Due to the RD in MHD turbulence, the magnetic energy spectrum peaks at the driving scale of MHD turbulence, where 
the turbulent energy and magnetic energy are in equipartition. 
We note, that the effect of magnetic reconnection on turbulent dynamo was also discussed in \citet{KulA92}. They adopted Petschek's model for reconnection \citep{Petschek:1964}, and thus the peak scale of magnetic energy spectrum is much less than that in \citet{XL16}, which cannot account for the large correlation scale of observed magnetic fields. 

\begin{figure*}[htbp]
\centering
\subfigure[$P_m=1$]{
   \includegraphics[width=0.45\textwidth]{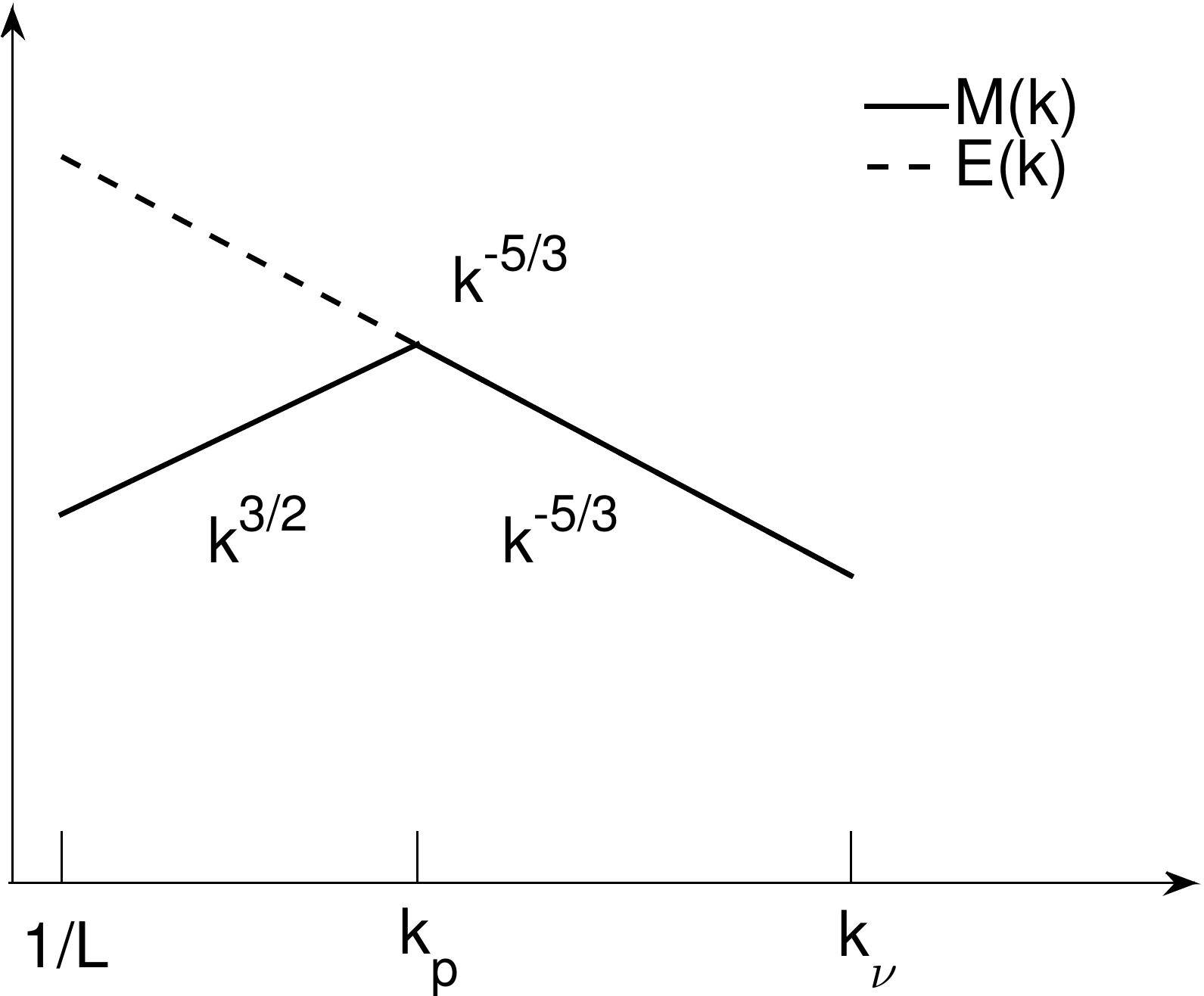}\label{fig: pmlt}}
\subfigure[$P_m>1$]{
   \includegraphics[width=0.45\textwidth]{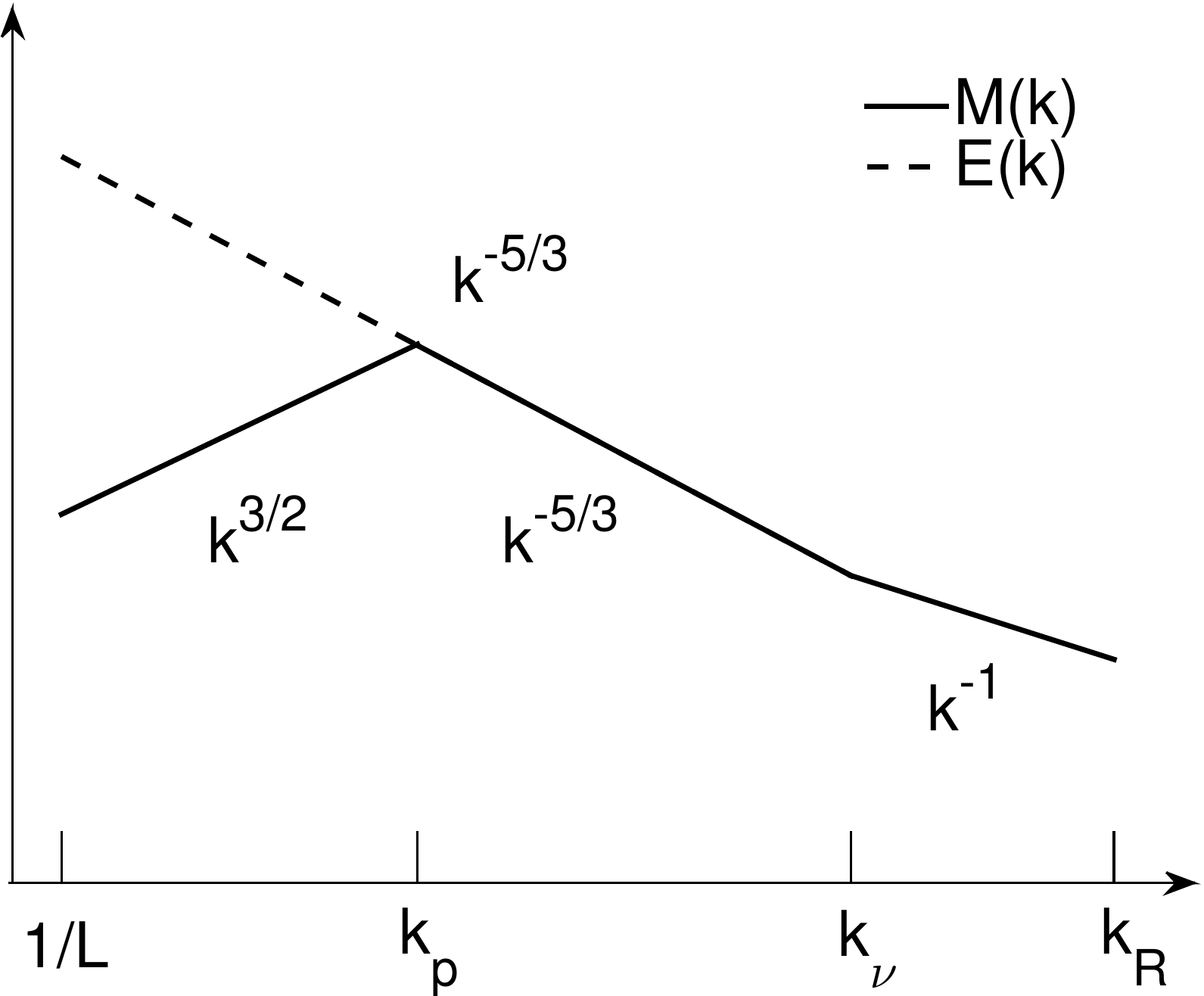}\label{fig: pmht}}
\subfigure[$P_m=1$, figure. 5.1 in \citet{Bran05}]{
   \includegraphics[width=0.45\textwidth]{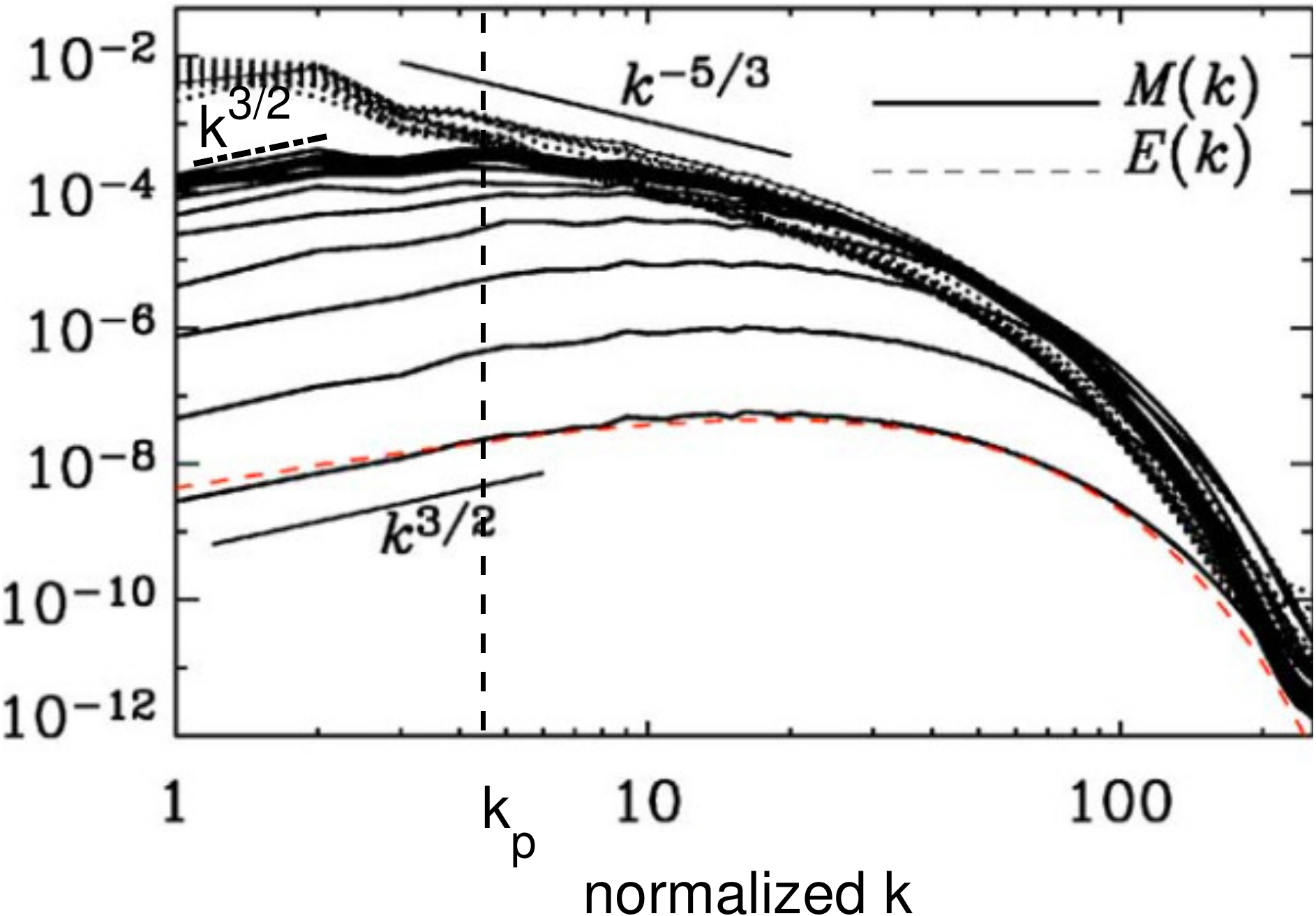}\label{fig: pmls}}
\subfigure[$P_m=50$, figure. 5.2 in \citet{Bran05}]{
   \includegraphics[width=0.45\textwidth]{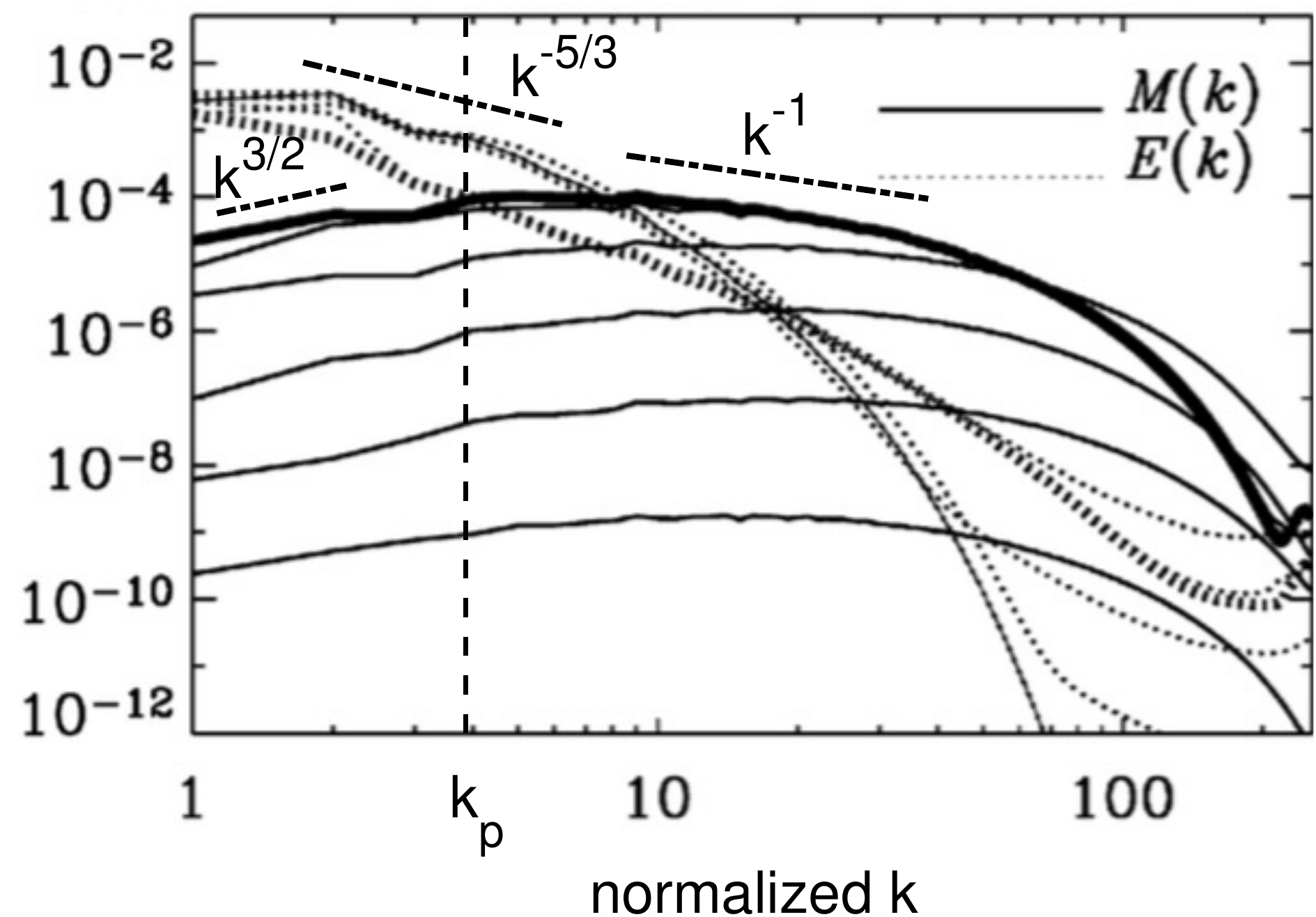}\label{fig: pmhs}}
\caption{ Upper panel: sketches of the magnetic (solid line) and turbulent (dashed line) energy spectra for the nonlinear dynamo.  
Lower panel: the original figures are taken from \citet{Bran05} (under license \href{https://creativecommons.org/licenses/by-nc-sa/3.0/}{CC BY-NC-SA}). 
The added dash-dotted lines indicate analytical spectral scalings, 
and the vertical dashed line indicates the peak scale of magnetic energy spectrum.
From \citet{XL16}, \copyright~AAS. Reproduced with permission. }
\label{fig: sket}
\end{figure*}

This peak scale increases with the growth of magnetic energy
\citep{XL16}, 
\begin{equation}
   k_p =  \Big[k_0^{-\frac{2}{3}} + \frac{3}{19} \epsilon^\frac{1}{3} (t - t_0)\Big]^{-\frac{3}{2}}.
\end{equation}
Here $k_0$ is the wavenumber where the spectrum peaks 
at the beginning of nonlinear dynamo at $t=t_0$. 
On smaller length scales, magnetic fields are dynamically important and results in the scale-dependent anisotropy of turbulence. 
On the other hand, the RD allows efficient diffusion and thus turbulent motions of magnetic fields. 
Therefore, instead of having a folded structure with magnetic field reversals only at the dissipation scale, 
the magnetic fields amplified by the nonlinear dynamo are turbulent magnetic fields with field reversals on all length scales within the 
inertial range of turbulence. 
Fig. \ref{fig: sket} illustrates a comparison between the analytical scalings of magnetic energy spectrum obtained by
\citet{XL16}
and numerical measurements in 
\citet{Bran05}
in cases with different magnetic Prandtl number $P_m$.
We see that the nonlinear dynamo theory based on the RD well explains the behavior of nonlinear dynamo presented in numerical simulations.

\begin{figure*}[htbp]
\centering
\subfigure[First star]{
   \includegraphics[width=0.48\textwidth]{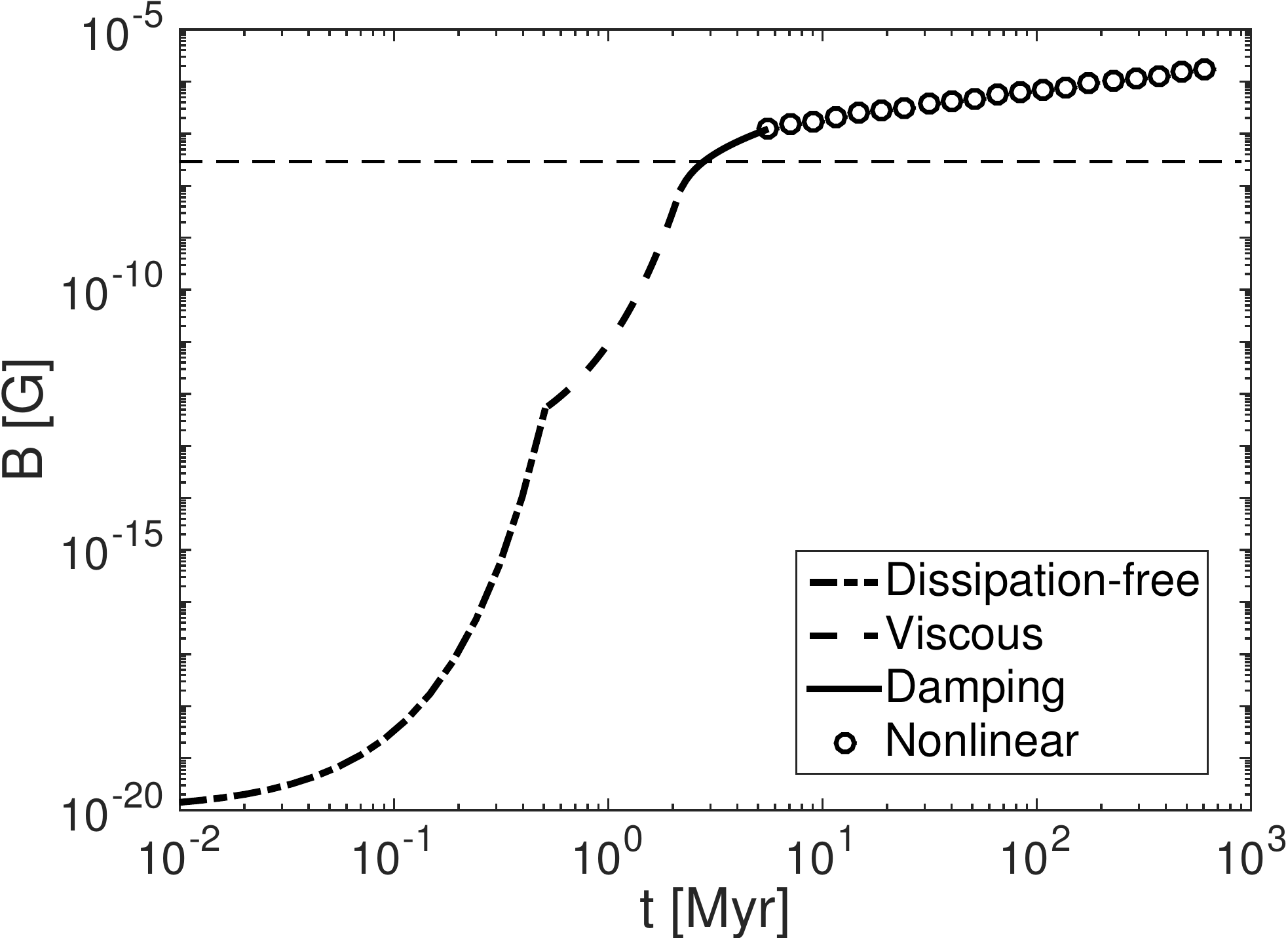}\label{fig: fsb}}
\subfigure[First star]{
   \includegraphics[width=0.48\textwidth]{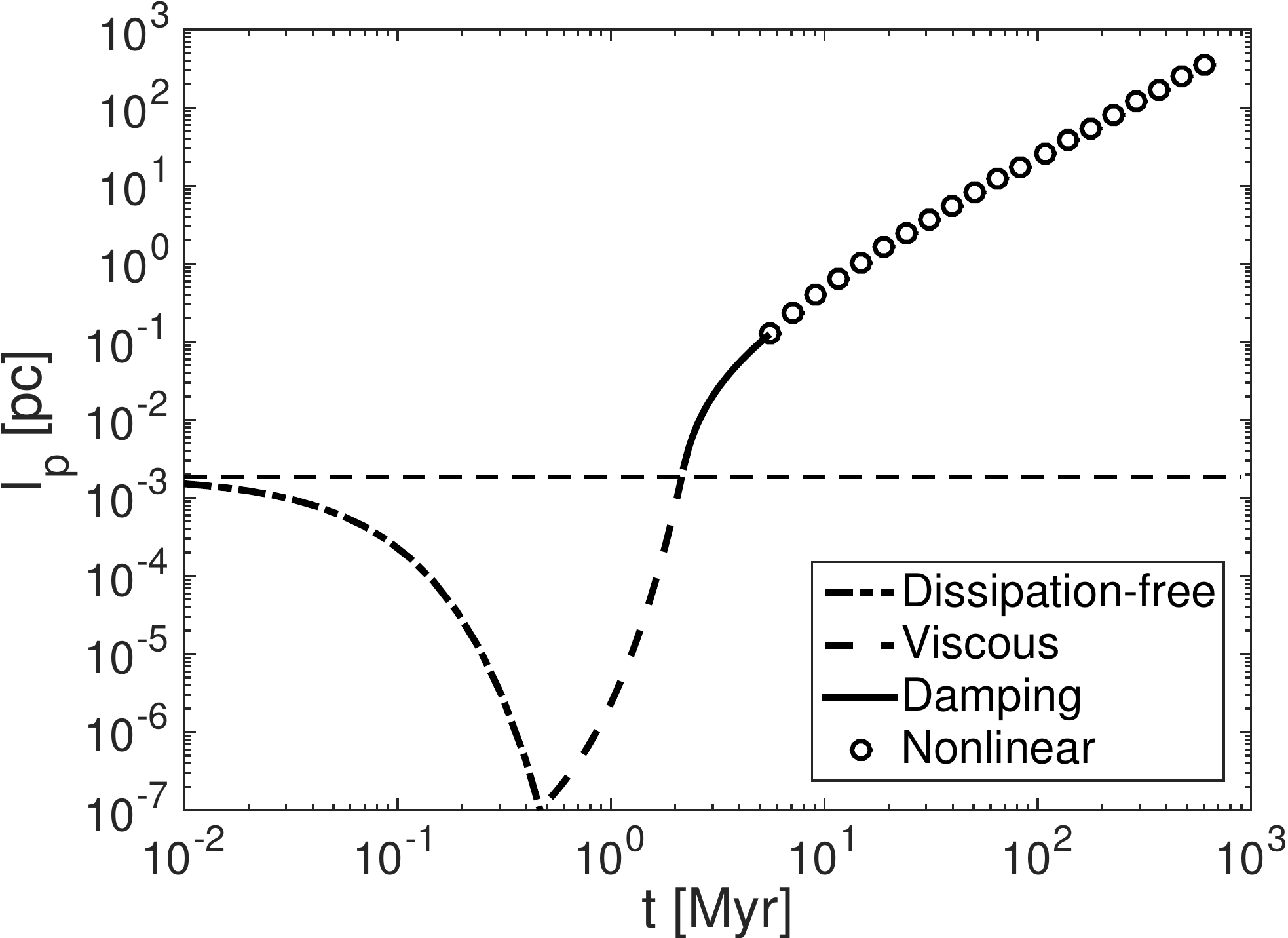}\label{fig: fss}} 
\caption{ The time evolution of the magnetic field strength and the correlation length of magnetic field 
during the formation of the first stars.  
From \citet{XL16}, \copyright~AAS. Reproduced with permission.}
\label{fig: firs}
\end{figure*}

The inefficiency of nonlinear dynamo has important astrophysical implications for the evolution and importance of magnetic fields in both 
early and present-day universe. 
In earlier studies on the turbulent dynamo during the primordial star formation, 
e.g., \citet{SchoSch12}, 
the RD was not taken into account. 
This leads to an efficient nonlinear dynamo and 
generation of dynamically significant magnetic fields over a short timescale. 
\citet{XL16} 
applied the nonlinear dynamo theory including the RD 
and 
examined the dynamo evolution of magnetic fields in environments like the first stars. 
As shown in Fig. \ref{fig: firs}, among different evolutionary stages, the nonlinear dynamo has the longest timescale due to the effect of RD. 
Since the resulting dynamo timescale is longer than the system's free-fall time, 
it is unlikely that the dynamo-amplified magnetic fields can regulate the gravitational collapse. 

\begin{figure*}[htbp]
\centering
\subfigure[Time evolution of $B$]{
   \includegraphics[width=0.48\textwidth]{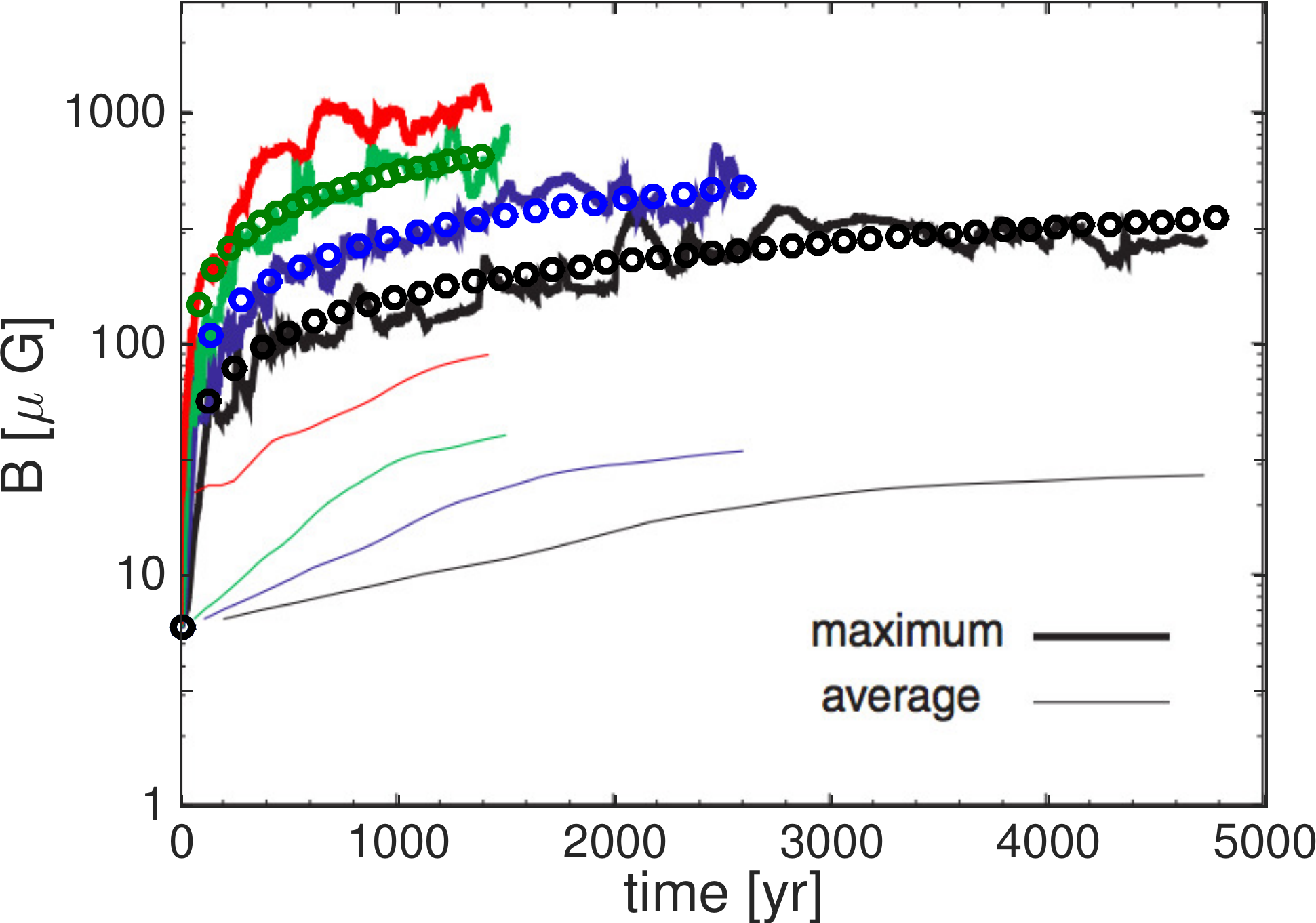}\label{fig: sh1}}
\subfigure[Spatial profile of $B$]{
   \includegraphics[width=0.48\textwidth]{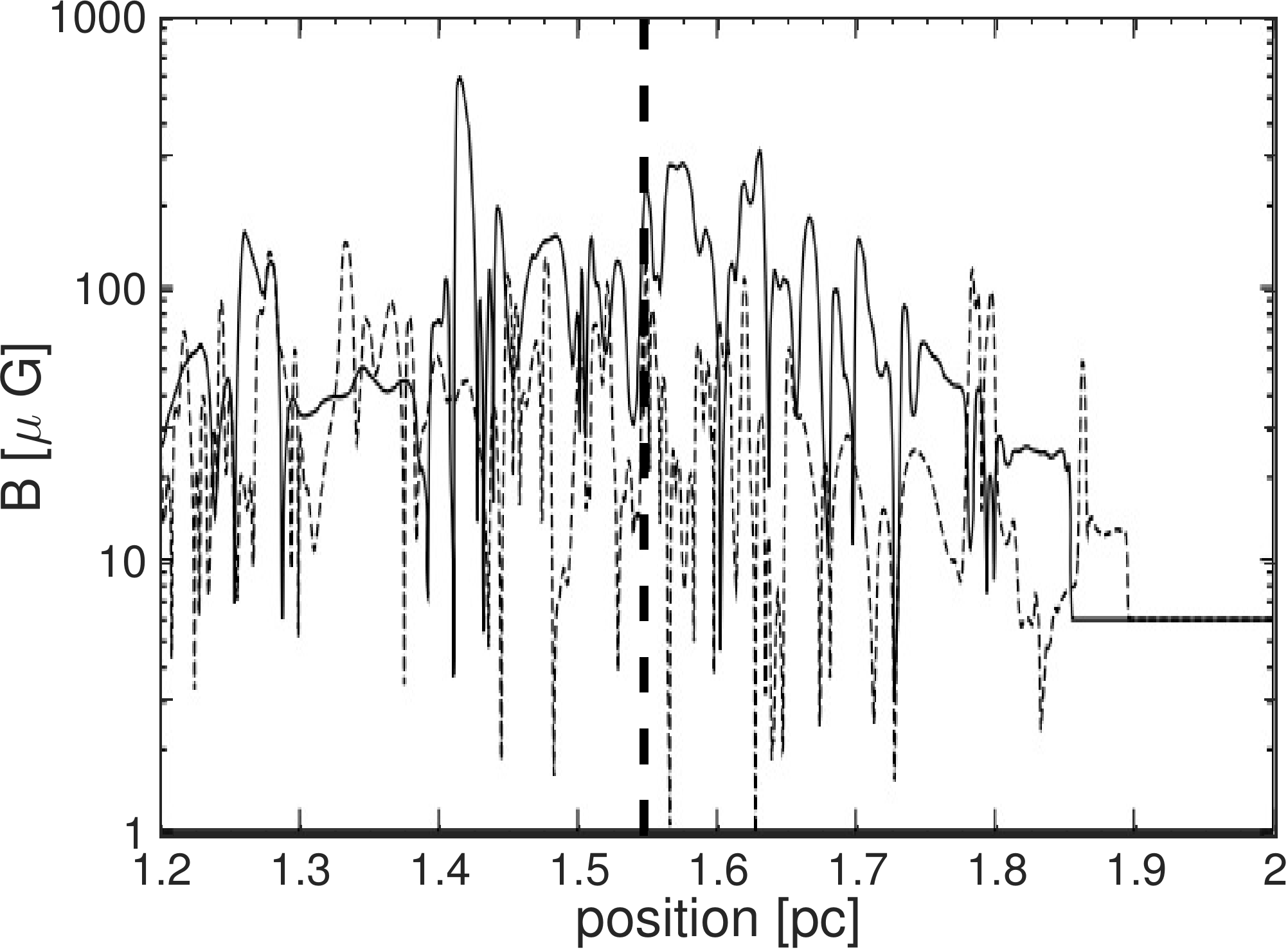}\label{fig: sh2}}  
\caption{Magnetic field amplification in the postshock medium. 
Compared with the numerical measurements by  
\citet{Ino09},
the circles in (a) and the vertical dashed line indicating the position for the maximum field strength 
in (b) are analytical results 
derived by \citet{XuLazarian:2017} using the nonlinear dynamo theory based on the RD. From \citet{XuLazarian:2017}, \copyright~AAS. Reproduced with permission.
}
\label{fig: dynsh}
\end{figure*}

Dynamo amplification of magnetic fields is also expected in supernova shocks to account for the strong magnetic fields required 
for the acceleration of Galactic cosmic rays. 
\citet{Xu2017}
analytically modeled the nonlinear dynamo in the downstream region of a supernova shock. 
Fig. \ref{fig: dynsh} displays a comparison between their analytical results and the numerical measurements in MHD simulations of a strong shock wave by 
\citet{Ino09}. 
By adopting the same turbulence parameters used in the simulations, 
\citet{Xu2017} 
found that the nonlinear dynamo theory based on the RD well explains the 
linear-in-time growth of magnetic energy and the amplification of magnetic field strength by two orders of magnitude seen in the simulations 
(Fig. \ref{fig: sh1}). 
Due to the inefficiency of the nonlinear dynamo, the maximum field strength can only be reached at a distance on the order of $0.1$~pc
behind the shock front (Fig. \ref{fig: sh2}). 
All these findings are consistent with the properties of magnetic fields implied by the X-ray hot spots observed in supernova remnants 
\citep{Uch08}.
It suggests the nonlinear dynamo involving the RD as the origin of strong magnetic fields in supernova remnants.

\subsection{Reconnection diffusion and star formation}
\label{ssec: redsf}

The Richardson dispersion describes the dynamics of particles and magnetic field lines from the point of view of Lagrangian approach. At the same time, for many problems the description of the effects induced by turbulent eddies in the laboratory system is preferable. This is Eulerian approach to the dynamics of the diffusion of magnetic field and this process was termed ``reconnection diffusion''
\citep{Lazarian:2005,LazarianYan:2014,LEC12}. The name is designed to stress the importance of the process of magnetic reconnection that enables the diffusion process.
Naturally, reconnection diffusion is intrinsically related to the Richardson diffusion in time and space that we described in 
\S \ref{ssec: rdffvio}. 
In many respects, the process has similarities to the normal turbulent diffusion in hydrodynamics, which is not so surprising in view of the fact that the dynamics of eddies perpendicular to magnetic field is similar to the Kolmogorov turbulence. Indeed, such eddies depend on mixing of fluid motions perpendicular to the local direction of the magnetic field (see LV99) and the process takes place only if reconnection events happen through every eddy turnover. For small scale eddies magnetic field lines are nearly parallel. Therefore, when they intersect, the pressure gradient is not $V_A^2/l_{\|}$ but rather $(l_{\bot}^2 /l_{\|})V_A^2$. This happens
since only the energy of the component of the magnetic field that is not shared is available to drive the outflow. At the same time, the characteristic length contraction of a given field line
due to reconnection between adjacent eddies is $l_{\bot}^2 /l_{\|}$. Taken together, this gives an effective ejection rate of $V_A /l_{\|}$ . Since 
the width of the diffusion layer over the length $l_{\|}$ is $l_{\bot}$ one can write the mass conservation in a form by $V_{rec} \approx V_A (l_{\bot}/l_{\|})$. This provides the reconnection rate $V_A /l_{\|}$ , which coincides with the nonlinear cascade rate on the scale $l_{\|}$.

\begin{figure}
\centering
\includegraphics[width=0.48\textwidth]{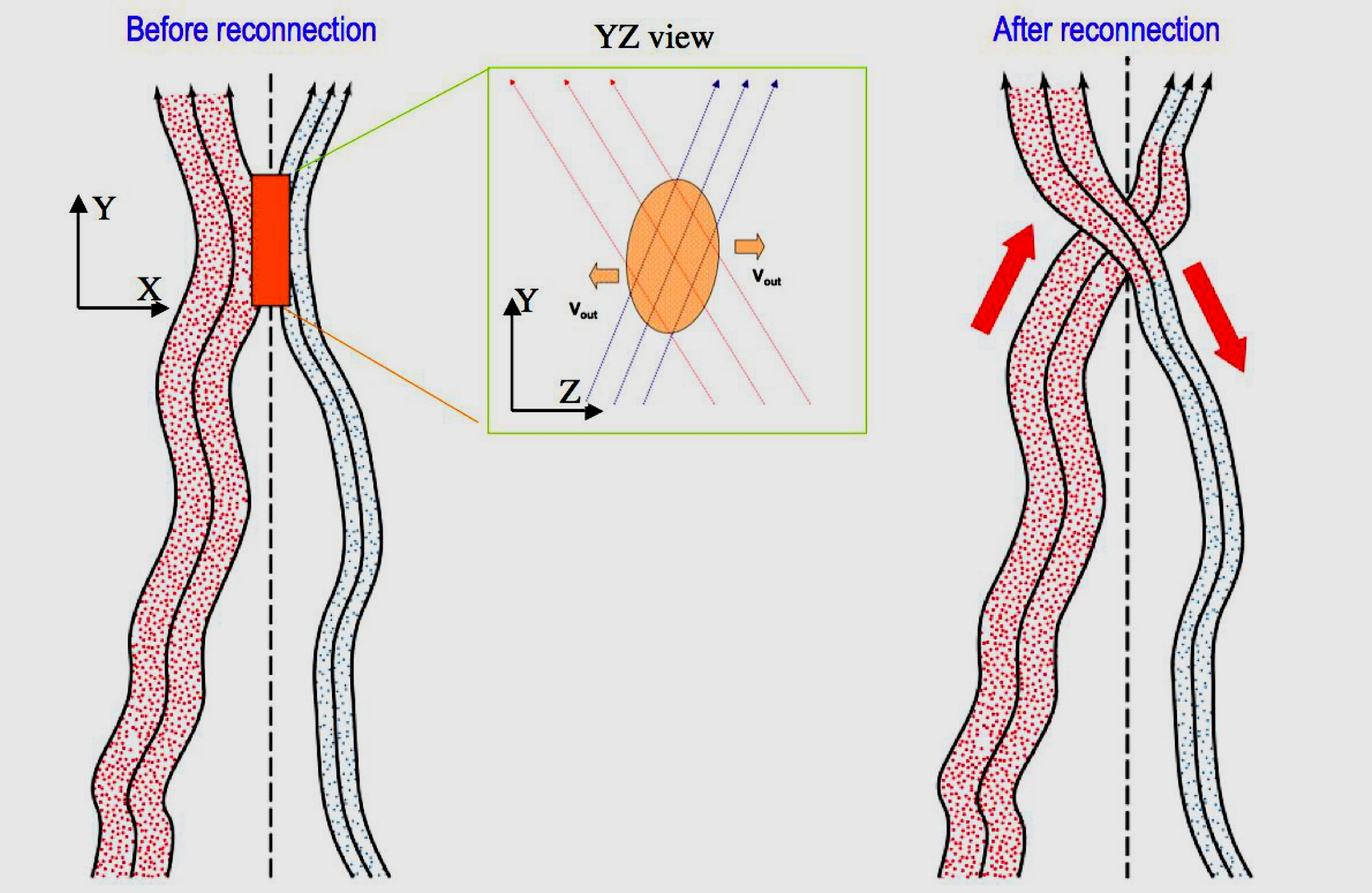}
\caption{Motion of matter in the process of reconnection diffusion. 3D magnetic flux tubes get into contact and after reconnection plasma streams along magnetic field lines. {\it Left}: XY projection before reconnection, upper panel shows that the flux tubes are at angle in X-Z plane. {\it Right}: after reconnection. From \cite{LEC12}, \copyright~AAS. Reproduced with permission.}
\label{fig29}
\end{figure}

The description above provides an introduction to the mathematical framework of turbulent diffusivity in a homogeneous magnetized fluid.
The standard theory of star formation is based on the assumption that magnetic field lines preserve their identify and the diffusion of charged particles perpendicular to magnetic field lines is restricted. As a result, the mass loading of magnetic field lines does not change. However, LV99 model
 suggests that the standard assumptions are not applicable if magnetized fluids are turbulent. As a result, in the presence of MHD turbulence, the diffusion of plasma perpendicular to magnetic field is inevitable.
We shall illustrate the reconnection diffusion showing how it allows plasma to move perpendicular to the mean inhomogeneous magnetic field (see Figure \ref{fig29}). 
This is relevant to star formation. Consider two magnetic flux tubes with entrained plasmas that intersect each other at an angle. Due to reconnection the identity of magnetic field lines changes. Consider a situation that before the reconnection event the plasma pressures $P_{plasma}$ in the tubes are different, but the total pressure $P_{plasma}+P_{magn}$ is the same for two tubes. This is a situation of a stable equilibrium. If plasmas are partially ionized,  ambipolar diffusion, i.e. slow diffusion of neutrals in respect to magnetic field can make gradually smoothen the magnetic field pressure gradients. In the presence of turbulence a different  much faster process related to turbulent reconnection takes place.

Magnetic field lines in the presence of turbulence are not parallel. Such field lines can reconnect and do reconnect all the time in a turbulent flow. The process of reconnection connects magnetic fields with different mass loading and plasma pressures. As a result, plasmas stream along magnetic field lines to equalize the pressure. In the process, the portions of magnetic flux tubes with higher magnetic pressure expand as plasma pressure increases due to the flow of plasma along magnetic field lines. The entropy of the system increases as magnetic and plasma pressures become equal through the volume. In other words, a process of the diffusion driven by turbulent reconnection takes place and this process does not depend on the degree of the ionization of the matter. In the absence of gravity, the outcome of the reconnection diffusion is to make magnetic field and plasmas more homogeneously distributed. Both motions of plasmas along the flux tubes and the exchange of parts of the flux tubes between different eddies contributes to the diffusion.

\begin{figure}
\centering
\includegraphics[width=0.48\textwidth]{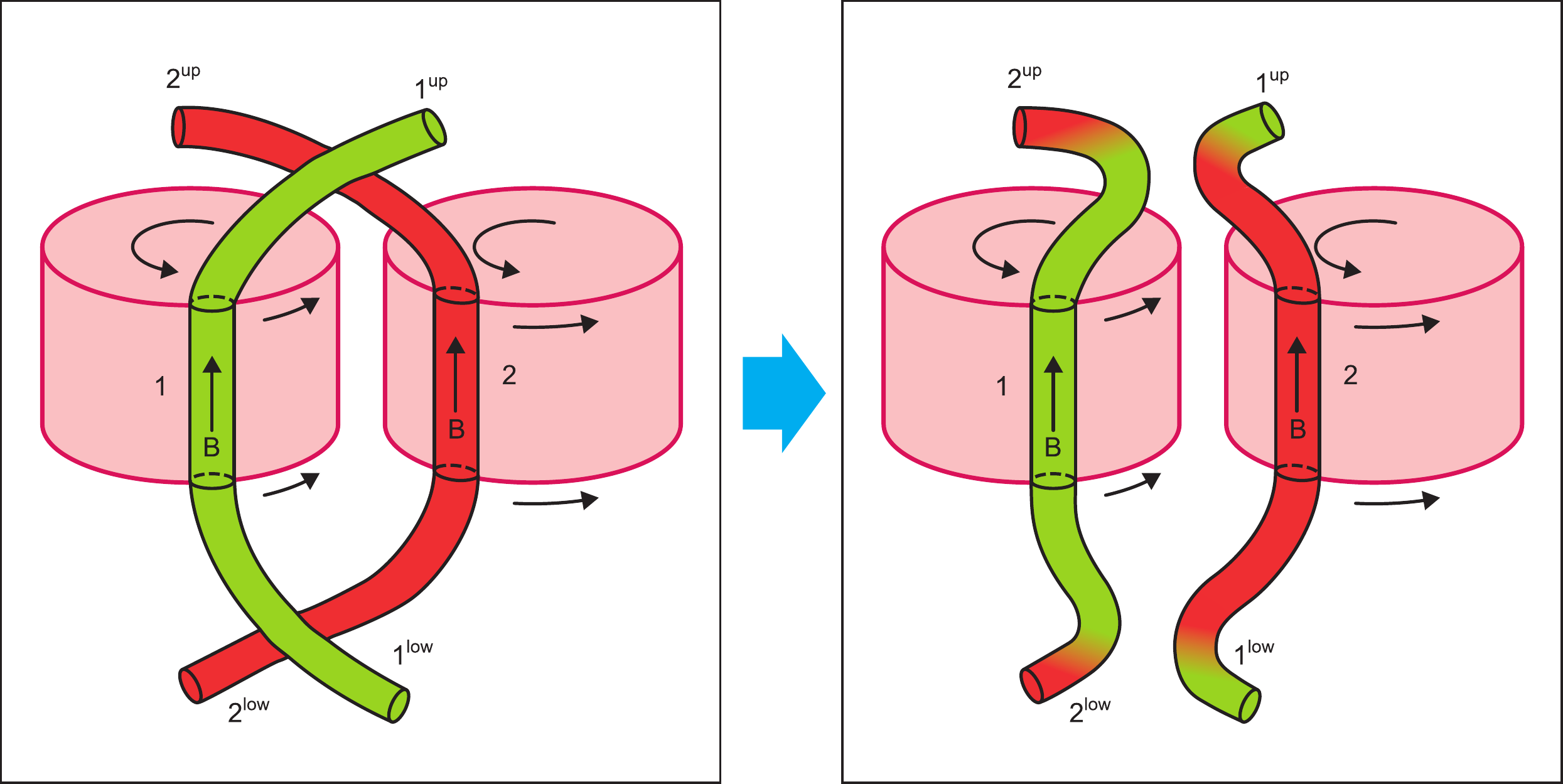}
\caption{Reconnection diffusion: exchange of flux with entrained matter. Illustration of the mixing of matter and magnetic fields due to reconnection as two flux tubes of different eddies interact. Only one scale of turbulent motions is shown. In real turbulent cascade such interactions proceed at every scale of turbulent motions. From \cite{LEC12}, \copyright~AAS. Reproduced with permission.}
\label{fig30}
\end{figure}

To get a clear picture of what is going on, consider a toy model of two adjacent magnetized eddies (see Figure \ref{fig30}). Magnetic flux tubes moving with the eddies reconnect and exchange plasmas and magnetic fields. This induces turbulent diffusion of both magnetic field and plasmas. In the case illustrated by Figure \ref{fig30}, the densities of plasma within the flux tubes can be different and the reconnection and creates a new flux tubes along which plasma redistributes due to the pressure difference. This process induces the diffusion of plasma perpendicular to the mean magnetic field. In reality, the process above happens at every scale and for turbulence with the extended inertial range, the shredding of the columns of plasmas with different density proceeds at all turbulence scales. The time scale of the eddy turnover in GS95 turbulence is $l/v_l\sim l^{2/3}$ and therefore, assuming the decreasing the size of an eddy in the process of cascading by a factor of 2, one can present the total time for the cascade from $l$ to the Kolmogorov dissipation scale as a geometric pregression.  Thus the full cascading takes place in about one eddy turnover time. Therefore the speed of plasma motions along the magnetic field is, in most cases, not important for the diffusion. 

As we discussed in \S \ref{sec:level1}  the magnetic flux freezing is not applicable to turbulent fluids. The gross violation of the flux freezing starts at the scales larger than the critical scale of the 
turbulent damping eddies. As we discussed in \S \ref{sec:turbulence3} in partially ionized gas, the corresponding scale perpendicular damping scale can be significantly larger than the resistive scale. This, however, does not
change the dynamics of turbulence at scales larger than $l_{\bot, crit}$. This fact follows both from observations and simulations (see 
\cite{Brandenburg2013}). Therefore the
process of reconnection diffusion is supposed to proceed in partially ionized gas and it will be governed by the dynamics of the large eddies in the MHD regime. Nevertheless, at scales smaller that $l_{\bot, crit}$ the evolution will be different. The structures at these scales may survive longer as the mixing is suppressed. This can provide an explanation for the existence of the tiny scale structures in HI observed 
\citep{Heiles1997}.

\begin{figure}
\centering
\includegraphics[width=0.48\textwidth]{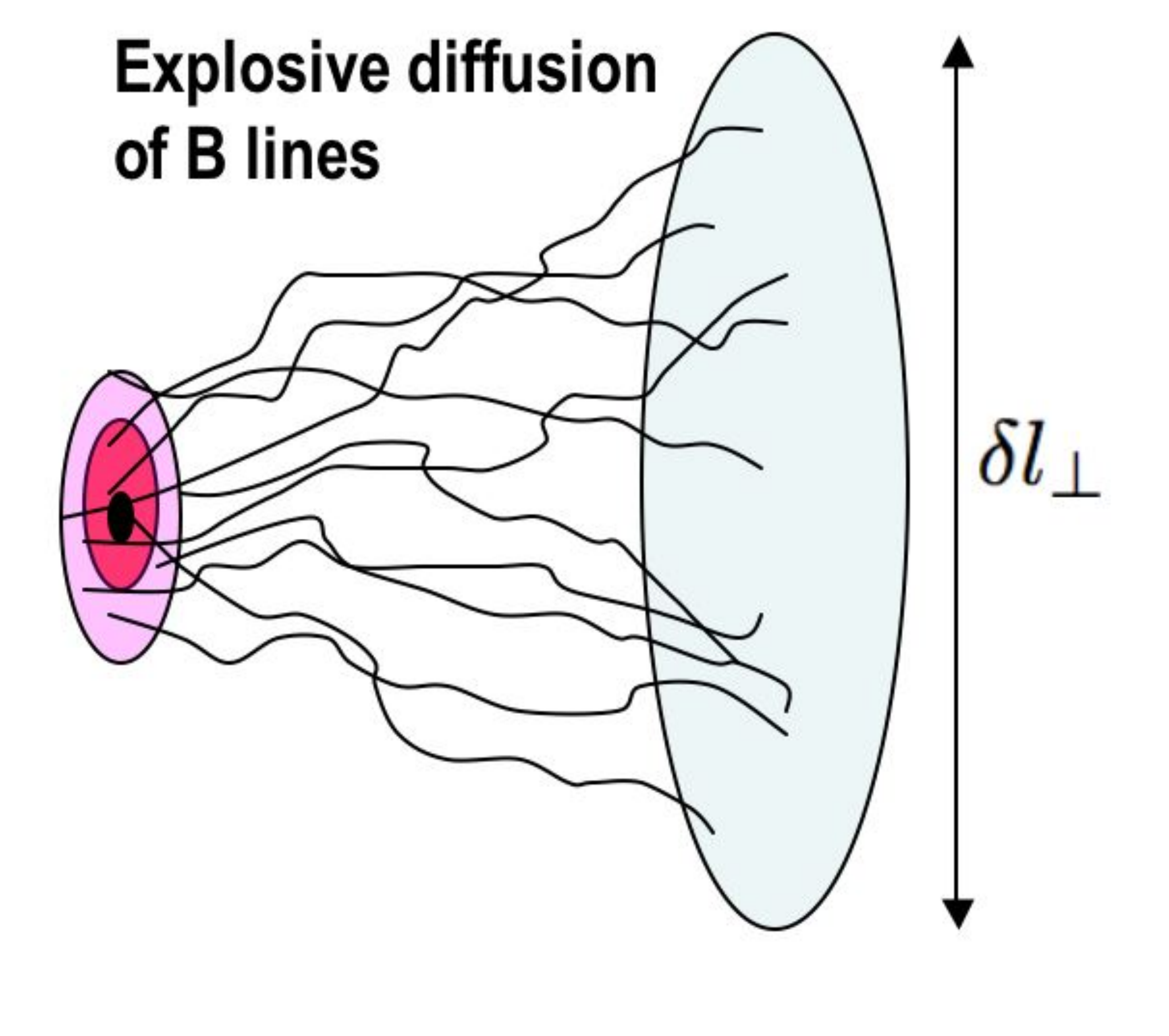}
\caption{Particle tracing magnetic field lines may start at different initial locations shown as coaxial ellipsoids. However after a period of time the field line spread over a larger volume and the final position of the field lines does not correlate with their initial position.  From \cite{LEC12}, \copyright~AAS. Reproduced with permission.}
\label{fig6}
\end{figure}

\begin{figure}
\centering
\includegraphics[width=0.48\textwidth]{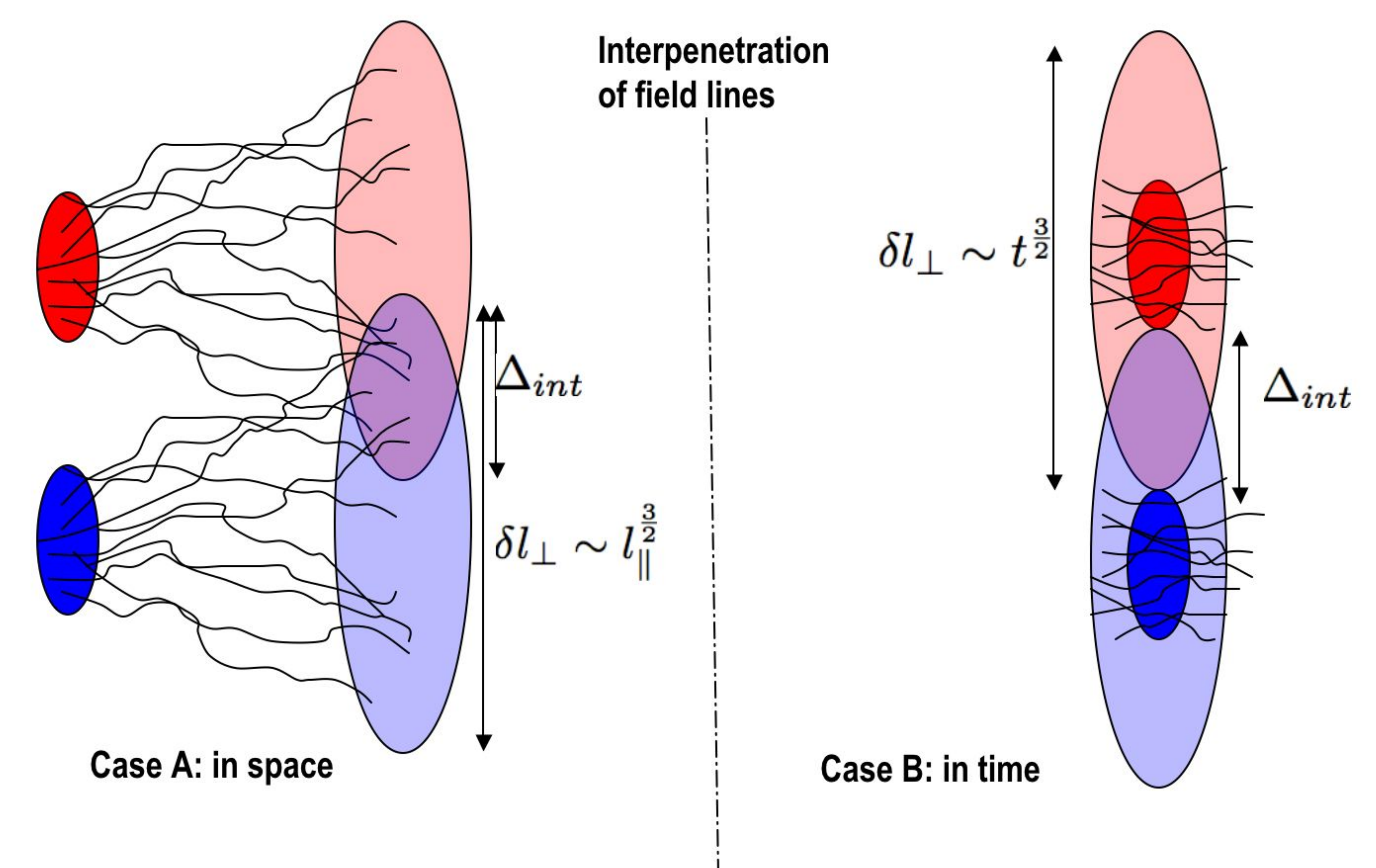}
\caption{Microscopic physical picture of reconnection diffusion. Magnetized plasma from two regions is spread by turbulence and mixed up over $\Delta$. {\it Left}: Description of the process in terms of field wandering in space. {\it Right}: Description of the spread in time. From \cite{LEC12}, \copyright~AAS. Reproduced with permission.}
\label{fig7}
\end{figure}

We should stress that the reconnection diffusion on the scales less than the turbulence injection scale is very closely related to the Richardson dispersion in time and space that we discussed in \S \ref{ssec: rdffvio}. Figure \ref{fig6} illustrates the Richardson dispersion in space, namely, the process of mixing up magnetic field lines originating from different spatial locations. It also illustrates the loss of the Laplacian determinism for magnetic field lines. In analogy with the example above, the final line spread $l_{\bot}$  does not depend on the initial separation of the field lines. The same loss of magnetic field identity is happening if the diffusion of magnetic field lines is traced in time. This is illustrated in Figure \ref{fig7}.

It is obvious that the accelerated separation of magnetic field lines is impossible to visualize without breaking a magnetic field line and reconnecting it with the neighboring field lines. Thus while instantaneously one can visualize a field line, the next moment the identify of this field line is entirely different, which is at odds with the classical concept of flux freezing. In fact, as it is discussed in ELV11, already the resistive term induces stochasticity associated with Ohmic diffusion. Therefore the definition of the magnetic field line on scales affected by resistivity gets not deterministic. In fact, introduces stochasticity and makes deterministic definition of field lines problematic. The determinism fails completely in the presence of turbulence, however. This was demonstrated explicitly in numerical calculations that we discuss in \S \ref{ssec: doffvio}.

The two types of the dispersion are not equivalent, however. The difference is obvious for the problem of the propagation of cosmic rays. There, following magnetic field lines is important for describing cosmic ray perpendicular diffusion that arises from fast moving particles that follow essentially an instantaneous realization of a magnetic field line (see 
\cite{LazarianYan:2014}). The effects of the time dependent Richardson dispersion are negligible in this context.

The concept of reconnection diffusion has been applied to different aspects of star formation, from the loss of magnetic support of molecular clouds to the solving the problem of magnetic breaking of accretion disks \cite[see][]{Lazarian:2014,Gonzalez-Casanova_etal:2016}  and ref. therein. These applications are easy to understand. The entire theory of star formation was developed assuming that the magnetic field are perfectly frozen into ionized component of the partially ionized gas. This was the basic assumption of both the classical studies \cite[see][]{MestelSpitzer:1956, Mestel:1966} and the elaborate theories that followed \cite[see][]{Shu_etal:1987, Mouschovias:1991, Nakano_etal:2002, Shu_etal:2004, Mouschovias_etal:2006}. Within this approach,
magnetic fields slow down and even prevent star formation
if the media is sufficiently magnetized. With the Alfven theorem
taken for granted,
the change of the flux to mass ratio within forming dense clumps happens due to neutrals
which do not feel magnetic field directly, but only through
ion-neutral interactions. In the presence of gravity, neutrals
get concentrated towards the center of the gravitational potential
while magnetic fields resist compression and therefore gradualy 
leave the forming protostar \cite[e.g.][]{Mestel:1965a, Mestel:1965b}. As a result, star formation  gets inefficient for magnetically dominated
(i.e. subcritical) clouds. This was viewed as a good news, as low efficiency of star formation
corresponds to observations \cite[e.g.][]{ZuckermanEvans:1974}. Therefore the star formation controlled by ambipolar diffusion became a standard paradigm. On closer examination, one can see that ambipolar diffusion does not solve all the
problems related to star formation. For instance, for clouds dominated by gravity, i.e. supercritical clouds, ambipolar diffusion is too slow and 
magnetic fields do not have time to leave the cloud. Therefore for supercritical clouds magnetic
field are expected be dragged into the star, forming stars with
magnetizations much in excess of the observed ones \cite[see][]{Galli_etal:2006, Johns-Krull:2007}.

Serious problems the traditional paradigm faces also with other processes associated with star formation, 
 e.g. the formation of the accretion disks
around forming stars.  In fact, the paradigm has been challenged by observations \cite[][see \citeauthor{Crutcher:2012}, \citeyear{Crutcher:2012}, for a review]{TrolandHeiles:1986, Shu_etal:2006, Crutcher_etal:2009, Crutcher_etal:2010} and this is suggestive that one cannot really describe the processes of star formation erroneously assuming that magnetic field are frozen in within turbulent fluids. It is important to note that turbulence has already become an essential part of the field of star formation picture \cite[see][]{Vazquez-Semadeni_etal:1995, Ballesteros-Paredes_etal:1999, Elmegreen:2000, Elmegreen:2002, McKeeTan:2003, ElmegreenScalo:2004, MacLowKlessen:2004, McKeeOstriker:2007} but the treatment of the turbulent magnetic fields stayed
within the flux freezing paradigm.

Without going into details, we can state, that the process of turbulent reconnection induces the reconnection diffusion that is faster than the ambipolar diffusion. This can be illustrated by an intentionally simplified arguments. As the most mass in molecular clouds is neutral gas, the diffusion coefficient for the fluid can be estimated as $\nu_n\sim v_n l_n$, where $v_n$ and $l_n$ are the velocity and mean free path of neutrals, respectively. If turbulence is present, the Reynolds number at the scale $l$, i.e. $Re_l\approx v_l l/\nu_n$ is larger than 1. It means that the turbulent diffusion coefficient $\kappa\sim v_l l$ is larger than the molecular diffusion coefficient $\nu$ \cite[see a more rigorous discussion in][]{Lazarian:2014}. As a result, one can expect that in turbulent cloud, the loss of magnetic flux will be driven through the reconnection diffusion as suggested in \cite{Lazarian:2005}. The numerical simulations, e.g. in \cite{Santos-Lima_etal:2010}, support this idea \cite[see also][]{Leao_etal:2013}.

Another important problem that has been solved using the reconnection diffusion concept was the so-called ``magnetic breaking catastrophe'' \cite[see][]{Shu_etal:2006}. The nature of the problem is easy to understand. If the interstellar magnetic field is frozen the matter forming the accretion disk, the forming Keplerian disk is subject to magnetic torques and expected to lose its angular momentum fast. In \cite{Gonzalez-Casanova_etal:2016} it was shown that magnetic reconnection changes the picture of star formation by both inducing the reconnection diffusion of magnetic field and changing the magnetic field configuration around the disk. The configuration of the magnetic field that connects the disk with the ambient interstellar media is frequently called ``split monopole'' as magnetic field from the ambient interstellar medium converge to the accretion disk, which size is very small compared to the size of the molecular cloud. In this configuration the magnetic field of different directions, i.e. converging from the media to the accretion disk and diverging from the accretion disk to the media come close together.  The reconnection changes part of the split monopole configuration to the dipole one. As a result, decoupling the magnetic field from the ambient interstellar or molecular cloud gas takes place. This significantly decreases the magnetic torques acting on the disk. The numerical simulations in \cite{Santos-Lima_etal:2014, Gonzalez-Casanova_etal:2016} show that the formed disks correspond to observations. 

Incidentally, turbulent reconnection and therefore reconnection diffusion does violate helicity conservation (LV99). This was demonstrated numerically in \cite{Bhat_etal:2014}. This once again proves that any attempts to substitute reconnection diffusion with effective ``turbulent resistivity'' or ``eddy resistivity'' are grossly incorrect (see more in \S \ref{sec:alsternatives3}).

\subsection{First Order Fermi acceleration of energetic particles by turbulent reconnection} 
\label{sec:implications3}

Magnetic reconnection was discussed in the context of particle acceleration before the LV99 model was introduced (e.g., \cite{Litvinenko:1996,ShibataTanuma:2001,Zenitani2001}). However, this process was intrinsically inefficient as it was based on the Ohmic-induced electric field of current sheets present within the models of reconnection known at that time.
The sweet-Parker reconnection is known to be extremely slow in most of astrophysical circumstnaces and unable to transfer any appreciable part of energy from magnetic field to energetic particles. At the same time, for Petschek reconnection, the fraction of magnetic energy that is available for particle acceleration in the small fraction of the magnetic volume occupied by the Ohmic diffusion region goes to zero as the resistivity goes to zero. Thus for both processes the Ohmic electric field acceleration is negligible.

\begin{figure*}
\centering
\includegraphics[width=0.48\textwidth]{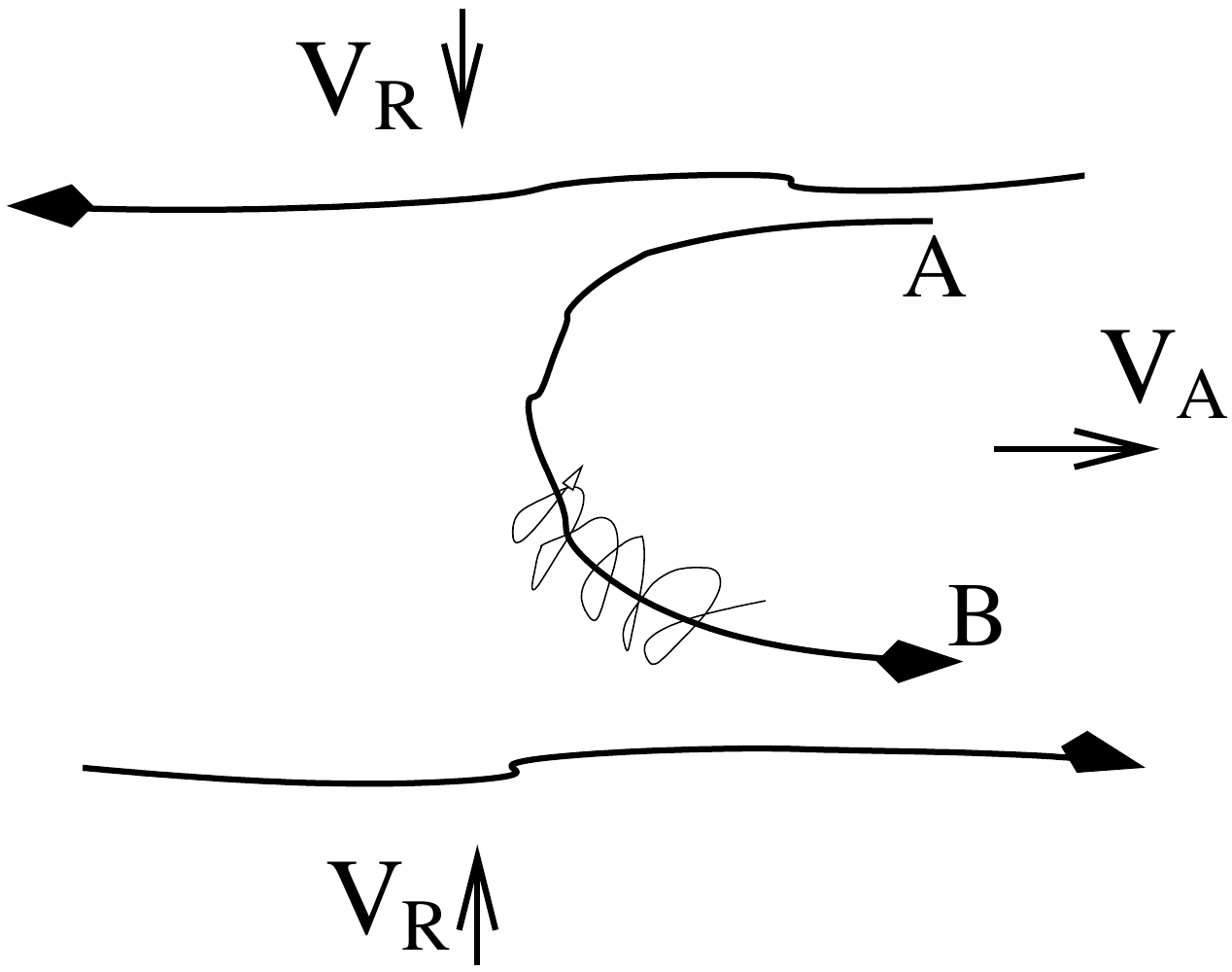}
\includegraphics[width=0.48\textwidth]{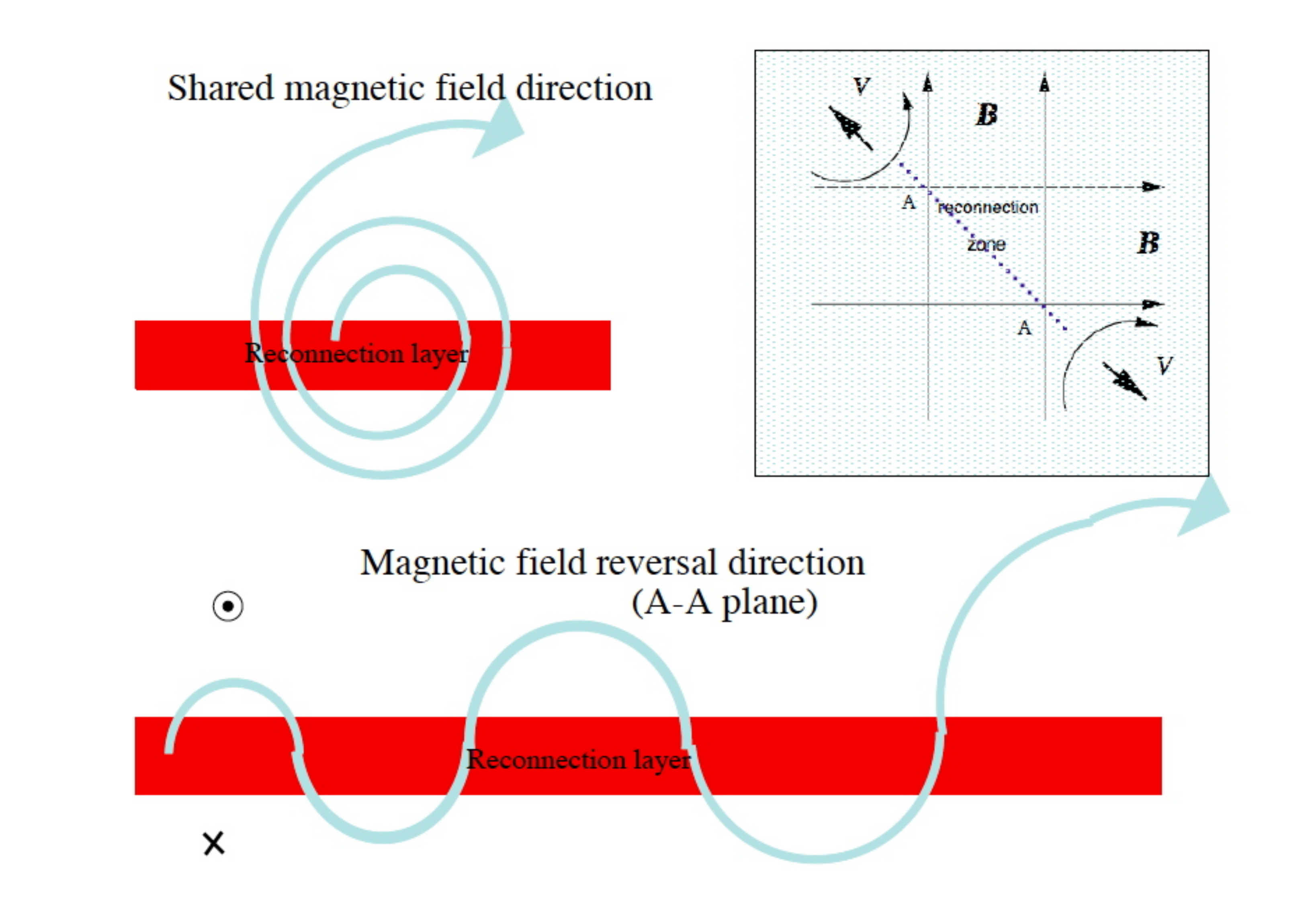}
\caption{{\it Left}: Shrinking magnetic loop within the turbulent reconneciton region with a particle that gains higher $p_{\|}$ as it follows the magnetic field.  From \citet{Lazarian:2005}.  {\it Right}: Particles with a sufficiently large Larmor radius interacts with converging magnetized flow and increases its perpendicular momentum $p_{\bot}$ as a result. The current sheet of the reconnection layer is denoted by red color. Due to magnetic reconnection the fluxes are moving towards each other with the velocity $V_{rec}$. As a result, a charged particle gets deflected by these converging magnetic mirrors and increases its $p_\bot$. The dynamics of a particle is different if viewed in the parallel the guide component of magnetic field and from the direction parallel to the plasma outflow, i.e. parallel to the A-A line in the upper right subpanel. In the former direction the components of magnetic field of the upper and lower magnetic fluxes are parallel. Therefore the particle gyrates in this field. The Larmor radius increases due to the $p_\bot$ increase and the particle trajectory presents a spiral with increasing radius as shown in the upper left subpanel. The observer viewing the reconnection layer perpendicular to the guide field, i.e. along the A-A direction of the upper right subpanel, reports field reversal at the reconnection layer with magnetic field being towards the observer above the reconnection layer and away from the observer below the reconnection layer. For this magnetic field configuration the particle samples magnetic field at both sides of the magnetic reversal and it changes the direction of its gyration as it crosses current sheet. As a result, the particle moves along the open trajectory perpendicular to the A-A line. The amplitude of its oscillations are increasing as $p_\bot$ increases as a result of the First Order Fermi acceleration.  Reprinted by permission from Springer Nature Customer Service Centre GmbH: Springer Nature, Space Science Reviews, \citet{Lazarian2012}, \copyright~2012.}
\label{fig31}
\end{figure*}

The acceleration process that is induced by the LV99-type of reconnection is based on the electric field arising from the fluid motions, i.e. $\sim v\times B$, i.e. disregards the Ohmic-induced electric field.In fact, the testing of particle acceleration within a region of turbulent reconnection in \citet{Kowal_etal:2012b} was performed making the Ohmic-induced electric field equal to zero.

The corresponding motions originally considered are illustrated by left panel of Figure \ref{fig31} that is taken from \citet{Lazarian:2005}.  Magnetic reconnection results in shrinking of magnetic loops and the charged particles entrained over magnetic loops get accelerated. One way to describe the acceleration is to consider energetic particles that move along the magnetic loop without collisions. In this setting the second adiabatic invariant $\int_a^b p_{\|}ds$ is conserved \cite[see][]{AnosovFavorskii:1987}. There $p_{\|}$ is the component of parallel momentum of a particle and the integration is performed along the magnetic loop with magnetic mirrors at points $a$ and $b$. It is obvious that as the loop shrinks the parallel momentum of the particle is bound to increase. 

Turbulent magnetic reconnection is associated with the process of decreasing of the overall length of magnetic field lines. Therefore in the reconnection region magnetic loops are systematically shrinking. Thus we expect to have a systematic increase of the parallel momentum of the particles trapped over magnetic field lines within the extended turbulent reconnection region (see the illustration of LV99 reconnection in the lower panel of Fig. \ref{fig:recon}).

There are several important points that should be made at this point. First of all, since the rate of turbulent reconnection is regulated only by the level of turbulence, for the sufficiently high level of turbulence the transfer of energy from the magnetic one is fast. Therefore for a significant amount of energy is available for the acceleration of particles per unit time. Moreover, the turbulent reconnection is a volume filling process and therefore the aforementioned shrinking of magnetic field lines is taking place over a significant layer of the thickness $\Delta$ (see Fig. \ref{fig:recon}). In addition,the LV99 reconnection operates in incompressible fluid and therefore the acceleration does not depend on the fluid compressibility.

As the particles are trapped in the magnetic field loop, one can consider that the volume available for such particles decreases as the magnetic field line shrink in the process of turbulent reconnection. This type of argument relate this type of acceleration to the traditional acceleration processes, e.g. shock acceleration, that require fluid compressibility.

Most important, it is obvious that the described process is the First Order Fermi accceleration process. Indeed, particles in the turbulent reconnection region are {\it systematically gaining energy}. Therefore the turbulent reconnection can be as efficient in terms of particle acceleration as the classical shock acceleration \citep{Krymskii:1977, Bell:1978}.

While Figure \ref{fig31} illustrates the gist the process of the physical essence of the First Order Fermi acceleration in turbulent reconnection regions, the quantitative description of the process is an still an issue of debates. The non-trivial issue for the description is for the process it is essential that the particle distribution function is anisotropic in terms of the particle momentum. Fast particle scattering that isotropize the momentum distribution decrease the efficiency of acceleration in the turbulent reconnection regions. In fact, the effect of the decreased acceleration efficiency in the presence of collisions is described in \cite{ChoLazarian:2006} for a similar process that increases only one component of particle angular momentum. 

In  \citet{deGouveiaDalPinoLazarian:2005} (henceforth GL05) the description of the acceleration was obtained assuming that the shrinking of magnetic field tubes can be approximated by particle bouncing between the 
mirrors created by the not-reconnected magnetic flux. These mirrors are approaching each other with the velocity $V_{rec}$. In this model the increase of the particle energy due to the increase of $p_{\|}$ arising due to particle reflections is somewhat similar to the First Order Fermi acceleration of particles in shocks. 

Further studies showed that additional processes that also increase the perpendicular component of angular momentum $p_\bot$ are also possible in turbulent reconnection regions. For instance, an illustration of this process from \citet{Lazarian2012} is presented in the right panel of  Figure \ref{fig31}. There the particles with sufficiently large Larmor radius interact over its gyroperiod both with the shared component of magnetic field (guide field) of the two reconnecting fluxes and with the reversing components of the two fluxes. The reversing component induces the particle motion along the open trajectory in the direction of the guide field. As the magnetic fluxes converge with $V_{rec}$ a particle gets increase of $\sim \frac{V_{rec}}{c} p_\bot$ over one gyration in the total magnetic field. Due to the magnetic field reversal, the direction of gyration changes which results in the motion of accelerating particles along the guide field. Moving this way a particle can either escape from the reconnection region or be reflected back. Magnetic inhomogeneities can reflect the particle back without changing the cosine of its pitch angle. 

Looking at Figure \ref{fig:recon} one may erroneously conclude that for the increase of $p_\bot$ through the process described above the particle radius should exceed $\Delta$. In fact, Figure \ref{fig:recon} is an idealized schematic aimed only to reflect the effect magnetic field wandering for the widening of the reconnection region. As it was discussed in LV99, the actual structure of turbulent reconnection layer is much more complex. In 3D reconnected magnetic loops move in the opposite directions and cross each other creating turbulent reconnection layers at smaller scales. These new layers, in their turn, present the microcosm of turbulent reconnection events. Therefore, we expect to see the hierarchy of turbulent reconnection layers with the decreasing thickness. These layers can increase $p_\bot$ of particles with different Larmor radii. 

It is important that both the increase of $p_{\|}$ and $p_\bot$ in reconnection regions happens for particle that do not experience scattering. The latter is a relatively slow process that limits the efficiency of the textbook version of the First Fermi acceleration in shocks. If scattering is not required the process of particle acceleration is getting much faster and therefore much more efficient. This makes us confident in the astrophysical importance of the First Order Fermi acceleration in turbulent magnetic reconnection. 

Naturally, the processes of ``collisionless'' acceleration in turbulent reconnection layers are complex. Therefore we view the calculations of the particle spectra in the pioneering study by \citet{deGouveiaDalPinoLazarian:2005} only as the first attempt to deal quantitatively with this interesting problem. The differences of the particle acceleration in tearing reconnection and turbulent reconnection were discussed in \cite{Beresnyak:2017} with more recent attempts to quantify the acceleration arising due to 3D turbulent reconenction that can be found in \cite{BeresnyakLi:2016}. 
We believe that more quantitative studies in this direction are necessary. 

\begin{figure}
\centering
\includegraphics[width=0.48\textwidth]{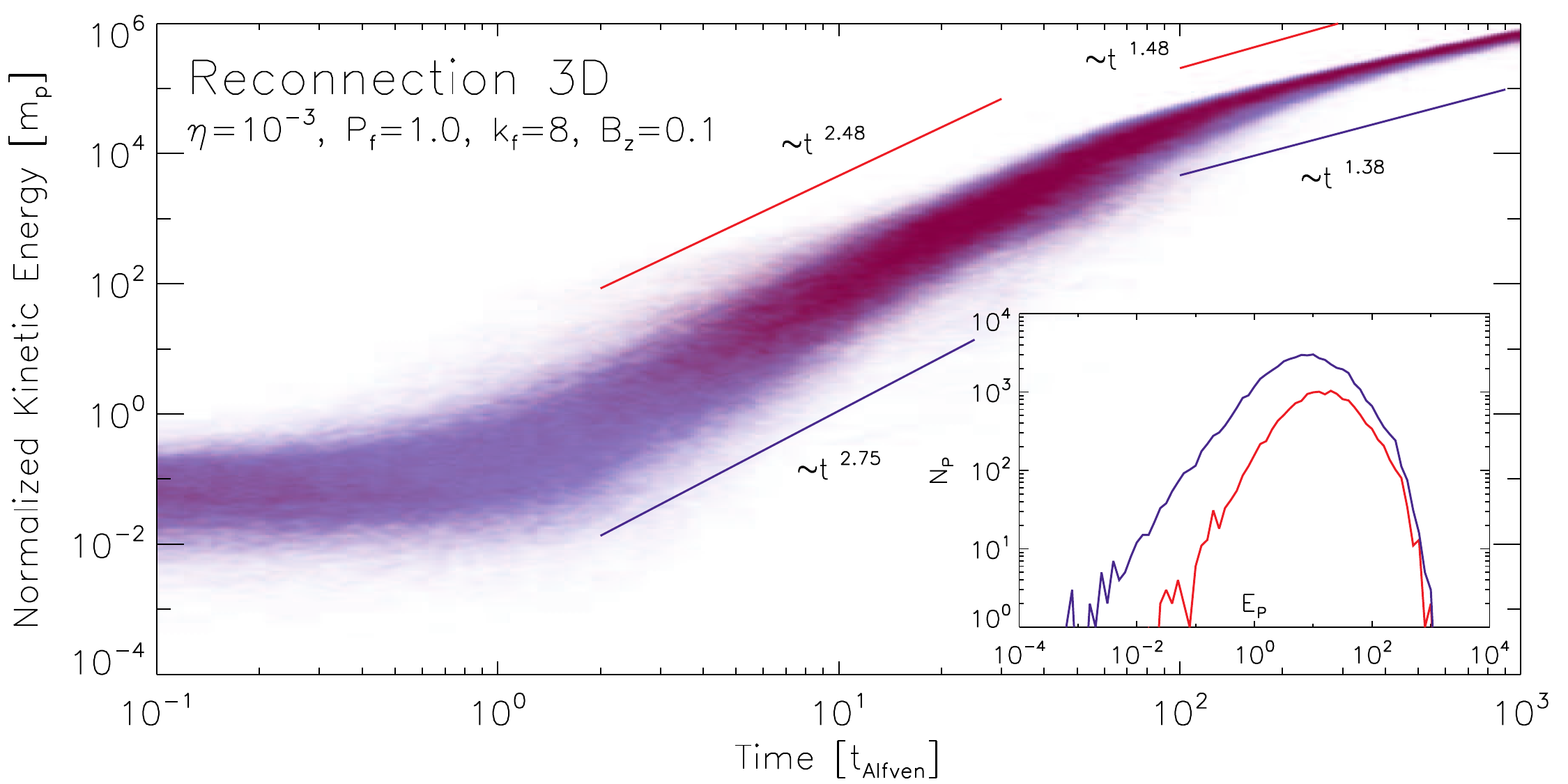}
\caption{Particle kinetic energy distributions for 10,000 protons injected in the simulated turbulent reconnection region. The colors indicate which velocity component is accelerated (red or blue for parallel or perpendicular, respectively). The energy is normalized by the rest proton mass. Reprinted figure with permission from \citet{Kowal_etal:2012b}. Copyright (2012) by the American Physical Society.}
\label{fig32}
\end{figure}

Testing of particle acceleration in a large-scale current sheet with embedded turbulence to make reconnection fast was performed in \citet{Kowal_etal:2012b} and its results are presented in Figure \ref{fig32}. The simulations were performed considering 3D MHD simulations of reconnection with the injection of 10,000 test particles.  The exponential growth of particle energy was reported with the acceleration rate is $\sim E^{-0.4}$ \citep{Khiali_etal:2015}. Both the parallel and perpendicular components of the particle momentum was growing, as it is shown in the subpanel of Figure \ref{fig32}. This suggests a rather complex process of acceleration presumably involving both processes depicted on the panels of Figure \ref{fig31}.

The process of turbulent reconnection requires 3D settings, while in 2D the acceleration of magnetic reconnection happens through the tearing instability. In this case the acceleration shown in the left panel of Figure \ref{fig31} still can happen, but in this degenerate case instead of 3D loops one has to deal with closed islands. The case of acceleration in islands was easier to demonstrate numerically and it was discussed for the 2D tearing reconnection in the subsequent paper by \citet{Drake_etal:2006}. Being published in Nature and appealing to the popular in the community tearing reconnection the paper had more impact that either GL05 or \citet{Lazarian:2005} publications. 

The acceleration arising from the 2D tearing reconnection can also be explained on the basis of the second adiabatic invariant. However, the deficiency of the process of acceleration in 2D compared to 3D is self-evident. The 2D magnetic islands are closed structures and the shrinking of magnetic field lines within these islands is limited. The islands can get from elliptical to circular and then further shrinking gets impossible. Therefore, unless the islands undergo a merger, the acceleration process within an individual islands stops as it gets circular. On the contrary, in 3D turbulent reconnection the loops can shrink all the way with no constraints arising from the artificial dimensional constraints. In fact, later research \cite[see][]{Dahlin_etal:2014, Dahlin_etal:2015, Dahlin_etal:2016, Dahlin_etal:2017} has shown that when the problem of tearing is considered in the space of correct dimensions, i.e. in 3D, the acceleration efficiency is increasing. In 3D islands become the contracting loops as in the original suggestion shown in the left panel of Figure \ref{fig31}.  

The contracting islands increase $p_\|$. The increase of $p_\bot$ as shown in the right panel of Figure \ref{fig31} as the 3D loops undergo intersect. However, as this process takes place for fluids of low viscosity the transition to full turbulence is expected as we discussed in \S \ref{sec:level2b}. In other words, we expect the acceleration within 3D reconnection layers to be a generic one. At the same time proving this with PIC simulations may not be easy. The serious deficiency of present day PIC simulations is that they face problems reproducing high Re number behavior of magnetized fluid in the reconenction regions. Therefore, such simulations, unlike the MHD ones, do not show trully turbulent outflows and, as a result, they are missing the key ingredient of realistic astrophysical reconnection, i.e. turbulence. It is important to understand that in the absence of turbulence the 3D reconnection does not feel additional degrees of freedom and therefore shows the 2D reconnection features.

Since it introduction, the new mechanism of acceleration by reconnection has been invoked in many studies (see 
\cite{Jaroschek_etal:2004,ZenitaniHoshino:2008,Zenitani_etal:2009,Cerutti_etal:2013,Cerutti_etal:2014,SironiSpitkovsky:2014,Werner_etal:2014,Guo2015}, see also \cite{HoshinoLyubarsky:2012} for a review). In particular, it was discussed in the context of acceleration of energetic particles in relativistic environments, like pulsars (e.g. \cite{Cerutti_etal:2013,Cerutti_etal:2014,SironiSpitkovsky:2014,UzdenskySpitkovsky:2014}) and relativistic jets of active galactic nuclei (AGNs) (e.g. \cite{Giannios:2010}). However, in most of the studies, the authors referred to 2D tearing reconnection, as the accepted process. But as we discussed above, turbulent reconnection is the necessary feature of realistic astrophysical settings and therefore the original acceleration process by 3D reconnection is more relevant to appeal in these studies.

The process of fast turbulent reconnection acceleration (TRA) is expected to be widespread. In particular, it has been discussed in \citet{LazarianOpher:2009} as a cause of the anomalous cosmic rays observed by Voyagers and in \citet{LazarianDesiati:2010} as a source of the observed cosmic ray anisotropies.
Turbulent reconnection was considered for the environments around of black hole sources \cite[GL05,][]{deGouveiaDalPino_etal:2010b, Kadowaki_etal:2015, Singh_etal:2015, Khiali_etal:2015, KhialideGouveiaDalPino:2015, delValle_etal:2016, Singh_etal:2018, Kadowaki_etal:2019}. The aforementioned studies are based on non-relativistic turbulent reconnection. However, as we discuss in \S \ref{sec:special} the relativistic and non-relativistic reconnection processes are rather similar. 


\subsection{Stochastic acceleration in MHD turbulence}
\label{stochastic_accel}

In the previous section we have discussed the acceleration of energetic particles in the large scale reconnection layers. However, as we discussed through the entire review, magnetic reconnection ubiquitous in turbulent fluids. For instance, the dynamics of magnetic field lines depicted in Figure \ref{fig11} is only possible through the unstoppable sequence of magnetic reconnection events changing the topologies of the magnetic field lines. The reconnection at all scales that happens over one eddy turnover time is picture of magnetic reconnection that was advocated in LV99 and that has been supported by more sophisticated theoretical arguments since then (see \S \ref{sec:violation1}). As a result, we expect the magnetic reconnection to be present through the entire turbulent volume and be an intrinsic, essential part of the turbulent cascade in magnetized fluids. 

Magnetic reconnection shrinks magnetic field lines and releases the energy stored in magnetic field. For the inertial range turbulence, the energy is being transferred to kinetic energy at the comparable scale. This kinetic energy bends the neighboring magnetic field flux, increasing magnetic field energy. In the turbulence with steady driving the transfer from magnetic energy to kinetic via the reconnection and its transfer via turbulent dynamo compensate each other. It is evident, that in the regions of magnetic reconnection magnetic field lines shrink and can accelerate particles, while in the regions of dynamo particles lose energy. This provides stochastic acceleration, but with a peculiar one. In particular, the increase of a particle energy as it get accelerated in a reconnection region can be significant. This is especially evident in superAlfvenic turbulence where extended reconnecting regions are produced. In other words, the process of acceleration is a random walk process, but with substantial increments in energy. In \citet{BruLaz16} this process was applied to the acceleration of cosmic rays in galaxy clusters and was termed 1.5 Fermi acceleration. Later this 1.5 Fermi acceleration was also employed  
to account for the acceleration of electrons in Gamma-ray bursts 
\citep{XuZ17,Xuy18}
and pulsar wind nebulae 
\citep{XuK19},
which successfully explains the features of their synchrotron spectra.

\subsection{Flares and bursts of reconnection}
\label{bursts}

In magnetically dominated media the release of magnetic energy via reconnection must result in the outflow that induces turbulence in astrophysical high Reynolds number plasmas. This, in its turn, inevitably increases the reconnection rate further enhancing turbulence. As a result, we get a reconnection instability as was discussed in LV99 and elaborated in \citet{LazarianVishniac:2009} (see more details in \cite{Lazarian:2005}). The applications of the theory range from solar flares to gamma ray bursts (GRBs) \cite[see][]{Lazarian_etal:2003}. This idea became the basis of the Internal-Collision-induced MAgnetic Reconnection 
and Turbulence (ICMART) model by \citet{ZhangYan:2011}, who showed that such a model can overcome several difficulties
of the traditional internal shock model. 
\citep{Ree94,Kob97,Dai98,Ghi00,Ku08} and can well interpret the lightcurves and spectra of GRBs 
\citep{ZhZ14,Uhm14,XuZ17,Xuy18}.

\begin{figure}[htbp]
\centering
\includegraphics[width=0.48\textwidth]{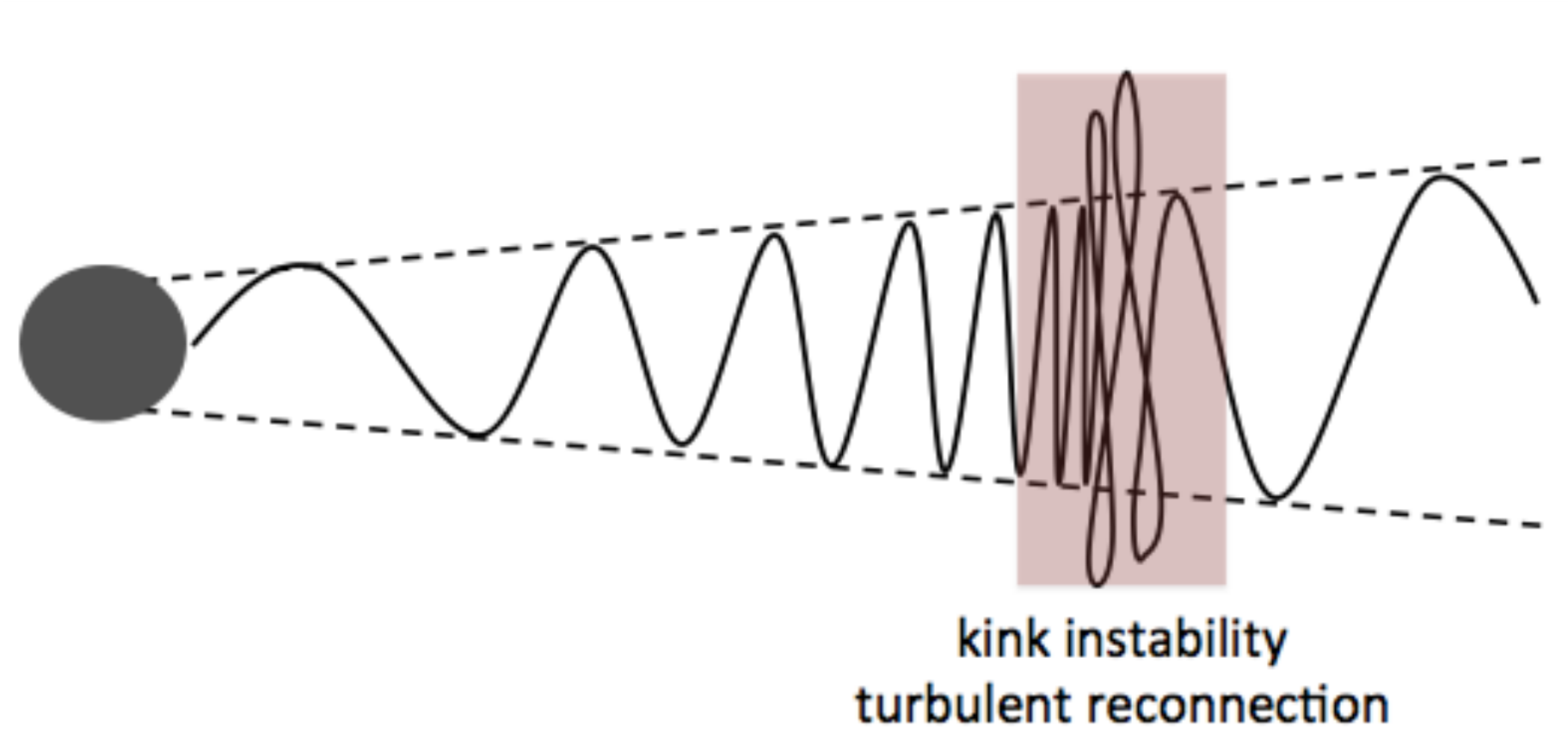}
\caption{Illustration of the kink instability in the GRB jet. From \citet{LazZ18}.}
\label{fig33}
\end{figure}

ZY11 suggested that the magnetic field reversals required to trigger ICMART events may be achieved through
internal collisions among high-$\sigma$ blobs.  \citet{Den15} performed  relativistic MHD
numerical simulations of collisions of magnetic blobs, and reported significant magnetic dissipation
 with an efficiency above $30\%$. However, the simulations were on the global scale and no detailed
turbulent reconnection was observed. \citet{LazZ18} used the advances of relativistic turbulent reconnection 
to improve the ICMART model. The authors also modified the ICMART model by identifying the kink
instability as the most probable mechanism of creating the magnetic configurations prone to reconnection and 
triggering the turbulent reconnection (see Figure \ref{fig33}).

The difference of the model in \citet{LazZ18} from other kink-driven models of GRBs (e.g. 
\cite{DrenkhahnSpruit:2002,GianniosSpruit:2006,Giannios:2008,McKinneyUzdensky:2012})
is that the kink instability also induces turbulence 
\citep{GalsgaardNordlund:1997,GerrardHood:2003}, 
which drives magnetic fast reconnection. The new model gets support with the simulations in \citet{2016ApJ...824...48S}.
There 
the kink instability was numerically studied in jets and the regions of kink were shown to coincide with the sites of magnetic reconnection.
The simulations were performed using the ideal MHD, but the authors authors appealed to LV99 reconnection model of reconnection to
claim that the reconnection that they observe is not a numerical artifact, but the actual process of magnetic reconnection. The improved
ICMART model was shown to provide a good agreement with observations, including the lightcurves, spectra and polarization. 

Energy release and particle acceleration arising from turbulent magnetic reconnection associated with the environment of black hole sources, e.g. active galactic nuclei (AGNs) and galactic black hole binaries (GHBs) has been also explored.This research was initiated in GL05, where it was proposed that turbulent reconnection takes place between the magnetic field lines arising from the inner accretion disk and the magnetosphere of the BH. Later, \citet{Kadowaki_etal:2015} revisited the aforementioned model and extended the study to explore also the gamma-ray flare emission arising from these sources. This study confirmed the dependence reported in GL05, namely, that if observed fast reconnection is driven by turbulence, there is a correlation between the calculated fast magnetic reconnection power and the BH. In fact, this dependence was show in \citet{Kadowaki_etal:2015} to span over the range of  mass varying by a factor $10^{10}$. It is interesting that this can explain not only the observed radio, but also the gamma-ray emission from GBHs and low luminous AGNs (LLAGNs). This match was confirmed for an extensive sample of more than 230 sources which include those of the so called fundamental plane of black hole activity \citep{Merloni_etal:2003} high luminous AGNs whose jet points to the line of sight) and GRBs does not follow the same trend as that of the low luminous AGNs and GBHs, suggesting that the observed radio and gamma-ray emission in these cases is not produced at the cores of the sources. The study by \citet{Singh_etal:2015} suggests that the results in \citet{Kadowaki_etal:2015} are robust and depend on turbulent reconnection properties rather than on the details of the accretion physics. 

\subsection{Implications of turbulent reconnection: summary}

Magnetic reconnection is an part and parcel of the strong Alfvenic turbulent cascade. The reconnection happens at all scales of the cascade. The implications of this are really numerous and only some of them have been explored so far.

Changing the conventional flux freezing paradigm, turbulent reconnection induces the process of reconnection diffusion, i.e. the process of diffusion of magnetic field and matter with the rates that are determined by the level of turbulence. As a result, the loss of magnetic flux in the process of star formation is significantly modified. It becomes independent of the ionization rate of the turbuletn fluid, which is a big change in understanding of star formation. Reconnection diffusion is important for the collapse of molecular clouds as well as for the formation of protoplanetary disks. The calculations that account for the reconnection diffusion can explain observations that are puzzling otherwise.

Reconnection diffusion exceeds the magnetic diffusion arising from microscopic plasma effects and resistivity. Therefore it is also an essential process for the magnetic field generation by turbulence, i.e. for turbulent dynamo. The predictions of the dynamo theory that accounts for reconnection diffusion correspond well to numerical simulations.

During magnetic reconnection magnetic lines shrink. Therefore energetic particles entrained on such magnetic field lines are getting the boost for their parallel component of momentum. This is the First order Fermi acceleration process and it takes place during the reconnection of oppositely directed large scale fluxes. Only 3D turbulent reconnection represents this process correctly. In comparison, the 2D tearing reconnection produces 2D islands, which are degenerate configurations arising only due to the artificially reduced dimension of the process. The 2D closed magnetic islands induce the acceleration that is less efficient compared to the acceleration to the  magnetic loops formed in 3D case.

If turbulence is just driven in the volume and no large scale magnetic field reversals are induced, the particle acceleration induced by reconnection still take place. However, the overall acceleration is stochastic:  particles are accelerated in the regions where magnetic field shrinks due to reconnection to the regions where magnetic field lines stretch due to turbulent dynamo. The two processes are intrinsic to the MHD turbulent cascade. 

Turbulent magnetic reconnection can be successfully applied to explain different observational phenomena. This is a big mostly unexplored field. The results obtained through applying the theory of turbulent reconnection to AGNs, GRBs etc. are encoraging and call for more research efforts in studying the astrophysical implications of turbulent reconnection.

\section{Alternative ideas and misconceptions in the literature}
\label{sec:alternatives}

\subsection{Definitions of magnetic reconnection and the understanding of stochastic flux freezing}

As we mentioned earlier, for decades magnetic reconnection was viewed as something very special and exceptional, necessarily associated with bursts and flares of magnetic energy release. In fact, a notion that magnetic reconnection requires magnetically-dominated low-beta plasmas can still can be occasionally found in the literature. Thus we mention possible definitions of reconnection.

Resistive diffusion is a process at the base of the classical Sweet-Parker reconnection. It definitely changes the magnetic connectivity of plasma elements, but the rate of this reconnection is slow, i.e. it strongly depends on the resistivity. Therefore for most astrophysical settings the Sweet-Parker reconnection rate is negligibly slow. 

Fast reconnection require some additional effects that allow the magnetic connectivity of plasma elements to proceed at rates independent on resistivity. In 1999 two ideas were competing: the plasma effects, in particular Hall-effect, creating X-point Petschek-type reconnection and 3D turbulence. By 2019 the first idea is no more popular and substituted by the tearing idea. At the same time, more evidence, as we have discussed in the review, accommulated in support of LV99 turbulent reconnection model. 

The reconnection is currently understood as an entire suite of reconnection phenomena, each of which has great physical significance, e.g.:
\begin{enumerate}
\item Field-line slippage or reconnection diffusion, which is the basic process underlying all others. This is an important process,  e.g., in star formation and in creating deviations from the Parker-spiral model. 
\item Topological reconnection, or the change in topology of the field-line passing through a given plasma element. An example is diffusion of plasma across a separatrix surface, as when solar-wind plasma leaks into the magnetosphere. In 3D turbulence this actually happens by the Forbes-Priest ''magnetic flipping'' phenomena, not by diffusion across separatrices \citep{PriestForbes:2000}.
\item  Magnetic energy release, in particular Alfv\'enic jets driven by field-line tension of topologically reconnected lines. This is just the classic flare-type reconnection. Of course, it requires reconnection in both senses 1) and 2).   
\end{enumerate}

As we discussed in \S \ref{ssec: rdffvio} and 
\S \ref{sec:level1} the classical flux freezing is violated in the turbulent fluids, where magnetic reconnection happens through the entire turbulent volume at all scales. However, this does not mean that suddenly the fluid getting highly resistivive or/and that the notion of magnetic field line completely loses its meaning. A common misunderstanding seems to be that turbulent violation of flux-feezing is an “all or nothing” thing. In reality, it was predicted back in LV99 that the reconnection rate depends upon the turbulence intensity, i.e. magnitude of the turbulent velocity fluctuations relative to the local mean Alfven speed. As a result, in low Alfven Mach number turbulence, i.e. for $M_A\ll 1$, the flux is ``nearly frozen'' \cite[see][]{Eyink_etal:2011}. For instance, by analysing the spacecraft data Eyink (2015) estimated that that the turbulent breakdown of flux-freezing in the solar wind was of the order 5\%, exactly what was 
required to explain the observed deviations from the Parker spiral model. Fluids with low $M_A$ provide a gradual transition to the settings where the conventional flux freezing is approximately true. 

\subsection{Critical importance of 3D simulations}

Most of the reconnection modeling is currently done using 2D PIC simulations. This is usually justified by the higher resolution that is available for such simulations compared to their 3D counterparts. We feel that these simulations are missing the physics essential in the large scale astrophysical reconnection processes. 

The modeling of a physical process using in a reduced dimensions is always a tricky and somewhat dangerous exercise. In many instances hydrodynamics and MHD flows radically change their properties if instead of 3D, the dimensions are reduced to 2D. The direction of energy flow in hydrodynamic turbulence is a notorious example of how much the reduction of the dimensions to 2D can change the physics of the process. Therefore it is essential that modeling in 2D is not taken for granted in any research, unless there are solid physical arguments presented that the reduced dimension does not change the essence of the process.

Magnetic reconnection is not an exception from the general rule above which is vivid for the case of turbulent reconnection. Note, that the 2D picture of Sweet-Parker reconnection represents the actual 3D physics well exactly as the fluids are assumed laminar, i.e. the fluid motions are very strongly restricted. As it was shown in LV99, the 3D effect of magnetic field wandering radically changes the nature of magnetic reconnection making it fast. Independent arguments on the importance of 3D geometry for magnetic reconnection are provided by Boozer (2018). 

The nature of MHD turbulence in 2D and 3D are different (see Eyink et al. 2011). The effect of magnetic field wandering induced by Alfvenic turbulence that is at the core of fast turbulent reconnection in 3D is not present in 2D. Therefore the issue of whether the 2D magnetic reconnection in the presence of turbulence is fast \cite[see][]{Loureiro_etal:2009, Kulpa-Dybel_etal:2010} is of pure academic interest. Even if one can prove that magnetic field wondering in 2D can proceed through magnetic nulls that are present at high Reynolds numbers, the actual 3D physics of turbulent reconnection is, nevertheless, is radically different. 

In addition, the development of turbulence, its evolution and its effects are radically different in 2D and 3D reconnection layers. Similarly, the development of instabilities and their interplay with turbulence also depend whether the reconnection layer is 3D or 2D. For instance, the tearing instability in 2D is accepted to be the dominant process driving magnetic reconnection. It was demonstrated that in 2D the reconnection proceeds in the plasmoid regime for all numerically explored Lundquist numbers and no MHD turbulence is generated. In this respect one should remember that not every chaotic evolving pattern is turbulence. Turbulence requires a cascade of energy. 2D tearing produces an inverse cascade of merging loops, which is different from MHD turbulence. On the contrary, as we discussed in \S \ref{sec:testing1}, 3D magnetic reconnection produces turbulence with the spectrum and anisotropy expected for MHD turbulence (see \S \ref{sec:turbulence2}). Moreover, the turbulence suppresses tearing and changing the nature of reconnection. Therefore, as we discuss below, 
 the 2D modeling of tearing reconnection  does not provide a correct  picture of the actual astrophysical reconnection. 
 
 The difficulty of studying 3D reconnection is the significant computational cost of such simulation. Therefore, the we are only approaching to the state that PIC simulations can resolve the initial stages of turbulence development in reconnection layers (see \S \ref{sec:kinetic}). However, these numerical difficulties do not justify making conclusions about astrophysical reconnection based on high resolution 2D PIC simulations. The two type of simulations sample different physics. As we demonstrated in \S \ref{sec:testing} and \S \ref{sec:special} the testing of 3D reconnection is already feasible with MHD codes. The applicability of MHD treatment for magnetic reconnection at scales much larger than the relevant plasma scales is demonstrated e.g. in \S \ref{sec:level4}. The numerical simulations discussed in \S \ref{Hall} (see also Kowal et al. 2009) also support this conclusion. 
 As a result, MHD modeling of 3D reconnection at present is the most realistic approach for exploring astrophysical reconnection.
 
 We would not judge to what extend the 2D PIC simulations reproduce the physics of plasma-scale reconnection relevant, e.g. magneotspheric reconnection. The  3D effects may be important if magnetic field wandering induced by whistler or kinetic Alfven turbulence (see \S \ref{whistler}) plays an important role for such reconnection.

\subsection{Comparison of turbulent and tearing reconnection models}
\label{sec:alternatives1}

Turbulent reconnection emerged at the time when the competing explanation for fast reconnection was the 
regular X-point Petsheck model. It is obvious that the latter reconnection scheme is difficult to realize in any realistic astrophysical
setting as the X-point configuration of the reconnection region would have to be preserved over the astronomically large
scale $L$ that enters the definition of the Lunquist number $S$ in spite of the action of external forces acting on it 
in the media. In fact, further research has shown that, even in the absence of external forces, a global X-point configuration 
collapses. Instead, the tearing stochastic scheme, which is closer to spirit to LV99, emerges in 2D configuration studies 
\cite[see][]{Loureiro_etal:2007, Uzdensky_etal:2010}.

Tearing reconnection was discussed in the literature prior to the more recent surge of interest to the process. The most notable contribution was done by Syrovatskii and his group 
\cite[see][]{Syrovatskii:1981}. 
It was also quantitatively discussed in LV99 and was shown to be a subdominant process compared to turbulent reconnection. Nevertheless, this is an important
process that can play the role of a trigger for a more robust turbulent reconnection process in the settings when the reconnecting magnetic fields are mostly laminar. The mechanism 
has been widely recognized by the community more recently 
\citep{Biskamp1986,ShibataTanuma:2001,Daughton_etal:2006,Daughton_etal:2009b,Daughton_etal:2011,Dau14,Fer12,Loureiro_etal:2007,Loureiro2012,Lapenta:2008,Bhattacharjee_etal:2009,Cassak_etal:2009,Huang2010,HuangBhattacharjee:2016,She10,Uzdensky_etal:2010,Huang2011,Bar11,Shen11,Tak13,Wy14}. 
The important feature recently capitalized upon is that in 2D tearing reconnection can proceed at smaller scales at higher rates. Magnetic bubbles so-produced undergo a cascade of merging events producing larger and larger bubbles, the size of which eventually determines the thickness of the outflow region. Unlike the X-point Petsheck 
model, the tearing reconnection is more difficult to quench by random external forcing as the formed magnetic bubbles, which are also called plasmoids, resist the external pressure that attempts to chock
the outflow. Conceptually, tearing reconnection, which invokes a cascade of mergers that increases the size of plasmoids, is closer to turbulent reconnection. 

In 2D MHD numerical simulations the transfer from Sweet-Parker to tearing reconnection was observed in the two dimensional current sheet starting with a particular Lundquist number larger than $S\approx 10^4$.
The resulting reconnection does not depend on the fluid resistivity. The study of tearing momentarily eclipsed the earlier mainstream research of the reconnection community, which attempted to explain fast reconnection appealing to the collisionless plasma effects that were invoked to stabilize the Petschek-type X point configuration for reconnection 
\citep{Shay1998,Drake:2001,Drake_etal:2006}.
Abandoning the artificially extended X point configurations was the natural and right step. Indeed, the stability of X- points in the situation of realistic astrophysical forcing was very dubious, as was pointed out e.g. in LV99. In a sense, tearing introduced stochasticity into reconnection and brought the mainstream models of magnetic reconnection towards the LV99 model. The tearing reconnection presents volume-filling Y-type reconnection with broad outflow similar to that in LV99 and in contrast to the Petschek X-point reconnection with a very localized region where reconnection takes place. It also allowed some ideas developed within the framework of LV99 theory to be transferred to the tearing mechanism. For instance, the model of First order Fermi acceleration of cosmic rays in turbulent reconnection \citep{deGouveiaDalPinoLazarian:2005,Lazarian:2005} was successfully applied within the tearing mechanism \citep{Drake_etal:2006}.

 The problems of transferring the ideas obtained in 2D to actual 3D settings of any realistic reconnection even without turbulence were discussed in a series of recent papers by \cite{Boozer:2019a, Boozer:2019b, Boozer:2019c}. Here we focus, however, on the reconnection in astrophysical environments where turbulence is a natural component that is extremely difficult to avoid in any realistic setting. Turbulence makes the 2D and 3D even more different as the nature of MHD turbulence in 3D is very different.
 
 For instance, arguments why LV99 model may not be applicable to 2D configurations were provided in \citet{Eyink_etal:2011}. Therefore, it is possible that in 2D tearing is the dominant mechanism but in our 3D world the situation appears very different. Numerical tests in a number of recent studies \citep{Kowal_etal:2012a, Eyink_etal:2013, Oishi_etal:2015, Beresnyak:2017, Kowal_etal:2017, Takamoto:2018} show that both in non-relativistic and relativistic MHD simulations the transfer to turbulent reconnection is inevitable even if the reconnection starts from the laminar state. In particular, in \S \ref{sec:testing1} we provide evidence that Kelvin-Hemholtz instability dominates over tearing instability in the outflow region. One may argue that the tearing instability can play the role of initiating the reconnection, but the subsequent process proceeds as driven by turbulence.  
From the theoretical point of view, the dominance of turbulent reconnection in astrophysics seems inevitable. 
In realistic 3D configurations the thicker outflows induced by plasmoid/tearing reconnection inevitably induce turbulence.
 Indeed, the mass conservation
constraint requires that in order to have fast reconnection one has to increase
the outflow region thickness $\Delta$ in proportion to $L_x$, which means 
proportional to the Lundquist number $S$. The Reynolds number $Re$ of the
outflow --- approximately $\Delta V_A /\nu$ where $\nu$ is viscosity --- grows also as $S$. As $Re$ increases the flow gets turbulent. 
Once the shearing rate introduced by eddies gets larger than the rate of the tearing instability growth, the instability should be suppressed. 

Potentially, one can increase the domain of tearing reconnection in 3D by increasing the Prandtl number $Pt = \eta/\nu$  of turbulence. Indeed,  as we discussed earlier,
the mass conservation constraint in the case of fast reconnection dictates the increase of the outflow region thickness in proportion to the extent of the zone of the magnetic field contact, which for the Y-type tearing reconnection is proportional to $S$. This provides $Re \approx Pt^{-1}S$ and requires very high computational effort to test the high $Pt$ self-driven reconnection. Note, that the increase of the numerical efforts grows in proportion to $S^4$, i.e. 3 dimensions + time. This is a steep dependence that frustrates the efforts to increase the Lundquist numbers significantly.
However, this is only applicable if there 
is no external turbulence driving, which is rather unrealistic in astrophysical environments. If turbulence 
is present, the turbulent reconnection proceeds along the lines discussed in  \S \ref{sec:model}. 

We believe that the reason that the tearing reconnection model is so popular within the reconnection community is that in 3D the PIC simulations do not yet have enough particles to reproduce turbulent fluid behavior in the reconnection zone. Therefore for most PIC simulations the regime of MHD-like turbulent reconnection is very difficult to achieve. Nevertheless,  in \S \ref{sec:kinetic} we reported the first results supporting turbulent reconnection that were obtained using a PIC code. We expect to see more PIC results relevant to MHD turbulent reconnection in the future. 

What we say about tearing/plasmoid reconnection should not be interpreted that that this type of reconnection is never present in 3D configurations. Indeed, if the initial magnetic field configuration is not turbulent it takes time given by Eq. (\ref{tcas}) for turbulent to develop. Thus we can expect to have observe the transient 3D tearing accompanied with the ejection of plasmoids if the reconnecting fluxes were not turbulent initially \cite[see][]{Huang_etal:2015} However, the reconnection is expected to proceed along the turbulent reconnection path at later times \cite[see][]{Kowal_etal:2017, Kowal_etal:2019}.

\subsection{Objections to the concept of ``reconnection-mediated turbulence''}
\label{sec:alternatives2}

As discussed above, tearing/plasmoid instability of current sheets is one possible route to turbulence that can enhance magnetic reconnection rates. It has also long been understood that such a tearing instability could occur for microscale current sheets in
MHD turbulence. For example, LV99 in their Appendix C considered this possibility and concluded that tearing instability would affect negligibly their estimates for turbulent reconnection rates in the inertial range and at larger scales. Recently, however, an extensive literature has developed suggesting that reconnection in MHD turbulence must necessarily be induced by tearing instability, \cite[see][]{Loureiro:2017aa,Bol17,Lou17,Wal18, Mallet_etal:2017a, Mallet_etal:2017b, Com18,vech2018magnetic}. The authors explore the subject of tearing instability and its possible modification of the properties of MHD turbulence. The conclusion of those works is that, at extremely high kinetic and magnetic Reynolds numbers, the microscale current sheets do become tearing unstable and that this instability affects the turbulence at scales smaller than a critical scale $\lambda_c,$ somewhat larger than the dissipation scale of turbulence. This new ``tearing-mediated regime'' 
of MHD turbulence does not modify the inertial-range properties of MHD turbulence at scales $\gg\lambda_c,$  
although the authors of the above works 
emphasize that the range of scales between $\lambda_c$
and the resistive scale (or the ion gyroradius in collisionless plasma turbulence) can be sizable in practice. The detailed conclusions of these papers depend upon extrapolating some properties of laminar tearing instability to turbulent environments, which might be questioned. In fact, We expect to see the suppression of the tearing instability by turbulence as it is described in \S \ref{sec:testing1}. However, the essential picture in this literature is fully consistent with the earlier conclusions of LV99, since GS95 scaling would not be altered at 
scales $>\lambda_c.$ In fact, the conclusions of LV99 do not dependent upon GS95 scaling but hold in a qualitatively similar form for any other spectrum of MHD turbulence 
(see LV99, Appendix D). The possible change of the turbulence spectrum that the authors discuss does not affect turbulent reconnection physics over the wide inertial range of scales $>\lambda_c.$ The possible effects of dynamical alignment \cite[][see more discussion of the effect and its testing in \citeauthor{BeresnyakLazarian:2019}, \citeyear{BeresnyakLazarian:2019}]{Boldyrev:2006, BeresnyakLazarian:2006} do not change the model of turbulent reconnection either. 


There is a serious misconception, however, that has been promulgated by the aforementioned works, e.g.
\cite{Loureiro:2017aa,Bol17,Lou17,Wal18, Mallet_etal:2017a, Mallet_etal:2017b,Com18,vech2018magnetic}. This definitely requires our comment in view of our discussion of flux freezing in 
 \S \ref{sec:level1}. The possible effects of tearing instability discussed above have also been described as ``reconnection-mediated turbulence'' or a modification 
of MHD turbulence ``due to reconnection'' at scales smaller than $\lambda_c.$ There is an implicit assumption in such statements that fast magnetic reconnection 
does not occur at inertial-range scales larger than this critical value and that it only happens at scales $\lambda_c$ due to tearing instability.
For example, 
\cite{Loureiro:2017aa} 
regard it as a major surprise that in MHD turbulence ``the anisotropic, current-sheetlike eddies become the sites of magnetic reconnection 
{\it before} [their emphasis] the formal Kolmogorov dissipation scale is reached'' and they argue that such fast reconnection sets in only because 
``Sweet-Parker current sheets above a certain critical aspect ratio ... are violently unstable to the formation of multiple magnetic islands, or plasmoids.'' 
In our opinion, both of these assumptions are false. First, there are many other viable pathways to MHD turbulence and fast reconnection besides tearing instability. 
More importantly, reconnection occurs at {\it all scales} in MHD turbulence, for magnetic eddies of {\it all sizes} (see more discussion in \S \ref{insight}). 

As we have discussed in e.g. in \S \ref{sec:level1} and \S \ref{insight}, that fast reconnection at all scales is part and parcel of the MHD turbulent cascade and it is absolutely 
essential for the dynamical consistency of MHD turbulence. There are also many observable consequences, such as deviations from the Parker spiral 
model \citep{Eyink2015} and reconnection jets/exhausts seen at all scales in the solar wind \citep{gosling2012magnetic}.
In fact, it has recently been conceded by one of the authors of 
\cite{Mallet_etal:2017a, Mallet_etal:2017b}
that ``magnetic fields reconnect at every scale over a time comparable with the correlation time associated with 
this scale (55--57) and thus, cannot preserve their identity over more than one parallel correlation scale of the Alfv\'enic turbulence'' 
\cite[][p. 5]{Mey19}.
This physical effect underlies the new theory proposed in 
\cite{Mey19} 
for fluidization of compressible modes in collisionless plasma turbulence. We therefore believe that statements 
about modifications of MHD turbulence ``by reconnection'' reflect not just a poor choice of terminology but in fact promote a real misunderstanding of the physics of turbulent magnetic reconnection. We therefore urge the community to adopt the accurate terminology of ``tearing-mediated MHD turbulence'' for such possible small-scale effects, as in the recent papers 
\citep{Wal18,Com18}, and not the physically misleading language of ``reconnection-mediated turbulence''. 
MHD turbulence always involves magnetic reconnection and at all scales (see \S \ref{ssec: rdffvio}, \S \ref{insight}). 

A separate issue is whether the theoretical treatment of ``tearing-mediated turbulence'' is accurate in the cited papers 
\citep{Loureiro:2017aa,Bol17,Lou17,Wal18, Mallet_etal:2017a, Mallet_etal:2017b, Com18,vech2018magnetic}. Indeed these various works reach 
somewhat divergent conclusions among themselves. \cite{Wal18} present numerical evidence in a special 2D setting that tearing 
instability can compete with nonlinear evolution and \cite{vech2018magnetic} interpret a break observed around $0.1-1$ Hz 
in solar wind magnetic energy spectra in terms of such effects. Existing theoretical treatments, however, 
assume that tearing rates are the same as for laminar current sheets and are unaffected by the 
surrounding turbulent environment. The validity of this assumption is not so clear.  We presented evidence 
in  \S \ref{sec:testing1}  that tearing instability is subdominant and suppressed even for the case of 3D freely-evolving 
reconnection. 

We note that another questionable point of the ``tearing-mediated turbulence'' that this model is based on the assumption that the dynamical alignment steadily increases with the decrease of the scale of turbulent motions over the entire turbulence inertial range it gets so large that the tearing instability gets efficient. This idea of ever increasing dynamical alignment, however, is challenged in a number of papers \cite[see][and ref. therein]{BeresnyakLazarian:2019} through the analysis of the high resolution numerical data. In fact, the data available to us indicates that the dynamical alignment is a transient effect at the vicinity of the injection scale and therefore it cannot proceed to the scales at which the tearing instability is expected. In other words, numerical simulations testify that the turbulence scaling that we described in \S \ref{sec:turbulence2} are correct. These scaling corresponds to MHD turbulence without any dynamical alignment over the ineritial range. At the same time, the predictions for the spectral slope and and the change of anisotropy with the scale predicted in the turbulence model with the dynamical alignment \citep{Boldyrev:2006} are inconsistent with numerical simulations \cite[see][]{Beresnyak:2014, BeresnyakLazarian:2019}. This surely leaves a lot of questions, e.g what the cause of the dynamical alignment that is seen over a limited range of scales is not yet clear.\footnote{One can argue that if this effect is not related to turbulence driving, it could be that other instabilities, e.g. Kelvin-Helmholtz one, stop this alignment before the tearing instability can become efficient.} However, the discussion of these issues is far beyond the scope of our review.

Naturally, more work will be required to properly test the theories of ``tearing-mediated turbulence'' and 
their underlying assumptions. We feel that it is important to stress that while the hypothetical tearing range is a reconnection range, the turbulence inertial-range at larger scales is also a reconnection range. The latter encompasses a much broader range of scales. For instance, in the solar wind the range of scales that is relevant to reconnection of magnetic structures is much larger than ion scales, including events of the type extensively catalogued in \cite{gosling2012magnetic}.

In any case, it is clear that tearing instability is not necessary for fast reconnection
in MHD turbulence. Reconnection rates independent of resistivity are observed at all inertial-range scales in 
current 3D MHD turbulence simulations (see \S \ref{ssec: doffvio})  and in simulations of large-scale magnetic reconnection in the presence of driven turbulence (see \S \ref{sec:testing1}), 
but the Lundquist numbers of microscale current sheets in those simulations are far below the critical values
$\sim 10^4$ or higher required for tearing instability. Even at the astronomically high Reynolds numbers that would be 
required to make microscale current sheets tearing unstable, the effects of such tearing will be limited to 
tiny length scales  $<\lambda_c$ and irrelevant for the astrophysically significant reconnection at much larger scales.

 \subsection{Objections to the concept of turbulent resistivity and the mean-field approach to reconnection}
 \label{sec:alsternatives3}
 
 LV99 and further papers on turbulent reconnection predict dramatic changes of the dynamics of magnetic fields in turbulent fluids compared to their laminar counterparts. These changes sometimes are erroneously associated with the concept of ``turbulent resistivity''. We strongly object to this association and claim that the description of magnetic phenomena based on the  ``turbulent resistivity'' idea is misleading and has {\it fatal} problems.
 
It is obvious that, if we know the reconnection rate, e.g. from LV99, then an eddy-resistivity can always be adjusted and tuned by hand to achieve the known required rate.
But this is engineering with little scientific justification. While the tuned reconnection rate will be correct by construction, this unphysical model will make other
predictions that will be wrong. For instance, the required large eddy-resistivity will smooth out all turbulent magnetic structure that are below an ``eddy-resistive scale''
$\ell_{eddy}$. This is in contrast to the physical reality, where turbulence produces strong small-scale inhomogeneities, such as current sheets, down to the dissipation
micro-scale. Naturally, field-lines in the flow smoothed by assumed eddy-resistivity will not show the super-diffusive Richardson-type separation at scales below
$\ell_{eddy}$. These are just a few examples how the wrong concept of ``eddy resistivity'' misrepresents the reality of magnetic turbulence. As a result, no correct
understanding of particle transport/scattering/acceleration in the turbulent reconnection zone is possible. In addition, it is possible to show that in the case of
relativistic reconnection, turbulent resistivities will introduce acausal, faster than light propagation effects.

The important point to remember, as stressed in the classical monograph of \cite{tennekes1972first} on fluid turbulence, is that the concept of ``turbulent diffusivity'' 
is both a very crude approximation and, also, ``a property of the flow, not of the fluid''.  One must coarse-grain the turbulent flow at some length-scale 
$\ell$ in order to see such effects in the averaged equations for the larger eddies.  As we discussed in \S \ref{renormalizing}, 
coarse-graining the MHD equations 
by eliminating modes at scales smaller than some length $\ell$ will indeed introduce a ``turbulent electric field'', i.e. an effective field induced by motions 
of magnetized eddies at smaller scales. However, as is well known in the fluid turbulence community, the resulting turbulent transport is not ``down-gradient''  and 
cannot be accurately represented by an enhanced diffusivity. This is what \cite{tennekes1972first} call the ``gradient-transport fallacy''. 
The physical reason arises from the fact that turbulence lacks separation in scales. This makes it impossible 
to use a simple ``eddy-resistivity'' description of the process. As a result, energy is not only absorbed by the smaller eddies, but also supplied by them, a phenomenon called 
``backscatter''. One should also remember that the turbulent electric field both destroys and creates magnetic flux. In a steady state MHD turbulence the processes of 
annihilation of magnetic field by reconnection and creation of magnetic field by turbulent dynamo are in balance. In the language of cascades, the turbulent electric field
creates both the forward cascade of magnetic energy and the inverse cascade of magnetic helicity. 
 
Attempts are often made to incorporate these effects into the description of MHD turbulence and reconnection by the ``mean-field theory'' that was originally 
developed in an attempt to understand turbulent magnetic dynamos.  In mean-field electrodynamics, the effects of turbulence are described using parameters such 
as anisotropic turbulent magnetic diffusivity 
$\beta_{ijkl}$ and magnetic $\alpha$-tensor $\alpha_{ij}$ so that the ``turbulent electric field'' becomes $\alpha_{ij}\overline{B}_j + \beta_{ijkl}\partial_k\overline{B}_l$.
Here the underlying assumption is that there is a wide separation $L_\nabla\gg \ell_T$ between the scale of variation $L_\nabla$ of mean-fields
and the size $\ell_T$ of the largest turbulent eddies, with $\overline{\bB}$ representing the field averaged over turbulent ensembles or spatially 
coarse-grained over some (arbitrary) length-scale $\ell$ satisfying  $L_\nabla\gg \ell\gg \ell_T.$
This mean-field approach was used in \cite{Guo_etal:2012} to formulate a model of turbulent reconnection alternative to LV99. 
Another model of turbulent reconnection based on the mean-field approach is presented in \cite{HigashimoriHoshino:2012}.
However, the obtained expressions for fast reconnection, unlike those in LV99, are grossly contradicted by the numerical tests of turbulent reconnection 
that we discussed in  \S \ref{sec:testing1}. This is not surprising, because the mean-field theory's fundamental assumption of scale-separation   
is badly violated in MHD turbulence. 

There is, in addition, a much deeper conceptual problem with attempts to explain fast reconnection by averaging fields over space-time or over 
ensembles and by invoking an enhanced resistivity. Such averaging is a purely passive operation on the turbulent fields that, obviously, 
can alter none of the objective physics but merely changes the mathematical description. The results depend entirely upon how  
one performs the averages.  {\it It is not possible to claim an ``explanation'' of the rapid pace of magnetic reconnection using such an approach, 
unless it is demonstrated that the reconnection rates obtained in the theory are strictly independent of the length and timescales of the averaging.}  
This is the idea of ``renormalization group invariance'' that we invoked in our discussion in  \S \ref{sec:level1}. Such invariance is never
demonstrated in applications of mean-field electrodynamics to turbulent flows without scale-separation. On the contrary, it is by systematically 
applying this principle in our coarse-graining RG analysis that we derive the necessary conditions for fast reconnection independent of any microscopic plasma non-idealities (see \S \ref{sec:level4}).  

Likewise, the stochastic flux-freezing that we discussed in  \S \ref{flux-freezing},  a concept intrinsically connected to fast turbulent reconnection, 
is very different from the dissipation of magnetic field by resistivity. The latter can be modelled in a Lagrangian description by field-line   
stochastic Brownian motion with dispersion growing $\propto \eta t$ in time. By contrast, the stochastic flux-freezing observed at very high Reynolds 
numbers corresponds to field-line dispersion growing by Richardson super-ballistic separation or, assuming GS95 scaling, 
separation $\propto \varepsilon t^3.$  Other crucial properties are also quite distinct. For example, it is well-known that magnetic 
helicity undergoes an inverse cascade but no forward cascade in MHD turbulence and is thus very well-conserved 
\citep{berger1984rigorous,aluie2017coarse}. This conservation is the cornerstone of the theory of large-scale astrophysical dynamos, as shown by 
\citet{Vishniac2001}.
Richardson dispersion of magnetic field-lines  is entirely compatible with conservation of magnetic helicity,
whereas enhanced turbulent resistivity would dissipate magnetic helicity. 
  

We can summarize that ``eddy resistivity'' is a very misleading concept that has no sound scientific basis whatsoever for MHD 
turbulence. As a result, it cannot be applied with any confidence to astrophysical problems. Analogous criticisms apply to the  
hyper-resistivity concept 
\citep{BhattacharjeeHameiri:1986,Hameiri1987,Strauss:1986,DiamondMalkov:2003},
which has also 
been invoked to explain fast reconnection within the context of mean-field electrodynamics. Apart from the aforementioned problems 
of using the mean-field approach while dealing with reconnection, we would like to point out that the derivation of hyper-resistivity 
is questionable from a different point of view. The required parallel electric field is derived from magnetic helicity conservation. Integrating 
the resulting expression by parts one derives a term proportional to the magnetic helicity current which looks like an effective resistivity. 
The derivation uses, however, several implicit assumptions. Critical among them is the assumption that the magnetic helicity 
of mean fields and of small scale, statistically stationary turbulent fields are separately conserved, up to tiny resistivity effects. This disregards the 
magnetic helicity fluxes through open boundaries that are essential for stationary reconnection.

 \subsection{On the ``turbulent ambipolar diffusion'' idea}
 \label{sec:alternatives4}
 
We note that the concept of reconnection diffusion in partially ionized gas (see \S \ref{ssec: redsf}) should not be mixed up with that of ``turbulent ambipolar diffusion'' 
  \citep{FatuzzoAdams:2002,Zweibel:2002}
 that was also developed to explain magnetic flux removal from molecular clouds. The latter concept is based on the idea that turbulence can create gradients of neutrals and those can accelerate the overall pace of ambipolar diffusion. The questions that naturally arise are (1) whether this process can proceed without magnetic reconnection and (2) what is the role of ambipolar diffusion in the process. 
 
 \cite{Heitsch2004}
 performed numerical simulations with 2D turbulent mixing of a layer with magnetic field perpendicular to the layer and reported fast diffusion that was of the order of turbulent diffusivity number $V_L L$, independent of ambipolar diffusion coefficient. However, this sort of mixing can happen without reconnection only in a degenerate case of 2D mixing with exactly parallel magnetic field lines. In any realistic 3D case turbulence will bend magnetic field lines and the mixing process does inevitably involve reconnection. Therefore the 3D turbulent mixing in magnetized fluid must be treated from the point of view of reconnection theory. 
 
 If reconnection is slow, the process in 
 \cite{Heitsch2004} cannot proceed due to the inability of magnetic field lines to freely cross each other (as opposed to the 2D case!). This should arrest the mixing and makes the conclusions obtained in the degenerate 2D case inapplicable to the 3D diffusion. If, however, reconnection is fast as predicted in LV99, then the mixing and turbulent diffusion will take place. However, such a diffusion is not expected to depend on the ambipolar diffusion processes, which is, incidentally, in agreement with results in 
 \cite{Heitsch2004}. 
 
In short, the answers to the questions above are that turbulent diffusion in
partially ionized gas is (1) impossible without fast turbulent reconnection
and (2) independent of ambipolar diffusion physics. In this situation we
believe that it is misleading to talk about ``turbulent ambipolar diffusion''
in any astrophysical 3D setting. In fact, the actual diffusion in turbulent
media is controlled by magnetic reconnection and is independent of ambipolar
diffusion process!

\subsection{On the universal $0.1 V_A$ reconnection rate}

For decades the reconnection research has been motivated mostly by Solar observations and there for explaining a number of the observed phenomena the reconnection rate of $\sim 0.1 V_A$ was required. Therefore providing models that would explain this ``Holy Grail'' number became the motivation for an extensive research effort. Therefore when this number was achieved, first with the Hall-MHD 2D reconnection simulations, and later, with 2D PIC tearing mode simulations for some researchers this signified the final successful resolving the long standing problem of magnetic reconnection. 

The reality, however, is different. The astrophysical reconnection does show a variety of reconnection rates, which are rather difficult to accommodate if one assumes that a universal microphysical plasma process regulates the physics of the reconnection. First of all, the evidence is accumulating that the Solar flares (see \S ref {sec:observations1}) and other explosive reconnecton phenomena (see \S \ref {sec:implications3}) require both periods of fast and slow reconnection. The deviations of magnetic field from the Parker spiral (see \S {sec:observations3}) and the reconnection diffusion of magnetic field demonstrated in numerical simulations (see \S \ref{ssec: doffvio}) definitely do not fit either into the paradigm of the universal reconnection at the $0.1 V_A$ or the requirement of the plasma effects to be essential for fast magnetic reconnection.

The correspondence of the canonical value of $0.1 V_A$ reported in the PIC simulations and realistic Solar physics settings was recently questioned in \cite{Beresnyak:2018} (see \S \ref{Hall}). The author pointed out that the reconnection sheet in Solar corona is $3\times 10^4$ ion skin depths. His simulations showed the significant decrease of the reconnection rate as the width of the reconnection layer was increasing from the value of less than $10$, the latter range being typical for most of the present-day PIC studies. In other words, the correspondence of the rates obtained in the PIC simulations with the ``desired'' $0.1 V_A$ value is coinsidental. 

The turbulent reconnection theory does not predict a single preferred value for the magnetic reconnection rate. On the contrary, the turbulent reconnection rate is being controlled by the level of turbulence. This turbulence may be driven by external processes or by reconnection itself or both externally and self-driven. In fact, we expect the role of self-driving to increase as the plasma beta is decreasing and there is more energy available to drive turbulence. 

The predicted rate of reconnection within one eddy turnover time is sufficient for explaining the observations that we are aware of. At the same time, the estimates in \cite{Eyink_etal:2011} show that for calm low-turbulence regions of Solar corona the rate of turbulent reconnection is formally less that that induced by the Sweet-Parker process. 

As we discussed in this review, the process of turbulent reconnection is intrinsically  related to the violation of the classical flux freezing based on the famous \cite{Alfven:1942} theorem. Instead we have the so-called ``stochastic flux freezing'' (see \S \ref{flux-freezing}), which is a special way of flux freezing violation in turbulent fluids. The latter is necessary to make the description of magnetized turbulent fluid self-consistent. In other words, one should not consider magnetic reconnection  as the process that involves annihilation of nearly anti-parallel magnetic flux tubes and that is accompanied by a tangible energy release. Yes, the latter is how the reconnection problem was initially formulated more than half a century ago. By now this  picture that is motivated by Solar flares is just one of the incarnations of the ubiquitous reconnection process that takes place in astrophysical turbulent fluids.

As a result, the idea of universal $0.1 V_A$ rate of magnetic reconnection is ill founded and it contradicts to the theoretical expectations, numerical testing as well as to the available observational data. The actual astrophysical reconnection rate can be both larger and smaller than this ``preferred'' rate.

 \section{Past, present and future of turbulent reconnection theory}
 \label{sec:future}
 
 It has been 20 years since the introduction of the turbulent reconnection model by LV99. By now the models that the LV99 model was competing at its start, such as the 2D X-point Hall MHD reconnection model, have been abandoned by the community. The currently popular model of plasmoid or tearing reconnection shares common features with the original LV99 idea. In fact, we believe that the tearing may be important for the onset of turbulent reconnection in the situation where magnetic fields are initially not sufficiently turbulent. At the same time, the numerical simulations with both non-relativistic and relativistic MHD codes testify that the steady-state 3D reconnection is generically turbulent (see \S \ref {sec:testing1},\S \ref{sec:level2b}, \S \ref {sec:special2}).

 Since its establishment, the idea of 3D turbulent reconnection has become an essential part of a bigger picture of the magnetic field dynamics in turbulent fluids. Deep connections have been identified between the turbulent reconnection and theories of MHD turbulence (see \S \ref{sec:level1}, \S \ref{ssec: doffvio}), turbulent dynamo (see \S \ref{sec:implications1}), magnetic field stochasticity (see \S \ref{ssec: rdffvio}), etc., which have shown that turbulent reconnection is an intrinsic part of magnetic field dynamics in realistic magnetized turbulent fluids.
 
 Being the fundamental property of turbulent and magnetized flows, the concept of turbulent reconnection has induced significant changes in understanding of fundamental astrophysical processes, including cosmic ray propagation and acceleration (see \S \ref{sec:implications3}), as well as the star formation (see \S \ref{ssec: redsf}). Naturally, the work in terms of exploring astrophysical consequences of turbulent reconnection are expected to bring more fruits, as many brunches of astrophysical theories were developed based on the assumption of classical flux freezing concept. It has been well established by now that the concept is not applicable to turbulent fluids (see \S \ref{ssec: doffvio}).
 
 The turbulent reconnection described in LV99 is ubiquitous in magnetized fluids on scales where the MHD approximation is valid. Recent research shows that it is applicable to both turbulent partially ionized and fully ionized plasmas, as well as to turbulent relativistic and non-relativistic fluids (see \S \ref{sec:special}). On smaller scales, for instance in magnetospheric research, the magnetic reconnection is different. Some ideas of turbulent reconnection can be carried over to this regime, e.g. in the form of whistler turbulence rather than MHD turbulence being the driver of the flux freezing violation (see \S \ref{whistler}). At the same time, one should clearly realise that the regime where the thicknesses of reconnection layers are comparable to the relevant plasma scales is very different from the typical settings of large-scale astrophysical reconnection. 
 
 The scale separation is frequently discussed as the biggest challenge of magnetic reconnection. This problem is indeed present in some models of reconnection, e.g. the classical Sweet-Parker reconnection. As a result, in the latter case the magnetic reconnection is slow. In turbulent fluids magnetic reconnection takes place on all scales, and therefore talking ``reconnection-mediated turbulence'' by assuming that magnetic reconnection in turbulent fluids happens only at some chosen scale, e.g. the scale of tearing, is not meaningful (see \S \ref{sec:alternatives2}). 
 Turbulent reconnection provides much required solid theoretical foundations for the MHD modeling of astrophysical processes taking place in realistically turbulent fluids when the scales involved much larger that the ion gyroradius. Indeed, magnetic reconnection is an essential feature of astrophysical flows and if properly resolving plasma scales were essential for the large-scale dynamics, the modeling of most astrophysical processes would be unrealistic. Indeed, is not feasible and is not feasible in any foreseable future to model with PIC simulations star formation, evolution of magnetized interstellar medium in galaxies and intracluster medium in clusters of galaxies! The turbulent reconnection justifies MHD modeling of various astrophysical processes provided that at the scales that are studies the fluids remain turbulent. 
 
 To explain the observed properties of magnetized fluids, in the absence of understanding the nature of turbulent reconnection and flux freezing violation that it entails, a number of ``engineering'' concepts, e.g.,  ``turbulent resistivity'', ``hyper-resistivity'', have been introduced in the astrophysical literature. These concepts have poor theoretical foundations and are frequently in direct contradictions with numerical simulations and observations. On the contrary, the theory of turbulent reconnection provides a way to properly describe the processes in turbulent astrophysical fluids (see \S \ref{sec:alsternatives3}). 
 
 The above statements about the universality and power of turbulent reconnection theory should not be understood in the way that we already know all the answers. In fact, there are still a number of fundamental issues to be addressed. For instance, the studies of turbulent relativistic reconnection are still at their infancy (see \S \ref{sec:turbulence4}). Many issues are still unclear in terms of turbulence in collisionless magnetized plasmas and the reconnecton that accompanies this turbulence (see \S \ref{sec:level4}). The combined theoretical and numerical efforts are required to achieve the progress along these directions.

 While turbulent reconnection in MHD approximation correctly addresses the large scale behavior of astrophysical magnetic fields, it is important to achieve the description of reconnection processes on all scales. At the moment, the turbulent reconnection studies do not have much overlap with the research in magnetized plasmas. We expect that the gap between the two fields to decrease in future. Reconnection at the scales where both collisionless effects and turbulent dynamics are important present the next theoretical, observational and numerical challenge, the challenge that we are starting to address.
 
 The field of reconnection is the vast field of research. Apparently, different types of reconnection takes place in different environments. For instance, Sweet-Parker reconnection has its own domain, but at high Lundquist numbers its nature changes. Based on the 2D numerical study one could assume that it transforms to the tearing/plasmoid reconnection. However, the 3D studies testify that the generic reconnection model is the turbulent one. This still leaves some parameter space for plasmoid reconnection. For instance, we believe that the plasmoid stage is important for the initial stages of reconnection when the reconnecting fluxes are laminar at the beginning. Similarly, the extensive magnetospheric reconnection research demonstrates that its physics is very different from the aforementioned types of reconnection. This is not surprising as the characteristic widths of magnetispheric current sheets are comparable with the characteristic plasma scales, i.e. the ion inertial length. Therefore, there is no surprise that collisionless plasma effects are getting important for the magnetospheric reconnection. Some ideas of turbulent reconnection can be transferred to this special small-scale reconnection. For instance, we discussed in the review the effect of opening up of the reconneciton outflow arising from field wandering that induced by whistler of kinetic Alfven modes. At the same time, it is clear that the reconnection on astrophysical scales is the domain where the MHD approximation is valid and therefore the LV99 turbulent reconnection is applicable. Further research should define better the conditions and regimes for different types of reconnection. 

\begin{acknowledgments}
AL acknowledges the support of NSF AST 1816234 and NASA TCAN AAG1967 grants. 
GK acknowledges the support of the Brazilian agency CNPq (no. 304891/2016-9). The final work on the review was done at Kavli Institute for hospitality during the Astroplasma19 event and AL thanks Anatoly Spitkovsky for organizing  productive discussions on MHD turbulence and turbulent reconnection during this event. AL thanks Andrey Beresnyak for fruitful discussions. HL gratefully acknowledges the support of DoE/OFES and LANL/LDRD programs. 
Figures in Sec. VI based on the PIC simulations were made by F. Guo and P. Kilian.   
SX acknowledges the support for Program number HST-HF2-51400.001-A provided by NASA through a grant from the Space Telescope Science Institute, which is operated by the Association of Universities for Research in Astronomy, Incorporated, under NASA contract NAS5-26555. We acknowledge exchanges with Nuno Loureiro and Alex Schekochihin on the nature of magnetic reconnection in a turbulent cascade as well as with Jim Drake on the field wandering in the whislter turbulence. ETV and AJ thank the AAS for support. We thank the anonymous referee for the numerous useful suggestions that improved the quality of our presentation.
\end{acknowledgments}

\bibliography{main}

\end{document}